\documentstyle[11pt,axodraw]{report}
\begin{document}
%%%%%%%%%%%%%%%%%%%%%%%%%%%%%%%%%%%%%%%%%%%%%%%%%%%%%%%%%%%%%%%%%
\def\beq{\begin{eqnarray}}    %%%  begequation/eqnarray
\def\eeq{\end{eqnarray}}      %%%  endequation/eqnarray
 \def\bea{\begin{eqnarray}}
 \def\eea{\end{eqnarray}}
 \def\ul{\underline}
 \def\ni{\noindent}
 \def\nn{\nonumber}
 \def\wt{\widetilde}
 \def\wh{\widehat}
 \def\Tr{\mbox{Tr}\ }
 \def\arccosh{\mbox{acrcosh}\ }

 %%%%%    SPECIAL SYMBOLS
 \def\tr{\,\mbox{tr}\,}                  %%% trace
 \def\Tr{\,\mbox{Tr}\,}                  %%% Trace
 \def\Res{\,\mbox{Res}\,}                %%% Residue
 \renewcommand{\Re}{\,\mbox{Re}\,}       %%% Real
 \renewcommand{\Im}{\,\mbox{Im}\,}       %%% Imaginary
 \def\lap{\Delta}                        %%% Laplacian
 \def\pa{\partial}                       %%% partial derivative

%%%%%%%%%%%%%%%%%%%%%%%%%%%%%%%%%%%%%%%%%%%%%%
%%%%   \def\Box{\nabla}
%%%%   \def{\bar \Box}{\bar{\vphantom{I}\Box}}
%%%%%%%%%%%%%%%%%%%%%%%%%%%%%%%%%%%%%%%%%%%%%%

\def\sTr{\,\mbox{Str}\,}                %%% SuperTrace
\def\lap{\Delta}                        %%% Laplacian
\def\sla{\!\!\!\slash}
%%%%%%%% for example, p\sla %%%%%%%%%%%%%%% Slash

%%%%%    GREEK ALPHABET
\def\al{\alpha}
\def\be{\beta}
\def\ch{\chi}
\def\ga{\gamma}
\def\de{\delta}
\def\vp{\varepsilon}
\def\ep{\epsilon}
\def\ze{\zeta}
\def\io{\iota}
\def\ka{\kappa}
\def\la{\lambda}
\def\na{\nabla}
\def\ro{\varrho}
\def\si{\sigma}
\def\om{\omega}
\def\ph{\varphi}
\def\ta{\tau}
\def\th{\theta}
\def\te{\vartheta}
\def\up{\upsilon}
\def\Ga{\Gamma}
\def\De{\Delta}
\def\La{\Lambda}
\def\Si{\Sigma}
\def\Om{\Omega}
\def\Te{\Theta}
\def\Th{\Theta}
\def\Up{\Upsilon}

\thispagestyle{empty}
%%%%%%%%%%%%%%%%%%%%%%%%%%%%%%%%%%%%%%%%%%%%%%%%
\begin{center}

{\large\sc Physical Aspects of the Space-Time Torsion}
\vskip 2mm

{I.L. Shapiro}\footnote
{On leave from 
Department of Mathematical Analysis, Tomsk 
State Pedagogical University, Russia} 
\vskip 2mm

{\small
\it Departamento de F{\'\i}sica, Universidade Federal de Juiz 
de Fora, CEP: 36036-330, MG, Brazil}

{\small
\it Tel/Fax: (55-32)-3229-3307/3312. $\,\,\,$ E - mail:  
shapiro@fisica.ufjf.br}
\vskip 2mm

\end{center} 
%%%%%%%%%%%%%%%%%%%%%%%%%%%%%%%%%%%%%%%%%%%%
\vskip 6mm

\noindent
{\bf Abstract.} 
$\,\,$ We review many quantum aspects of torsion theory and 
discuss the possibility of the space-time torsion to exist 
and to be detected. The paper starts, in Chapter 2, with a
pedagogical introduction to the classical gravity with torsion, 
that includes also interaction of torsion with matter fields.
Special attention is paid to the conformal properties of the
theory. In Chapter 3, the renormalization of quantum theory of 
matter fields and related topics, like renormalization group,
effective potential and anomalies, are considered. Chapter 4 is 
devoted to the action of spinning and spinless particles in 
a space-time with torsion, and to the discussion of  
possible physical effects generated by the background torsion. 
In particular, we review the upper bounds for the magnitude of 
the background torsion which are known from the literature.  
In Chapter 5, the comprehensive study of the possibility of a
theory for the propagating completely antisymmetric torsion 
field is presented. It is supposed that the propagating field 
should be quantized, and that its quantum effects must be 
described by, at least, some effective low-energy quantum field 
theory. We show, that the propagating torsion may be consistent 
with the principles of quantum theory only in the case when 
the torsion mass is much greater than the mass of the heaviest 
fermion coupled to torsion. Then, universality of the 
fermion-torsion interaction implies that torsion itself has a 
huge mass, and can not be observed in realistic experiments. 
Thus, the theory of quantum matter fields on the classical 
torsion background can be formulated in a consistent way, while 
the theory of dynamical torsion meets serious obstacles. In 
Chapter 6, we briefly discuss the string-induced torsion and 
the possibility to induce torsion action and torsion itself 
through the quantum effects of matter fields.
\vskip 6mm
\vskip 6mm



%%%%%%%%%%%%%%%%%%%%%%%%%%%%%%%%%%%%%%%%%%%%
\newpage
\noindent
{\Large\bf Content:}
\vskip 6mm

%%%%%%%%%%%%%%%%%%%%%%%%%%%%%%%%%%%%%%%%%%%%%%%%%%%%%%%%%%%%%%%%
%%%%%%%%%%%%%%%%%%%%%%%%%%%%%%%%%%%%%%%%%%%%%%%%%%%%%%%%%%%%%%%%
{\large\bf 1.$\,\,$Introduction.}
%%%%%%%%%%%%%%%%%%%%%%%%%%%%%%%%%%%%%%%%%%%%%%%%%%%%%%%%%%%%%%%%

%%%%%%%%%%%%%%%%%%%%%%%%%%%%%%%%%%%%%%%%%%%%%%%%%%%%%%%%%%%%%%%%%%%
%%%%%%%%%%%%%%%%%%%%%%%%%%%%%%%%%%%%%%%%%%%%%%%%%%%%%%%%%%%%%%%%%%%
\vskip 3mm 
{\large\bf 2.$\,\,$Classical torsion.}
%%%%%%%%%%%%%%%%%%%%%%%%%%%%%%%%%%%%%%%%%%%%%%%%%%%%%%%%%%%%%%%%%%%

%%%%%%%%%%%%%%%%%%%%%%%%%%%%%%%%%%%%%%%%%%%%%%%%%%%%%%%%%%%%%%%%%%%
\vskip 1mm \noindent
{2.1$\,\,$Definitions, notations and basic concepts.}
%%%%%%%%%%%%%%%%%%%%%%%%%%%%%%%%%%%%%%%%%%%%%%%%%%%%%%%%%%%%%%%%%%%

%%%%%%%%%%%%%%%%%%%%%%%%%%%%%%%%%%%%%%%%%%%%%%%%%%%%%%%%%%%%%%%%
\vskip 1mm \noindent
{2.2$\,\,$Einstein-Cartan theory and non-dynamical torsion.}
%%%%%%%%%%%%%%%%%%%%%%%%%%%%%%%%%%%%%%%%%%%%%%%%%%%%%%%%%%%%%%%%

%%%%%%%%%%%%%%%%%%%%%%%%%%%%%%%%%%%%%%%%%%%%%%%%%%%%%%%%%%%%%%%%
\vskip 1mm \noindent
{2.3$\,\,$Interaction of torsion with matter fields.}
%%%%%%%%%%%%%%%%%%%%%%%%%%%%%%%%%%%%%%%%%%%%%%%%%%%%%%%%%%%%%%%%

%%%%%%%%%%%%%%%%%%%%%%%%%%%%%%%%%%%%%%%%%%%%%%%%%%%%%%%%%%%%%%%%%%%
\vskip 1mm \noindent
{2.4$\,\,$Conformal properties of torsion.}
%%%%%%%%%%%%%%%%%%%%%%%%%%%%%%%%%%%%%%%%%%%%%%%%%%%%%%%%%%%%%%%%%%%

%%%%%%%%%%%%%%%%%%%%%%%%%%%%%%%%%%%%%%%%%%%%%%%%%%%%%%%%%%%%%%
\vskip 1mm \noindent
{2.5$\,\,$Gauge approach to gravity. Higher derivative
gravity theories with torsion.}
%%%%%%%%%%%%%%%%%%%%%%%%%%%%%%%%%%%%%%%%%%%%%%%%%%%%%%%%%%%%%%

%%%%%%%%%%%%%%%%%%%%%%%%%%%%%%%%%%%%%%%%%%%%%%%%%%%%%%%%%
\vskip 1mm \noindent
{2.6$\,\,$An example of the possible effect of 
classical torsion.}
%%%%%%%%%%%%%%%%%%%%%%%%%%%%%%%%%%%%%%%%%%%%%%%%%%%%%%%%%

%%%%%%%%%%%%%%%%%%%%%%%%%%%%%%%%%%%%%%%%%%%%%%%%%%%%%%%%%%%%
%%%%%%%%%%%%%%%%%%%%%%%%%%%%%%%%%%%%%%%%%%%%%%%%%%%%%%%%%%%%
\vskip 3mm {\large\bf 3.$\,\,$
Renormalization and anomalies in curved space-time 
with torsion.}
%%%%%%%%%%%%%%%%%%%%%%%%%%%%%%%%%%%%%%%%%%%%%%%%%%%%%%%%%%%%

%%%%%%%%%%%%%%%%%%%%%%%%%%%%%%%%%%%%%%%%%%%%%%%%%%%%%%%%%%%%
\vskip 1mm \noindent
{3.1$\,\,$ General description of renormalizable theory.}
%%%%%%%%%%%%%%%%%%%%%%%%%%%%%%%%%%%%%%%%%%%%%%%%%%%%%%%%%%%%

%%%%%%%%%%%%%%%%%%%%%%%%%%%%%%%%%%%%%%%%%%%%%%%%%%%%%%%%%%%%%
\vskip 1mm \noindent
{3.2$\,\,$ One-loop calculations in the vacuum sector.}
%%%%%%%%%%%%%%%%%%%%%%%%%%%%%%%%%%%%%%%%%%%%%%%%%%%%%%%%%%%%%

%%%%%%%%%%%%%%%%%%%%%%%%%%%%%%%%%%%%%%%%%%%%%%%%%%%%%%%%%%%
\vskip 1mm \noindent
{3.3$\,\,$One-loop calculations in the matter fields 
sector.}
%%%%%%%%%%%%%%%%%%%%%%%%%%%%%%%%%%%%%%%%%%%%%%%%%%%%%%%%%%%

%%%%%%%%%%%%%%%%%%%%%%%%%%%%%%%%%%%%%%%%%%%%%%%%%%%%%%%%%%%%
\vskip 1mm \noindent
{3.4$\,\,$Renormalization group and universality 
in the non-minimal sector.}
%%%%%%%%%%%%%%%%%%%%%%%%%%%%%%%%%%%%%%%%%%%%%%%%%%%%%%%%%%%%

%%%%%%%%%%%%%%%%%%%%%%%%%%%%%%%%%%%%%%%%%%%%%%%%%%%%%%%%%%%%
\vskip 1mm \noindent
{3.5$\,\,$Effective potential of 
scalar field in the space-time with torsion. 
Spontaneous symmetry
breaking and phase transitions induced by curvature and torsion.}
%%%%%%%%%%%%%%%%%%%%%%%%%%%%%%%%%%%%%%%%%%%%%%%%%%%%%%%%%%%%

%%%%%%%%%%%%%%%%%%%%%%%%%%%%%%%%%%%%%%%%%%%%%%%%%%%%%%%%%%%%
\vskip 1mm \noindent
{3.6$\,\,$Conformal anomaly in the spaces with torsion. 
Trace anomaly and modified trace anomaly.}
%%%%%%%%%%%%%%%%%%%%%%%%%%%%%%%%%%%%%%%%%%%%%%%%%%%%%%%%%%%%

%%%%%%%%%%%%%%%%%%%%%%%%%%%%%%%%%%%%%%%%%%%%%%%%%%%%%%%%%%%%
\vskip 1mm \noindent
{3.7$\,\,$Integration of conformal anomaly
and anomaly-induced effective actions of vacuum.
Application to inflationary cosmology.}
%%%%%%%%%%%%%%%%%%%%%%%%%%%%%%%%%%%%%%%%%%%%%%%%%%%%%%%%%%%%

%%%%%%%%%%%%%%%%%%%%%%%%%%%%%%%%%%%%%%%%%%%%%%%%%%%%%%%%%%%
\vskip 1mm \noindent 
{3.8 $\,\,$Chiral anomaly in the spaces with torsion.
Cancellation of anomalies.}
%%%%%%%%%%%%%%%%%%%%%%%%%%%%%%%%%%%%%%%%%%%%%%%%%%%%%%%%%%%%%%

%%%%%%%%%%%%%%%%%%%%%%%%%%%%%%%%%%%%%%%%%%%%%%%%%%%%%%%%%%%%%%
%%%%%%%%%%%%%%%%%%%%%%%%%%%%%%%%%%%%%%%%%%%%%%%%%%%%%%%%%%%%%%
\vskip 3mm 
{\large\bf 4.$\,\,$Spinning and spinless particles and the 
possible effects on the classical background of torsion.}
%%%%%%%%%%%%%%%%%%%%%%%%%%%%%%%%%%%%%%%%%%%%%%%%%%%%%%%%%%%%%

%%%%%%%%%%%%%%%%%%%%%%%%%%%%%%%%%%%%%%%%%%%%%%%%%%%%%%%%%%%%%%%%%%
\vskip 1mm \noindent
{4.1$\,\,$Generalized Pauli equation with torsion.}
%%%%%%%%%%%%%%%%%%%%%%%%%%%%%%%%%%%%%%%%%%%%%%%%%%%%%%%%%%%%%%%%%%

%%%%%%%%%%%%%%%%%%%%%%%%%%%%%%%%%%%%%%%%%%%%%%%%%%%%%%%%%%%%%
\vskip 1mm \noindent
{4.2$\,\,$Foldy-Wouthuysen transformation with torsion.}
%%%%%%%%%%%%%%%%%%%%%%%%%%%%%%%%%%%%%%%%%%%%%%%%%%%%%%%%%%%%%

%%%%%%%%%%%%%%%%%%%%%%%%%%%%%%%%%%%%%%%%%%%%%%%%%%%%%%%%%%%%%%%%%%%%
\vskip 1mm \noindent
{4.3 
$\,\,$Non-relativistic particle in the external torsion field.}
%%%%%%%%%%%%%%%%%%%%%%%%%%%%%%%%%%%%%%%%%%%%%%%%%%%%%%%%%%%%%%%%%%%%

%%%%%%%%%%%%%%%%%%%%%%%%%%%%%%%%%%%%%%%%%%%%%%%%%%%%%%%%%%%%%
\vskip 1mm \noindent
{4.4 $\,\,$Path-integral approach for the relativistic particle
   with torsion.}
%%%%%%%%%%%%%%%%%%%%%%%%%%%%%%%%%%%%%%%%%%%%%%%%%%%%%%%%%%%%%

%%%%%%%%%%%%%%%%%%%%%%%%%%%%%%%%%%%%%%%%%%%%%%%%%%%%%%%%%%%%%%%%%
\vskip 1mm \noindent
{4.5 $\,\,$Space-time trajectories for the spinning and spinless
  particles in an external torsion field.}
%%%%%%%%%%%%%%%%%%%%%%%%%%%%%%%%%%%%%%%%%%%%%%%%%%%%%%%%%%%%%%%%%

%%%%%%%%%%%%%%%%%%%%%%%%%%%%%%%%%%%%%%%%%%%%%%%%%%%%%%%%%%%%%%%%%%%%%%%%
\vskip 1mm \noindent
{4.6 $\,\,$Experimental constraints for the constant background
torsion.}
%%%%%%%%%%%%%%%%%%%%%%%%%%%%%%%%%%%%%%%%%%%%%%%%%%%%%%%%%%%%%%%%%%%%%%%%

%%%%%%%%%%%%%%%%%%%%%%%%%%%%%%%%%%%%%%%%%%%%%%%%%%%%%%%%%%%%%%%%%%%%%%%%
%%%%%%%%%%%%%%%%%%%%%%%%%%%%%%%%%%%%%%%%%%%%%%%%%%%%%%%%%%%%%%%%%%%%%%%%
\vskip 3mm {\large\bf 5.$\,\,$
The effective quantum field theory approach for the 
dynamical torsion.}
%%%%%%%%%%%%%%%%%%%%%%%%%%%%%%%%%%%%%%%%%%%%%%%%%%%%%%%%%%%%%%%%%%%%%%%%

%%%%%%%%%%%%%%%%%%%%%%%%%%%%%%%%%%%%%%%%%%%%%%%%%%%%%%%%%%%%%%%%%%%%%%%%
\vskip 1mm \noindent
{5.1 $\,\,$Early works on the quantum gravity with torsion.}
%%%%%%%%%%%%%%%%%%%%%%%%%%%%%%%%%%%%%%%%%%%%%%%%%%%%%%%%%%%%%%%%%%%%%%%%

%%%%%%%%%%%%%%%%%%%%%%%%%%%%%%%%%%%%%%%%%%%%%%%%%%%%%%%%%%%%%%%%%%%%%%%%
\vskip 1mm \noindent
{5.2 $\,\,$General note about the effective approach to torsion.}
%%%%%%%%%%%%%%%%%%%%%%%%%%%%%%%%%%%%%%%%%%%%%%%%%%%%%%%%%%%%%%%%%%%%%%%%

%%%%%%%%%%%%%%%%%%%%%%%%%%%%%%%%%%%%%%%%%%%%%%%%%%%%%%%%%%%%%%%%%%%%%%%%
\vskip 1mm \noindent
{5.3 $\,\,$Torsion-fermion interaction again:
Softly broken symmetry associated with torsion and
the unique possibility for the low-energy torsion action.}
%%%%%%%%%%%%%%%%%%%%%%%%%%%%%%%%%%%%%%%%%%%%%%%%%%%%%%%%%%%%%%%%%%%%%%%%

%%%%%%%%%%%%%%%%%%%%%%%%%%%%%%%%%%%%%%%%%%%%%%%%%%%%%%%%%%%%%%%%%%%%%%%%
\vskip 1mm \noindent
{5.4 $\,\,$Brief review of the possible torsion effects in  
high-energy physics.}
%%%%%%%%%%%%%%%%%%%%%%%%%%%%%%%%%%%%%%%%%%%%%%%%%%%%%%%%%%%%%%%%%%%%%%%%

%%%%%%%%%%%%%%%%%%%%%%%%%%%%%%%%%%%%%%%%%%%%%%%%%%%%%%%%%%%%%%%%%%%%%%%%%
\vskip 1mm \noindent
{5.5 $\,\,$First test of consistency: loops in the fermion-scalar 
systems break unitarity.}
%%%%%%%%%%%%%%%%%%%%%%%%%%%%%%%%%%%%%%%%%%%%%%%%%%%%%%%%%%%%%%%%%%%%%%%%

%%%%%%%%%%%%%%%%%%%%%%%%%%%%%%%%%%%%%%%%%%%%%%%%%%%%%%%%%%%%%%%%%%%%%%%%
\vskip 1mm \noindent
{5.6 $\,\,$Second test: problems with the quantized
fermion-torsion systems.}
%%%%%%%%%%%%%%%%%%%%%%%%%%%%%%%%%%%%%%%%%%%%%%%%%%%%%%%%%%%%%%%%%%%%%%%%

%%%%%%%%%%%%%%%%%%%%%%%%%%%%%%%%%%%%%%%%%%%%%%%%%%%%%%%%%%%%%%%%%%%%%%
\vskip 1mm \noindent
{5.7 $\,\,$Interpretation of the results: do we have a chance to meet
propagating torsion?}
%%%%%%%%%%%%%%%%%%%%%%%%%%%%%%%%%%%%%%%%%%%%%%%%%%%%%%%%%%%%%%%%%%%%%%

%%%%%%%%%%%%%%%%%%%%%%%%%%%%%%%%%%%%%%%%%%%%%%%%%%%%%%%%%%%%%%%%%%%%%%%%
\vskip 1mm \noindent 
{5.8 $\,\,$What is the difference with metric?}
%%%%%%%%%%%%%%%%%%%%%%%%%%%%%%%%%%%%%%%%%%%%%%%%%%%%%%%%%%%%%%%%%%%%%%%%

%%%%%%%%%%%%%%%%%%%%%%%%%%%%%%%%%%%%%%%%%%%%%%%%%%%%%%%%%%%%%%%%%
%%%%%%%%%%%%%%%%%%%%%%%%%%%%%%%%%%%%%%%%%%%%%%%%%%%%%%%%%%%%%%%%%%
\vskip 3mm 
{\large\bf 6.$\,\,$Alternative approaches: induced torsion.}
%%%%%%%%%%%%%%%%%%%%%%%%%%%%%%%%%%%%%%%%%%%%%%%%%%%%%%%%%%%%%%%%%%

%%%%%%%%%%%%%%%%%%%%%%%%%%%%%%%%%%%%%%%%%%%%%%%%%%%%%%%%%%%%%%%%%%
\vskip 1mm \noindent
{6.1$\,\,$ Is that torsion induced in string theory?}
%%%%%%%%%%%%%%%%%%%%%%%%%%%%%%%%%%%%%%%%%%%%%%%%%%%%%%%%%%%%%%%%%%

%%%%%%%%%%%%%%%%%%%%%%%%%%%%%%%%%%%%%%%%%%%%%%%%%%%%%%%%%%%%%%%%%%
\vskip 1mm \noindent
{6.2 $\,\,$ Gravity with torsion induced by quantum effects 
of matter.}
%%%%%%%%%%%%%%%%%%%%%%%%%%%%%%%%%%%%%%%%%%%%%%%%%%%%%%%%%%%%%%%%%%

%%%%%%%%%%%%%%%%%%%%%%%%%%%%%%%%%%%%%%%%%%%%%%%%%%%%%%%%%%%%%%%%%%
%%%%%%%%%%%%%%%%%%%%%%%%%%%%%%%%%%%%%%%%%%%%%%%%%%%%%%%%%%%%%%%%%%
\vskip 3mm  {\large\bf 7.$\,\,$Conclusions.} 
%%%%%%%%%%%%%%%%%%%%%%%%%%%%%%%%%%%%%%%%%%%%%%%%%%%%%%%%%%%%%%%%%%
\newpage

%%%%%%%%%%%%%%%%%%%%%%%%%%%%%%%%%%%%%%%%%%%%%%%%%%%%%%%%%%%%%%%%
%%%%%%%%%%%%%%%%%%%%%%%%%%%%%%%%%%%%%%%%%%%%%%%%%%%%%%%%%%%%%%%%
\chapter{Introduction.}
%%%%%%%%%%%%%%%%%%%%%%%%%%%%%%%%%%%%%%%%%%%%%%%%%%%%%%%%%%%%%%%%

The development of physics, until recent times, went from experiment 
to theory. New theories were created when the previous ones did not 
fit with some existing phenomena, or when the theories describing 
different classes of phenomena have shown some mutual contradictions. 
At some point this process almost stopped. However, recent data on the 
neutrino oscillations should be, perhaps, interpreted in such a way 
that the Minimal Standard Model of particle physics does not 
describe the full spectrum of existing particles. On the other
hand, one can mention the supernova observational evidence for a 
positive cosmological constant and the lack of the natural explanation
for the inflation. This probably indicates that the extension
of the Standard Model must also include gravity. 
The desired fundamental theory is expected to provide the solution 
to the quantum gravity problem, hopefully explain the observable 
value of the cosmological constant and maybe even predict the 
low-energy observable particle spectrum. The construction  
of such a fundamental theory meets obvious difficulties: besides 
purely theoretical ones, there is an extremely small link with the 
experiments or observations. Nowadays, the number of theoretical 
models or ideas has overwhelming majority over the number of their 
possible verifications. In this sense, today the theory is very 
far ahead of experiment. 

In such a situation, when a fundamental theory is unknown or it 
can not be verified, one might apply some effective approach and 
ask what could be the traces of such 
a theory at low energies. In principle, there can be two kinds of 
evidences: new fields or new low energy symmetries. 
The Standard Model is composed by three
types of fields: spinors, vectors and scalars. On the other hand, 
General Relativity yields one more field -- metric, which describes the 
properties of the space-time. Now, if there is some low-energy 
manifestation of the fundamental theory, it 
could be some additional characteristics of the space-time, different 
from the fields included into the Standard Model. 
One of the candidates for this
role could be the space-time torsion, which we are going to discuss in this 
paper. Torsion is some independent characteristic of the space-time, 
which has a very long history of study (see \cite{hehl} for the 
extensive review and references, 
mainly on various aspects of classical gravity theory with torsion).
In this paper we shall concentrate on the quantum
aspects of the theory, and will look at the problem, mainly, from 
quantum point of view.  Our purpose will be to apply 
the approach which is standard in the high-energy physics when things 
concern the search for some new particle or interaction. 
One has to formulate the 
corresponding theory in a consistent way, first at the level of lower 
complexity,
and then investigate the possibility of experimental manifestations. 
After that, 
it is possible to study more complicated models. Indeed, for the case of 
torsion, which has not been ever observed, the study of experimental 
manifestations 
reduces to the upper bounds on the torsion parameters from various 
experiments.
Besides this principal line, the extensive introduction 
to the gravity with torsion will be given in Chapter 2.

For us, the simplest level of the torsion theory will be the 
classical background for quantum matter fields. As we shall 
see, such a theory can be formulated in a consistent way. The 
next level is, naturally, the theory of dynamical (propagating) 
torsion, which should be considered in the same manner as metric 
or as the constituents of the Standard Model. We shall present 
the general review of the original publications \cite{betor,guhesh} 
discussing the restrictions in implementing torsion into a gauge 
theory such as the Standard Model.  
By the end of the paper, we discuss the possibility of torsion 
induced in string theory and through the quantum effects of 
matter fields.

\vskip 10mm

%%%  \newpage
%%%%%%%%%%%%%%%%%%%%%%%%%%%%%%%%%%%%%%%%%%%%%%%%%%%%%%%%%%%%%%%%%%%
%%%%%%%%%%%%%%%%%%%%%%%%%%%%%%%%%%%%%%%%%%%%%%%%%%%%%%%%%%%%%%%%%%%
\chapter{Classical torsion.}
%%%%%%%%%%%%%%%%%%%%%%%%%%%%%%%%%%%%%%%%%%%%%%%%%%%%%%%%%%%%%%%%%%%

This Chapter  mainly contains introductory material,  
which is necessary for the next Chapters, where the quantum aspects
of torsion will be discussed.  

%%%%%%%%%%%%%%%%%%%%%%%%%%%%%%%%%%%%%%%%%%%%%%%%%%%%%%%%%%%%%%%%%%%
\section{Definitions, notations and basic concepts.}
%%%%%%%%%%%%%%%%%%%%%%%%%%%%%%%%%%%%%%%%%%%%%%%%%%%%%%%%%%%%%%%%%%%

Let us start with the basic notions of gravity with torsion.
In general, our notations correspond to those in 
\cite{thesis,book}. The metric $g_{\mu\nu}$ and torsion 
$T^\alpha_{\cdot\;\beta\gamma}$ are independent characteristics 
of the space-time. In order to understand
better, how the introduction of torsion becomes possible, let us
briefly review the construction of covariant derivative in 
General Relativity. We shall mainly 
consider the algebraic aspect of the covariant derivative.
For the geometric aspects, related to the notion of 
parallel transport,
the reader is referred, for example, to \cite{hehl}. 

The partial derivative of a scalar field is a covariant vector 
(one-form). However,
the partial derivative of any other tensor field does not form
a tensor. But, one can add to the partial derivative some
additional term such that the sum is a tensor. 
The sum of
partial derivative and this additional term is called covariant
derivative. For instance, in the case of the (contravariant) 
vector $A^\al$ the covariant derivative looks like
\beq
\na_\be\,A^\al = \pa_\be\,A^\al + \Gamma_{\be\ga}^\al \,A^\ga\,,
\label{1.1}
\eeq
where the last term is a necessary addition. The covariant 
derivative (\ref{1.1}) is a tensor if and only if the affine 
connection $\Ga^\al_{\be\ga}$ transforms in a special non-tensor 
way. The rule for constructing the covariant derivatives of other 
tensors immediately follows from the following facts:

i) The product of co- and contravariant vectors 
$A^\al$ and $B_\al$  should be a scalar. 
Then 
$$
\na_\be\,(A^\al B_\al) = \pa_\be\,(A^\al B_\al)
$$ 
and therefore
\beq
\na_\be\,B_\al = \pa_\be\,B_\al - \Gamma_{\be\al}^\ga \,B_\ga\,.
\label{1.2}
\eeq

ii) In the same manner, one can notice that the contraction of any 
tensor with some appropriate set of vectors is a scalar, and arrive
at the standard expression for the covariant derivative of an arbitrary
tensor 
\beq
\na_\be\,T^{\al_1 ...}_{\,\,\,\,\ga_1 ...} = 
\pa_\be\,T^{\al_1 ...}_{\,\,\,\,\ga_1 ...} 
+ \Gamma_{\be\la}^{\al_1} \,T^{\la ...}_{\,\,\,\,\ga_1 ...} + ... 
- \Gamma_{\be\ga_1}^\tau \,\,T^{\al_1 ...}_{\,\,\,\,\tau ...} - ...\,\,. 
\label{1.3}
\eeq
Now, (\ref{1.1}) and (\ref{1.2}) become particular cases of 
(\ref{1.3}). At this point, it becomes
clear that the definition of $\Ga^\al_{\be\ga}$ contains, from the
very beginning, some ambiguity. Indeed, (\ref{1.3}) remains a tensor
if one adds to $\Ga^\al_{\be\ga}$ {\it any} tensor $C^\al_{\be\ga}$:
\beq
\Ga^\al_{\be\ga} \to \Ga^\al_{\be\ga} + C^\al_{\be\ga}\,.
\label{1.4}
\eeq
A very special choice of $\Ga^\al_{\be\ga}$, which is used in
General Relativity, appears as a consequence of two requirements:
\vskip 1mm

i) symmetry $\,\Ga^\al_{\be\ga} = \Ga^\al_{\ga\be}\,$ and

ii) metricity of the covariant derivative $\na_\al\,g_{\mu\nu} = 0$.
\vskip 1mm

\noindent 
If these conditions are satisfied, one can apply (\ref{1.3}) and 
obtain the unique solution for $\,\Ga^\al_{\,\,\be\ga}$:
\beq
{\Gamma}^\alpha_{\;\beta\gamma} =
\left\{ ^{\al}_{\be\ga}\right\} = 
\frac12\,g^{\al\la}\,\left(
\pa_\be\,g_{\la\ga} + \pa_\ga\,g_{\la\be} - \pa_\la\,g_{\be\ga} \right)\,.
\label{1.5}
\eeq
The expression (\ref{1.5}) is called Christoffel symbol, 
it is a particular case of the affine connection. 
Indeed, (\ref{1.5}) is a very important object, because it 
depends on the metric only.  (\ref{1.5}) it is the simplest 
one among all possible affine connections. It is very useful to 
consider (\ref{1.5}) as some ``reference point'' for all the 
connections. Other connections can be considered as  (\ref{1.5}) 
plus some additional tensor as in  (\ref{1.4}). It is easy to
prove that the difference between any two connections is a tensor.

When the space-time is flat, the metric and the expression 
(\ref{1.5}) depend just on the choice of the 
coordinates, and one can choose them in such a
way that $\,\left\{ ^{\al}_{\be\ga}\right\}\,$ vanishes 
everywhere. On the contrary, if we consider, as in (\ref{1.4}),  
\beq
{\wh \Ga}^\al_{\be\ga} = \left\{ ^\al_{\be\ga}\right\} 
+ C^\al_{\,\cdot\,\be\ga}\,,
\label{1.6}
\eeq
than the tensor $\, C^\al_{\,\cdot\,\be\ga}\,$ (and, consequently, 
the whole connection $\,{\wh \Ga}^\al_{\be\ga}\,$) can not be 
eliminated by a choice of the coordinates. Even if one takes 
the flat metric, the covariant derivative based on 
$\,{\wh \Ga}^\al_{\be\ga}\,$ does not reduce to the coordinate 
transform of partial derivative. Thus, the introduction of an 
affine connection different from Christoffel symbol means that the 
geometry is not completely described by the metric, but has another, 
absolutely independent characteristic -- tensor $C^\al_{\be\ga}$. 
The ambiguity in the definition of $\Ga^\al_{\be\ga}$ is very 
important, for it enables one to introduce gauge fields 
different from gravity, and thus describe various interactions. 

In this paper we shall consider the particular choice of the 
tensor $\,C^\al_{\be\ga}$. Namely, we suppose that the affine 
connection $\,\tilde{\Gamma}^\alpha_{\;\beta\gamma}\,$ is 
not symmetric:
\beq
\tilde{\Gamma}^\alpha_{\,\beta\gamma} 
- \tilde{\Gamma}^\alpha_{\,\gamma\beta} 
= T^\alpha_{\,\cdot\,\beta\gamma}
\neq 0\,.
\label{tor}
\eeq
At the same time, we postulate that the corresponding 
covariant derivative satisfies the metricity condition 
$\tilde{\nabla}_\mu g_{\alpha\beta} = 0$
\footnote{The breaking of this condition means that one adds one more
tensor to the affine connection. This term is called non-metricity,
and it may be important, for example, in the consideration of the 
first order formalism for General Relativity. However, 
we will not consider the theories with non-metricity here.}.
The tensor $T^\alpha_{\;\cdot\,\beta\gamma}$ is called torsion. 

Below, we use the notation (\ref{1.5}) for the Christoffel symbol, 
and the notation with tilde for the connection with torsion and for 
the corresponding
covariant derivative. The metricity condition enables one to express
the connection through the metric and torsion in a unique way as
\beq
\tilde{\Gamma}^\alpha_{\;\beta\gamma} 
= {\Gamma}^\alpha_{\;\beta\gamma} + K^\alpha_{\cdot\;\beta\gamma}\,,
\label{con}
\eeq
where 
\beq
K^\alpha_{\;\cdot\beta\gamma} = \frac{1}{2}
\left( T^\alpha_{\;\cdot\beta\gamma} -
T^{\;\alpha}_{\beta\cdot\gamma} - 
T^{\;\alpha}_{\gamma\cdot\beta} \right)
\label{1.7}
\eeq
is called the contorsion tensor.
The indices are raised and lowered by means of the metric.
It is worthwhile 
noticing that the contorsion is antisymmetric in the first
two indices: $K_{\al\be\ga}=-K_{\be\al\ga}$, while torsion 
$\,T^\alpha_{\;\cdot\beta\gamma}\,$ 
itself is antisymmetric in the last two indices.  

The commutator of covariant derivatives in the space with torsion 
depends on the torsion and on the curvature tensor. 
First of all, consider the commutator acting on the 
scalar field $\,\ph$. We obtain
\beq
\left[{\tilde{\na}}_\al\,,\,{\tilde{\na}}_\be\right]\,\ph
= K^\la_{\,\cdot\,\al\be}\,\pa_\la\ph\,,
\label{scalar}
\eeq
that indicates to a difference with respect to the commutator 
of the covariant derivatives $\,\na_\al\,$ based on the Christoffel 
symbol (\ref{1.5}). In the case of a vector, after some simple 
algebra we arrive at the expression
\beq
\left[{\tilde{\na}}_\al\,,\,{\tilde{\na}}_\be\right]\,P^\la = 
T^\tau_{\cdot\al\be}\,{\tilde{\na}}_\tau\,P^\la 
+ {\tilde{R}}^\la_{\,\cdot\tau\al\be}\,P^\tau\,, 
\label{comu}
\eeq
where $\,{\tilde{R}}^\la_{\,\cdot\,\tau\al\be}\,$ is the curvature 
tensor in the space with torsion: 
\beq
{\tilde{R}}^\la_{\,\cdot\tau\al\be} = 
\pa_\al\,\tilde{\Gamma}^\la_{\cdot\tau\beta}
- \pa_\be\,\tilde{\Gamma}^\la_{\cdot\tau\al} 
+ \tilde{\Gamma}^\la_{\cdot\ga\al}\,\tilde{\Gamma}^\ga_{\cdot\tau\be}
- \tilde{\Gamma}^\la_{\cdot\ga\be}\,\tilde{\Gamma}^\ga_{\cdot\tau\al}\,.
\label{curva}
\eeq
Using (\ref{scalar}), (\ref{comu}) and that the product 
$\,P^\la B_\la\,$ is a scalar, one can easily derive the commutator 
of covariant derivatives acting on a 1-form $B_\la$ and then calculate
such a commutator acting on any tensor. In all cases the commutator 
is the linear combination of curvature (\ref{curva}) and torsion.  

The curvature (\ref{curva}) 
can be easily expressed through the Riemann tensor
(curvature tensor depending only on the metric), 
covariant derivative $\,\na_\al\,$ (torsionless) and contorsion as
\beq
{\tilde{R}}^\la_{\cdot\,\tau\al\be} = R^\la_{\cdot\,\tau\al\be} 
+ \na_\al\,K^\la_{\cdot\tau\be} - \na_\be\,K^\la_{\cdot\tau\al} + 
K^\la_{\cdot\ga\al}\,K^\ga_{\cdot\tau\be}-
K^\la_{\cdot\ga\be}\,K^\ga_{\cdot\tau\al}\,.
\label{riemann}
\eeq
Similar formulas can be written for the Ricci tensor 
and for the scalar curvature with torsion:
\beq
{\tilde{R}}_{\tau\be} = 
{\tilde{R}}^\al_{\,\cdot\,\tau\al\be} = R_{\tau\be} 
+ \na_\la\,K^\la_{\cdot\tau\be} - \na_\be\,K^\la_{\cdot\tau\la} + 
K^\la_{\cdot\ga\la}\,K^\ga_{\cdot\tau\be} -
K^\la_{\cdot\tau\ga}\,K^\ga_{\cdot\la\be}\,.
\label{ricci}
\eeq
(notice it is not symmetric) and 
\beq
{\tilde {R}} = g^{\tau\be}\,{\tilde{R}}_{\tau\be}
= R + 2\,\na^\la\,K_{\cdot\,\la\tau}^\tau 
-K_{\tau\la\,\cdot}^{\,\,\,\,\,\,\la}\,
K^{\tau\ga}_{\cdot\,\cdot\,\,\ga} 
+ K_{\tau\ga\la}\,K^{\tau\la\ga}\,.
\label{R}
\eeq
\vskip 2mm
It proves useful to divide torsion into 
three irreducible components:

i) the trace vector $T_{\beta} = T^\alpha_{\,\cdot\,\beta\alpha}\,\,,$

ii) the (sometimes, it is called pseudotrace) axial vector
$\;S^{\nu} = \epsilon^{\alpha\beta\mu\nu}T_{\alpha\beta\mu}\;$ and

iii) the tensor
$\;q^\alpha_{\,\cdot\,\beta\gamma}\;$, 
which satisfies two conditions $q^\alpha_{\;\cdot\,\beta\alpha} = 0$ 
and $\epsilon^{\alpha\beta\mu\nu}q_{\alpha\beta\mu} =0$

\noindent
Then, the torsion field can be expressed through these new fields as
\footnote{In the most of this paper, we consider the four-dimensional 
space-time. More general, $n$-dimensional formulas concerning 
classical gravity with torsion can be found in Ref. \cite{anhesh}.
More detailed classification of the torsion components can be 
found in Ref. \cite{class}.}
\beq
T_{\alpha\beta\mu} = \frac{1}{3} 
\left(T_{\beta}\,g_{\alpha\mu} - T_{\mu}\,g_{\alpha\beta}\right) 
- \frac{1}{6}\, \varepsilon_{\alpha\beta\mu\nu}\,S^{\nu} 
+ q_{\alpha\beta\mu}\,.
\label{irr}
\eeq
Using the above formulas, it is not difficult to express the 
curvatures (\ref{riemann}), (\ref{ricci}), (\ref{R}) 
through these irreducible components. We shall 
write only the expression for scalar curvature, which 
will be useful in what follows
\beq
 {\tilde R}= R - 2\na_\al \, T^\al - \frac{4}{3}\, T_\al\,T^\al 
+ \frac{1}{2}\,q_{\alpha \beta \gamma} \,q^{\alpha \beta \gamma} 
+ \frac{1}{24}\, S^\al\,S_\al\,.
\label{irred}
\eeq

%%%%%%%%%%%%%%%%%%%%%%%%%%%%%%%%%%%%%%%%%%%%%%%%%%%%%%%%%%%%%%%%
\section{Einstein-Cartan theory and non-dynamical torsion}
%%%%%%%%%%%%%%%%%%%%%%%%%%%%%%%%%%%%%%%%%%%%%%%%%%%%%%%%%%%%%%%%

In order to start the discussion of gravity with torsion, we 
first consider a direct generalization of General Relativity, 
which is usually called Einstein-Cartan theory. Indeed, our 
consideration will be very brief. For further information
one is recommended to look at the review \cite{hehl}. Our first 
aim is to generalize the Einstein-Hilbert action 
$$
S_{EH} = - \,\frac{1}{\ka^2}\,\int d^4x\,\sqrt{-g}\,R\,
$$ 
for the space with torsion. It is natural to substitute scalar 
curvature $R$ by (\ref{irred}), despite the change of the 
coefficients for the torsion terms in (\ref{irred}) can not be 
viewed as something wrong. As we shall see later, in quantum 
theory the action of gravity with torsion is induced with the 
coefficients which are, generally, different from the ones in 
(\ref{EC}). The choice of the volume element $dV_4$ should be done 
in such a manner that it transforms like a scalar and also reduces 
to the usual $d^4x$ for the case of a flat space-time and 
global orthonormal coordinates. Since we have two independent 
tensors: metric and torsion, the correct transformation 
property could be, in principle, 
satisfied in infinitely many ways. For example, one
can take $dV_4 = d^4x\,\sqrt{-g}$ as in General Relativity, or
$dV_4 = d^4x\,\sqrt{det (S_\mu T_\nu - S_\mu T_\nu)}$,  
or choose some other form. However, if we request that the 
determinant becomes $d^4x$ in a flat space-time limit with 
zero torsion, all the expressions similar to the last one
are excluded. In this paper we postulate, as usual, that the 
volume element in the 
space with torsion depends only on the metric 
and hence it has the form $dV_4 = d^4x\,\sqrt{-g}$.
Then, according to (\ref{irred}), the most natural 
expression for the action of gravity with torsion will be
\beq
S_{EC} = - \frac{1}{\ka^2}\,\int d^4x\,\sqrt{-g}\,{\tilde R} 
= - \frac{1}{\ka^2}\,\int d^4x\sqrt{-g}\left(
R - 2\na_\al T^\al - \frac{4}{3}
 T^2_\al + \frac{1}{2} q^2_{\alpha \beta \gamma} +
 \frac{1}{24} S^2_\al \right)\,.
\label{EC}
\eeq
The second term in the last integrand is a
total derivative, so it does not affect the equations of 
motion, which 
have the non-dynamical form $T^\al_{\,\cdot\,\be\ga}=0$.
Therefore, on the mass shell the theory (\ref{EC}) is 
completely equivalent to General Relativity. 
The difference appears when we add the external source for 
torsion. Imagine that torsion is coupled to some matter 
fields, and that the action of these fields depends on 
torsion in such a way that it contains the term
\beq
S_m = \int d^4x\,\sqrt{-g}\,\,K^\al_{\,\cdot\,\be\ga}\,
\Sigma_\al^{\,\cdot\,\be\ga}\,
\label{spinor-source}
\eeq
where the tensor $\,\Sigma_\al^{\,\cdot\,\be\ga}\,$ is constructed 
from the matter fields (it is similar to 
the dynamically defined Energy-Momentum tensor) but may also depend 
on metric and torsion. One can check that, for the Dirac fermion 
minimally coupled to torsion, the $\,\Sigma_\al^{\,\cdot\,\be\ga}\,$ 
is nothing but the expression for the spin tensor of this field. 
One can use this as a hint and choose
\beq
\Sigma_\al^{\,\cdot\,\be\ga} = 
\frac{1}{\sqrt{-g}}\,\frac{\de S_m}{\de K^\al_{\,\cdot\,\be\ga}}
\label{spindefin}
\eeq
as the dynamical definition of the 
spin tensor for the theory with the classical action $S_m$. 
Unfortunately, in some theories this formula gives the result 
different from the one coming from the Noether theorem, and 
only for the minimally coupled Dirac spinor (see the next section)
the result is the same. Next, since there is no experimental 
evidence for torsion, we can safely suppose it to be very weak.
Then, as an approximation, $\,\Sigma_\al^{\,\cdot\,\be\ga}\,$ 
can be considered independent of torsion. 
In this case, the equations following from the 
action $S_{EC}+S_m$ have the structure 
\beq
K \, \sim \, \ka^2 \Sigma \,\sim\, 
\frac{1}{M_p^2} \,\cdot \Sigma\,, 
\label{conta1}
\eeq
where $\,M_p=1/\ka\,$ is the Planck mass.
Then, torsion leads to the contact spin-spin interaction 
with the classical potential 
\beq
V(\Sigma) \sim  \frac{1}{M_p^2} \,\cdot \Sigma^2 
\label{conta2}
\eeq
Some discussion of this contact interaction can be found in 
\cite{hehl}. In section 2.6 we shall provide an example, 
illustrating the
possible importance of this interaction in the Early Universe.
Since the last expression  (\ref{conta2}) contains 
a $\,{1}/{M_p^2}\approx 10^{-38}\,GeV^{-2}\,$ factor, it can only
lead to some extremely weak effects at low energies. Therefore,
the effects of torsion, in the Einstein-Cartan theory, are 
suppressed by the torsion mass which is of the Planck order 
$M_p$. Even if one introduces kinetic terms for the
torsion components, the situation would remain essentially 
the same, as far as we consider low-energy effects. 

An alternative possibility is to suppose that  
torsion is light or even massless. In this case torsion can 
propagate, and there would be a chance to meet really independent 
torsion field. The review of theoretical limitations on this 
kind of theory \cite{betor,guhesh}
is one of the main subject of the present review
(see Chapter 4). These limitations 
come from the consistency requirement for the effective 
quantum field theory for such a ``light'' torsion.

%%%%%%%%%%%%%%%%%%%%%%%%%%%%%%%%%%%%%%%%%%%%%%%%%%%%%%%%%%%%%%%%
\section{Interaction of torsion with matter fields}
%%%%%%%%%%%%%%%%%%%%%%%%%%%%%%%%%%%%%%%%%%%%%%%%%%%%%%%%%%%%%%%%

In order to construct the actions of matter fields in an external
gravitational field with torsion we impose the principles of 
locality and general covariance. Furthermore, in order to preserve 
the fundamental features of the original flat-space theory, 
one has to require the symmetries of a given theory (gauge 
invariance) in flat space-time to hold for the theory in curved 
space-time with torsion. It is also natural to forbid the 
introduction of new parameters with the dimension of inverse 
mass. This set of conditions enables one to construct the 
consistent quantum theory of matter fields on the classical 
gravitational background with torsion. The form of the action
of a matter field is fixed except the values of some new 
parameters (nonminimal and vacuum ones) which remain arbitrary. 
This procedure which we have described above, leads to the 
so-called non-minimal actions.

Along with the nonminimal scheme,
there is a (more traditional) minimal one. According to it
the partial derivatives $\pa_\mu$ are substituted by the
covariant ones $\tilde{\nabla}_\mu$, the flat
metric $\eta_{\mu\nu}$ by $g_{\mu\nu}$ and the
volume element $d^4x$ by the covariant expression $d^4x\sqrt{-g}$.
We remark, that the minimal scheme gives, for the case of 
the Einstein-Cartan theory, the action (\ref{EC}), 
while the non-minimal scheme would make the coefficients 
at the torsion terms arbitrary. 

We define the minimal generalization of the action for the scalar 
field as
\beq
S_0=\int d^4x\,\sqrt{-g}\left\{\,
\frac12\,g^{\mu\nu}\,\pa_\mu\ph\,\pa_\nu\ph
- V(\ph)\,\right\}\,.
\label{scalar1}
\eeq
Obviously, the last action
does not contain torsion. One has to notice some peculiar 
property of the last statement. If one starts from the 
equivalent flat-space expression 
$$
S_0\,=\,\int d^4x\,\left\{\,- \frac12\,\ph \,\pa^2\,\ph
- V(\ph)\,\right\}
$$
then the generalized action does contain torsion. This can be 
easily seen from the following simple calculation:
\beq
{\tilde \na}^2 \ph =
g^{\mu\nu}\,{\tilde \na}_\mu\,{\tilde \na}_\nu \ph = 
\Box \ph + T^\mu\,\pa_\mu\ph\,.
\label{funnynotice}
\eeq
Thus, at this point, the minimal scheme contains a small 
ambiguity which can be cured through the introduction of 
the non-minimal interaction.  
For one real scalar, one meets five possible nonminimal 
structures (compare to the expression (\ref{EC})) 
$\,\ph^2 P_i$, where \cite{bush2}
\beq
P_1 = R, \,\,\,\,\,\,
P_2 = \na_\al\,T^\al, \,\,\,\,\,\,
P_3 = T_\al\,T^\al, \,\,\,\,\,\,
P_4 = S_\al\,S^\al, \,\,\,\,\,\,
P_5 = q_{\al\be\ga}\,q^{\al\be\ga}\,.
\label{Pi}
\eeq
Correspondingly, there are
five nonminimal parameters $\xi_1\,...\,\xi_5$.
The general non-minimal free field action has the form
\beq
S_0=\int d^4\sqrt{-g}\,\left\{\,
\frac12\,g^{\mu\nu}\,\na_\mu\ph\,\na_\nu\ph
+ \frac12\,m^2\,\ph^2 + \frac12\,\sum_{i=1}^{5}\xi_i\,P_i\,\ph^2
\,\right\}\,.
\label{scalar1-nm}
\eeq

A more complicated scalar content gives rise to more 
nonminimal terms \cite{bush1}.
In particular, for the complex scalar field $\phi$ one can 
introduce, into the covariant Lagrangian, 
the following additional term 
$$
\De L(\phi, \phi^\dagger)
= i\,\xi_0\,T^\mu\,\left(\,\phi^\dagger\cdot\pa_\mu\,\phi
 - \pa_\mu\phi^\dagger\cdot\phi\,\right)\,.
$$
In the case of a scalar $\ph$ coupled to a pseudoscalar $\chi$, 
there are other possible non-minimal terms:  
$$
\De L(\ph, \chi)
=\frac12\,\xi^\prime_0\,S^\mu\,\left(\,\ph\,\pa_\mu\,\chi
 - \chi\,\pa_\mu\,\ph\,\right) + \ph\,\chi\sum_{j=6}^{10}
\xi_j\,D_j\,,
$$
where 
$$
D_6=\na_\mu\,S^\mu \,, \,\,\,\,\,\,\,\,
D_7=T_\mu\,S^\mu \,, \,\,\,\,\,\,\,\,
D_8=\ep_{\mu\nu\al\be}\,S^\mu\,q^{\nu\al\be} \,, 
$$$$
D_9=\ep_{\mu\nu\al\be}\,
q_\la^{\,\,\,\,\,\mu\nu}\,q^{\la\al\be}\,\,\,\,\,\,\,\,
D_{10}=\ep_{\mu\nu\al\be}\,
q_{\,\cdot\,\la\,\cdot\,}^{\mu\,\,\,\,\,\nu}\,q^{\al\la\be}\,.
$$
More general scalar models can be treated in a similar way. 
%%%%%%%%%%%%%%%%%%%%%%%%%%%%%%%%%%%%%%%%%%%%%%%%%%%%%%%%%%%%%
\vskip 4mm

For the Dirac spinor
\footnote{Various aspects of the minimal fermion-torsion 
interaction have been considered in many papers, e.g. 
\cite{datta,aud,hehl}.}, 
the minimal procedure leads to the 
expression for the hermitian action
\beq
S_{\frac12 , min} = \frac{i}{2}\,\int d^4x\sqrt{g}\,\left(\,
{\bar \psi}\,
\ga^\al \,\tilde{\na}_\al\psi -
 \tilde{\na}_\al{\bar \psi} 
\,\ga^\al \,\psi -2im {\bar \psi} \psi\,  \right)\,.
\label{dirac1}
\eeq
Here $\ga^\mu = e^\mu_a\,\ga^a$, where $\,\ga^a\,$
is usual (flat-space) $\,\gamma$-matrix, and $\,e^\mu_a\,$
is tetrad (vierbein) defined through the standard relations
$$ 
e^\mu_a\cdot e_{\mu b} = \eta_{ab}\,,\,\,\,\,\, 
e^\mu_a\cdot e^{\nu a} = g^{\mu\nu}\,,\,\,\,\,\, 
e_\mu^a\cdot e_{\nu a} = g_{\mu\nu}\,,\,\,\,\,\, 
e_\mu^a\cdot e^{\mu b} = \eta^{ab}\,.
$$
The covariant derivative of a Dirac spinor $\,\tilde{\na}_\al\psi$ 
should be defined to be consistent with the covariant derivative of  
tensors. We suppose that 
\beq
\tilde{\na}_\mu\psi = \pa_\mu\psi + \frac{i}{2}\,
\tilde {w}^{ab}_\mu\,\, \si_{ab}\, \,\psi\,,
\label{spinor11}
\eeq
where $\,\tilde {w}^{ab}_\mu\,$ is a new object 
which is usually called spinor connection, and 
$$
\si_{ab} = \frac{i}{2}\,\left(\ga_a\ga_b - 
\ga_b \ga_a\right). 
$$
The conjugated expression is
\beq
\tilde{\na}_\mu\,{\bar \psi} = \pa_\mu\,{\bar \psi}
 - \frac{i}{2}\,{\bar \psi}\,\,
\tilde {w}^{ab}_\mu\,\,\si_{ab}\,.
\label{spinor12}
\eeq
Now, we consider how the covariant derivative acts on the 
vector $\,{\bar \psi}\ga^\al\psi\,$. As we already learned
in section 2.1, if the connection provides the proper 
transformation
law for the vector, it does so with any tensor. Therefore, 
the only one equation for the spinor connection 
$\,\tilde {w}^{ab}_\mu\,$ is
\beq
\tilde{\na}_\mu\,\left(\,{\bar \psi}\,\ga^\al \,\psi
\,\right) = {\pa}_\mu\,\left(\,{\bar \psi}\,\ga^\al 
\,\psi \,\right) + \tilde{\Ga}^\al_{\cdot\,\la\mu}\,
\left(\,{\bar \psi}\,\ga^\al \,\psi\,\right)
= {\na}_\mu\,\left(\,{\bar \psi}\,\ga^\al 
\,\psi \,\right) + K^\al_{\cdot\,\la\mu}\,
\left(\,{\bar \psi}\,\ga^\al \,\psi\,\right)\,.
\label{spinor13}
\eeq
Replacing (\ref{spinor11}) and (\ref{spinor12}) into
(\ref{spinor13}), after some algebra, we arrive at the 
formula for the spinor connection
\beq
\tilde {w}_{\mu ab} = {w}_{\mu ab} + \frac14\,
K^\al_{\cdot\,\la\mu}\,\left(\,e^\la_a\,e_{b\al}
- e^\la_b\,e_{a\al}\,\right)\,,
\label{connection}
\eeq
where 
\beq
w_{\mu ab}= \frac14\,\left(\,e_{b\al} \pa_\mu \, e^\al_a
- e_{a\al} \pa_\mu \, e^\al_b\,\right)
+ \frac14\,\Ga^\al_{\la\mu}\,\left(\,
e_{b\al} \,e^\la_a - e_{a\al}\,e^\la_b \,\right)
\label{spin-connect}
\eeq
is spinor connection in the space-time without torsion.

Substituting (\ref{connection}) into (\ref{dirac1}), and 
performing integration by parts, we arrive at two equivalent 
forms for the spinor action
$$
S_{\frac12 , min} = i\,\int d^4x\sqrt{g}
\,\,{\bar \psi}\,\left(\,
\ga^\al\,\tilde{\na}_\al - \frac12\,\ga^\al\,T_\al - im 
\,\right)\,\psi =
$$\beq
= i\,\int d^4x\sqrt{g}\,{\bar \psi}\,\left(\,
\ga^\al\,\na_\al  - \frac{i}{8}\,\ga^5\ga^\al\,S_\al - im
\,\right)\,\psi\,,
\label{dirac2}
\eeq
where we use standard representation for the Dirac matrices,
such that $\ga^5=-i\ga^0\ga^1\ga^2\ga^3$ and $(\ga^5)^2=1$.
Also, $\ga^5=\ga_5$. 
The first integral is written in terms of the covariant 
derivative $\,\tilde{\na}_\al\,$ with torsion, while 
the last is expressed through the torsionless covariant 
derivative  $\,\na_\al$. Indeed the last form is
more informative, for it tells us that only the axial vector
$\,S_\mu\,$ couples to fermion, and the other two
components:
$\,T_\mu\,$ and $\,q^\al_{\cdot\,\be\ga}\,$ completely 
decouple. It is important to notice, that in the first 
of the integrals (\ref{dirac2}) the $\,\ga^\al\,T_\al$-term
has an extra factor of $\,\,i\,\,$ as compared to the 
electromagnetic (or any vector) field. This indicates, 
that if being taken separately from the first term 
$\,\ga^\al\,\tilde{\na}_\al$, the $\,\ga^\al\,T_\al$-term
would introduce an imaginary part into the spinor action,
and therefore it has no sense. Of course, the same concerns 
the term $\,i\int {\bar \psi} \ga^\al{\tilde \nabla}_\al \psi$. 
The imaginary terms, coming from the two parts,
cancel, and give rise to a Hermitian action for the spinor, 
equivalent to (\ref{dirac1}) or to the second integral in 
(\ref{dirac2}).   

The non-minimal interaction is a bit more complicated.
Using covariance, locality, dimension, and 
requesting that the action does not break parity
one can construct only two non-minimal (real, of course)
terms with the structures already known from (\ref{dirac2}). 
\beq
S_{\frac12, non-min} \,=\, i\,\int d^4x\sqrt{g}\,\,{\bar \psi}\,
\Big(\, \ga^\al\,\na_\al  + \sum_{j=1,2} \eta_j\,Q_j - im
\,\Big)\,\psi\,.
\label{dirac2-nm}
\eeq
with 
$$
Q_1 = i\ga^5\,\ga^\mu\, S^\mu\,,\,\,\,\,\,\,\,\,
Q_2 = i \ga^\mu\, T^\mu
$$
and two arbitrary 
non-minimal parameters $\eta_1,\,\eta_2$. The minimal theory
corresponds to $\eta_1 = -\frac18,\,\,\,\eta_2 = 0$. Let us again notice 
that the $T_\mu$-dependent term in (\ref{dirac2-nm}) is different 
from the last term in the first representation in (\ref{dirac2}).
In (\ref{dirac2-nm}) all the terms are real. We observe, that 
the interaction of the torsion trace $T_\mu$ with fermion is
identical to the one of the electromagnetic field. Therefore, in some
situations when torsion is considered simultaneously with the external 
electromagnetic field $\,A_\mu\,$, 
one can simply redefine $\,A_\mu\,$ such that the torsion trace
$\,T_\mu\,$ disappears.

Consider the symmetries of the 
Dirac spinor action non-minimally coupled to the vector
and torsion fields
\beq
S_{1/2}= i\,\int d^4x\,{\bar \psi}\, \left[
\,\ga^\mu \,\left( {\pa}_\mu + ieA_\mu + i\,\eta\,\ga_5\,S_\mu\,\right)
- im \,\right]\,\psi\,.
\label{diraconly}
\eeq
As compared to (\ref{dirac2-nm}), here we have changed the 
notation for the
nonminimal parameter of interaction between spinor fields and the
axial part $S_\mu$ of torsion: $\eta_1 \to \eta$. Furthermore, 
we used the possibility  to redefine the external
electromagnetic potential $A_\mu$ in such a way that it absorbs
the torsion trace $T_\mu$.

The new interaction with torsion does not spoil
the invariance of the action under usual gauge transformation:
\beq
\psi' = \psi\,e^{\al(x)}
,\,\,\,\,\,\,\,\,\,\,\,\,\,\,
{\bar {\psi}}' = {\bar {\psi}}\,e^{- \al(x)}
,\,\,\,\,\,\,\,\,\,\,\,\,\,\,
A_\mu ' = A_\mu - {e}^{-1}\, \pa_\mu\al(x)\,.
\label{trans1}
\eeq
Furthermore,
the massless  part of  the action (\ref{diraconly})
is invariant under the transformation in which the axial vector 
$\,S_\mu\,$ plays the role of the gauge field
\beq
 \psi' = \psi\,e^{\ga_5\be(x)}
,\,\,\,\,\,\,\,\,\,\,\,\,\,\,
{\bar {\psi}}' = {\bar {\psi}}\,e^{\ga_5\be(x)}
,\,\,\,\,\,\,\,\,\,\,\,\,\,\,
S_\mu ' = S_\mu - {\eta}^{-1}\, \pa_\mu\be(x)\,.
\label{trans}
\eeq
Thus, in the massless sector of the theory one faces generalized
gauge symmetry depending on the scalar, $\al(x)$, and pseudoscalar,
$\be(x)$, parameters of the transformations, while the
massive term is not invariant under the last transformation.

%%%%%%%%%%%%%%%%%%%%%%%%%%%%%%%%%%%%%%%%%%%%%%%%%%%%%%%%%%%%%%%%%%%
Consider whether the massless vector field might 
couple to torsion. Here, one has to use the principle of 
preserving the symmetry. The minimal interaction
with torsion breaks the gauge invariance for the vector 
field, since
$$
\tilde F_{\mu\nu} = \tilde{\na}_\mu\,A_\nu - 
\tilde{\na}_\nu\,A_\mu = F_{\mu\nu} 
+ 2\,A_\la\,K^\la_{\cdot\,\,[\mu\nu]}
%%%%%%%%%%  \label{vector-min}
$$
is not invariant. The possibility to modify the gauge transformation 
in the theory with torsion has been studied in \cite{novello,rr}. 
In this paper we opt to keep 
the form of the gauge transformation unaltered, and postulate that 
the gauge vector does not couple to torsion. The reasons for
this choice is the following. First of all, when one is 
investigating the quantum field theory in external torsion field, 
it is natural to separate the effects of external field from the 
purely matter sector. Thus, the modification of the gauge 
transformation does not fit with our approach. Furthermore,
the most important part $\,S_\mu\,$ of the torsion tensor does
not admit the fine-tuning of the gauge transformation. In other words,
for the most interesting case of purely antisymmetric torsion 
it is not possible to save gauge invariance for the vector 
coupled to torsion in a minimal way. 

Let us consider the non-minimal interaction for the special 
case of an abelian gauge vector. One can introduce several 
nonminimal terms which do not break the gauge invariance:
$\,\int d^4x\sqrt{-g}\,F_{\mu\nu}\,K^{\mu\nu}$. 
The most general form of $\,K^{\mu\nu}$ is: 
$$
K^{\mu\nu} = \th_1\ep^{\mu\nu\al\be}\,T_\al\,S_\be + 
\th_2\ep^{\mu\nu\al\be}\,\pa_\al\,S_\be +
\th_3\ep^{\mu\nu\al\be}\,q^\la_{\cdot\,\,\al\be}\,S_\la +
$$
\beq
\th_4\,q^\la_{\cdot\,\,\mu\nu}\,T_\la +
\th_5\,(\pa_\mu\,T_\nu- \pa_\mu\,T_\nu)+
\th_6\,\pa_\la\,q^\la_{\cdot \,\mu\nu}\,.
\label{vector-nm}
\eeq

For the non-abelian vector, the non-minimal structures 
like (\ref{vector-nm}) are algebraically impossible. In the Standard 
Model, where all vectors are non-abelian, the non-minimal 
terms (\ref{vector-nm}) do not exist.

And so, we have constructed the actions of free scalar, spinor and 
vector fields coupled to torsion. In general, there are two types 
of actions: minimal and non-minimal. As we shall see in the next 
sections,  the non-minimal interactions with spinors and scalars 
provide certain advantages
at the quantum level, for they give the possibility to construct  
renormalizable theory \cite{bush1}. 
 
%%%%%%%%%%%%%%%%%%%%%%%%%%%%%%%%%%%%%%%%%%%%%%%%%%%%%%%%%%%%%%%%%%%
\section{Conformal properties of torsion}
%%%%%%%%%%%%%%%%%%%%%%%%%%%%%%%%%%%%%%%%%%%%%%%%%%%%%%%%%%%%%%%%%%%

Conformal symmetry with torsion 
has been studied in many papers (see, for example,
\cite{obukhov,bush1,anhesh} and references therein). 
Here, we shall 
summarize the results obtained in \cite{bush1,anhesh}. 

For the torsionless theory the conformal transformation of the 
metric, scalar, spinor and vector fields take the form:
\beq
g_{\mu\nu}\to g^\prime_{\mu\nu} = g_{\mu\nu}\,e^{2\si}\,,\,\,\,\,\,\,\,
\ph \to \ph^\prime =\ph \,e^{-\si}\,,\,\,\,\,\,\,\,
\psi \to \psi^\prime =\psi \,e^{-3/2\,\si}\,,\,\,\,\,\,\,\,
A_\mu\to A_\mu^\prime = A_\mu\,,
\label{semmetr}
\eeq
where $\,\si = \si(x)\,$. In the absence of torsion, the actions 
of free fields are invariant if they are massless, besides in the
scalar sector one has to put $\,\xi =\frac16$. Indeed, the 
interaction terms of the gauge theory (gauge, Yukawa and 4-scalar 
terms) are conformally invariant. 

The problem is to define the conformal transformation for torsion, 
such that the free actions formulated in the previous section 
would be invariant for these or that values of the nonminimal 
parameters. It turns out that there are three different ways
to choose the conformal transformation for torsion. 
\vskip 2mm

%%%%%%%%%%%%%%%%%%%%%%%%%%%%%%%%%%%%%%%%%%%%%%%%%%%%%%%%%%%%
{\it i) Week conformal symmetry} \cite{bush1}.
Torsion does not
transform at all: $T^\la_{\cdot\,\mu\nu} \to 
{T^\prime}^\la_{\cdot\,\mu\nu} = T^\la_{\cdot\,\mu\nu}$.
The conditions of conformal symmetry are absolutely the 
same as in the torsionless theory. 
\vskip 2mm
%%%%%%%%%%%%%%%%%%%%%%%%%%%%%%%%%%%%%%%%%%%%%%%%%%%%%%%%%%%%

{\it ii) Strong conformal symmetry} \cite{bush1}. 
In this version, torsion transforms as: 
\beq
T^\la_{\cdot\,\mu\nu} \to 
{T^\prime}\,^\la_{\cdot\,\mu\nu} = T^\la_{\cdot\,\mu\nu} 
+ \om\,\left(\de_\nu^\la\,\pa_\mu - \de_\mu^\la\,\pa_\nu
\right)\si(x).
\label{strongconf}
\eeq
This transformation includes an arbitrary parameter 
\footnote{Instead of introducing an arbitrary numerical 
parameter $w$, one could replace, in the transformation rule for 
torsion (\ref{strongconf}), the parameter $\,\si\,$ for 
some other, independent parameter. This observation
has been done in 1985 by A.O. Barvinsky and V.N. Ponomaryev 
in the report on my PhD thesis \cite{thesis}.}, $\,\,\,\om$. 
Indeed the above 
transformation means that only the torsion trace transforms
$$ %%%\beq 
T_\mu \to T^\prime_\mu = T_\mu + 3\,\om\,\pa_\mu\,\si(x).
$$  %%%%%%\label{traceconf}
%%%%%% \eeq
Other components of torsion remain inert under (\ref{strongconf}).
For any value of $\,\om\,$, the free actions (\ref{scalar1-nm})
and (\ref{dirac2-nm}) are invariant if they depend only on the 
axial vector $S_\mu$ and tensor $q^\la_{\cdot\,\mu\nu}$, but 
not on the trace $T_\mu$. Of course, this is quite natural, 
because only $T_\mu$ transforms. The restrictions  
imposed by the symmetry are $\,\xi_2=\xi_3=\eta_2=0$.
The immediate result of the 
strong conformal symmetry is the modified Noether 
identity, which now reads
\beq
\frac{\de S}{\de g_{\mu\nu}}\,\de g_{\mu\nu} +
\sum\,\,\frac{\de S}{\de \Phi}\,\de \Phi +
\frac{\de S}{\de T^\al_{\cdot\,\be\ga}}
\,\de T^\al_{\cdot\,\be\ga} = 0\,,
\label{Noether2}
\eeq
where $\,\Phi\,$ is the full set of matter fields.
Using the equations of motion, the
definition of the Energy-Momentum Tensor 
$\,\,T_{\mu\nu}\,=\,-\,
\frac{2}{\sqrt{-g}}\,\,\frac{\de S}{\de g^{\mu\nu}}$,
and the definition of the Spin Tensor given in 
(\ref{spindefin}), we get
$$
T^{\mu\nu}\,\de g_{\mu\nu} + \left(\,\Si_\la^{\cdot\,\mu\nu} 
- \frac12\,\Si_{\cdot\,\cdot\,\la}^{\mu\nu}\,\right)\,\de 
T^\la_{\cdot\,\mu\nu} = 0\,,
$$
that gives  
\beq
2T^\mu_\mu - 3\,\om\,\na_\mu\,\Si_\nu^{\cdot\,\mu\nu} = 0\,.
\label{Noether22}
\eeq
Along with the standard relation for the
trace of the Energy-Momentum Tensor $\,T^\mu_\mu = 0$, 
here we meet an additional 
identity $\,\na_\mu\,\Si_\nu^{\cdot\,\mu\nu} = 0$. It is
interesting that this identity is exact, for it is not violated 
by the anomaly on quantum level. In order to understand this, 
we remember that the $\,T_\mu$--dependence is purely
non-minimal, and if it does not exist in classical theory, 
it can not appear at the quantum level. 
\vskip 2mm

%%%%%%%%%%%%%%%%%%%%%%%%%%%%%%%%%%%%%%%%%%%%%%%%%%%%%%%%%%%%%%%%%%%%
{\it iii) Compensating conformal symmetry \cite{obukhov,anhesh}.}
This is the most interesting and complicated version of 
the conformal transformation. 

Let us consider the massless scalar,
non-minimally coupled to metric and torsion (\ref{scalar1-nm}).
One can add to it the $\la\ph^4$-term, without great changes in 
the results. Then  
 \bea
 S =\int d^4x \sqrt{-g}\; \left\{
 \, \frac{1}{2} \;g^{\mu\nu}\partial_{\mu}\ph\;
\partial_{\nu}\ph +  \frac{1}{2}\;\sum_{i=1}^{5}\,\xi_i P_i \ph^2
 - \frac{\la}{4!}\,\ph^4 \, \right\}\,.
 \label{n7}
 \eea
where $P_i$ were defined at (\ref{Pi}). 

 The equations of motion for the torsion tensor can
 be split into three independent equations
 written for the components
 $T_\al,\,S_\al,\,q_{\al\be\ga}$; they yield:
 \beq
 T_\al=\frac{\xi_2}{\xi_3}\,\cdot\, \frac{\na_\al\ph}{\ph}
 \,,\,\,\,\,\,\,\,\,\,\,\,\,
 \,S_\mu = q_{\al\be\ga} = 0\,.
 \label{n9}
 \eeq
 Replacing these expressions back into the action (\ref{n7}),
 we obtain the on-shell action
 \bea
 S = \int d^4x \sqrt{-g}\; \left\{
 \frac{1}{2} (1-\frac{\xi^2_2}{\xi_3})\;
 g^{\mu\nu}\partial_{\mu}\ph \;
 \partial_{\nu}\ph +  \frac{1}{2}\; \xi_1 \ph^2R
 -\frac{\la}{4!}\ph^{4} \right \}\,,
 \label{onshell}
 \eea
 that can be immediately reduced to the torsionless
 conformal action
 \bea
 S = \int d^4x \sqrt{-g}\; \left\{
 \frac{1}{2}\; g^{\mu\nu}\partial_{\mu}\ph \;
 \partial_{\nu}\ph  +  \frac{1}{12}\; \ph^2\,R
 -\frac{\la^ \prime}{4!}\ph^{4} \right \}\,,
 \label{sem-torcao}
 \eea
 by an obvious change of variables, whenever the non-minimal 
 parameters satisfy the condition
 \beq
 \xi_1=\frac{1}{6}\,
 \left(1-\frac{\xi^2_2}{\xi_3}\right)\,
 \label{n8}
 \eeq
and some obvious relation between $\la$ and $\la^\prime$.
Some important observation is in order.
It is well-known, that the conformal action (\ref{sem-torcao})
is classically equivalent to the Einstein-Hilbert action of 
General Relativity, but with the opposite sign \cite{deser,conf}. 
In order to check this, we take such a wrong-sign action:
 \beq
 S_{EH}[g_{\mu\nu}] =
 \int d^4x \sqrt{- {\hat g}}\;
 \left\{ \, \frac{1}{\ka^2}\, {\hat R} + \La  \, \right\} \cdot
 \label{0.1}
 \eeq
 This action depends on the metric $\,{\hat g}_{\mu\nu}$.
  Performing conformal transformation
 $\,{\hat g}_{\mu\nu} = g_{\mu\nu}\cdot e^{2\si(x)}$,
 we use the standard relations between geometric
 quantities of the original and transformed metrics:
 \beq
 \sqrt{-{\hat g}} = \sqrt{-g}\;e^{4\si}\,, \;\;\;\;\;\;\;\;\;
 {\hat R} = e^{-2\si}\left[R -
 6\Box\si - 6(\na\si)^2 \right] \cdot
 \label{n1}
 \eeq
 Substituting (\ref{n1}) into (\ref{0.1}), after
 integration by parts, we arrive at:
 $$ %%%%%%   \bea
 S_{EH}[g_{\mu\nu}] =
 \int d^4x \sqrt{-g}\,
 \left\{ \frac{6}{\ka^2}\,e^{2\si}\,(\na\si)^2
 +\frac{e^{2\si}}{\ka^2}\,R + \La e^{4\si}\right \},
 $$ %%%%%%%  \label{n2}
 %%%%%%%%%%  \eea
 where 
 $(\na \si)^2 = g^{\mu\nu}\partial_\mu\si \partial_\nu\si $.
 If one denotes
 $$ %%% \beq
 \ph =  e^{\si}\,\cdot\,
 \sqrt{ \frac{12}{\ka^2} } \,,
 $$ %%% \label{n3}
 %%%    \eeq
 the action (\ref{0.1}) becomes
 \beq
 S=\int d^4x \sqrt{-g} \left\{
 \frac{1}{2}\,(\na\ph)^2+\frac{1}{12}R\ph^2+\La \left(\,
 \frac{\ka^2}{12}\,\right)^2\cdot
 \ph^ 4 \right \}\,, 
 \label{n5}
 \eeq
that is nothing but (\ref{sem-torcao}).  And so, 
the metric-scalar theory described by the action of eq. (\ref{n5})
and the metric-torsion-scalar theory (\ref{n7}) with the
constraint (\ref{n8}) are equivalent to 
the General Relativity with cosmological constant.

 One has to notice that the first two theories exhibit 
 an extra local
 conformal symmetry, which compensates an extra (with respect to
 (\ref{0.1})) scalar degree of freedom. Moreover, (\ref{n5}) is a
 particular case of a family of
 similar actions, linked to each other
 by the reparametrization of the scalar or (and) the conformal
 transformation of the metric \cite{conf}. The symmetry
 transformation which leaves the action (\ref{n5}) stable is
 \beq
 g^{\prime}_{\mu\nu}=g_{\mu\nu}\cdot e^{2\rho(x)}
 \,,\, \, \, \, \, \, \, \, \, \, \,
 \ph^{\prime} = \ph \cdot e^{-\rho(x)}\,.
 \label{nnn}
 \eeq
 The version of the
 Brans-Dicke theory with torsion (\ref{n7})
 is conformally equivalent to General Relativity (\ref{0.1}) 
 provided
 that the new condition (\ref{n8}) is satisfied and there are only
 external conformally covariant sources for
 $\,S_\mu$, $\,{q^{\alpha}}_{\,\cdot\,\beta\gamma}\,$
 and for the transverse component of $\,T_{\al}\,$. Such a sources
 do not spoil the conformal symmetry.

 Now, we can see that 
 the introduction of torsion provides some theoretical advantage.
 If we start from the positively defined gravitational action
 (\ref{0.1}), the sign of the scalar action (\ref{n5}) should
 be negative, indicating to the well known instability of the
 conformal mode of General Relativity (see, for example, 
 \cite{hawking-foam} and also \cite{wet,mod}
 for the recent account of this problem and further references).
 It is easy to see that the metric-torsion-scalar theory may be
 free of this problem, if we choose
 $\,\frac{\xi^2_2}{\xi_3}-1 > 0$. In this case one meets 
 the equivalence of the
 positively defined scalar action (\ref{n7}) to the
 action (\ref{0.1}) with the negative sign. The negative
 sign in (\ref{0.1}) signifies, in turn, the positively defined
 gravitational action. Without torsion one can achieve positivity
 in the gravitational action only by the expense of taking the 
 negative kinetic energy for the scalar action in (\ref{n5}).

 The equation of motion (\ref{n9}) for $T_\al$ 
 may be regarded as a constraint that fixes
 the conformal transformation for this vector to be consistent
 with the one
 for the metric and scalar. Then, instead of (\ref{nnn}), one has
 \beq
 g^{\prime}_{\mu\nu}=g_{\mu\nu}\cdot e^{2\rho(x)}\,,\, \, \, \, \,
 \, \, \, \, \, \,
 \ph^{\prime} = \ph \cdot e^{-\rho(x)}
 \,,\, \, \, \, \, \, \, \, \, \, \,
 T^{\prime}_\al =
 T_\al - \frac{\xi_2}{\xi_3}\cdot\partial_\al\rho(x)
 \label{mmm}
 \eeq
 The on-shell equivalence  
 can be also verified using the equations of motion \cite{anhesh}.
 It is easy to check, by direct inspection,
 that even off-shell, the theory with
 torsion (\ref{n7}), satisfying the relation (\ref{n8}), may be
 conformally invariant whenever we define the transformation
 law for the torsion trace according to (\ref{mmm}):
 and also postulate that the other pieces of torsion:
 $S_\mu$ and ${q^\al}_{\,\cdot\,\be\ga}$, do not transform.
 The quantities $\,\sqrt{-g}\,$ and $\,R\,$ transform as in
 (\ref{n1}).
 One may introduce into the action other conformal invariant
 terms depending on torsion. For instance:
 $$ %%% \beq
 S = - \frac{1}{4}\, \int d^4x\,\sqrt{-g}\,\,T_{\al\be}\, 
 T^{\al\be},
 $$ %%% \label{stressT}
 %%%%%% \eeq
 where $T_{\al\be}=\partial_\al T_\be-\partial_\be T_\al$.
 In the spinor sector one has to request $\eta_2=0$ in 
 (\ref{dirac2-nm}), as it was for the strong conformal symmetry. 
 It is important to notice, for the future,
 that on the quantum level
 this condition does not break renormalizability, even if 
 $\,\xi_{2,3}$ are non-zero.

  %%%%%%%%%%%%%%%%%%%%%%%%%%%%%%%%%%%%%%%%%%%%%%%%%%%%%%%%%%%%%%%
 One can better understand the equivalence between General Relativity 
 and conformal
 metric-scalar-torsion theory (\ref{n7}), (\ref{n8})
 after presenting an alternative form for the symmetric action.
 All torsion-dependent terms in (\ref{n7}) 
 may be unified in the expression
 \beq
 {\cal P}  = -\frac16\,\frac{\xi^2_2}{\xi_3}\,R +
 \xi_2\,(\na_\mu T^\mu) +
 \xi_3\,T_\mu T^\mu +
 \xi_4\, S_\mu^2
 + \xi_5\,q_{\mu\nu\la}^2 \,.
 \label{P}
 \eeq
 It is not difficult to check that the transformation law
 for this new quantity is especially simple: 
 $\,\,{\cal P}^\prime = e^{-2\rho}\,{\cal P}$.
 Using new quantity, the conformal invariance of the
 action becomes obvious:
 \beq
 S_{inv} = \int d^4x \sqrt{-g}\; \left\{
 \,\frac{1}{2} \;g^{\mu\nu}\partial_{\mu}\ph\;
 \partial_{\nu}\ph +  \frac{1}{12}\;R\ph^2
 + \frac12\,{\cal P} \,\ph^2 \right\}\,.
 \label{inv}
 \eeq

In order to clarify the role of torsion 
in our conformal model, we construct one more representation 
for the 
metric-scalar-torsion action with local conformal symmetry.
Let us  start, once again, from the action (\ref{n7}), (\ref{n8})
and perform only part of the transformations (\ref{mmm}): 
\beq 
\ph\rightarrow  \ph^{\prime} = \ph \cdot e^{-\rho(x)}
\,,\, \, \, \, \, \, \, \, \, \, \, T_\al \rightarrow
T^{\prime}_\al = T_\al - 
\frac{\xi_2}{\xi_3}\cdot\partial_\al\rho(x)\,.
\label{partofmmm} 
\eeq 
Of course, if we supplement
(\ref{partofmmm}) by the transformation of the metric, we arrive
at (\ref{mmm}) and the action does not change. On the other hand,
(\ref{partofmmm}) alone might lead
to an alternative conformally equivalent description of the
theory. Taking $\,\rho\,$ such that
$$
\ph \cdot e^{-\rho(x)} =
\frac{12}{\ka^2}\,\left(1-\frac{\xi^2_2}{\xi_3}\right)
 = const\,,
$$
we obtain, after some algebra, the following action:
\beq
 S = \frac{1}{\ka^2}\,\int d^4x \sqrt{-g}\;
\Big\{\,R 
+ \frac{3}{\ka^2 (1-{\xi^2_2}/{\xi_3})}\,
\Big[
\xi_4 S_\mu^2
+ \xi_5 q_{\mu\nu\la}^2
+ \xi_3 \Big(T_\al-\frac{\xi_2}{\xi_3}\,\na_\al\ln\ph
\Big)^2\,\Big]\,\Big\}\,.
\label{oneq}
\eeq
This form of the action does not contain interaction between
curvature and the scalar field. At the same time, the latter 
is present until we use the equations of motion
(\ref{n9}) for torsion.
Torsion trace looks here like a Lagrange multiplier, and only
using torsion equations of motion, one can obtain the action 
of GR. It is clear that one can arrive at the same action
(\ref{oneq}), making the transformation of the metric
as in (\ref{mmm}) instead of (\ref{partofmmm}).

 To complete this part of our consideration, we mention that
 the direct generalization
 of the Einstein-Cartan theory including an extra scalar may be
 conformally equivalent to General Relativity, provided
 that the non-minimal parameter takes an appropriate
 value. To see this, one uses the relation (\ref{irred})
  and replace it into the "minimal" action
 \beq
 S_{ECBD} =
 \int d^4x \sqrt{-g}\, 
 \left\{ \frac{1}{2}\,g^{\mu\nu}\partial_\mu\ph\,
 \partial_\nu\ph+\frac{1}{2}\,\xi\,{\wt R}\ph^2 \right\} \,.
 \label{mini}
 \eeq
 It is easy to see that the condition (\ref{n8}) is satisfied
 for the special value $\,\,\xi =  \frac13$,
 contrary to the famous $\xi = \frac16$ in the
 torsionless case. The effect of changing conformal value of
 $\xi$ due to the non-trivial transformation of torsion
 has been discussed in \cite{payo} and \cite{anhesh} (see 
 also further references there).

%%%%%%%%%%%%%%%%%%%%%%%%%%%%%%%%%%%%%%%%%%%%%%%%%%%%%%%%%%%%%%
\section{Gauge approach to gravity. Higher derivative
gravity theories with torsion}
%%%%%%%%%%%%%%%%%%%%%%%%%%%%%%%%%%%%%%%%%%%%%%%%%%%%%%%%%%%%%%

There are many good reviews on the gauge approach to gravity
(see, for example, \cite{hehl}), and since the aim of the present 
paper is to treat torsion from the field-theoretical point of
view, we shall restrict ourselves to a brief account of the results 
and some observations. 

In section 2.1 we have
introduced covariant derivative and found, that this can be done
in different ways, because one can add to the affine connection any 
tensor $C^\al_{\,\cdot\,\be\ga}$ (see eq. (\ref{1.6})). Any such 
extension of the affine connection is 
related to some additional physical field, exactly because 
$C^\al_{\,\cdot\,\be\ga}$ is a tensor and can not be removed by a
coordinate transformation. It was already mentioned in section 2.1, 
that the introduction of covariant 
derivative is related to the general coordinate transformations. 
The aim of the gauge approach to gravity is to show that the 
same construction, including the metric and the covariant derivative 
based on an arbitrary connection, can be achieved through the local 
version of the Lorentz-Poincare symmetry. If one requests
the theory to be invariant with respect to the Poincare group with 
the infinitesimal parameters depending on the space-time point,
one has to introduce two compensating fields: vierbein $e_\al^a$ 
and some independent spinor connection $W^{bc}_\mu$  
\cite{utiama,kibble,hehl,hehl-review} (see also \cite{kubo,book}
and \cite{tse82,niew,ivansard,pobaob,GSW} for alternative
considerations). In this way, one naturally arrives at the gauge 
approach to gravity. One of the important applications of this 
approach is the natural and compact formulation of simple 
supergravity \cite{dezu}, where the supersymmetric generalization 
of the Einstein-Cartan theory emerges. 

The gauge approach, exactly as the one of the section 2.1,
does not provide any reasonable restrictions on $W^{bc}_\mu$, and 
one has to introduce (or not) these restrictions additionally. 
In this paper we suppose that the covariant derivative possess
metricity $\na_\al e^\mu_a=0$. Then $W^{bc}_\mu$ becomes 
${\tilde w}^{bc}_\mu$ - spinor connection with torsion. 
The interesting question to 
answer is whether the description of the gravity with torsion 
in terms of the variables ($e_\al^a,\,{\tilde w}^{bc}_\mu$) 
is equivalent to the description in terms of the variables
($g_{\mu\nu},\,T^\al_{\,\cdot\,\be\ga}$). 

One can make the following observation. The first set 
($e_\al^a,\,{\tilde w}^{bc}_\mu$) corresponds to the first order 
formalism, while the second set ($g_{\mu\nu},\,T^\al_{\,\cdot\,\be\ga}$)
to the second order formalism. The origin of this is that in the
last case the non-torsional part of the affine connection 
is a function of the metric, while, within the gauge approach, 
the variables ($e_\al^a,\,{\tilde w}^{bc}_\mu$) 
are mutually independent {\large\it completely}. 
One has to notice that, in gravity, 
the equivalence between the first order and second order 
formalism is a subtle matter. The simplest situation is the following. 
One takes the action
\beq
S_{1} = \int d^4x\sqrt{-g}\,g^{\mu\nu}\,R_{\mu\nu}(\Ga)\,, 
\label{1st}
\eeq
where $R_{\mu\nu}(\Ga)=\pa_\la\Ga^\la_{\mu\nu}-
\pa_\mu\Ga^\la_{\nu\la} + \Ga^\la_{\mu\nu}\Ga^\tau_{\la\tau}-
\Ga^\la_{\mu\tau}\Ga^\tau_{\la\nu}$ depends on the 
connection $\Ga^\la_{\mu\nu}$ which is independent 
on the metric $g_{\mu\nu}$. Then the equations for these two 
fields 
$$
\frac{\de S_{1}}{\de g_{\mu\nu}}=0
\,,\,\,\,\,\,\,\,\,\,\,\,\,\,\,\,\,\,\,
\frac{\de S_{1}}{\de \Ga^\la_{\mu\nu}}=0
$$
lead to the conventional Einstein equations and also to the 
standard expression for the affine connection (\ref{1.5}).
In this case the first order formalism is  
equivalent to the usual second order formalism.
However, this is not true if one chooses some
other action for gravity. For instance, introducing higher
derivative terms or adding to (\ref{1st}) additional terms 
depending on the non-metricity, one can indeed lose classical 
equivalence between two 
formalisms\footnote{We remark that on the quantum level there
is no equivalence even for the (\ref{1st}) action 
\cite{diplom}.}. For the general action, the only possibility to 
link the connection with the metric is to impose the 
metricity condition. 

Let us come back to our case of gravity with torsion.
From the consideration above it is clear that the descriptions
in terms of the variables ($e_\al^a,\,{\tilde w}^{bc}_\mu$) 
and ($g_{\mu\nu},\,T^\al_{\,\cdot\,\be\ga}$) can not be
equivalent unlike we work with the Einstein-Cartan action
(\ref{1st}).
But, the non-equivalence comes only from the usual difference 
between first and second order formalisms, and has nothing 
to do with torsion. In order to have 
comparable situations, we have to replace the first set by 
($e_\al^a,\,\De w^{bc}_\mu$), where 
$\De w^{bc}_\mu = {\tilde w}^{bc}_\mu - w^{bc}_\mu$ and 
$w^{bc}_\mu$ depends on the vierbein through (\ref{spin-connect}). 
The equivalence between the sets 
($g_{\mu\nu},\,T^\al_{\,\cdot\,\be\ga}$) and ($e_\al^a,\,\De w^{bc}_\mu$)
really takes place and this
can be checked explicitly. First of all, one has to establish 
the equivalence (invertible relation) between metric and vierbein. 
This can be achieved by deriving
\beq 
\frac{\de g_{\mu\nu}}{\de e_\al^a} = 2\,\de^\al_{(\mu}\,e_{\nu ) a}
\,\,\,\,\,\,\,\,\,\,\,\,{\rm and}\,\,\,\,\,\,\,\,\,\,\,\,\,
\frac{\de e_\be^b} {\de g_{\mu\nu}} = \frac12\,e^{b (\mu}\,\de^{\nu )}_\be\,.
\label{metric-vierbein}
\eeq
In a similar fashion, one can calculate the derivatives 
\beq 
\frac{\de \De w_\mu^{ab}}{\de T_{\la\rho\si}} 
= \frac12\,e^{\la [b}\,e^{a]\,[\rho}\,e^{\si ]}_\mu 
+\frac14\,\de^\la_\mu\,e^{a\,[\rho}\,e^{\si]\,b}
\,\,\,\,\,\,\,\,\,\,\,{\rm and}\,\,\,\,\,\,\,\,\,\,\,
\frac{\de T_{\la\rho\si}}{\de \De w_\mu^{ab}}
= 4 \,e_{\la [a}\,e_{b][\si}\,\de^\mu_{\rho ]} \,.
\label{torsion-connect}
\eeq
Thus, the transformation from one set of variables 
($g_{\mu\nu},\,T^\al_{\,\cdot\,\be\ga}$) to another one
($e_\al^a,\,\De w^{bc}_\mu$)
is non-degenerate and two (second order in the torsion-independent part)
descriptions are equivalent. If some statement about torsion is true for
the ($g_{\mu\nu},\,T^\al_{\,\cdot\,\be\ga}$) variables, 
it is also true for the ($e_\al^a,\,\De w^{bc}_\mu$) variables, and 
{\large\it v.v.} 

Now, since we established the equivalence between different 
variables, we can try to formulate some torsion theories 
more general than the one based on the Einstein-Cartan action.  
On the classical level the consideration can be based only 
on the general covariance and other symmetries.
Let us restrict ourselves to the local actions only. 

As we have already learned, there are three possible
descriptions of gravity with torsion: 
\vskip 1mm
a) In terms of the metric $g_{\mu\nu}$, torsion 
$T^\al_{\,\cdot\,\be\ga}$ and Riemann curvature 
$R^\la_{\,\cdot\,\al\be\ga}$. When useful, torsion 
tensor can be replaced by its irreducible components
$T_\al,\,S_\be,\,q^\la_{\,\cdot\,\ga\tau}$.
\vskip 1mm
b) In terms of the metric $g_{\mu\nu}$, torsion 
$T^\al_{\,\cdot\,\be\ga}$ and curvature (\ref{curva}) 
with torsion
${\tilde R}^\la_{\,\cdot\,\al\be\ga}$. The relation 
between two curvature tensors,with and without torsion, 
is given by (\ref{riemann}).
\vskip 1mm
c) In terms of the variables ($e_\al^c,\,{\tilde w}^{ab}_\mu$) 
and the corresponding curvature 
$$
R^{ab}_{\,\,\,\,\,\,\mu\nu} = \pa_\mu\,{\tilde w}^{ab}_\nu
-  \pa_\mu\,{\tilde w}^{ab}_\nu 
+ {\tilde w}^{ac}_\mu \, {\tilde w}^b_{c\nu}
 -{\tilde w}^{ac}_\nu \, {\tilde w}^b_{c\mu}\,.  
$$
\vskip 1mm
We consider the possibility $a$) as the most useful one and 
will follow it in this paper, in particular for the 
construction of the new actions. 

The invariant local action can be always
expanded into the power series in derivatives of the 
metric and torsion. It is natural to consider torsion 
to be of the same order as the affine connection, that 
is $\,T\sim \pa g$. Then, in the second order (in the 
metric derivatives) we find just those terms which were
already included into the Einstein-Cartan action (\ref{EC}),
but possibly with other coefficients. In the next
order one meets numerous possible structures of the 
mass dimension 4, which were 
analyzed in Ref. \cite{chris}. Indeed, this action,
which includes more than 100 dynamical terms, and many
surface terms, does not look attractive for 
deriving physical predictions of the theory. 
It is important that this general action, and many its 
particular cases, describe torsion dynamics. In fact, all 
interesting and physically important cases, like the 
action of vacuum for the quantized matter fields, 
second order (in $\alpha^\prime$) string effective action, 
possible candidates for the torsion action \cite{betor}
are nothing but the particular cases of the bulky action
of \cite{chris}. In what follows we consider some of the 
mentioned particular cases of \cite{chris}. 
Some other torsion actions
which will not be presented here: the general fourth derivative 
actions 
with absolutely antisymmetric torsion, with and without 
local conformal symmetry, were described in \cite{book}. 

%%%%%%%%%%%%%%%%%%%%%%%%%%%%%%%%%%%%%%%%%%%%%%%%%%%%%%%%%
\section{An example of the possible effect of 
classical torsion}
%%%%%%%%%%%%%%%%%%%%%%%%%%%%%%%%%%%%%%%%%%%%%%%%%%%%%%%%%

There is an extensive bibliography on different aspects 
of classical gravity with torsion. We are not going to review 
these publications here, just because our main target is the
quantum theory. However, it is worthwhile to present 
a short general remark and consider a simple but interesting 
example.

The expression ``classical action of torsion'' can be used only 
in some special sense. In the Einstein-Cartan theory, with or 
without matter, torsion does not have dynamics and therefore 
can only lead to the contact interaction between spins. On the 
other hand, the spin of the particle is essentially quantum 
characteristic. Therefore, the classical torsion can be 
understood only as the result of a semi-classical approximation 
in some quantum theory. Now, without going into the details, 
let us suppose that such an approximation can be
done and consider its possible effects. The most natural
possibility is the application to early cosmology, which has
been studied long ago (see, for example, discussion in 
\cite{hehl}). Here, we are going to consider this issue in 
a very simple manner. 

One can suppose that in the Early Universe, due to quantum
effects of matter, the average spin (axial) current is nonzero. 
Let us demonstrate that this might lead to a non-singular 
cosmological solution. For simplicity, we suppose that torsion is
completely antisymmetric and that there is a 
conformally constant spinor current
\beq
J^\mu \,=\, <{\bar \psi}\ga^5\ga^\mu\psi> \,. 
\label{global-spin-cur}
\eeq
The Einstein-Cartan action
(\ref{EC}), with this additional current is \cite{bonsh}
\footnote{The one-loop quantum calculations in this model, 
and the construction of the on-shell 
renormalization group, has been performed in \cite{bonsh}.}
\beq
S = \int d^4x\sqrt{-g}\,\left[\,-\frac{1}{16\pi G}\,(
\,R+\th\,S_\mu\, S^\mu\,) + S_\mu\,J^\mu\,\right]\,.
\label{global}
\eeq
We have included an arbitrary coefficient $\th$ into the 
Einstein-Cartan action, but it could be equally well included 
into the definition of the global current (\ref{global-spin-cur}). 
It is worth mentioning that at the quantum level the introduction 
of such coefficient is justified. Since torsion does not
have its own dynamics, on shell it is simply expressed through
the current
\beq
S^\mu = \frac{8\pi G}{\th}\,J^\mu\,.
\label{global-torsion}
\eeq
Replacing (\ref{global-torsion}) back into the action
(\ref{global}) we arrive at the expression 
\beq
S = \int d^4x\sqrt{-g}\,\left[\,-\frac{1}{16\pi G}\,R
+ \frac{4\pi G}{\th}\,J_\mu\,J^\mu\,\right]\,,
\label{new-global}
\eeq
which resembles the Einstein-Hilbert action with the cosmological
constant. However, the analogy is incomplete, because the
square of the current $\,J^\mu\,$ has conformal properties
different from the ones of the cosmological constant. 
Consider, for the sake of simplicity, the conformally flat metric
$$
g_{\mu\nu} = \eta_{\mu\nu}\cdot a^2(\eta)\,,
$$
where $\,\eta\,$ is the conformal time. According to
(\ref{global-spin-cur}), the current $\,J^\mu\,$ has to be
replaced by $\,J^\mu = a^{-4}(\eta)\cdot {\bar J}^\mu$, where
$\,{\bar J}^\mu\,$ is constant. Denoting
$$
\frac{32}{3}\,\,\frac{\pi G^2}{\th}\,\,
\eta_{\mu\nu}\,{\bar J}^\mu \,{\bar J}^\nu = K = const\,, 
$$
we arrive at the action and the corresponding equation 
of motion for $a(\eta)$:
\beq
S \,=\, -\, \frac{3}{8\pi G}\,
\int d\eta \int d^3x \,\left[\,(\na a)^2 - \frac{K}{a^2}
\,\right]\,\,\,;\,\,\,\,\,\,\,\,\,\,\,\,\,\,\,\,\,\, 
\,\,\,\,\,\,\,\,\,\,\,\,\,\,\,\,\,\,\,\,\,
\frac{d^2a}{d\eta^2}=\frac{K}{a^3}\,.
\label{vse-global}
\eeq
This equation can be rewritten in terms of physical time $\,t\,$
(where, as usual, $\,a(\eta)d\eta=dt$) as
$$
a^2{\ddot a} + a {\dot a}^2 = Ka^{-3}\,.
$$
After the standard reduction of order, the integral solving this
equation is written in the form
\beq
\int \frac{a^2\,da}{\sqrt{Ca^2-K}} \,=\, t-t_0\,,
\label{intergral-global}
\eeq
where $\,C\,$ is the integration constant. The last integral
has different solutions depending on the signs of $K$ and $C$.
Consider all the possibilities:  
\vskip 2mm

\noindent
1) The global spinor current is time-like and $K>0$. 
Then, the eq. (\ref{intergral-global}) shows that: 
i) C is positive, and ii) $a(t)$ has
minimal value $\,a_0=\sqrt{K/C} > 0$. Thus, the presence of the 
global time-like
spinor current, in the Einstein-Cartan theory, prevents the
singularity. Indeed, since such a global spinor current can
appear only as a result of some quantum effects, one can consider
this as an example of quantum elimination of the Big Bang
singularity. The singularity is prevented, in this example, at
the scale comparable to the Planck one. This is indeed natural,
since the dimensional unity in the theory is the Newton constant. 
The dimensional considerations \cite{hehl} show that in the
Einstein-Cartan theory the effects of torsion become relevant
only at the Planck scale. Finally, the explicit solution of the 
equation (\ref{intergral-global}) has the form
\beq
arccosh\left(\sqrt{\frac{C}{K}}\,a\right)
+ a\,\sqrt{a^2 - \frac{K}{C}}
=\frac{2C^{3/2}}{K}\,(t-t_0)\,,
\label{otvet-global}
\eeq
where $\,C\,$ is an integration constant. The value of $\,C\,$ 
can be easily related to the minimal possible value of $a$. 
In the long-time limit we meet the asymptotic behaviour 
$\,a \sim t^{2/3}$. The importance of torsion,
in the Einstein-Cartan theory, is seen only at small 
distances and times and for the scale factor comparable to 
$a_0=\sqrt{K/C}$. At this scale torsion prevents singularity
and provides the cosmological solution with bounce.
\vskip 2mm

\noindent
2) The spinor current is space-like and $K<0$. 
Then, for any value of $C$, there are singularities. 
In the case of positive $C$ the solution is
\beq
\frac{a}{C}\,\sqrt{1+\frac{C}{|K|}a^2}
- \frac{|K|^{1/2}}{C^{3/2}}\,{\rm ln}
\left[\,\sqrt{\frac{C}{|K|}}\,a
+ \sqrt{1+\frac{C}{|K|}\,a^2}\,\right]  
= 2\,(t-t_0)\,,
\label{otvet-global2}
\eeq
while in case of negative $C$ the the solution is
\beq
- \frac{a}{|C|}\,\sqrt{1-\left|\frac{C}{K}\right|\,a^2}
+ \left|\frac{K}{C^3}\right|^{1/2}\,{\rm arcsin}\,
\left(\left|\frac{C}{K}\right|^{1/2}\,a\right)= 2\,(t-t_0)\,,
\label{otvet-global3}
\eeq
and for $C=0$ it is the simplest one 
$$
a(t) = \Big[\,3\,|K|\,(t-t_0)\,\Big]^{1/3} \sim t^{1/3}\,.
%%%%  \label{c=0}
$$
\vskip 2mm

\noindent
3) The last case is when the spinor current is light-like and 
$\,K=0$. Then, $\,C>0\,$ and the solution is 
\beq
a(t) = \Big[\,2\,\sqrt{C}\,(t-t_0)\,\Big]^{1/2} \sim t^{1/2}\,.
\label{K=0}
\eeq
This is, of course, exactly the same solution as one meets 
in the theory without torsion. Light-like spin vector 
decouples from the conformal factor of the metric. 
\vskip 1mm

The above solutions are, up to our knowledge, new 
(see, however, Ref.'s \cite{s1,s2,s3} where other,
similar, non-singular solutions were obtained) and
may have some interest for cosmology. 
We notice that the second-derivative inflationary models 
with torsion have attracted some interest recently,
in particular they were used for the analysis of 
the cosmic perturbations \cite{palle,garcia}.

\vskip 10mm
%%% \newpage
%%%%%%%%%%%%%%%%%%%%%%%%%%%%%%%%%%%%%%%%%%%%%%%%%%%%%%%%%%%%
%%%%%%%%%%%%%%%%%%%%%%%%%%%%%%%%%%%%%%%%%%%%%%%%%%%%%%%%%%%%
\chapter{Renormalization and anomalies in curved space-time 
with torsion.}
%%%%%%%%%%%%%%%%%%%%%%%%%%%%%%%%%%%%%%%%%%%%%%%%%%%%%%%%%%%%

The classical theory of torsion, which has been reviewed 
in the previous Chapter, is not really consistent, unless 
quantum corrections are taken into account. 
The consistency of a quantum theory usually 
includes such requirements as unitarity, renormalizability
and the conservation of fundamental symmetries on the 
quantum level. In many cases, these requirements help to 
restrict the form of the classical theories, and thus improve 
their predictive power, even in the classical framework. 

The condition of unitarity is relevant for the  
propagating torsion. But, for the
study of the quantum theory of matter on classical 
curved background with torsion, it is useless.
Therefore we have to start by formulating the 
renormalizable quantum field theory of the matter fields 
on curved background with torsion and related issues
like anomalies.

There is an extensive literature devoted to the quantum 
field theory in curved space-time (see, for example,  books
\cite{birdav,grmamo,fulling,book} and references therein). 
In the book \cite{book} the quantum theory on curved 
background with torsion has been also considered. We shall 
rely on the formalism developed in 
\cite{bush1,thesis,bush-Izv,sh-alexeev,bosh-pot,buodsh,book,anhesh},
and consider some additional applications
\footnote{In part,
we repeat here the content of Chapter 4 of \cite{book}, but some
essential portion of information was not known at the time when 
\cite{book} was written, or has not been included into that 
edition.}.

%%%%%%%%%%%%%%%%%%%%%%%%%%%%%%%%%%%%%%%%%%%%%%%%%%%%%%
\section{General description of renormalizable theory}
%%%%%%%%%%%%%%%%%%%%%%%%%%%%%%%%%%%%%%%%%%%%%%%%%%%%%%

Let us start out with some gauge theory (some version of 
SM or GUT) which is
renormalizable in flat space-time, and describe its 
generalization for the curved background with torsion. 
The theory includes spinor, vector and scalar fields 
linked by gauge, Yukawa and 4-scalar interactions, and is 
characterized by gauge invariance and maybe by some other 
symmetries. It is useful to introduce, from the very beginning, 
the non-minimal interactions between matter fields and 
torsion. One can notice, that the terms 
describing the matter self-interaction 
have dimensionless couplings and hence they can not 
(according to our intention not to introduce the inverse-mass 
dimension parameters), be affected by torsion. Thus, the  
general action can be presented in the form \cite{bush1}:
$$
S = \int d^4x\sqrt{g}\,
\left\{ -\frac14\,\left(G_{\mu\nu}^a\right)^2 
+\frac12\,g^{\mu\nu}\,{\cal D}_\mu\phi\,{\cal D}_\nu\phi 
+\frac12\,\left(\sum \xi_i\,P_i + M^2\right) \phi^2 
- V_{int}(\phi) +
\right.
$$
\beq
\left.
+i{\bar \psi} \left(\ga^\al \,{\cal D}_\al +\sum \eta_j\,Q_j - 
im + h\phi \right)\psi \, \right\}\,\, + \,\, S_{vac}\,,
\label{GUT}
\eeq
where ${\cal D}$ denotes derivatives which are covariant 
with respect to both gravitational and gauge fields but do not 
contain torsion. $\xi_iP_i$ and $\eta_jQ_j$ are non-minimal 
terms described in section 2.3.
The last term in (\ref{GUT}) represents the 
vacuum action, which is a necessary element of the renormalizable 
theory. We shall discuss it below, especially in the next section.

The action of a renormalizable theory must include all the terms 
that can show up as counterterms. So, let us investigate 
which kind of counterterms one can meet in the matter fields 
sector of the theory with torsion. We shall consider 
both general non-minimal theory (\ref{GUT}) and its particular 
minimal version. Some remark is in order. The general 
consideration of renormalization in curved space-time, based 
on the BRST symmetry, has been performed in \cite{tmf,book}. 
The generalization to the theory with 
torsion is straightforward and it is not worth to present it 
here. Instead, we are going to discuss the renormalization 
in a more simple form, using 
the language of Feynman diagrams, and also will refer to the
general statements about the renormalization of the 
gauge theories in presence of the background fields
\cite{dewitt,aref,kallosh}.

The generating functional of the Green functions, in the 
curved space-time with torsion, can be postulated in the form:
\beq
Z[J,\,g_{\mu\nu},\,T^\al_{\cdot\,\be\ga}]
\,=\, {\cal N}\,\int d\Phi\,
{\rm exp}\,\left\{\,i\,S[\Phi,g,T] \,+\, i\phi\,J
\right\}\,,
\label{Z}
\eeq
where $\Phi$ denotes all the matter (non-gravitational) fields 
$\phi\,\,$ (with spins $\,0,\,1/2,\,1\,$) and the Faddeev-Popov ghosts 
$\,c,{\bar c}$. $J$ are the external sources for the matter 
fields $\,\phi$. In the last term, in the exponential, we are using 
condensed (DeWitt) notations.  ${\cal N}=Z^{-1}[J=0]\,$ 
is the normalization factor.

Besides the source term, (\ref{Z}) depends 
on the external fields $g_{\mu\nu}$ and $T^\al_{\cdot\,\be\ga}$. 
One has to define how to modify the 
perturbation theory in flat space-time so that it 
incorporates the external fields. The corresponding 
procedure is similar to that for the purely metric background.
One has to consider the metric as a sum of 
$\,\eta_{\mu\nu}\,$ and of the perturbation $h_{\mu\nu}$ 
$$
g_{\mu\nu}=\eta_{\mu\nu}+h_{\mu\nu}\,.
$$ 
Then, we 
expand the action $S[\Phi,g,T]$ such that the propagators 
and vertices of all the fields (quantum and background) are 
the usual ones in the flat space-time. The internal lines 
of all the diagrams are only those of the matter fields, 
while external lines 
are both of matter and background gravitational 
fields (metric $\,h_{\mu\nu}\,$ and torsion). As a result,
any flat-space digram gives rise to the infinite set of  
diagrams, with increasing number of the background fields
tails. An example of such set is depicted at Fig. 1. 
%%%%%%%%%%%%%%%%%%%%%%%%%%%%%%%%%%%%%%%%%%%%%%%%%%%%%%%%
%%%% \newpage

\vskip 1mm
\noindent
%%%%%%%%%%%%%%%%%%%%%%%%%%%%%%%
\begin{picture}(120,120)(0,0)
\Line(0,25)(25,25)\BCirc(50,25){25}\Line(75,25)(100,25) 
\Vertex(75,25){3} \Vertex(25,25){3} 
\Text(110,25)[c]{$\to$}
\end{picture}
%%%%%%%%%%%%%%%%%%%%%%%%%%%%%%%
\begin{picture}(120,120)(0,0)
\Line(0,25)(25,25)\BCirc(50,25){25}\Line(75,25)(100,25) 
\Vertex(75,25){3} \Vertex(25,25){3} 
\Text(110,25)[c]{$+$}
\end{picture}
%%%%%%%%%%%%%%%%%%%%%%%%%%%%%%%
\begin{picture}(120,120)(0,0)
\Line(0,25)(25,25)\BCirc(50,25){25}\Line(75,25)(100,25) 
\Vertex(75,25){3} \Vertex(25,25){3} 
\Photon(50,50)(25,90){2}{8}\Photon(50,50)(40,100){2}{8}
\Photon(50,50)(75,90){2}{8}
\DashCArc(50,50)(25,70,95){2}\Vertex(50,50){3}
\Text(110,25)[c]{$+$}
\end{picture}
%%%%%%%%%%%%%%%%%%%%%%%%%%%%%%%
\\
%%%%%%%%%%%%%%%%%%%%%%%%%%%%%%%
\begin{picture}(120,120)(0,0)
\Text(-10,25)[c]{$+$}
\Line(0,25)(25,25)\BCirc(50,25){25}\Line(75,25)(100,25) 
\Vertex(75,25){3} \Vertex(25,25){3} 
\Photon(25,25)(-10,55){2}{8}\Photon(25,25)(0,65){2}{8}
\Photon(25,25)(15,75){2}{8}
\DashCArc(25,25)(25,100,125){2}\Vertex(25,25){3}
\end{picture}
%%%%%%%%%%%%%%%%%%%%%%%%%%%%%%%
%% \\
%%%%%%%%%%%%%%%%%%%%%%%%%%%%%%%
\begin{picture}(120,120)(0,0)
\Text(-10,25)[c]{$+$}
\Line(0,25)(25,25)\BCirc(50,25){25}\Line(75,25)(100,25) 
\Vertex(75,25){3} \Vertex(25,25){3} 
\Photon(25,25)(-10,55){2}{8}
\Photon(25,25)(0,65){2}{8}\Photon(25,25)(15,75){2}{8}
\DashCArc(25,25)(25,100,125){2}\Vertex(25,25){3}
\Photon(50,50)(25,90){2}{8}\Photon(50,50)(40,100){2}{8}\Photon(50,50)(75,90){2}{8}
\DashCArc(50,50)(25,70,95){2}\Vertex(50,50){3}
\end{picture}
%%%%%%%%%%%%%%%%%%%%%%%%%%%%%%%
\begin{picture}(120,120)(0,0)
\Text(-10,25)[c]{$+$}
\Line(0,25)(25,25)\BCirc(50,25){25}\Line(75,25)(100,25) 
\Vertex(75,25){3} \Vertex(25,25){3} 
\Photon(50,50)(25,90){2}{8}\Photon(50,50)(40,100){2}{8}
\Photon(50,50)(75,90){2}{8}
\DashCArc(50,50)(25,70,95){2}\Vertex(50,50){3}
\Photon(75,25)(80,65){2}{8}\Photon(75,25)(90,65){2}{8}
\Photon(75,25)(110,60){2}{8}
\DashCArc(75,25)(25,45,70){2}\Vertex(75,25){3}
\Text(110,25)[c]{$\,\,\,+\,\,\,\,\,\,...$}
\end{picture}
%%%%%%%%%%%%%%%%%%%%%%%%%%%%%%%
\vskip 3mm

\noindent
{\small\it Figure 1. $\,\,$ 
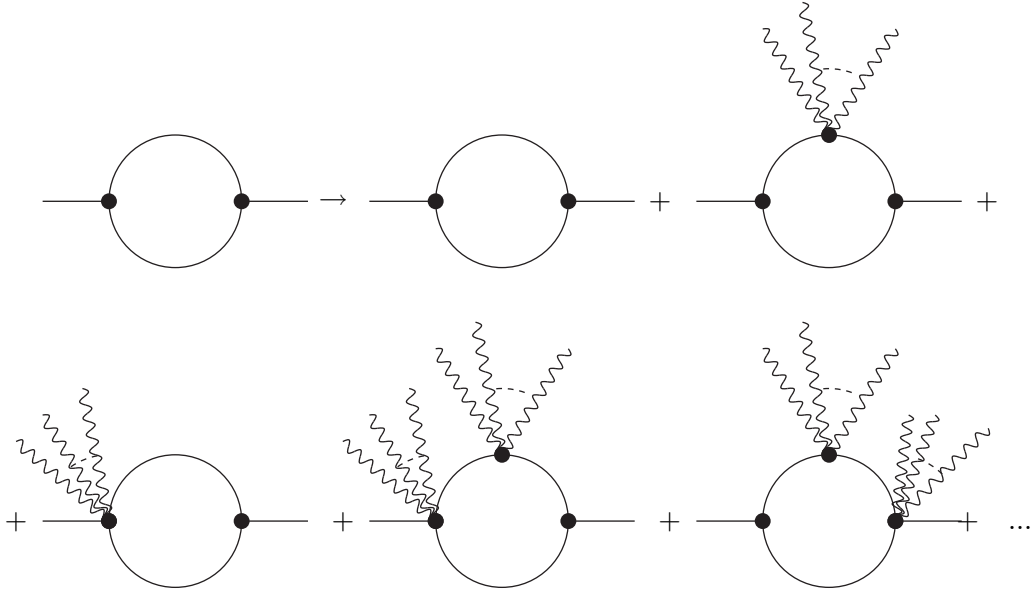
The straight lines correspond to the matter (in this case scalar with 
$\,\lambda\varphi^3\,$ 
interaction) field, and wavy lines to the external 
field (in this case metric). A single diagram in flat space-time
generates an infinite set of families of diagrams in curved space-time.
The first of these generated diagrams is exactly the one in the
flat space-time, and the rest have external gravity lines.}
\vskip 1mm

Let us now remind three relevant facts. 
%%%%%%% *********
{\large\it First}, 
when the number of 
vertices increases, the superficial degree of divergence
for the given diagram may only decrease. Therefore, the 
insertion of new vertices of interaction with the background 
fields 
$\,\,g_{\mu\nu},\,T^\al_{\cdot\,\be\ga}$ can not increase 
the degree
of divergence. In other words, for any flat-space diagram, 
all generated diagrams with gravitational external tails 
have the same or smaller index of divergence than the original 
diagram. 
%%%%%%% *********
{\large\it Second}, since we are working with the 
renormalizable theory, the number of the divergent $n$-loop 
diagrams, in flat space-time, is finite. 
As a result, after generating the diagrams with external 
gravity (metric and torsion) tails, we meet a finite number 
of the {\it families} of divergent diagrams at any loop order. 
Furthermore, 
including an extra vertex of interaction with 
external field one can convert the quadratically divergent 
diagram into a logarithmically divergent one. For example, 
the quadratically divergent diagram of Fig. {\it 2a} generates
the logarithmically divergent ones of Fig. {\it 2b}. The
diagrams from the  Fig. {\it 2b} give rise to the 
$R\ph^2$-type counterterm. Similarly, the diagrams 
of Fig. 2c and Fig. 2d produce
 $\,{\bar \psi}\ga^5\ga^\mu S_\mu \psi\,$ and $\,\ph^2S^2\,$-type
counterterms.

\vskip 3mm
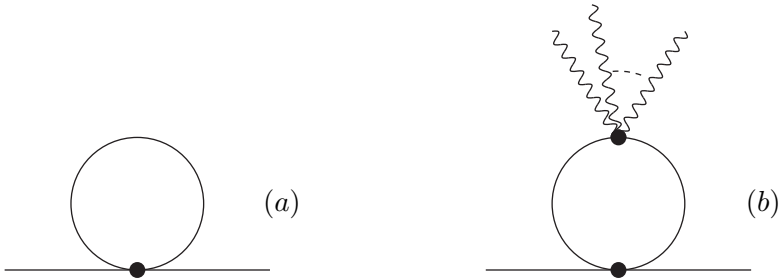
\noindent
%%%%%%%%%%%%%%%%%%%%%%%%%%%%%%%
\begin{picture}(120,120)(0,0)
\Line(0,0)(100,0) \BCirc(50,25){25} \Vertex(50,0){3}
\Text(105,25)[c]{$(a)$}
\end{picture}
$\,\,\,\,\,\,\,\,\,\,\,\,\,\,$
$\,\,\,\,\,\,\,\,\,\,\,\,\,\,$
%%%%%%%%%%%%%%%%%%%%%%%%%%%%%%%
\begin{picture}(120,120)(0,0)
\Photon(50,50)(25,90){2}{8}\Photon(50,50)(40,100){2}{8}
\Photon(50,50)(75,90){2}{8}
\DashCArc(50,50)(25,70,95){2}\Vertex(50,50){3}
\Line(0,0)(100,0) \BCirc(50,25){25} 
\Vertex(50,0){3} \Vertex(50,50){3}
\Text(105,25)[c]{$(b)$}
\end{picture}
%%%%%%%%%%%%%%%%%%%%%%%%%%%%%%%
\\
%%%%%%%%%%%%%%%%%%%%%%%%%%%%%%%
\begin{picture}(120,120)(0,0)
\DashLine(0,0)(100,0){4} \CArc(50,0)(25,0,180) 
\Vertex(25,0){3}\Vertex(75,0){3}\Vertex(50,0){3}
\Line(48,0)(48,-50)\Line(48,-50)(52,-50)\Line(52,0)(52,-50)
\Text(105,25)[c]{$(c)$}
\end{picture}
$\,\,\,\,\,\,\,\,\,\,\,\,\,\,$ $\,\,\,\,\,\,\,\,\,\,\,\,\,\,$
%%%%%%%%%%%%%%%%%%%%%%%%%%%%%%%
\begin{picture}(120,120)(0,0)
\Line(0,0)(25,0)\DashCArc(50,0)(25,0,360){3}\Line(75,0)(100,0) 
\Line(48,-25)(48,-75)\Line(48,-75)(52,-75)\Line(52,-25)(52,-75)
\Line(48,25)(48,75)\Line(48,75)(52,75)\Line(52,25)(52,75)
\Vertex(25,0){3}\Vertex(75,0){3}\Vertex(50,-25){3}\Vertex(50,25){3}
\Text(105,25)[c]{$(d)$}
\end{picture}
%%%%%%%%%%%%%%%%%%%%%%%%%%%%%%%
\vskip 30mm
%% \newpage

\noindent
{\small\it Figure 2}. 
{\small\it 
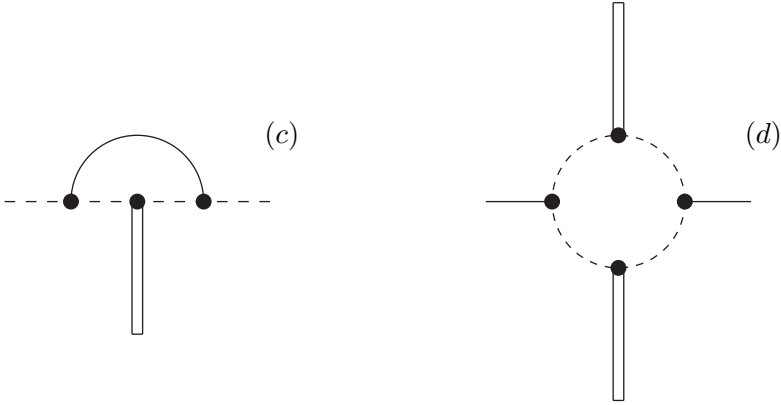
(a) Quadratically divergent graf for the $\,\lambda\varphi^4$-theory.}
\\
{\small\it (b)
The example of logarithmically divergent graf generated by 
the graf at Fig. 2a and the procedure presented at Fig. 1.
This diagram contributes to the $\,\,R\varphi^2 $-type counterterm.}
\\
{\small\it (c)
Dashed lines represent spinor, continuous lines represent
scalar and double line -- external torsion $S_\mu$. This 
diagram gives rise to the 
${\bar \psi}\,\gamma^5\gamma^\mu S_\mu\psi$-type counterterm.}
\\
{\small\it (d) 
This diagram produces $\,\,\varphi^2 S^2$-type counterterm.}
\vskip 3mm

%%%%%%% *********
{\large\it Third}, there
are general proofs \cite{volatyu} that the divergences of a 
gauge invariant theory can be removed, at any loop, by the 
gauge invariant and local counterterms\footnote{In 
case of the diffeomorphism invariance, this
can be also proved using the existence of the explicitly 
covariant perturbation technique based on the local momentum 
representation and Riemann normal coordinates. 
For example, in \cite{parker-toms} this technique has been 
described in details and applied to the extensive one-loop 
calculations. In the case of gravity with torsion, the local 
momentum representation have been used in Ref. \cite{cogzer}.}.
Indeed these 
theorems apply only in the situation when there is no anomaly. 
In the present case we have regularizations (say, dimensional 
\cite{hooft,leibr}, or properly used higher derivative 
\cite{slavnov, asorey}) which preserve, on the quantum level,
both general covariance and gauge invariance of the model. 
Thus, we are in a position to use general covariance and 
gauge invariance for the analysis of the counterterms.
Anomalies do not threaten these symmetries, for in the 
four-dimensional space-time there are no gravitational 
anomalies. 

Taking all three points into account, we arrive at the following 
conclusion. The counterterms of the theory in an external 
gravitational background with torsion have the same dimension 
as the counterterms for the corresponding theory in flat 
space-time. These counterterms possess general covariance and
gauge invariance, which are the most important symmetries of
the classical action. 

At this stage one can
explain why the introduction of the non-minimal interaction 
between torsion and matter (spin-1/2 and spin-0 fields) is so 
important. The reason is that the appearance of the
non-minimal 
counterterms is possible, for they have proper symmetries 
and proper dimensions. Let us imagine that we have started 
from the
minimal theory, that is take $\,\eta_1=\frac18,\,\eta_2=0\,$
and $\,\xi_{1,2,3,4,5}=0.$ Then, the classical action depends on 
the metric
$g_{\mu\nu}$ and on the axial vector component $S_\mu$ of torsion.
Thus, the vertices of interaction with these two fields will 
modify the diagrams and one can expect that the counterterms
depending on $g_{\mu\nu}$ and $S_\mu$ will appear. According 
to our analysis these counterterms should be of three possible
forms (see Figures 2b -- 2d): 
$$
\int d^4x\,\sqrt{-g}\,S_\al\,S^\al\,\ph^2
\,,\,\,\,\,\,\,\,\,\,\,\,\,
\int d^4x\,\sqrt{-g}\,R\ph^2
\,,\,\,\,\,\,\,\,\,\,\,\,\,
\int d^4x\,\sqrt{-g}\,{\bar \psi}\,\ga^5\ga^\al S_\al\,\psi\,,
$$
and therefore these three structures should be included into 
the classical action in order to provide renormalizability. 
Therefore, the essential nonminimal interactions with 
torsion are the
ones which contain the torsion pseudotrace $\,S_{\mu}$.
If the space-time possesses torsion, the non-minimal parameters 
$\eta_1$ and $\xi_4$ have the same status as the $\xi_1$
parameter 
has for the torsionless theory. Of course, $\xi_1$ remains
to be essential -- independent of whether torsion is present. 

The special role of the two parameters $\,\eta_1,\,\xi_4\,$, 
as compared to others: $\eta_2$ and $\xi_{2,3,5}$, 
is due to the fact that minimally only $\,S_{\mu}$-component 
of torsion interacts with matter fields. It is remarkable
that not only spinors but also scalars have to interact 
with torsion if we are going to have a renormalizable 
theory. 

The terms which describe interaction 
of matter fields with $T_\mu$ and $q^\al_{\,\cdot\,\be\ga}$ 
components of torsion, can be characterized as purely 
non-minimal. One can put parameters 
$\,\xi_{2,3,4},\,\eta_2$ to be zero simultaneously without
jeopardizing the renormalizability. Indeed, if the 
$\,\eta_2$-term is included, it is necessary to introduce 
also the $\,\xi_{2,3}$-type terms. In the case of abelian 
gauge theory with complex scalars one may need to 
introduce some extra non-minimal terms \cite{bush2}
(see also sections 2.3 and 3.3). 

Besides the non-minimal terms, one can meet 
the vacuum structures which satisfy the conditions 
of dimension and general covariance. The action of 
vacuum depends exclusively on the gravitational fields 
$\,\,g_{\mu\nu}\,$ and $\,T^\al_{\cdot\,\be\ga}$.
Hence, the corresponding counterterms result from the
diagrams which have only the external tails of 
these fields. The most general form of the vacuum action 
for gravity with torsion has been constructed in \cite{chris}. 
This action satisfies the conditions of covariance and 
dimension, but it is very bulky for it contains 168 terms 
constructed from curvature, torsion and their derivatives. 
Using the torsionless curvature, one can distinguish 
the terms of the types 
$$
R_{...}^2 \,,\,\,\,\,\,\,\,\,\,\,\, R_{...}T^2
\,,\,\,\,\,\,\,\,\,\,\,\,R_{...}\na T
\,,\,\,\,\,\,\,\,\,\,\,\,
T^2\na T\,,\,\,\,\,\,\,\,\,\,\,\,T^4
$$
plus total derivatives. 

It turns out, that the number of necessary 
terms can be essentially reduced without giving up the 
renormalizability. At the one-loop level, we meet just an 
algebraic sum of the closed loops of free vectors,
fermions and scalars, and only the last two kind of 
fields contribute to the torsion-dependent vacuum sector. 
Therefore, calculating closed scalar and spinor loops 
one can fix the necessary form of the classical action of 
vacuum, such that this action is sufficient for 
renormalizability but does not contain any unnecessary 
terms. Since we are considering the renormalizable 
theory, the divergent vertices in the matter field 
sector are local and have the same algebraic structure 
as the classical action. For this reason, the structures 
which do not emerge as the one-loop vacuum counterterms, will 
not show up at higher loops too. Hence, one can 
restrict the minimal necessary form of the vacuum  
action, using the one-loop calculations.

%%%%%%%%%%%%%%%%%%%%%%%%%%%%%%%%%%%%%%%%%%%%%%%%%%%%%%%%%%%%%
\section{One-loop calculations in the vacuum sector}
%%%%%%%%%%%%%%%%%%%%%%%%%%%%%%%%%%%%%%%%%%%%%%%%%%%%%%%%%%%%%

In this section we derive the one-loop 
divergences for the free matter fields in an external 
gravitational field with torsion. As we already learned 
in the previous sections, only scalar and spinor fields 
couple to torsion, so we restrict the 
consideration by these fields.

For the purpose of one-loop calculations we shall consistently 
use the Schwinger-DeWitt technique. One can find the review of 
this method, 
its generalizations and developments and the list of 
many relevant references in \cite{dewitt,bavi,vilk,avram}. 
Also, in Chapter 5, some new application of this technique will 
be given. Now we need just a simplest version of the 
Schwinger-DeWitt technique.
The one-loop contribution to the effective action 
$\,{\bar \Ga}^{(1)} = \frac{i}{2}\,{\rm Tr}\ln\,{\hat H}\,$
has the following integral representation
\beq
{\bar \Ga}^{(1)} \,=\, -\, \frac{i}{2}\,{\rm Tr}\,
\int_0^\infty\,\frac{ds}{s}\,\,
\frac{ i\,{\cal D}^{1/2}(x,x^\prime)}{(4\pi i\,s)^{n/2}}
\,\,{\rm exp}\,\left\{\,
{-ism^2+\frac{i}{2s}\,\si(x,x^\prime)}\right\}
\,\,\sum_{k=0}^{\infty}\,(is)^k{\hat a}_k(x,x^\prime)\,,
\label{a2..}
\eeq	
where $\,\si(x,x^\prime)\,$ is the world function (geodesic distance 
between two close points, $\,\si=\frac12\,\na_\mu\si\na^\mu\si\,$)
and $\,{\cal D}^{1/2}(x,x^\prime)\,$ is the Van Vleck-Morette 
determinant
$$
{\cal D}^{1/2}(x,x^\prime)\,=\,
\Big|\,{\rm det}\Big(\,
-\frac{\pa^2 \si}{\pa x^\mu\,\pa x^\nu}\,\Big)\,\Big|\,.
$$
$n$ is the parameter of the dimensional regularization.
The details about the dimensional regularization in the 
Schwinger-DeWitt technique can be found in \cite{bavi}. 

For the minimal differential operator 
\beq
{\wh H}={\wh 1}\,{\Box}\,+\,2\,{\wh h}^\la\,\na_\la\,+\,
{\wh \Pi}\,
\label{minoper}
\eeq 
acting on the fields of even Grassmann parity, 
the divergent part of the functional trace (\ref{a2..})
is a factor of the coincidence limit of the trace 
$$
\tr \lim_{x^\prime \to x}{\hat a}_2(x,x^\prime) 
$$
of the second coefficient of the Schwinger-DeWitt 
expansion. Direct calculation yields \cite{dewitt}
$$
\Ga^{(1)}_{div}({\wh H})
= \frac{i}{2}\,\Tr\ln\,\left.\Big(\,-\,\frac{{\wh H}}{\mu^2}\,
\Big)\right|_{div} 
= - \frac{\mu^{n-4}}{\vp}\,\int d^nx\sqrt{-g}\,\tr\,\left[\,
\frac{\wh 1}{180}\left(R_{\mu\nu\al\be}^2 - R_{\mu\nu}^2
+ {\Box}\,R\right) + 
\right.
$$
\beq 
\left.
+ \frac16\,{\Box}{\wh P} 
+ \frac12\,{\wh P}\cdot{\wh P} 
+ \frac{1}{12}\,{\wh S}_{\mu\nu}\cdot{\wh S}^{\mu\nu}\,\right]\,,
\label{a2}
\eeq
where $\,\vp = (4\pi)^2\,(n-4)$ is the parameter of 
dimensional regularization, $\mu$ is the dimensional parameter, 
${\wh 1}$ is the identity matrix in the space of the given fields,
$$
{\wh P} = {\wh \Pi} + \frac{\wh 1}{6}\,R - \na_\al\,{\wh h}^\al
- {\wh h}^\al\,{\wh h}_\al
$$
and
$$
{\wh S}_{\mu\nu} = 
(\na_\nu\,\na_\mu - \na_\mu\,\na_\nu)\,{\wh 1} +
\na_\nu\,{\wh h}_\mu - \na_\mu\,{\wh h}_\nu + 
{\wh h}_\nu\,{\wh h}_\mu - {\wh h}_\mu\,{\wh h}_\nu\,.
$$
One has to notice that the last formula is nothing but 
the commutator of the covariant derivatives 
$$
D_\al = \na_\al + {\wh h_\al}. 
$$
Of course, the expression (\ref{a2}) can be written 
in terms of the covariant derivative with torsion
${\tilde \na}_\al$, but it is useful to separate 
the torsion dependent terms. The last observation is that  
for the operator (\ref{minoper}) acting on the fields
of odd Grassmann parity, the expression (\ref{a2})
changes its sign. In a complicated situations with the 
operators of mixed Grassmann parity (like that 
we shall meet in Chapter 5) it is useful 
to introduce special notation $\,Str\,$ for the supertrace.  

Let us first consider the calculation of divergences 
for the especially simple case of free
scalar field. The one-loop divergences are given by  
eq. (\ref{a2}), where
$$
 {\wh H}_{sc} \,=\,-\, 
\frac12\,\frac{\de^2\,S_{0}}{\de^2\ph}
\,=\,  {\Box} - m^2 - \sum \xi_i\,P_i\,.
$$
Here we use the notation (\ref{scalar1-nm}) 
of the previous Chapter. 
Applying (\ref{a2}), one immediately obtains
$$
\Ga^{(1)}_{div}({\rm scalar}) = -\,\frac{i}{2}\,\Tr\ln\, 
\Big(\,\frac{{\wh H}_{sc}}{\mu^2} \,\Big)\,\Big|_{div} =
$$
\beq 
=\,-\,\frac{\mu^{n-4}}{\vp}\,\int d^nx\sqrt{-g}\,\left[\,
\frac{1}{180}\left(R_{\mu\nu\al\be}^2 - R_{\mu\nu}^2
+ {\Box}\,R\right) + \frac16\,{\Box}{\wh P} 
+ \frac12\,{\wh P}^2 \,\right]\,,
\label{scalardivs}
\eeq
where 
$$
{\wh P} = \frac16\,R - \sum_i\xi_i\,P_i - m^2\,.
$$
As it was already mentioned above, in order to provide 
renormalizability one has to include into the classical 
action of vacuum all the structures that can appear as 
counterterms. For the scalar field 
on the external background of gravity with torsion,
the list of the integrands of the vacuum action consists 
of $\,R_{\mu\nu\al\be}^2\,$ and  $\,R_{\mu\nu}^2\,$, 
five total derivatives $\,{\Box}P_i$, ten products 
$\,P_i\,P_j\,$ and six mass dependent terms: 
$\,m^4\,$ and $\,m^2\,P_i$. The total number of necessary
vacuum structures is 23, and 7 of them are total 
derivatives. 
This number of 23 can be compared, from one side, 
with the 6 terms 
$$
R_{\mu\nu\al\be}^2,\,\,\,R_{\mu\nu}^2,\,\,\,\,
R^2,\,\,\,\,\Box R, \,\,\,\,m^4\,\,\,\,\,m^2R
$$ 
which emerge
in the torsionless theory, and from the other side,
with the 168 algebraically possible covariant terms 
constructed from curvature, torsion 
and their derivatives \cite{chris}.

It is sometimes useful to have another basis for the 
torsionless fourth derivative terms. We shall use the 
following notations:
$$
C^2 = C_{\mu\nu\al\be}C^{\mu\nu\al\be} =
R_{\mu\nu\al\be}R^{\mu\nu\al\be} - 2 \,R_{\al\be}R^{\al\be} +
\frac13\,R^2
$$
for the square of the Weyl tensor, which is conformal invariant 
at four dimensions and
$$
E = 
R_{\mu\nu\al\be}R^{\mu\nu\al\be} - 
4 \,R_{\al\be}R^{\al\be} + R^2
$$
for the integrand of the Gauss-Bonnet topological term. 
The inverse relations have the form
\beq
R_{\mu\nu\al\be}^2 = 2C^2-E+\frac13\, R^2
\,\,\,\,\,\,\,\,\, {\rm and} \,\,\,\,\,\,\,\,\,
R_{\mu\nu}^2 = \frac12\,C^2-\frac12\,E+\frac13 R^2\,.
\label{basis}
\eeq
%%%%%%%%%%%%%%%%%%%%%%%%%%%%%%%%%%%%%%%%%%%%%%%%%%%%

Let us now consider the fermionic determinant, which 
has been studied by many authors (see, for example, 
\cite{gold,kimura,obukhov-spectr,buodsh,cogzer,gus}).
One can perform the calculation by writing the action
through the covariant derivative without torsion 
\cite{obukhov-spectr,buodsh}. So, we start 
from the general non-minimal action (\ref{diraconly}).
The divergent contribution from the single fermion 
loop is given by the expression
\beq
\Ga_{div} [g, A, S] = - i\Tr\ln {\hat H}\Big|_{div}\,,
\label{efac}
\eeq
where
$$
{\hat H} =
i\ga^\al \,\left({\cal D}_{\al} - im \right)
\,\,\,\,\,\,\,\,\,\,\,\,\,\,\,\,\,\,
{\rm and}\,\,\,\,\,\,\,\,\,\,\,\,\,\,\,
{\cal D}_\al =
\na_\al + ieA_\al + i\eta\ga^5\,S_\al
$$
is generalized covariant derivative. For the massless theory 
this covariant derivative respects general covariance, the
abelian gauge symmetry (\ref{trans1}) and the 
additional gauge symmetry (\ref{trans}). In the massive case this last
symmetry is softly broken, but the above notation is still useful.

In order to calculate functional determinant (\ref{efac}), 
we perform the transformation which preserves covariance 
with respect to the derivative ${\cal D}_\al$. First observation is 
that, by dimensional reasons, the (\ref{efac}) is even in 
the mass $m$.
In other words, (\ref{efac}) does not change if we replace $m$ 
by $-m$. It proves useful to introduce the conjugate derivative,
$D^*_\mu = \pa_\mu + ieA_\al - i \eta\ga^5 S_\mu\,$ and the
conjugated operator
$\,{\hat H}^*=i\ga^\al \,\left({\cal D}_{\al} + im \right)$. 
Then, we can perform the transformation:
$$
\Ga_{fermion} = - \frac{i}{2} \,Tr\ln\,{\hat H}\cdot{\hat H}^* =
 - \frac{i}{2} \,Tr\ln\,\{- \ga^\mu D_\mu\ga^\nu D_\nu - m^2\} =
$$
\beq
= -
\frac{i}{2} \,Tr\ln\,\{\,- (\ga^\mu\ga^\nu D^*_\mu D_\nu + m^2)\,\}\,.
\label{kvadrat}
\eeq
After a simple algebra, one can write two useful forms for the last
operator: the non-covariant (with respect to $\,{\cal D}_\al\,$):
$$
-\,{\hat H}\cdot{\hat H}^* = \na^2 + R^\mu\na_\mu + \Pi\,,
$$
with (here $\si^{\mu\nu}=e^\mu_a\,e^\mu_b\,\si^{ab}
=\frac{i}{2}\,[\ga^\mu\ga^\nu-\ga^\nu\ga^\mu]\,$) 
$$
R^\mu = 2ie\,A_{\mu} + 2\eta\,\si^{\mu\nu}\,S_{\nu}\,\ga_5\,,
$$
$$
\Pi =  ie\nabla^{\mu}A_{\mu} + i\eta\ga^5\,\pa_\mu\,S^{\mu}
-e^2\,A^{\mu}A_{\mu} 
+\frac{ie}{2}\,\ga^{\mu}\ga^{\nu}\,F_{\mu\nu}-\frac{1}{4}\,R + m^2+
$$
\beq
+\frac{i}{2}\,\eta\ga^{\mu}\ga^{\nu}\ga_5 S_{\mu\nu}
+\eta^2 \,S_{\mu}S^{\mu} + 2ie\eta\si^{\mu\nu}\ga^5\,A_{\mu}S_{\nu}\,;
\label{eta5}
\eeq
where $\,S_{\mu\nu} = \pa_\mu S_\nu - \pa_\nu S_\mu$,
$\,\,F_{\mu\nu}=\pa_\mu A_\nu - \pa_\nu A_\mu\,$ 
and covariant
\beq
-\,{\hat H}\cdot{\hat H}^* = D^2 + E^\mu D_\mu + F,
\label{eta55}
\eeq
\beq
{\rm with }\,\,\,\,\,\,\,\,\,\,\,\,\,\,
E^\mu =  -2i\,\eta\ga_\nu\ga^\mu\ga^5\,S^{\nu}
\,,\,\,\,\,\,\,\,\,\,\,\,\,\,\,
F = m^2 - \frac14\,R 
+ \frac{ie}{2}\,\ga^{\mu}\ga^{\nu}\,F_{\mu\nu}
+ \frac{i}{2}\,\eta\ga^5\,\ga^{\mu}\ga^{\nu}\,S_{\mu\nu}\,.
\label{cova}
\eeq
The intermediate expressions, for the covariant version, are 
\beq
\wh{P}=i\eta\ga^5\,\nabla_{\mu}S^{\mu} - 2\eta^2S_{\mu}S^{\mu}
+ \frac{ie}{2}\,\ga^{\mu}\ga^{\nu}\,F_{\mu\nu}-\frac{1}{12}R+m^2 
\label{PPPP}
\eeq
and 
\beq
\wh{S}_{\al\be} & = & \frac{1}{4}\,\ga^{\rho}\ga^{\la}\,R_{\be\al\rho\la}
-ieF_{\al\be}+\eta\si_{\al\nu}\ga^5\nabla_{\be}S^{\nu}
- \eta\si_{\be\nu}\ga^5\nabla_{\al}S^{\nu} + \nonumber \\
& + & (\ga_{\al}\ga_{\mu}\ga_{\be}\ga_{\nu}
-\ga_{\be}\ga_{\mu}\ga_{\al}\ga_{\nu})\,\eta^2\,S^{\mu}S^{\nu}\,. 
\label{S}
\eeq

The divergent part of (\ref{kvadrat}) can be easily evaluated 
using the general formula (\ref{a2}).
After some algebra we arrive at the following divergences:
$$
{\bar \Ga}_{div}^{(1)}({\rm Dirac \,\,spinor})  
= \frac{\mu^{n-4}}{\vp}\,\int d^n x \sqrt{-g}\;\left\{\, 
\frac{2}{3}e^2F_{\mu\nu}^2+ \frac{2}{3}\eta ^2S_{\mu\nu}^2
- 8\,m^2\,\eta^2S_{\mu}S^{\mu}
-\frac{1}{3}m^2R+2m^4+
\right.
$$
\beq
\left.
+\frac{1}{72}\,R^2- \frac{1}{45}R_{\mu\nu}^2
-\frac{7}{360}\,R_{\mu\nu\rho\la}^2
-\frac{4}{3}\,\eta^2\,\Box (S^{\mu}S_{\mu})
+\frac{4}{3}\,\eta^2\,\nabla_{\mu}(S^{\nu}\nabla_{\nu}S^{\mu}
-S^{\mu}\nabla_{\nu}S^{\nu})-\frac{1}{30}\Box R\right\}\,.
\label{contra}
\eeq
%%%%%%%%%%%%%%%%%%%%%%%%%%%%%%%%%%%%%%%%%%%%%%%%%%%%%%%%%%%%%

The form of the divergences (\ref{contra}) is related to 
the symmetry transformation (\ref{trans}). For instance, 
diffeomorphism and gauge invariance (\ref{trans1}) are preserved. 
The one-loop divergences contain the $S_{\mu\nu}^2$-term, that 
indicates that in the massless theory the symmetry (\ref{trans}) 
is also preserved. And the appearance of the massive divergent 
$\,m^2S^2\,$ term reveals that the symmetry under transformation 
(\ref{trans}) is softly broken by the fermion mass. The symmetry 
(\ref{trans}) and the form of divergences (\ref{contra}) will be 
extensively used later on, in Chapter 5, when we try to formulate 
the consistent theory for the propagating torsion. Indeed, it 
is very important that the longitudinal 
$\,\left(\pa_\nu S^\nu\right)^2$-term is absent 
in (\ref{contra}), for it would break the symmetry (\ref{trans}).

The one-loop divergences coming from the fermion sector, 
add some new necessary terms to the vacuum 
action. The terms which were not necessary for the 
scalar case, but appear from spinor loop are (remind that
$T_\mu$ is hidden inside $A_\mu$):
\beq
S_{\mu\nu}S^{\mu\nu}\,,\,\,\,\,\,
T_{\mu\nu}T^{\mu\nu}\,,\,\,\,\,\,
\na_\mu(S_\nu\na^\nu S^\mu - S^\mu\na_\nu S^\nu)\,,
\label{extravac!!!}
\eeq
where we denoted, as in Chapter 2,
 $\,\,T_{\mu\nu} = \pa_\mu T_\nu - \pa_\nu T_\mu$.
Finally, for the theory including scalars, gauge vectors
and fermions, the total number of necessary vacuum structures
is 26, and 7 of them are surface or topological terms. 
One can see, that in such theory 142 of the algebraically 
possible counterterms \cite{chris} never show up, most of 
them depend on the $\,q^\al_{\,\cdot\,\be\ga}\,$-component of 
torsion. 

%%%%%%%%%%%%%%%%%%%%%%%%%%%%%%%%%%%%%%%%%%%%%%%%%%%%%%%%%%%
\section{One-loop calculations in the matter fields 
sector.}
%%%%%%%%%%%%%%%%%%%%%%%%%%%%%%%%%%%%%%%%%%%%%%%%%%%%%%%%%%%

The one-loop calculation in the matter fields sector
in curved space-time with torsion is very important and
it has been carried out for various gauge models 
\cite{bush1,bush-Izv,bush2,njl,sh-alexeev}. 

The first such calculation was done in 
\cite{bush1,bush-Izv} for several $SU(2)$ models. The
renormalization of the same models in flat space-time has 
been studied earlier in \cite{votyu}, where they were taken
as examples of theories with the asymptotic freedom 
in all (gauge, Yukawa and scalar) couplings taking place
on the special solutions of the renormalization group 
equations. Later on, the same models have been used in 
\cite{buod} for the first calculation of the one-loop 
divergences and the study of the renormalization group for 
the complete gauge theory in an external gravitational field. 
The one-loop renormalization in the torsion-dependent
non-minimal sector \cite{bush1} possesses some universality, 
and this motivated consequent calculations performed in 
\cite{bush1,bush-Izv,bush2,sh-alexeev} for the abelian, 
$O(N)$ and $SU(N)$ 
models with various field contents. Here, we present only a 
brief account of these calculations, so that the origin of 
the mentioned universality becomes clear.

As we have already learned, there is some difference in the
interaction with torsion for abelian and non-abelian gauge 
models. In the last case, the massless vector field does not 
interact with torsion at all, while in the first the 
non-minimal interaction (\ref{vector-nm}) is yet possible. 
Therefore, besides being technically simpler,
the abelian gauge theory is somehow more general and we 
shall use it to discuss the details of the one-loop 
renormalization. We mention, that the one-loop calculation 
for the abelian model was first performed in \cite{obupon}, 
but has been correctly explained only in \cite{bush2}. 

Let us consider the abelian gauge model including gauge,
Yukawa and
four-scalar interactions. The classical action is 
some particular case of (\ref{GUT}), it is given by 
the sum of nonminimal matter actions 
(\ref{scalar1-nm}), (\ref{dirac2-nm}), (\ref{vector-nm}), 
interaction terms and the action of vacuum with all 
26 terms described in the previous section. 

$$
S=\int d^4x\sqrt{-g}\,\Big\{\,
-\frac14\,F_{\mu\nu}\,F^{\mu\nu} + K^{\mu\nu}\,F_{\mu\nu} +
$$
$$
+ \frac12\,g^{\mu\nu}\,\pa_\mu\ph\,\pa_\nu\ph
+ \frac12\,m^2\,\ph^2 
+ \frac12\,\sum_{i=1}^{5}\xi_i\,P_i\,\ph^2
- \frac{1}{24}\,f\,\ph^4 +
$$
\beq
+ i\,{\bar \psi}\,\Big(\,
\ga^\al\,\na_\al + ie\,\ga^\al\,A_\al 
- im -i\,h\,\ph + \sum_{j=1,2} \eta_j\,Q_j 
\,\Big)\,\psi\Big\}\,\, + \,\,S_{vac}\,.
\label{abelian+action}
\eeq

For the sake of the one-loop calculations 
one can use the same formula (\ref{a2}). 
However, since we are going to consider, simultaneously, 
the fields with different Grassmann parity, this formula 
must be modified by replacing usual traces, $Tr$ and $tr$, 
by the supertraces, $sTr$ and $str$. 

The decomposition of the fields into classical background 
$\,A_\mu$, $\ph$, ${\bar \psi}$, $\psi\,$
and quantum ones $\,a_\mu$, $\si$, ${\bar \chi}$, $\chi\,$ is 
performed in the following way:
\beq
A_\mu\to A_\mu + a_\mu\,,\,\,\,\,\,\,\,\,\,\,
\ph \to \ph + i\si\,,\,\,\,\,\,\,\,\,\,\,
{\bar \psi}={\bar \psi}+{\bar \chi}\,,\,\,\,\,\,\,\,\,\,\,
\psi=\psi+\chi\,,\,\,\,\,\,\,\,\,\,\,
\chi = -\frac{i}{2}\,\ga^\la\,\na_\la\eta\,.
\label{shiftabel}
\eeq
In the abelian theory, we can fix the gauge freedom by 
simple condition $\,\na_\mu\,a^\mu =0$. The 
Faddeev-Popov ghosts decouple from other quantum 
fields. They contribute only to the vacuum sector, 
which we already studied in the previous section. 
The resulting operator (\ref{minoper}) has the block 
structure which emerges from the bilinear expansion of the 
classical action \cite{bush2}
\beq
S^{(2)} = \frac12\,\int d^4x\sqrt{-g}\,\,\,(
a^\mu\,|\,\si\,|\,{\bar \chi})\,\,
\left( {\wh H} \right)\,\,
\left(\matrix{a^\nu \cr \si \cr \eta \cr}\right)\,.
\label{bilinear-abel}
\eeq
The details of the calculation can be found in 
\cite{obupon,bush1}. The matrix structure of the 
operators ${\wh P}$ and ${\wh S}_{\la\tau}$ is
the following:
$$
{\wh P} = \left(\matrix{  
  P_{11}     &  P_{12}   &  P_{13}   \cr
  P_{21}     &  P_{22}   &  P_{23}   \cr
  P_{31}     &  P_{32}   &  P_{33}   \cr}
\right)\,\,\,\,\,\,{\rm and}\,\,\,\,\,\,
{\wh S}_{\la\tau} = \left(\matrix{  
  S_{11, \la\tau} & S_{12, \la\tau} & S_{13, \la\tau}   \cr
  S_{21, \la\tau} & S_{22, \la\tau} & S_{23, \la\tau}   \cr
  S_{31, \la\tau} & S_{32, \la\tau} & S_{33, \la\tau}   \cr}
\right)\,.
$$
where the terms essential for our consideration are \cite{bush2}
$$
S_{31}=S_{32}=0\,,\,\,\,\,\,\,\,\,\,\,\,\,\,\,\,\,\,\,
P_{32}=2h\,\psi\,,\,\,\,\,\,\,\,\,\,\,\,\,\,\,\,\,\,\,
P_{31}=-2e\,\ga_\nu\psi\,,
$$$$
P_{13}=-\frac{ie}{2}\,\na_\rho {\bar \psi}\ga_\mu\ga^\rho
+ieh\,{\bar \psi}\ga_\mu \ph 
- \frac12\,e^2\,{\bar \psi}\ga_\mu A_\rho \ga^\rho 
- \frac{i}{4}\,e\,{\bar \psi}\ga_\mu\ga^\rho 
\sum \eta_j\,Q_j\, \ga_\rho\,,
$$
\beq 
P_{23}=\frac{ih}{2}\,\na_\rho {\bar \psi}\ga^\rho
-ih^2\,{\bar \psi} \ph 
+ \frac12\,eh\,{\bar \psi} A_\rho \ga^\rho 
+ \frac{i}{4}\,h\,{\bar \psi}\ga^\rho \sum 
\eta_j\,Q_j \ga_\rho\,.
\label{PS}
\eeq
Now, substituting ${\wh P}$ and ${\wh S}_{\la\tau}$
into (\ref{a2}), after some algebra we arrive at the 
divergent part of the one-loop effective action 
$$
{\bar \Ga}^{(1)}_{div} \,=\,-\,\frac{\mu^{n-4}}{\vp}\,
\int d^nx\sqrt{-g}\,\left\{\,
-\frac{2e^2}{3}\,F_{\mu\nu}\,F^{\mu\nu} 
+\frac{8e\eta_2}{3}\,F_{\mu\nu}\,\na^\mu T^\nu + 
2h^2\,g^{\mu\nu}\,\pa_\mu\ph\,\pa_\nu\ph + 
\right. 
$$
$$
\left.
+ \left(\, \frac{f^2}{8} - 2h^4 \, \right)\ph^4\,
+ \,\frac12\,\ph^2\,\left[\,-\left(\sum_{i}\xi_iP_i
-\frac16\,R \right)\,f 
+ \frac23\,h^2R -16h^2\,\eta^2_1S_\mu S^\mu \,\right]+
\right. 
$$
$$
\left.
+ i(2e^2+h^2)\,{\bar \psi}\ga^\mu \,
(\na_\mu + ieA_\mu -i\eta_2 T_\mu)\psi
+ (2e^2-h^2)\,\eta_1\,{\bar \psi}\ga^5\ga^\mu S_\mu\psi
+ (8e^2-2h^2)h\,{\bar \psi}\ph\psi
\right\} + 
$$
\beq
 + \,\,{\rm vacuum \,\,\, divergent \,\,\, terms}\,.
\label{div-abel}
\eeq
\vskip 2mm
The above expression is in perfect agreement with the general 
considerations of section 3.1. 
Let us comment on some particular points. 

1. One can see that the non-minimal divergences of the 
$\,{\bar \psi}\ga^5\ga^\mu S_\mu\psi\,$ and 
$\,\ph^2\,S_\mu S^\mu\,$-types really emerge, even if the
starting action includes only minimal interaction with
$\,\eta_{2}=\frac18\,,\,\,\eta_{2}=0\,$ and 
$\,\xi_{2,3,4,5} =0$. Thus, exactly as we have supposed, 
the nonminimal interaction of the $S_\mu$ - component of 
torsion with spinor and scalar is necessary for the 
renormalizability. The list of essential parameters includes 
$\xi_1, \eta_1, \xi_4$ parameters of the action which are 
not related to $T_\mu$ or $q^\al_{\cdot\,\be\ga}$. Another 
potentially important parameter is $\th_2$, because the 
corresponding topological counterterm (\ref{vector-nm}) can 
emerge at higher loops. 

2. All terms with $\,T_\mu$ and $q^\al_{\,\cdot\,\be\ga}\,$
components of torsion are purely nonminimal. Furthermore, 
the substitution $eA_\mu + \eta_2 T_\mu \to eA_\mu$
explains all details of the renormalization of the
parameter $\eta_2$. In particular, this concerns the
nonminimal interaction of the gauge vector $A_\mu$ with 
torsion trace $T_\mu$, which leads to the renormalization 
of the non-minimal parameter $\th_5$ in (\ref{vector-nm}).
It is worth mentioning that such a mixing is not possible 
for the non-abelian case \cite{bush1}.

3. One can observe some simple hierarchy of the parameters.
The non-minimal parameters $\,\xi_i,\eta_j,\th_k\,$ do 
not affect the renormalization of the coupling constants 
$e,h,f$ and masses. One can simply look at Figure 1 to 
understand why this is so. In turn, vacuum parameters do 
not affect the renormalization of neither coupling constants 
nor the non-minimal parameters $\xi_i,\eta_j,\th_k$. 
One important consequence of this is that the renormalization
group equations in the minimal matter sector are independent 
on external fields so that the renormalization of the couplings 
and masses is exactly the same as in the flat space-time.
The renormalization group in the mixed non-minimal sector
depends on the matter couplings, but does not depend
on the vacuum parameters. Finally, the renormalization  
in the vacuum sector depends, in general, on the nonminimal 
parameters. 

4. The last observation is the most complicated one. The 
contributions to the spinor sector may come only from the 
mixed sector of the products of the operators (\ref{PS}).
Now, since $S_{31}=S_{32}=0$, 
all the fermion renormalization comes from two traces: 
$$
tr\,(P_{13}\cdot P_{31}) \,\,\,\,\,\,\,\,\, {\rm and}                
\,\,\,\,\,\,\,\,\, tr\,(P_{23}\cdot P_{32}) \,.
$$
It is easy to see, that the arrangement of the 
$\gamma$-matrices in the expressions for  
$$
P_{13}\,,\,\,\,\,\,\,\,\,\,\,  P_{31}\,,\,\,\,\,\,\,\,\,\,\,  
P_{23}\,,\,\,\,\,\,\,\,\,\,\,  P_{32}
$$
is universal in the sense that it does not depend on the 
gauge group. For any non-abelian theory 
this arrangement is the same as for the simple abelian 
model under discussion. Therefore, in the fermionic sector,
the signs of the counterterms will always be equal to the 
ones we meet in the abelian model. The renormalization of the 
essential parameter $\eta_1$ has the form
\beq
\eta_1^0 = \eta_1\,\left(\,1 - \frac{C}{\vp}\,h^2\, \right)\,,
\label{renorm-eta1}
\eeq
with $\,C=2$ in the abelian case.
Using the above consideration we can conclude that the
renormalization of this parameter in an arbitrary 
gauge model will have the very same form (\ref{renorm-eta1}) 
with positive coefficient $\,C$. Of course, the value of 
this coefficient may depend on the gauge group. In 
the theory with several fermion fields the story becomes
more complicated, because there may be 
more than one non-minimal parameters $\eta_1$, and in 
general they may be different for different fields. The 
renormalization can mix these parameters, but the 
consideration presented above is a useful hint to the
sign universality of the $\beta$-function for $\eta_1$, 
which really takes place \cite{bush1,bush2,bush-Izv,sh-alexeev}. 

The details of the
calculations of the one-loop divergences in the variety
of $SU(2)$, $SU(N)$, $O(N)$ models including the finite
and supersymmetric models can be found in 
\cite{bush1,bush2,bush-Izv,sh-alexeev}, part of them was 
also presented in \cite{book}. Qualitatively
all these calculations resemble the sample we have 
just considered, so that all the complications come from the 
cumbersome group relations and especially from the 
necessity to work with many-fermion models. The results
are in complete agreement with our analysis, in particular
this concerns the universal sign of $C$ in (\ref{renorm-eta1}).

Let us now discuss another, slightly different, example.
Consider, following  \cite{njl}, the Nambu-Jona-Lasinio (NJL) 
model in curved space-time with torsion. 
This model is regarded as an effective theory of the SM
which is valid at some low-energy scale. If we are 
interested in the renormalization of the theory in 
an external gravitational field, then the NJL model may be 
regarded as the special case of the theory with
the  Higgs scalars \cite{hill&Salopek,mutodi}.
Our purpose is to study the impact of torsion 
for this effective theory.

Consider the theory of $N$ - component spin  ${1\over2}$ field
with four - fermion interaction in an external gravitational 
field with torsion. The action is
\beq
S_{njl} = \int d^4x\sqrt{-g}\,\left\{ 
L_{gb} + i{\bar{\psi}}{\gamma}^{\mu}(D_{\mu} - i{\eta}_1
{\gamma}^5 S_{\mu}){\psi} + G{({\bar{\psi}}_L {\psi}_R)}^2
\,\right\}  
\label{1}
\eeq
Here, $L_{gb}$ is the Lagrangian of the gauge boson field, 
$D_{\mu}$ is the covariant derivative with
respect to both general covariance and gauge invariance, 
$G$ is the
dimensional coupling constant. The above Lagrangian (\ref{1}) 
is direct generalization of the one of the paper 
\cite{hill&Salopek} for the case of gravity with torsion. 
The introduction of the nonminimal interaction with torsion 
reflects the relevance of such an interaction at high 
energies.

Introducing the auxiliary Higgs field $H$, one can cast the 
Lagrangian in the form
\beq
S_{njl} = \int d^4x\sqrt{-g}\,\left\{ 
L_{gb} + i{\bar{\psi}}{\gamma}^{\mu}(D_{\mu} - i{\eta}_1
{\gamma}^5 S_{\mu}){\psi} + ({\bar{\psi}}_L {\psi}_R H +
{\bar{\psi}}_R {\psi}_L H^\dagger) 
- m^2 H^\dagger H\,\right\}\,.          
\label{2}
\eeq
It is easy to see that, in this form, the theory (\ref{2}) 
is not renormalizable due to the divergences in the scalar 
and gravitational sectors. However, our previous analysis 
can be successfully applied here if we add to (\ref{2})
an appropriate action of the external fields. First we 
notice that the
scalar field is non-dynamical, so that both kind of 
divergences are similar to the vacuum ones in ordinary 
gauge theories in curved space-time. Then, one can provide the
renormalizability by introducing the Lagrangian $L_{ext}$ 
of external fields $\,H,\,H^\dagger,\,g_{\mu\nu},\,S_\mu\,$ 
into the action (\ref{2}).  

If we do not consider 
surface terms, the general form of $L_{ext}$ is:
\beq
L_{ext} = g^{\alpha\beta}{\cal{D}}_{\alpha}H {\cal{D}}_{\beta} H^\dagger
+ {\xi}_1R H^\dagger H +
{\xi}_4 H^\dagger H S_{\mu}S^{\mu} 
- {\lambda \over2}(H^\dagger H)^2 +
L_{vac}\,,
\label{3}
\eeq
where the form of $L_{vac}$ corresponds to the possible 
counterterms (\ref{contra}), as it has been discussed in 
the previous section. Some terms in (\ref{2}) 
were transferred to $L_{ext}$. This reflects 
their role in the renormalization.

When analyzing possible divergences of the theories 
(\ref{1}) and (\ref{2}), one meets a serious difference.  
The theory (\ref{1}) is not renormalizable, while the 
theory (\ref{2}), (\ref{3}) is. The reason is the use of the
fermion - bubble approximation. Then, (\ref{2}) becomes the 
theory of the free spinor field in the external background 
composed by gauge boson, scalar, metric and torsion fields.
All possible divergences in this theory emerge at the one-loop 
level, and only in the external field sector. 
In this sense, one can formulate the renormalizable NJL model 
in the external gravitational field with torsion.

The direct calculations give the
following result for the divergent part of the effective action (we omit
the gauge boson and surface terms):
$$
{\bar \Gamma}^{(1)}_{div} 
\,=\,-\, {2N\,\mu^{n-4} \over{\varepsilon}} \int d^nx \sqrt {-g}\, 
\Big\{
g^{\alpha\beta}{\cal{D}}_{\alpha}H {\cal{D}}_{\beta} H^\dagger +
{1\over6}\,R H^\dagger H +
$$
\beq
+ 4{\eta}_1^2 H^{+}H S_{\mu}S^{\mu} - (H^{+}H)^2 
- {1\over3}\,S_{\mu\nu}\,S^{\mu\nu} 
+{1\over{20}}C_{\mu\nu\alpha\beta}C^{\mu\nu\alpha\beta} \Big\}
\,+\,...\,\,.      
\label{njl}
\eeq
The first terms in (\ref{njl}) are the same as
in a purely metric theory \cite{hill&Salopek,mutodi},
while others are typical for the theory with external torsion.

%%%%%%%%%%%%%%%%%%%%%%%%%%%%%%%%%%%%%%%%%%%%%%%%%%%%%%%%%%%%
\section{Renormalization group and universality 
in the non-minimal sector}
%%%%%%%%%%%%%%%%%%%%%%%%%%%%%%%%%%%%%%%%%%%%%%%%%%%%%%%%%%%%

The renormalizability of the Quantum Field Theory in curved
background enables one to formulate the renormalization group 
equation for the effective action and parameters of the theory. 
The derivation of these equations in the 
space-time with torsion is essentially the same as for the 
purely metric background. Let us outline the  
formulation of the renormalization group \cite{tmf,book}.
The renormalized effective action 
$\,\Ga\,$ depends on the matter fields $\Phi$ (as before, 
we denote {\it all} kind of non-gravitational fields in 
this way), 
parameters $\,P\,$ (they include all couplings, masses, non-minimal 
parameters and the parameters of the vacuum action), external fields 
$\,g_{\mu\nu},\, T^\al_{\,\cdot\,\be\ga}$, dimensional parameter 
$\mu$ and the parameter of the dimensional regularization $\,n$. 
The renormalized effective action is equal to the bare one: 
\beq
\Ga\left[g_{\mu\nu}, T^\al_{\,\cdot\,\be\ga}, \Phi, P, \mu, n\right]
= \Ga_0\left[g_{\mu\nu}, T^\al_{\,\cdot\,\be\ga}, 
\Phi_0, P_0, n\right]
\label{bare-ren}
\eeq
Taking derivative with respect to $\mu$ we arrive at the equation
\beq
\left[\,\,\mu\frac{\pa}{\pa\mu} + \beta_P\frac{\pa}{\pa P}
+ \int d^nx\sqrt{-g}\,\gamma\Phi\,\frac{\de}{\de \Phi}\,\,\right]\,
\Ga\left[g_{\mu\nu}, T^\al_{\,\cdot\,\be\ga}, \Phi, P, \mu, n\right]
= 0\,.
\label{mu-indep}
\eeq
Here $\beta$ and $\ga$ functions are defined in a usual way:
\beq
\beta_P(n) = \mu\frac{\pa P}{\pa\mu} \,\,\,\,\,\,\,\,\,\,\,\,
{\rm and} \,\,\,\,\,\,\,\,\,\,\,\,
\ga(n)\,\Phi = \mu\frac{\pa \Phi}{\pa\mu}\,.
\label{bety}
\eeq
The conventional $n=4$ beta- and gamma-functions are defined 
through the limit $\,n \to 4$.
Using the dimensional homogeneity of the effective action
we get, in addition to (\ref{mu-indep}), the equation 
\beq
\left[\,\frac{\pa}{\pa t} + 
\mu\frac{\pa}{\pa\mu} + d_P\frac{\pa}{\pa P}
+ d_\Phi\int d^nx\sqrt{-g}\,\Phi\frac{\de}{\de \Phi}\right]\,
\Ga\left[g_{\mu\nu}, T^\al_{\,\cdot\,\be\ga}, \Phi, P, \mu, n
\,\right] = 0\,,
\label{dim-indep}
\eeq
where $d_P$ and $d_\Phi$ are classical dimensions of the parameters
and fields. Combining (\ref{mu-indep}) and (\ref{dim-indep}), 
setting $t=0$, replacing the operator 
$\,\,2g_{\mu\nu}\frac{\de}{\de g_{\mu\nu}}\Ga[g_{\mu\nu},...]\,\,$
by $\,\,\frac{\pa}{\pa t}\Ga[e^{-2t}g_{\mu\nu},...]$, and taking 
the limit $n\to 4$, 
we arrive at the final form of the renormalization group 
equation appropriate for the study of the short-distance 
limit: 
\beq
\left[\frac{\pa}{\pa t} - (\beta_P-d_P)\frac{\pa}{\pa P}
-(\ga - d_\Phi)\int d^nx\sqrt{-g}\,
\Phi\frac{\de}{\de \Phi}\right]\,\Ga\left[g_{\mu\nu} e^{-2t}, 
T^\al_{\,\cdot\,\be\ga}, \Phi, P, \mu \right] \,=\, 0\,.
\label{RGE}
\eeq
The general solution of this equation is 
\beq
\Ga\left[g_{\mu\nu} e^{-2t},  
T^\al_{\,\cdot\,\be\ga}, \Phi, P, \mu \right]
= \Ga\left[g_{\mu\nu}, T^\al_{\,\cdot\,\be\ga}, \Phi(t), 
P(t), \mu \right]\,,
\label{sol-RGE}
\eeq
where fields and parameters satisfy the equations
$$
\frac{d\Phi}{dt} = [\ga(t)-d_\Phi]\,\Phi
\,,\,\,\,\,\,\,\,\,\,\,\,\,\,\,\,\,\,\,\,\,\,\,
\Phi(0)=\Phi\,,
$$
\beq
\frac{d P}{dt} = \be_P(t)-d_P
\,,\,\,\,\,\,\,\,\,\,\,\,\,\,\,\,\,\,\,\,\,\,\, P(0)=P\,.
\label{RGE-const}
\eeq

In fact, torsion does not play much role in the 
above derivation, mainly because it does not transform under 
scaling. In fact, this is natural, because the physical 
interpretation of the UV limit in curved space-time is the 
limit of short distances. But, geometrically, the distance between 
two points does not depend on torsion, so it is not a big
surprise that torsion is less important than metric here. 

From eq. (\ref{sol-RGE}) follows that the investigation of the 
short-distance limit reduces to the analysis of 
the equations (\ref{RGE-const}). Our main interest will be the 
behaviour of the non-minimal and vacuum parameters related to 
torsion, but we shall consider the renormalization of other 
parameters when necessary.  

First of all, according to our previous discussion of the 
general features of the renormalization, the 
matter couplings and masses obey the same renormalization 
group equations as in the flat space-time. Furthermore, the 
running of  $\,\xi_1\,$ and those vacuum parameters, 
which are not related to torsion, satisfy the same equations 
as in the torsionless theory. Let us concentrate our attention 
on the parameters related to torsion.
Consider the most important equation for the 
non-minimal parameter $\eta_1$. Using the universality 
of its one-loop renormalization (\ref{renorm-eta1}), 
and the classical dimension (in $\,n\neq 4$) for the
Yukawa coupling 
$$
(2\pi)^2\,\frac{dh}{dt}=\frac{n-4}{2}\,h + \beta_h(4)\,,
$$ 
we can derive the universal form of the renormalization 
group equation at $n=4$ 
\beq
(4\pi)^2\,\frac{d\eta_1}{dt} = C\eta_1h^2\,.
\label{uni-RG}
\eeq
Here, according to (\ref{renorm-eta1}), the constant $C$ is 
positive, but its magnitude depends on the gauge group. 
For instance, in the case of abelian theory $C=2$, and for the 
adjoint representation of the $SU(N)$ group the value is
$C=1$ \cite{sh-alexeev}. The physical interpretation of the
universal running (\ref{uni-RG}) is obvious: the 
interaction of fermions with torsion becomes stronger in 
the UV limit (short distance limit in curved space-time). 
This provides, at the first sight, an attractive opportunity 
to explain very weak (if any) interaction between torsion and
matter fields. Unfortunately, the numerical effect of
this running is not sufficient. Let us consider the simplest 
case of the constant Yukawa coupling $h=h_0$. Then, 
replacing, as usual, $\frac{d}{dt}$ for $\mu\frac{d}{d\mu}$, 
the (\ref{uni-RG}) gives 

\beq
\frac{\eta_1(\mu)}{\eta_1(\mu_{UV})} =
\left(\frac{\mu}{\mu_{UV}}\right)^{Ch_0^2/(4\pi)^2}\,.
\label{behav}
\eeq
It is easy to see, that even the 50-order change of the 
scale $\mu$ changes $\eta_1$ for less than one order. 
The effect of (\ref{behav})
is stronger for the heavy fermions with larger 
magnitude of the Yukawa coupling. If we suppose, that all the 
fermions emerge after some superstring phase transition,
and that the high energy values of the non-minimal 
parameters are equal for all the spinors, the low-energy 
values of these parameters will not differ for more than 
2-3 times. Indeed, there may be some risk in the above 
statement, related to the non-perturbative low-energy 
effects of QCD. However, since the quarks are confined 
inside the nucleus, and there is no chance to observe their 
interaction with an extremely weak background torsion
(see the next Chapter for the modern upper bounds for
the background torsion), the effect of the $\eta_1 (t)$ 
running does not have much physical importance. 

In the non-minimal scalar sector we meet standard equation 
for the $\xi_1$ parameter, which does not depend on 
torsion \cite{bush1,book} 
\beq
(4\pi)^2\,\frac{d\xi_1}{dt} = \left(\xi_1-\frac16\right)\,
[k_1g^2+k_2h^2+k_3f]\,,
\label{uni-scal-xi1}
\eeq
where the magnitudes of $k_1,k_2,k_3$ depend on the gauge
group, but always $k_1<1$ and $k_{2,3}>0$. The equation 
for the scalar-torsion interaction parameter $\xi_4$ has
the form \cite{bush1}
\beq
(4\pi)^2\,\frac{d\xi_4}{dt} \, = \,
[k_1g^2+k_2h^2+k_3f]\,\xi_4\,-\,k_4h^2\,\eta_1^2\,,
\label{uni-scal-xi4}
\eeq
with $k_4>0$. The asymptotic behaviour of $\xi_1$ depends on 
the gauge group and on the 
multiplet composition of the model. For some models 
$\xi_1 \to 1/6$ in the UV limit $t\to \infty$ \cite{buod,yagu},
and one meets the asymptotic conformal invariance. 
For other theories, including the minimal $SU(5)$ GUT
\cite{parker-toms}, the asymptotic behaviour is the opposite:
$|\xi_1| \to \infty$ at UV. It is remarkable, that due to the 
non-homogeneous term in the beta-function (\ref{uni-scal-xi4})
for $\xi_4$, this parameter has an universal behaviour which does
not depend on the gauge group. It is easy to see, that in 
all cases
\beq
|\xi_4(t)| \to \infty \, \, \, \, \, \, \, \, \, \, \, \, 
{\rm at} \, \, \, \, \, \, \, \, \, \, \, \, t\to \infty\,.
\eeq
Consequently, the interaction of scalar with torsion also 
gets stronger at shorter distances, and weaker at long 
distances. Qualitatively the result is the same as for the 
parameter $\eta_1$.

Consider, for example, the $SU(2)$ gauge model with one charged 
scalar multiplet and two sets of fermion families in the
fundamental representation of the gauge group. In flat space-time, 
the $\beta$-functions for this model (which is quite similar 
to the SM) have been derived in Ref. \cite{votyu}. The 
Lagrangian of the renormalizable theory in curved space-time 
with torsion has the form \cite{bush1}:  
$$
  {\cal L} = -\frac14\,G^a_{\mu\nu}G^{a\,\mu\nu} 
+ g^{\mu\nu}\,(\pa_\mu\phi^{\dagger} 
+ \frac{ig}{2}\,\tau^a\,A^a_\mu\phi^{\dagger})\,(\pa_\mu\phi 
- \frac{ig}{2}\,\tau^a\,A^a_\mu\phi) 
- \frac{f}{8}\,(\phi^{\dagger}\phi)^2 + 
$$$$
+ \sum_{i=1}^{5}\,\xi_iP_i\,\phi^{\dagger}\phi 
+ i\xi_0\,T^\al (\phi^{\dagger} \pa_\al \phi 
- \pa_\al \phi^{\dagger}\cdot\phi) 
+ \sum_{k=1}^{m} \,i\,{\bar \chi}^{(k)}\Big(\,\ga^\mu\na_\mu
+ \sum_{j=1,2}\,\de_jQ_j\,\Big)\chi^{(k)} + 
$$
\beq
+ \sum_{k=1}^{m+n} \,i\,{\bar \psi}^{(k)}\Big(\,\ga^\mu\na_\mu
- \frac{ig}{2}\,\tau^a\,\ga^\mu A^a_\mu
+ \sum_{j=1,2}\,\eta_jQ_j\,\Big)\psi^{(k)} 
- h\sum_{k=1}^{m} \,\left(\,{\bar \psi}^{(k)}\chi^{(k)} \phi
+ \phi^{\dagger} {\bar \chi}^{(k)}\psi^{(k)}\,\right)\,.
\label{su(2)}
\eeq
Let us write, for completeness, the full set of the 
$\beta$-functions for the coupling constants and non-minimal 
parameters.   
\vskip 3mm

\noindent
i) Couplings \cite{votyu}: 
$$
(4\pi)^2\,\frac{dg^2}{dt} = - b^2g^2\,,\,\,\,\, \,\,\,\, \,\,\,\, 
\,\,\,\, \,\,\,\, \,\,\,\, \,\,\,\, \,\,\,\, 
b^2= - \frac{43-4(m+n)}{3}\,g^4\,,
$$
$$
(4\pi)^2\,\frac{dh^2}{dt} = (3+4m)h^4 - \frac92\,g^2 h^2\,,
$$
\beq
(4\pi)^2\,\frac{df}{dt} = 3f^2-32mh^4+9g^4+8mfh^2-9fg^2\,.
\label{coupl}
\eeq
Likewise all three $SU(2)$ models of Voronov and Tyutin 
\cite{votyu}, this one is asymptotically free in all effective 
couplings, but only in the special regime, when they are 
proportional to each other. The necessary condition of the 
asymptotic freedom in $g^2$ is $m+n\leq 10$ and for $h^2$ it is 
$m+n \geq 8$. The asymptotic freedom in $f(t)$ happens only 
for $m+n=10$ and $m=1$. 
On the special asymptotically free solutions of the renormalization 
group equations one meets the following behaviour:
\beq
g^2(t) = \frac{g^2}{1+b^2g^2t/(4\pi)^2}\,,
\,\,\,\,\,\,\,\,\,\,\,\,\,\,\,\,
h^2(t)=\frac12\,g^2
\,,\,\,\,\,\,\,\,\,\,\,\,\,\,\,\,\, 0 < f \leq g^2\,.
\label{couplsolu}
\eeq
%%%%%%%%%%%%%%%%%%%%%%%%%%%%%%%%%%%%%%
\vskip 3mm

\noindent
ii) Non-minimal parameters \cite{buod,bush1}:
$$
(4\pi)^2\,\frac{d\eta_1}{dt} = h^2(\eta_1+\de_1)\,,
\,\,\,\,\,\,\,\,\,\,\,\,\,\,\,\,
(4\pi)^2\,\frac{d\de_1}{dt} = 2h^2\eta_1\,,
$$
$$
(4\pi)^2\,\frac{d\eta_2}{dt} = h^2(\eta_2-\de_2-\frac12\,\xi_0)
\,,\,\,\,\,\,\,\,\,\,\,\,\,\,\,\,\,
(4\pi)^2\,\frac{d\de_2}{dt} = h^2(-2\eta_2 + \xi_0)\,,
$$$$
(4\pi)^2\,\frac{d\xi_0}{dt} = - 4m^2h^2(\de_2-\eta_2)\,,
$$
$$
(4\pi)^2\,\frac{d\xi_1}{dt} = \Big(\xi_1-\frac16\Big)\cdot A\,,
\,\,\,\,\,\,\,\,\,\,\,\,\,\,\,\,
(4\pi)^2\,\frac{d\xi_{2,5}}{dt} = A\xi_{2,5}\,,
\,\,\,\,\,\,\,\,\,\,\,\,\,\,\,\,{\rm where}
\,\,\,\,\,\,\,\,\,\,\,\,\,\,\,\,
A = \frac32\,f -\frac92\,g^2 + 4mh^2\,,
$$
\beq
(4\pi)^2\,\frac{d\xi_3}{dt} = A\,\xi_3  +4mh^2\,(\eta_2-\de_2)^2\,,
\,,\,\,\,\,\,\,\,\,\,\,\,\,\,\,\,\,
(4\pi)^2\,\frac{d\xi_4}{dt} = A\,\xi_4 - mh^2\,(\eta_1+\de_1)^2\,.
\label{non-min}
\eeq
The solutions of these equations are rather cumbersome
\cite{thesis} and we will not write them here, but only 
mention that they completely agree with the general analysis
given above (the same is true for all other known examples).
Finally, the asymptotic behaviour of the non-minimal parameters
is the following \cite{thesis,bush1}:
$$
\xi_1 - \frac16, \xi_0, \eta_2, \de_2 \to 0
\,,\,\,\,\,\,\,\,\,\,\,\,\,\,\,\,\, 
\eta_1, \de_1 \to \infty \cdot sgn(\de_1+2\eta_1)
\,,\,\,\,\,\,\,\,\,\,\,\,\,\,\,\,\, 
\xi_4 \to + \infty\,.
$$
\vskip 3mm

A special case is the renormalization group equations
for the NJL model (\ref{2}).
Since this theory can be viewed as the theory
of free spinor fields in an external scalar and gravitational 
fields, the exact $\beta$-functions coincide with the one-loop 
ones. The renormalization group equations for the essential 
effective couplings $\,\xi_4,\, \eta_1\,$ have the form:
$$
(4{\pi})^2 {d{\eta_1}^2 \over{dt}} = {8N\over3}{\eta_1}^4
$$
\beq
(4{\pi})^2 {d{\xi_4} \over{dt}} = 2N(\xi_4 - {8\over3}{\eta_1}^2)
\label{666}
\eeq
The analysis of these equations shows that
the strength of the interaction of spinor and scalar fields 
with torsion increases at short distances.

Let us now consider the renormalization group for the
parameters of the vacuum energy. If we write the vacuum action as
\beq
S_{vac} = \int d^4x\sqrt{-g}\,\sum_{k=1}^{26}\,p_k\,J_k\,,
\label{vacuum-torsion}
\eeq
where $p_k$ are parameters and $J_k$ - vacuum terms (for
instance, $J_1=C^2,\,J_2=E,\,...\,\,$ the one-loop divergences
will have the form 
\beq
\Ga_{div} = - \frac{\mu^{n-4}}{\ep}\,
\int d^nx\sqrt{-g}\,\sum_{k=1}^{26}\,\De_k\,J_k\,.
\label{divs-vac-torsion}
\eeq
Here $\De_k$ are the sums of the contributions from the 
free scalar, vector and spinor fields. The relations 
between renormalized and bare parameters have the form
$$
p^0_k = \mu^{n-4}\,\left[p_k + \frac{\De_k}{\ep} \right]\,.
$$
The $\be$-functions are derived according to the standard 
rule $\be_k= \mu\frac{d p_k}{d\mu}$. 
It is important, from the technical point of view, 
that the one-loop vacuum divergences depend only on the 
non-minimal parameters $\xi, \eta$, but not on the 
couplings $g,h,f$. The renormalization of the non-minimal 
parameters does not include the $\mu^{n-4}$ factor, and 
that is why we can derive universal expressions for the 
vacuum $\beta$-functions and renormalization group equations
\beq
\frac{dp_k}{dt}
=\be_k = (4-n) \,p_k - \frac{\De_k}{(4\pi)^2}
\,,\,\,\,\,\,\,\,\,\,\,\,\,\,\,\,\,\,\,\,\,\,
p_k(0)=p_{k0}\,.
\label{beta's-vac}
\eeq
Standard $n=4$ beta-functions can be obtained through the 
limit $n\to 4$. Since the total number of vacuum terms is 26, 
it does not have much sense to study the details of scaling 
behaviour for all of them. We shall just indicate some 
general properties. One can distinguish the $\De_k$
coefficients which are parameter independent (like the 
$\De_1, \De_2$), the ones which are proportional to the 
squares of the masses of scalars or spinors $m^2$, 
the ones which are proportional to the squares of the 
non-minimal parameters of the torsion-matter interaction
$\eta_j^2$ or $\xi_i^2$. 
In general, all these types of terms will have distinct 
asymptotic behaviour. Let us consider, for simplicity, only 
those parameters which are related to the completely 
antisymmetric torsion and correspond to the massless theory. 
For the same reason we can take some finite model, in 
which the beta-functions for masses equal zero (this
can be also considered as an approximation, because 
typically $\eta_1$ runs stronger than the mass). We can 
notice that if the scalar mass does not run, the behaviour 
of $\xi_4$ is $\xi_4(t)\sim \eta^2_2(t)$. Finally, for the 
torsion-independent terms we get the asymptotic behaviour 
$$
p_k(t) - p_{k0} \sim - \frac{\De_k}{(4\pi)^2}\,t
$$
while for the $\eta_1^2$-type terms the behaviour is
$$
p_k - p_{k0} \sim e^{2Ch^2t/(4\pi)^2} \sim \eta_1^2(t)  
$$
and for the $\xi_4^2$-type term 
$$
p_k - p_{k0} \sim e^{4Ch^2t/(4\pi)^2} \sim \eta_1^4(t)\,.  
$$
Thus, the running of the torsion-dependent vacuum terms
is really different from the running of the torsion-independent 
ones. For the last ones, we meet the usual power-like scaling, 
which may signify asymptotic freedom in UV or (for the 
massless theories) in IR -- this depends on the relative sign 
of $p_{k0}$ and $\De_k$. For the torsion-dependent terms
the behaviour is approximately exponential, exactly as
for the non-minimal parameter $\eta_1$. Indeed, the numerical 
range of the running is not too large, because of the 
$1/(4\pi)^2$-factor.

%%%%%%%%%%%%%%%%%%%%%%%%%%%%%%%%%%%%%%%%%%%%%%%%%%%%%%%%%%%%
\section{Effective potential of scalar field in the space-time 
with torsion. Spontaneous symmetry breaking and phase transitions 
induced by curvature and torsion}
%%%%%%%%%%%%%%%%%%%%%%%%%%%%%%%%%%%%%%%%%%%%%%%%%%%%%%%%%%%%

Let us investigate further the impact of the renormalization and 
renormalization group in the matter field sector of the effective 
action. We shall follow \cite{bosh-pot} and consider the 
effective potential of the massless 
scalar field in the curved space-time with torsion. The effective 
potential $V$ is defined as a zero-order term in the derivative 
expansion of the effective action of the scalar field $\ph$: 
\beq
\Ga [ \ph ] = \Ga_0 + \int d^4x\sqrt{-g}\,
\left\{\, - V(\ph) 
+ \frac12\,Z(\ph)\,g^{\mu\nu}\,\pa_\mu\ph\pa_\nu\ph \,+ \,...\,
\,\right\}\,,
\label{efpo}
\eeq
where the dots stand for higher derivative terms, and $\Ga_0$ 
is the vacuum effective action. We shall discuss the derivation 
of $\Ga_0$ in the next sections. 

The classical potential of the scalar field has the form
\beq
V_{cl} = a\,f\, \ph^4\, -\, 
\sum_{i=1}^{5}\,b_i\, \xi_i\, P_i\,\ph^2\,,
\label{clpo}
\eeq
where we have used the notations (\ref{Pi}). If the space-time
metric  is non-flat and the scalar field
couples to spinors through the Yukawa interaction, two of the 
non-minimal parameters $\xi_1$ and $\xi_4$ are necessary 
non-zero. Therefore, even for the flat space-time metric the
potential feels torsion through the parameter $\xi_4$.
Here we consider the general case of the curved metric and 
take all parameters $\xi_i$ arbitrary for the sake of 
generality. 

The quantum corrections to the classical potential (\ref{clpo})
can be obtained using the renormalization group method 
\cite{colwe,tmf,book,bosh-pot}. The renormalization group equation
for the effective potential follows from the renormalization group 
equation (\ref{mu-indep}) for the whole effective action. Since   
(\ref{mu-indep}) is linear, all terms in the expansion 
(\ref{efpo}) satisfy this equation independently. 
It is supposed that the divergences were already removed by the 
renormalization of the parameters, and therefore in this case 
one can put $n=4$ from the very beginning. Thus we get 
\beq
\left[\,\mu\frac{\pa}{\pa\mu} + \delta\,\frac{\pa}{\pa \al} 
+ \beta_P\frac{\pa}{\pa P}
+ \int d^4x\sqrt{-g}\,\gamma\,\ph\,\frac{\de}{\de \ph}\,\right]\,
V\left(g_{\mu\nu}, T^\al_{\,\cdot\,\be\ga}, \ph, P, \mu\,\right)
= 0\,.
\label{efpo-indep}
\eeq
Here $P$ stands, as before, for all the parameters of the theory:  
gauge, scalar and Yukawa couplings and non-minimal 
parameters $\xi_i$. 
$\al$ is the gauge fixing parameter corresponding to the term
${\cal L}_{gf} = \frac{1}{2\al}(\na_\mu A^\mu)^2$ and 
$\de$ - is the renormalization group function corresponding to $\al$. 

We shall solve (\ref{efpo-indep}) 
in the approximation $\ph^2 \gg \Big| P_i \Big|$ for all $P_i$, 
and neglect higher order terms. Physically, this 
approximation corresponds to the weakly oscillating metric 
and weak external torsion. In the gravitational field without 
torsion this approximation has been used  in \cite{ish}.
Similar method can be 
applied to the derivation of other terms in the effective action,
including higher order terms \cite{wolf} (see also \cite{book}).
The initial step is to write the effective potential in the 
form $V=V_1+V_2$, where $V_1$ does not depend on the external fields
$g_{\mu\nu},\,T^\al_{\,\cdot\,\be\ga}$, and 
$V_2= \sum_{i=1}^{5}\,V_{2i}\, P_i$. Since all $P_i$ are linear
independent, $\,V_{2i}\,$ must satisfy the equation 
(\ref{efpo-indep}) independently. 
Then this equation is divided into the following set of 
equations for $V_1,\,V_{2i}$:
\beq
({\cal D} - 4\ga)\,\frac{d^4\,V_1}{d\ph^4}=0\,,\,\,\,\,\,\,\,\,\,\,
\,\,\,\,\,\,\,\,\,\,
({\cal D} - 2\ga)\,\frac{d^2\,V_{2i}}{d\ph^2}=0\,,
\label{eqs}
\eeq
where
\beq
{\cal D} = - (1+\ga)\frac{\pa}{\pa t} 
+ \delta\,\frac{\pa}{\pa \al} + \beta_P\,\frac{\pa}{\pa P}\,,
\label{eqqs}
\eeq
and $\,\,t=(1/2)\,{\rm ln}\ph^2/\mu^2$. If we use the standard 
initial conditions \cite{book}
$$ %%% \beq
\frac{d^4\,V_1}{d\ph^4}\Big|_{t=0}\,=\,4\!\,af\,,
\,\,\,\,\,\,\,\,\,\,\,\,\,\,\,\,\,\,\,\,
\frac{d^2\,V_{2i}}{d\ph^2}\Big|_{t=0}\,=\,-2\,b_i\xi_i P_i\,,
$$ %%% \label{inicon1}
%%%%%% \eeq
the solution of the equations (\ref{eqs}) can be easily found 
in the form:
\beq
\frac{d^4\,V_1}{d\ph^4}\,=\,4\!\,af(t)\,\si^4(t)
\,,\,\,\,\,\,\,\,\,\,\,\,\,\,\,\,\,\,\,\,\,\,\,\,\,\,\,\,\,\,
\frac{d^2\,V_{2i}}{d\ph^2} \,=\,-2\,b_i\,\xi_i(t)\,\si^2(t)\,P_i\,,
\label{inicon}
\eeq
where
$$
\si(t) = \exp \left\{
\,-\int_0^t {\bar \ga}\,[P(t^\prime)]\,dt^\prime \right\}\,.
$$
The effective charges $P(t)\equiv (\,f(t),\,\xi_i(t),\,...\,)$ 
satisfy the renormalization group equations \cite{book}
\beq
{\dot P}(t) = {\bar \be}_P(t)\,,\,\,\,\,\,P(0)=P
\,,\,\,\,\,\,\,\,\,\,\,\,\,{\rm where}\,\,\,\,\,\,\,\,\,\,\,
{\bar \be}_P = \frac{\be_P}{1+\ga}
\,\,\,\,\,\,\,\,\,\,\,{\rm and}\,\,\,\,\,\,\,\,\,\,\,
{\bar \ga} = \frac{\ga}{1+\ga}\,.
\label{charges}
\eeq

In the one-loop approximation one has to take the linear dependence 
on $t$, so that $P(t)=P+\be_P\cdot t$. Then equations (\ref{inicon})
become
$$
\frac{d^4\,V_1}{d\ph^4}\,=\,4\!\,a \left[\,
f + \frac12\,(\be_f - 4\,f\ga)\,{\rm ln}\,\frac{\ph^2}{\mu^2}
\,\right] \,,
$$
\beq
\frac{d^2\,V_{2i}}{d\ph^2}\,=\,-\,2\,b_i\,
\left[\,\xi_i + 
\frac12\,(\be_{\xi_i} - 2\,\xi_i\ga)
\,{\rm ln}\,\frac{\ph^2}{\mu^2}\,\right]
\,.
\label{pochtisoluef}
\eeq
Integrating these equations with the renormalization (boundary) 
conditions
$$ %%%%  \beq
\frac{d^2\,V_1}{d\ph^2}\Big|_{\ph=0} \,=\, 0
\,,\,\,\,\,\,\,\,\,\,\,\,\,\,\,\,\,\,\,
\frac{d^4\,V_1}{d\ph^4}\Big|_{\ph=\mu} \,=\, 4\!\,a\,f
\,,\,\,\,\,\,\,\,\,\,\,\,\,\,\,\,\,\,\,
\frac{d^2\,V_{2i}}{d\ph^2}\Big|_{\ph=\mu}\,=\,-\,2\,b_i\xi_i\,P_i
\,,
$$ %%%%  \label{bondar}
%%%%%%%  \eeq
we arrive at the final expression for the one-loop effective
potential
\beq
V = a\,f\, \ph^4\, + A\,\Big(\,{\rm ln}\frac{\ph^2}{\mu^2}
-\frac{25}{6}\,\Big)\,\ph^4
- \,\sum_{i=1}^{5}\, \left[\,b_i \xi_i + B_i
\Big(\,{\rm ln}\frac{\ph^2}{\mu^2}-3\,\Big)\,\right]
\,P_i\,\ph^2\,,
\label{efpotya}
\eeq
with the coefficients
\beq
A = \frac{a}{2}\,( \be_f - 4\,f\ga ) 
\,\,\,\,\,\,\,\,\,\,\,\,\,\,\,\,\,
{\rm and} \,\,\,\,\,\,\,\,\,\,\,\,\,\,\,\,\,
B_i = \frac12\,b_i\,(\be_{\xi_i} - 2\,\xi_i\ga )\,.
\label{coefs}
\eeq
The eq. (\ref{efpotya}) is the general expression for the 
one-loop effective potential in the linear in $P_i$
approximation. One can substitute the values of $a$ and $b$ 
from some classical theory, together with the corresponding $\beta$ 
and $\gamma$ functions, and derive the quantum corrections using  
(\ref{efpotya}) and (\ref{coefs}). The gauge fixing dependence 
enters the effective potential through the $\gamma$-function
of the scalar field. We remark that, in general, the problem of 
gauge dependence of the effective potential is not simple to 
solve. It was discussed, for instance, in \cite{gauge-dep}. 
In the case of $f \sim g^4$ and $\xi_i \sim g^2$ the gauge fixing 
dependence goes beyond the one-loop approximation and the 
corresponding ambiguity disappears. These relations are direct 
analogs of the one which has been used in \cite{colwe} for the
similar effective potential on flat background without torsion. 

Let us present, as an example, the explicit expression for the 
effective potential for the theory (\ref{su(2)}) with $m+n=10$
and $m=1$. The necessary 
$\beta$-functions are given in (\ref{non-min}). The $\ga$-function
can be easily calculated to be 
$$
\ga(\al)=\frac34\,\al g^2 -\frac94\,g^2 + 2h^2\,.
$$
Then, using the general formulas (\ref{efpotya}), (\ref{coefs})
one can easily derive the effective potential 
for the theory  (\ref{su(2)}) 
\footnote{Some small misprints of \cite{bosh-pot} are corrected 
here.}.
$$
V = \frac{f}{8}\,(\phi^{\dagger}\phi)^2 
- \sum_{i=1}^{5}\,\xi_iP_i\,\phi^{\dagger}\phi 
+ \frac{3f^2+9g^4-32h^4-3\al fg^2}{(4\pi)^2}\,\phi^{\dagger}\phi\,
\Big(\,{\rm ln}\frac{|\phi|^2}{\mu^2}-\frac{25}{6}\,\Big) -
$$
$$
- \frac{1}{2(4\pi)^2}\,\phi^{\dagger}\phi\,
\Big(\,{\rm ln}\frac{|\phi|^2}{\mu^2}-3\,\Big) \,\cdot\,
\Big[\,
\frac32\,(f-\al g^2)\,\sum_{i=1}^{5}\,\xi_iP_i - \frac16\,
\left(\frac32\,f+ 4h^2-\frac92\,g^2\right)\,P_1 -
$$
\beq
- h^2(\de_2-\eta_2)^2\,P_3 - h^2(\de_1+\eta_1)^2\,P_4
\,\Big]\,.
\label{efposu2}
\eeq
\vskip 3mm

Consider, using the general expressions (\ref{efpotya}), 
(\ref{coefs}) the spontaneous symmetry breaking. It is easy to see 
that, for the $\,\sum_{i=1}^{5}\,b_i\xi_i\, P_i > 0$ case, the
classical potential (\ref{clpo}) has a minimum at 
\beq
\ph_0^2 = \frac{1}{2af}\,\sum_{i=1}^{5}\,b_i\xi_i\, P_i \,,
\label{ssb}
\eeq
so that the spontaneous symmetry breaking might 
occur, at the tree level, without having negative 
mass square. It is not difficult to take into 
account the quantum corrections to $\ph_0^2$, just solving 
the equations $\frac{\pa V}{\pa \ph}=0$ iteratively. For 
instance, after the first step one obtains
$$ %%%  \beq
\ph_1^2= \ph_0^2 
- \frac{A}{af}\,\ph^2_0\,\Big(\,{\rm ln}\frac{\ph_0^2}{\mu^2}
-\frac{11}{3}\,\Big) + \frac{1}{2af}\,
\sum_{i=1}^{5}\,B_i\, P_i\,\Big(\,{\rm ln}\frac{\ph_0^2}{\mu^2}
- 2\,\Big) \,.
$$ %%%  \label{ssb1}
%%%%%%  \eeq

One may consider a marginal case, when 
$\,\sum b_i\xi_i\, P_i < 0$, so that there is
no spontaneous symmetry breaking at the tree level. Suppose also 
that the $\xi_i \approx 0$, so that the
absolute value of the sum $\,\sum b_i\xi_i\, P_i$
is very small, and that $\,\sum b_i\,\beta_{\xi_i}\, P_i > 0$, 
such that the sign of $\,\sum b_i\xi_i\, P_i$ changes under the 
quantum corrections. Approximately, $B_i=\frac12\,b_i\,\beta_{\xi_i}$.
Then, from the equation $\frac{\pa V}{\pa \ph}=0$ one gets
\beq
\ph^2 - \sum_{i=1}^{5}\frac{b_i\,\beta_{\xi_i}}{2af}\,P_i
-\frac{11}{3}\,\frac{A}{af}\,\ph^2 + 
{\rm ln}\frac{\ph^2}{\mu^2}\,\Big(\,\frac{A}{af}\,\ph^2
-\frac{1}{4af}\,\sum_{i=1}^{5}\,b_i\,\beta_{\xi_i}\,P_i\Big)\,.
\label{ssb2}
\eeq
One can denote the positive expression
$$
\ph_0^2 = \frac{1}{2af}\,\sum_{i=1}^{5}\,b_i\,\beta_{\xi_i}\,P_i\,.
$$ 
Then, after the first iteration eq. (\ref{ssb2}) gives
\beq
\ph_1^2= \ph_0^2 + \frac{11}{3}\,\frac{A}{af}\,\ph^2_0
-{\rm ln}\frac{\ph_0^2}{\mu^2}\,
\Big(\,\frac{A}{af}\,\ph_0^2
-\frac{1}{4af}\,\sum_{i=1}^{5}\,b_i\,\beta_{\xi_i}\,P_i\,\Big)\,.
\label{ssb3}
\eeq
It is easy to see, that in this case the spontaneous symmetry 
breaking emerges only due to the quantum effects. In all the 
cases: classical or quantum, the effect of spontaneous symmetry 
breaking is produced by expressions like $b_i\,\xi_i\,P_i$
or $b_i\,\beta_{\xi_i}\,P_i$. In particular, the effect can be 
achieved only due to the torsion, without the Ricci curvature 
scalar $P_1=R$. 

Let us now investigate the possibility of phase transitions
induced by curvature and torsion. We shall be interested in the 
first order phase transitions, when the order parameter $<\ph>$
changes by jump. It proves useful to introduce the dimensionless
parameters $x=\ph^2/\mu^2$ and $y_i=P_i/\mu^2$. The equations
for the critical parameters $x_c,\,y_{ic}$ corresponding to the 
first order phase transition, are \cite{kirj}:
\beq
V(x_c,\,y_{ic})=0\,,\,\,\,\,\,\,\,\,\,\,\,\,\,\,\,\,\,\,\,\,
\frac{\pa V}{\pa x}\Big|_{x_c,\,y_{ic}}=0
\,,\,\,\,\,\,\,\,\,\,\,\,\,\,\,\,\,\,\,\,\,
\frac{\pa^2 V}{\pa x^2}\Big|_{x_c,\,y_{ic}} > 0\,.
\label{phase}
\eeq
These equations lead to the following conditions:
$$
2A^2=-\sum_{i=1}^{5}\,q_i\,\vp_i\,D_i \pm 
\left[\sum_{i,j}(D_iD_j-4A^2B_iB_j)\vp_i\vp_j q_i g_j\right]^{1/2}\,,
$$
\beq
af - \frac83\,A + A\,{\rm ln}x -\frac12\,
\sum_{i=1}^{5}\,B_i\vp_i q_i > 0\,,
\label{usl}
\eeq
where
\beq
D_i=Ab_i\xi_i-afB_i-\frac56\,Ab_i
\,,\,\,\,\,\,\,\,\,\,\,\,\,\,\,\,\,\,\,\,\,\,
\vp_i=sign P_i, 
\,,\,\,\,\,\,\,\,\,\,\,\,\,\,\,\,\,\,\,\,\,\,
q_i = \frac{y_{ic}}{x}\,.
\label{D_i}
\eeq
Besides (\ref{usl}), the quantities $q_i$ have to satisfy the 
conditions $0<q_i\ll 1$. The last inequality means that
our approximation $\ph^2\gg |P_i|$ is valid. 

In order to analyze the above conditions one has to implement such 
relations for the parameters that the result would be gauge fixing 
independent. Let us, for this end, take the relation 
$af = \frac{11}{3}\,A$,
as it has been done in \cite{colwe}. Then $f \sim g^4$. 
Consider the special case $|\xi_i|\ll g^2$. Then from (\ref{usl})
follows \cite{bosh-pot} 
$D_i\approx -\frac92\,AB_i$, $\,q_i\approx \frac{A}{4.3B_i\vp_i}$. 
We notice that in the first of (\ref{usl}) one has to take 
positive sign, otherwise the $q_i\ll 1$ condition does not hold.
In this approximation $A>0$, $B_i>0$, therefore one has to take 
all $\vp_i>0$. As we see, the theory admits the 
first order phase transition, which may be induced by curvature
and (or) torsion. In fact, there are other possibilities, for 
instance where all or part of the non-minimal parameters satisfy 
opposite relations  $|\xi_i|\gg g^2$ \cite{bosh-pot}. We 
will not present the discussion of these possibilities here. 
An important observation is that, in the point of minimum,
the effective potential generates the induced gravity with 
torsion: 
\beq
S_{ind} = - \int d^4x\sqrt{-g}\,V(\ph_c) =
- \int d^4x\sqrt{-g}\,\left\{\,\,
\La_{ind} - \frac{1}{16\pi G_{ind}}\,
\sum_{i}\,\th_i^{(ind)} P_i\,\,\right\}\,,
\label{ECind}
\eeq
with 
\beq
\La_{ind} = a\,f\, \ph_c^4\, 
+ A\,\left[\,{\rm ln}\frac{\ph_c^2}{\mu^2}
-\frac{25}{6}\,\right]\,\ph_c^4
\label{lambda}
\eeq
and 
\beq
- \frac{1}{16\pi G_{ind}}\,\th_i^{(ind)} 
\,= \,\sum_{i=1}^{5}\, \left[\,b_i \,\xi_i + B_i
\Big(\,{\rm ln}\frac{\ph_c^2}{\mu^2}-3\,\Big)\,\right]
\,\ph_c^2\,.
\label{indu}
\eeq
It is reasonable to take $\th_i^{(ind)}=1$, 
and then other $\th_i^{(ind)}$
will give the coefficients in the induced analog of the 
Einstein-Cartan action (\ref{EC}). 
Since these coefficients depend on the non-minimal 
parameters $\xi_i$, and 
these parameters have different scale dependence (see, for example, 
(\ref{non-min})), the coefficients of the induced action 
are, in general, 
different from the ones in (\ref{EC}), which correspond to the 
$\int\sqrt{-g}{\tilde R}$-type action of the Einstein-Cartan theory. 

%%%%%%%%%%%%%%%%%%%%%%%%%%%%%%%%%%%%%%%%%%%%%%%%%%%%%%%%%%%%
\section{Conformal anomaly in the spaces with torsion. 
Trace anomaly and modified trace anomaly}
%%%%%%%%%%%%%%%%%%%%%%%%%%%%%%%%%%%%%%%%%%%%%%%%%%%%%%%%%%%%

As we have already learned in Section 2.4, three different 
types of local conformal symmetry are possible in the theory
of gravity with torsion. Consequently, one  
meets different versions of conformal anomaly, which 
violates these symmetries at the quantum level. In this 
section, we
shall consider only the vacuum sector, and just notice 
that the trace anomaly in the matter field sector with 
torsion (which requires the renormalization
of composite operators similar to the one performed in 
\cite{brawn-collins}) has not been performed yet. On the
other hand, this anomaly would not be very different 
from the one in the purely metric theory, and
the vacuum effects look much more interesting. 

Let us start from the anomaly corresponding to the 
{\it week conformal symmetry} \cite{buodsh}. In this 
case, torsion does not transform and the Noether identity, 
in the vacuum sector, is just 
the same as in the purely metric gravity:
\beq
T_\mu^\mu = - \frac{2}{\sqrt{-g}}\,g_{\mu\nu}\,
\frac{\de S}{\de g_{\mu\nu}} = 0\,.
\label{trace-class-weak}
\eeq
The last identity indicates that the vacuum action can be 
chosen to be conformal invariant 
\footnote{In fact, this is true (exactly as in the 
purely metric theory) only in the one-loop approximation. 
At higher loops, the non-minimal parameter $\xi_1$ of the 
scalar-curvature interactions departs from the one-loop
conformal fixed point \cite{brv}. As a result, in order to 
preserve the renormalizability, one needs, strictly speaking, 
a non-conformal vacuum action. But, as it was noticed 
in \cite{brv}, the coefficients in front of the non-conformal
terms can be safely kept very small, and one can always
consider the conformal invariant vacuum action as a very 
good approximation. In order to avoid the discussion 
of this issue, we consider, in this section, the vacuum 
effects of the free matter fields.}. 
The vacuum action may include the non-conformal 
terms, but their renormalization is
not necessary in the case of conformal invariant free
massless fields. But these terms may be,
indeed, important from other points of view. 
In particular, one can include the Einstein-Cartan action
into $S_{vacuum}$, and treat the anomaly-induced 
effective action as quantum correction to the classical 
action of gravity with torsion at the very high energy, when 
the particle masses are negligible. 

Consider the anomaly in 
the identity (\ref{trace-class-weak}). We shall use the
dimensional regularization, which is the most useful for
this purpose \cite{ddi,duff}. The divergent part of the 
one-loop effective action that emerges after integrating 
over one real scalar and one Dirac spinor, have 
the form (\ref{scalardivs}) and (\ref{contra}). If we 
restrict consideration by the case of a purely 
antisymmetric torsion, the result for the 
$N_0$ real scalars, $N_{1/2}$ Dirac spinors and 
$N_1$ gauge bosons will be
$$
{\bar \Ga_{div}}(N_0, N_{1/2}, N_1) \,=\, 
-\, \frac{\mu^{n-4}}{\vp}\,\int d^n x\sqrt{-g}\,\Big\{ 
\,\left( \frac{N_0}{120} + \frac{N_{1/2}}{20}
+ \frac{N_1}{10} \right) C^2 -
$$
$$
- \left( \frac{N_0}{360} + \frac{11\,N_{1/2}}{360}
+ \frac{31\,N_1}{180}\right) E + 
 \left( \frac{N_0}{180} + \frac{N_{1/2}}{30}
- \frac{N_1}{10}\right) {\Box} R 
- \frac{2\eta^2\,N_{1/2}}{3}\,S_{\mu\nu}S^{\mu\nu} +
$$
$$
+\frac{N_0}{2}\,\xi^2\,(S^\mu S_\mu)^2
+\left(\,\frac{4N_{1/2}}{3}\,\eta^2-\frac{N_0}{6}\,\xi
\,\right){\Box}(S^\mu S_\mu) 
- \frac{4N_{1/2}}{3}\,\eta^2\,
\na_\mu(S_\nu\na^\nu S^\mu - S^\mu\na_\nu S^\nu) \}=
$$
\vskip 2mm
$$
=- \frac{\mu^{n-4}}{\vp}\,\int d^n x\sqrt{-g}\,\{ 
aC^2+bE+c{\Box}R+dS^2_{\mu\nu}+eS^4+f{\Box}S^2+
$$
\beq
+g\na_\mu(S_\nu\na^\nu S^\mu - S^\mu\na_\nu S^\nu)\,. 
\label{1d}
\eeq

The standard arguments show that the one-loop effective 
action of vacuum is conformal invariant before the 
local counterterm 
$\De S$ is introduced \cite{duff}. Consider the general 
expression for the one-loop effective action 
\beq
\Ga = S + {\bar \Ga} + \De S\,,
\label{total}
\eeq
where ${\bar \Ga}$ is the quantum correction to the 
classical action and $\De S$ is an infinite 
local counterterm 
which is called to cancel the divergent part of 
${\bar \Ga}^{(1)}$. Then, the anomalous trace is
\beq
T = <T_\mu^\mu> = - \frac{2}{\sqrt{-g}}\,g_{\mu\nu}
\frac{\de}{\de g_{\mu\nu}} {\bar \Ga}\Big|_{n=4}
= - \frac{2}{\sqrt{-g}}\,g_{\mu\nu}
\frac{\de}{\de g_{\mu\nu}} \De S\Big|_{n=4}
\label{trace}
\eeq
The most simple way of calculating this expression 
is to perform the local conformal transformation 
\beq
{g}_{\mu\nu} = {\bar g}_{\mu\nu}\cdot e^{2\si}
\,,\,\,\,\,\,\,\,\,\,\,\,
\si = \si(x)
\,,\,\,\,\,\,\,\,\,\,\,\, 
\det({\bar g}_{\mu\nu})=const
\label{conf}
\eeq
and use the identity
\beq
 - \frac{2}{\sqrt{-g}}\,g_{\mu\nu}
\frac{\de}{\de g_{\mu\nu}}\,A[g_{\mu\nu}]
= - \frac{1}{\sqrt{-{\bar g}}}\,e^{- 4\si}
\frac{\de}{\de \si}\,A[{\bar g}_{\mu\nu}\,e^{2\si}]
\,\Big|_{{\bar g_{\mu\nu}}\rightarrow g_{\mu\nu},
\si\rightarrow 0}\,.
\label{deriv}
\eeq
When this operator acts on 
$$
\De S = \frac{\mu^{n-4}}{\ep}\,
 \int d^nx\sqrt{-{\bar g}}\,e^{(n-4)\si}\cdot\biggl(
\,a{\bar C}^2 + ...\biggr)\,,
$$ 
the $\,1/(n-4)$-factor 
cancels and we immediately arrive at the expression
\beq
T = - \frac{1}{(4\pi)^2}\,
\biggr[\,aC^2+bE+c{\Box}R+dS^2_{\mu\nu}+eS^4+f{\Box}S^2
+g\na_\mu(S_\nu\na^\nu S^\mu - S^\mu\na_\nu S^\nu)\,\biggr]\,, 
\label{anomaly}
\eeq
with the same coefficients $a,b,...,g$ as in (\ref{1d}). 
The derivation of the anomaly for the general torsion case
can be done in the same way \cite{thesis}. The important 
thing is that for the case of the {\large\it week conformal 
symmetry} all components of torsion $S_\mu, T_\mu$
and $q^\al_{\,\cdot\be\ga}$ do not transform.
\vskip 2mm

The calculation of the anomaly for the case of {\it strong 
conformal symmetry} always reduces to the one for the 
{\it weak conformal symmetry}.
As it was already noticed in section 2.4, the Noether identity 
(\ref{Noether22}) separates into two independent identities: one 
of them is (\ref{trace-class-weak}), and second simply requests
that the actions does not depend on the torsion trace $T_\mu$. 
As it was mentioned in section 2.4, the second identity can not 
be violated by the anomaly, and we are left with 
(\ref{trace-class-weak}) and with the corresponding 
anomaly (\ref{anomaly}).
\vskip 2mm

Let us now consider the most interesting case of the 
{\it compensating conformal symmetry}, which 
will lead us to the modified trace anomaly. We shall
follow Ref. \cite{anhesh}. This version of conformal 
symmetry depends on the torsion trace $T_\mu$ and on the 
non-minimal interaction of this trace with 
scalar. In the spinor sector the symmetry requires
that there is no any interaction with $T_\mu$, so that
$\eta_2=0$. Therefore, we can restrict our consideration 
to the case of a single scalar field. 

It is easy to see that the Noether identity corresponding 
to the symmetry (\ref{mmm}) looks as follows: 
$$ %%%  \beq 
\,2 g_{\mu\nu}\,\frac{\de
S_t}{\de g_{\mu\nu}} +\frac{\xi_2}{\xi_3}\,\pa_\mu\,\frac{\de
S_t}{\de T_\mu} - \ph\,\frac{\de S_t}{\de \ph}  \,=\,0 \,.
$$ %%%  \label{neuth} 
%%%%%%  \eeq 
Then, due to the conformal invariance of the vacuum
divergences (coefficient at the $n=4$ pole), 
the vacuum action may be chosen in such a way
that 
\beq 
-\sqrt{-g}\,{\cal T}\, = 
\,2 g_{\mu\nu}\,\frac{\de S_{vac}}{\de g_{\mu\nu}}
+\frac{\xi_2}{\xi_3}\,\pa_\mu\,\frac{\de S_{vac}}{\de T_\mu}
  \,=\,0 \,.
\label{vacu}
\eeq
The new form (\ref{vacu}) of the conformal Noether
identity indicates the modification of the
conformal anomaly. In the theory under discussion, the
anomaly would mean $\,<{\cal T}>\,\neq\,0$ instead
of usual $\,<{T}^\mu_\mu>\,\neq\,0$. Therefore,
we have a special case here and one can not
directly use the relation between the one-loop
counterterms and the conformal anomaly derived above,
just because this relation does not
take into account the non-trivial transformation law
for the torsion field. 

One can derive this new anomaly directly, using 
the same method as before. However, it is possible to
find $\,<{\cal T}>\,$ in a more economic way, after 
performing a special decomposition of the background fields.
Let us try to change the background variables in such
a way that the transformation of torsion is absorbed by
that of the metric. The crucial observation is that ${\cal P}$, 
from (\ref{P}), transforms, under (\ref{mmm})
\footnote{As a
consequence, the action $\,\int\sqrt{-g}{\cal P}\phi^2\,$ 
is conformal invariant. This fact has been originally 
discovered in \cite{obukhov}.},
as ${\cal P}^\prime = {\cal P} \cdot e^{-2\si(x)}$. 
The non-trivial transformation of
torsion is completely absorbed by ${\cal P}$. 
Since ${\cal P}$ only depends on
the background fields, we can present it in any
useful form. One can imagine, for instance, ${\cal P}$
to be of the form $\,{\cal P}=g^{\mu\nu}\,\Pi_\mu\,\Pi_\nu\,$ 
where the vector $\,\Pi_\mu\,$
doesn't transform, exactly as the axial vector $S_\mu$. 
After that, the calculation readily reduces to the case of an 
antisymmetric torsion (\ref{anomaly}), described above. 

 For the single scalar, the 1-loop divergences have the form
 \bea
 \Ga^{(1)}_{div} = - \frac{\mu^{n-4}}{(4\pi)^2\,(n-4)}\,
 \int d^nx\sqrt{-g}\,\left\{\frac{1}{120}\,C^2-\frac{1}{360}\,E
 + \frac{1}{180}\,{\Box}R + \frac{1}{6}\,{\Box} {\cal P}
 + \frac{1}{2}\,{\cal P}^2 \right \}\,.
 \label{div}
 \eea
 Taking into account the arguments presented
 above, one can immediately cast the anomaly under the form
 \bea
 <{\cal T}> \,=\, - \frac{1}{(4\pi)^2}\,\left[
 \frac{1}{120}\,C_{\mu\nu\al\be}\,C^{\mu\nu\al\be}
 -\frac{1}{360}\,E
 + \frac{1}{180}\,{\Box}R + \frac{1}{6}\,{\Box} {\cal P}
 + \frac{1}{2}\,{\cal P}^2 \right]\,.
 \label{tracean}
 \eea

%%%%%%%%%%%%%%%%%%%%%%%%%%%%%%%%%%%%%%%%%%%%%%%%%%%%%%%%%%%%
\section{Integration of conformal anomaly
and anomaly-induced effective actions of vacuum.
Application to inflationary cosmology}
%%%%%%%%%%%%%%%%%%%%%%%%%%%%%%%%%%%%%%%%%%%%%%%%%%%%%%%%%%%%

One can use the conformal anomaly to restore the induced 
effective action of vacuum. This action can be regarded as 
a quantum correction to the classical gravitational action.
We notice, that the induced action proved to be the best 
tool in the analysis of anomaly, see e.g. \cite{balbi,anju},
including the theory with torsion \cite{buodsh}.

The equation for the finite part of the 1-loop correction 
${\bar \Ga}$
to the effective action can be obtained from anomaly.
Let us consider, first, the weak conformal symmetry. Then,
\beq
- \frac{2}{\sqrt{-g}}\,g_{\mu\nu}
\frac{\de\, {\bar \Ga} }{\de g_{\mu\nu}} \,=\, T\,.
\label{maineq}
\eeq
In the case of purely metric gravity this equation has 
been solved in \cite{rei,frts}. For the 
torsion theory with the weak conformal symmetry the 
solution has been found in \cite{buodsh} (see also 
\cite{book}). Finally, for the most complicated case of the 
compensating conformal symmetry, the problem has been 
solved in \cite{anhesh}.

We start from the case of purely antisymmetric torsion,
corresponding to the strong conformal symmetry. 
The simplest possibility in solving (\ref{maineq})
is to divide the metric in two parts: the conformal factor 
$\sigma(x)$ and the fiducial metric ${\bar g}_{\mu\nu}(x)$ with 
fixed determinant (\ref{conf}), 
and write the (\ref{mainequation}) via (\ref{deriv}).
Since torsion does not transform, we put $S_\mu = {\bar S}_\mu$. 
Then we get \cite{buodsh,book}
$$
{\bar \Ga} = S_c[{\bar g}_{\mu\nu},\,{\bar S}_\mu] + 
\frac{1}{(4\pi)^2}\,
\int d^4 x\sqrt{-{\bar g}}\,\{ 
a\si {\bar C}^2 + b\si ({\bar E}-\frac23 {\bar {\na}}^2
{\bar R}) + 2b\si{\bar \De}_4\si +
$$$$
+ d\si {\bar S}_{\mu\nu}^2 + e\si ({\bar S}_\mu 
{\bar S}^\mu)^2+(f+g/2){\bar S}^2({\bar \na}\si)^2 
+ g ({\bar S}^\mu {\bar \na}_\mu\si)^2 
- g{\bar \na}_\mu \si\,({\bar S}_\nu {\bar \na}^\nu {\bar S}^\mu
-{\bar S}^\mu  {\bar \na}_\nu {\bar S}^\nu)-    
$$
\beq
-f\,{\bar \na}_\mu \si\,{\bar \na}^\mu {\bar S}^2 
- \frac{1}{12}\,(c+\frac23 b)[{\bar R} - 6({\bar \na}\si)^2 - 
({\bar \Box} \si)]^2\,\} \,
+\,S_c[{\bar g}_{\mu\nu},{\bar S}_\mu]\,,
\label{quantum}
\eeq
where $S_c[{\bar g}_{\mu\nu},{\bar S}_\mu]\,$ 
is an unknown functional of 
the metric and torsion which serves as an 
integration constant for any solution of (\ref{maineq}).
Indeed, if one succeeds to rewrite (\ref{quantum}) in terms 
of the original variables $\,g_{\mu\nu},\,S_\mu$, the action
$S_c[{\bar g}_{\mu\nu},{\bar S}_\mu]\,$ must be 
replaced by an arbitrary conformal-invariant functional 
of these variables. It is, in principle, possible to proceed 
and, following \cite{rei}, derive the covariant form of the 
induced action (\ref{quantum}). This action contains, 
exactly as in the torsion-less case, the local and non-local 
pieces \cite{ddi,rei}.

The action  (\ref{quantum}), 
being the quantum correction to the 
Einstein-Cartan theory, can serve as a basis for the 
non-singular cosmological model with torsion. This model 
has been constructed in \cite{buodsh} (see also \cite{book}). 
Without going into technical details, we just summarize 
that, for the 
conformally flat metric $g_{\mu\nu} = \eta_{\mu\nu}\,a^2$
and isotropic torsion axial vector $S_{\mu}=(T,0,0,0)$,
the dynamical equations have approximate classical 
solution of the form (in physical time)
\beq
a(t)=a(0)\,e^{Ht}\,,\,\,\,\,\,\,\,\,\,\,\,\,\,\,\,\,
T(t)=T(0)\,e^{-2Ht}\,.
\label{nashi_starye_dobrye_dostijenia}
\eeq
This solution has an obvious physical interpretation:
torsion exponentially decreases during inflation, and 
that is why it is so weak today. Of course, this 
concerns only specific background torsion, but the
result is indeed relevant for the cosmological 
applications of torsion. 
%%%%%%%%%%%%%%%%%%%%%%%%%%%%%%%%%%%%%%%%%%%%%

Let us now derive the 
conformal non-invariant part of the effective action of 
vacuum, which is responsible for the modified conformal 
anomaly (\ref{tracean}). Taking into account our previous
treatment of the compensating conformal transformation of 
torsion, we consider 
it hidden inside the quantity ${\cal P}$ of eq. (\ref{P}),
and again imagine ${\cal P}$ to be of the form
$\,{\cal P}=g^{\mu\nu}\,\Pi_\mu\,\Pi_\nu\,$. Then,
the equation for the effective action
$\Ga [g_{\mu\nu},\Pi_\al]\,$ is
\footnote{We remark that this equation is valid only for
the "artificial" effective action
$\Ga [g_{\mu\nu},\Pi_\al]\,$, while the effective
action in original variables $g_{\mu\nu}, T^\al_{\be\ga}$
would satisfy the modified equation (\ref{vacu}). The
standard form of the equation (\ref{mainequation})
for the effective action is achieved only through
the special decomposition of the external fields.}
\beq
- \frac{2}{\sqrt{-g}}\,g_{\mu\nu} \frac{\de\,\Ga}{\de
g_{\mu\nu}} {} = <{\cal T}>\,.
\label{mainequation}
\eeq
In order
to find the solution for $\,\Ga$, we can factor out the conformal
piece of the metric ${g}_{\mu\nu} = {\bar g}_{\mu\nu}\cdot
e^{2\si}$, where ${\bar g}_{\mu\nu}$ has fixed determinant
and put $\,{\cal P} = {\bar {\cal P}}\cdot e^{-2\si(x)}$, that
corresponds to $\,{\bar \Pi}_\al = \Pi_\al$.
The result can be obtained directly from the
effective action derived in \cite{buodsh}, and we get
$$
{\Ga}\, = \,S_c[{\bar g}_{\mu\nu}; {\bar {\cal P}}]
\,- \,\frac{1}{12}\cdot\frac{1}{270 (4\pi)^2}\,\int d^4 x
\sqrt{-g (x)}\,R^2(x) + \frac{1}{(4\pi)^2}\,
\int d^4 x\sqrt{-{\bar g}}\,\left\{
\,\si\, \Big[ \frac{1}{120}\,{\bar C}^2 -
\right.
$$
\beq
\left.
- \frac{1}{360}\,({\bar E} - \frac23\,{\bar \na}^2 {\bar R})
+ \frac12\,{\bar {\cal P}}^2 ] +
 \frac{1}{180}\,\si {\bar \De}\si
-\frac16\,({\bar \na}_\mu\si)\,{\bar \na}^\mu {\bar {\cal P}}
+\frac16\,{\bar {\cal P}}({\bar \na}_\mu\si)^2\,\right\}\,,
\label{efac-conforma}
\eeq
where $S_c[{\bar g}_{\mu\nu}; {\bar {\cal P}}]\,$ is an unknown
functional of the metric  ${\bar g}_{\mu\nu}(x)$ and 
$\,{\bar {\cal P}}\,$,
which acts as an integration constant for any solution of
(\ref{mainequation}).

Now, one has to rewrite (\ref{efac-conforma}) in
terms of the original field variables, $g_{\mu\nu}, {T^\al}_{\be\ga}$.
Here, we meet a small problem, because we only have, for the moment,
the definition $\Pi_\al = {\bar \Pi}_\al$ for the artificial variable
$\Pi_\al$, but not for the torsion. Using the previous result
(\ref{mmm}), we can define
$$ 
 {T^\al}_{\be\ga} = {{\bar T}^\al}_{\,\,\be\ga}
-\frac{1}{3}\cdot\left[\,\de^\al_\ga\,\pa_\be\si -
\de^\al_\be\,\pa_\ga\si\,\right]\,, 
$$ 
where
$\,{{\bar T}^\al}_{\,\,\be\ga}\,$ is an arbitrary tensor. Also, we
call ${\bar T}^\al = {\bar g}^{\al\be}\,{\bar T}_\be$ etc. Now, we
can rewrite (\ref{efac-conforma}) in terms of 
metric and torsion components
$$ 
{\Ga}\, = \,S_c[{\bar g}_{\mu\nu}; {{\bar T}^\al}_{\,\,\be\ga}]
\,- \,\frac{1}{12}\cdot\frac{1}{270 (4\pi)^2}\,\int d^4 x \sqrt{-g
(x)}\,R^2(x) + 
$$$$ 
+ \frac{1}{(4\pi)^2}\, \int d^4 x\sqrt{-{\bar
g}}\,\left\{ \,+ \frac{1}{180}\,\si {\bar \De}\si +
\frac{1}{120}\,{\bar C}^2\,\si - \frac{1}{360}\,({\bar E} -
\frac23\,{\bar \na}^2 {\bar R})\,\si \right. 
$$$$ 
\left. +
\frac{1}{72}\,\si\,\left[ \,-\frac{\xi^2_2}{\xi_3}\,{\bar R} +
6\xi_2\,({\bar \na}_\mu {\bar T}^\mu) + 6\xi_3\,{\bar T}_\mu {\bar
T}^\mu  + 6\xi_4\, {\bar S}_{\mu}{\bar S}^{\mu} + 6\xi_5\,{\bar
q}_{\mu\nu\la}{\bar q}^{\mu\nu\la} \right]^2 + \right. 
$$ 
\beq
\left. \frac16\,\left[({\bar \na}^2\si\,+\, ({\bar
\na}_\mu\si)^2\,\right]\cdot \left[
\,-\frac{\xi^2_2}{\xi_3}\,{\bar R} + 6\xi_2\,({\bar \na}_\mu {\bar
T}^\mu) + 6\xi_3\,{\bar T}_\mu {\bar T}^\mu + 6\xi_4\, {\bar
S}_{\mu}{\bar S}^{\mu} + 6\xi_5\,{\bar q}_{\mu\nu\la}{\bar
q}^{\mu\nu\la} \right]\,\right\}\,. 
\label{efac-final} 
\eeq 
This
effective action is nothing but the generalization of the 
expression (\ref{quantum}) for the case of general
metric-torsion background and compensating conformal symmetry.
The curvature dependence in the last two terms appears
due to the non-trivial transformation law for torsion.
The physical interpretation of the action (\ref{efac-final})
coincide with the one of (\ref{quantum}) in case 
$T_\mu=q^\al_{\,\cdot\,\be\ga}=0$.  

From a technical point of view, eq. (\ref{efac-final}) is a very 
interesting example of an exact derivation of the anomaly-induced 
effective action for the case when the background 
includes, in addition to metric, another field with the
nontrivial conformal transformation. 

%%%%%%%%%%%%%%%%%%%%%%%%%%%%%%%%%%%%%%%%%%%%%%%%%%%%%%%%%%%
\section{Chiral anomaly in the spaces with torsion.
Cancellation of anomalies}
%%%%%%%%%%%%%%%%%%%%%%%%%%%%%%%%%%%%%%%%%%%%%%%%%%%%%%%%%%%%%%

Besides the conformal trace anomaly, in the theory with torsion
one can meet anomalies of other Noether identities. 
In particular, the systematic study of chiral anomalies has been 
performed in Refs. \cite{beja,adler-a,adba}. In many cases, due 
to the special content of the gauge theory, the anomaly cancel.
The most important particular example is the Standard Model
of particle physics, 
which is a chiral theory where left and right components of the 
spinors emerge in a different way. The violation 
of the corresponding symmetries could, in principle, lead to 
the inconsistency of the theory \cite{grja}. However, it does 
not happen in the SM, because the dangerous anomalies cancel.

The history of chiral anomaly in curved space-time with torsion 
started simultaneously with the derivation of divergences 
(always related to the $a_2(x,x^\prime)$-coefficient) for 
the corresponding fermion operator \cite{gold}. In this
paper, however, the explicit result has not been achieved 
because of the cumbersome way of calculation. Let us 
indicate only some of the subsequent calculations
\cite{kimura,obukhov-spectr,niya,yaki,cogzer1,coggia,doma1}
and other papers devoted to the closely related issues like index 
theorems, topological structures and Wess-Zumino \cite{wz}
conditions \cite{ombh,chza,mavr,yaj,PeeWa99}. We shall not 
go into details of these works but only present the most 
important and simple expression for the anomaly and give
a brief review of other results. 

For the massless Dirac fermion, there is the exact classical 
symmetry (\ref{trans}), and the corresponding Noether identity
is $\,\,\na_\mu\,J^\mu_5=0$, where (\ref{global-spin-cur})
$$
J^\mu_5 \,= \,{\bar \psi}\ga^5\ga^\mu\psi\,.
$$
The anomaly appears due to the divergences coming from the 
fermion loop. The mechanism of the violation of the Noether 
identity can be found in many books (for instance, in
\cite{collins,itszub}). However, the standard methods of 
calculating anomaly using Feynman diagrams are not very useful 
for the case of gravity with torsion. In principle, one can 
perform such calculations using the methods mentioned at the
beginning of this Chapter: either introducing external 
lines of the background fields, or using the local momentum 
representation. The most popular are indeed the functional 
methods (see, e.g. \cite{fuj}), which provide the covariance 
of the divergences automatically.

The investigation of the anomaly in curved space-time
using the so-called analytic regularization based on the 
Scwinger-DeWitt (Seeley-Minakshisundaram) expansion 
for the elliptic operator on the compact manifolds with 
positive-defined metric has been performed in \cite{romsch}. 
The anomaly can be calculated on the Riemann or Riemann-Cartan 
\cite{obukhov-spectr}) manifold with the Euclidean signature. 
The Euclidean rotation 
can be done in a usual way, and the result can be analytically 
continued to the pseudo-Euclidean signature. Then the vacuum 
average of the axial (spinor) current is 
$$ %%%% \beq
<J^\mu_5> \,=\, 
\frac{\int d{\bar \psi} d \psi\,J^\mu_5\,\,{\rm exp} 
\Big\{\,\int d^4x\sqrt{g}\,{\bar \psi} \ga^\mu D_\mu \psi\,\Big\}}
{\int d{\bar \psi} d \psi\,\,{\rm exp}\Big\{\,\int d^4x\sqrt{g}
\,{\bar \psi} \ga^\mu D_\mu \psi\,\Big\}}\,,
$$ %%%% \label{average}
%%%%%%% \eeq
where $D_\al = \na_\al + i\eta\ga_5\,S_\al$ is covariant 
derivative with the antisymmetric torsion. It can 
be easily reduced to the minimal covariant derivative
${\tilde \na}_\al = \na_\al - \frac{i}{8}\,\ga_5\,S_\al$,
but we will not do so, and keep $\eta$ arbitrary in order
to have correspondence with other sections of this Chapter. 

The vacuum average of the axial current divergence 
can be presented as \cite{romsch,obukhov-spectr} 
$$ %%%%%% \beq
<\na_\mu\, J^\mu_5> \,=\, \frac{\de}{\de \al(x)}\,
\frac{\int d{\bar \psi} d \psi\,\,{\rm exp} \Big\{\,
\int d^4x\sqrt{g}\,(\,{\bar \psi} \ga^\mu D_\mu \psi
-J^\mu_5\,\pa_\mu \al\,)\,\Big\}}
{\int d{\bar \psi} d \psi\,\,{\rm exp}\Big\{
\,\int d^4x\sqrt{g}\,
{\bar \psi} \ga^\mu D_\mu \psi\,\Big\}}\Big|_{\al=0}\,.
$$ %%%%%%%% \label{average-diver}
%%%%%%%%%%% \eeq
The analysis of \cite{romsch,obukhov-spectr} shows that
this expression is nothing but 
\beq
{\cal A} \,=\,
<\na_\mu\, J^\mu_5> \,=\, 2\,\lim_{x\to x^\prime}\,\tr \,\ga^5\,
a_2(x\,,\, x^\prime)\,,
\label{anom}
\eeq
where $a_2(x\,,\, x^\prime)$ is the second coefficient of the 
Schwinger-DeWitt expansion (\ref{a2..}). 
Applying this formula to the 
theory with torsion, that is using the expressions (\ref{a2}),
(\ref{P}), (\ref{S}), one can obtain the expression for the
anomaly in the external gravitational field with torsion
\cite{obukhov-spectr} 
\beq
{\cal A} \,=\,\frac{2}{(4\pi)^2}\,\Big[\,
\na_\mu {\cal K}^\mu 
+ \frac{1}{48}\,\vp_{\rho\si\al\be}
\,R^{\al\be}_{\,\cdot\,\cdot\,\mu\nu} \,R^{\mu\nu\rho\si}
+\frac16\,
\eta^2\,\vp^{\mu\nu\al\be}\,S_{\mu\nu}\,S_{\al\be}\,\Big]\,,
\label{anomal}
\eeq
where 
$$
{\cal K}^\mu \,=\,-\,\frac23\,\eta\,\Big(\,{\Box} 
+ 4\eta^2\,S_\la S^\la - \frac12\,R\,\Big)\,S^\mu\,.
$$
One can 
consider the anomalies of other Noether currents  
\cite{coggia} and obtain more general expressions.

From the physical point of view, the most important 
question is whether the presence of torsion preserves the
cancellation of anomalies in the matter sector. If this 
would not be so, the introduction of torsion could face 
serious difficulty. This problem 
has been investigated in \cite{coggia} and especially in 
\cite{doma1}. The result is that the presence of an
external torsion does not modify the anomalous divergence of 
the baryonic current
$\,\,J^\mu_B\,=\,\frac{1}{N_c}\,Q^\dagger\,\ga^\mu\,Q\,\,$
while it changes the leptonic current 
$\,\,J^\mu_L\,=\,L^\dagger\,\ga^\mu\,L\,\,$ by the expression
(\ref{anomal}). Independent on whether neutrinos are massless or 
massive, the cancellation of anomalies holds in the presence
of external torsion. In particular, weak external torsion does 
not affect the quantization of the SM charges \cite{doma1}. 
Therefore, the existence of the {\sl background} 
torsion does not lead to any inconsistency. On the other hand, the 
leptonic current gains some additional contributions, and this could,
in principle, lead to some effects like anisotropy of 
polarization of light coming from distant galaxies \cite{doma}.
However, taking the current upper bound on the background torsion 
from various experiments (see section 4.6)
one can see that the allowed magnitude of the background torsion 
is insufficient to explain the experimental
data which have been discussed in the literature \cite{nodrol}. 
%%%%%%%%%%%%%%%%%%%%%%%%%%%%%%%%%%%%%%%%%%%%%%%%%%%%%%%%%%%%%%

%%%%%%%%%%%%%%%%%%%%%%%%%%%%%%%%%%%%%%%%%%%%%%%%%%%%%%%%%%%%%%
%%%%%%%%%%%%%%%%%%%%%%%%%%%%%%%%%%%%%%%%%%%%%%%%%%%%%%%%%%%%%%
\chapter{Spinning and spinless particles and the 
possible effects on the classical background of torsion.}
%%%%%%%%%%%%%%%%%%%%%%%%%%%%%%%%%%%%%%%%%%%%%%%%%%%%%%%%%%%%%

The purpose of this Chapter
is to construct the non-relativistic approximation for the quantum
field theory on  torsion background and also develop the consistent
formalism for the  spinning and spinless particles on the 
torsion background. We shall follow the original papers
\cite{babush,rysh,gegi}. First we construct the non-relativistic 
approximation for the spinor field and particle, then use the
path integral method to construct the action of a relativistic 
particle. The action of a spin $1/2$
massless particle with torsion has been first established
in \cite{rumpf} on the basis of global supersymmetry. The same 
action has been rediscovered in \cite{RieHo96}, where it was
also checked using the index theorems. In \cite{PeeWa99}) 
the action of a massive particle has been obtained through the
squaring of the Dirac operator. This action does not possess
supersymmetry, and does not have explicit link to the supersymmetric
action of \cite{rumpf,RieHo96}.  In \cite{gegi} the action
of a spinning particle on the background of torsion and electromagnetic
field has been derived in the framework of the Berezin-Marinov
path integral approach. This action possesses local supersymmetry
in the standard approximation of weak torsion field and includes
the previous actions of \cite{rumpf,RieHo96,PeeWa99}) 
 as a limiting cases.  

In the last section of the Chapter we present 
a brief review of the possible physical effects and of the existing 
upper bounds on the magnitude of the background torsion from some
experiments. Since our purpose is to investigate the effects of 
torsion, it is better not to consider the effects of the metric. 
For this reason, in this and next sections we shall consider the
flat metric $g_{\mu\nu} = \eta_{\mu\nu}$.

%%%%%%%%%%%%%%%%%%%%%%%%%%%%%%%%%%%%%%%%%%%%%%%%%%%%%%%%%%%%%%%%%%
\section{Generalized Pauli equation with torsion}
%%%%%%%%%%%%%%%%%%%%%%%%%%%%%%%%%%%%%%%%%%%%%%%%%%%%%%%%%%%%%%%%%%

The Pauli-like equation with torsion has been (up to our knowledge)
first discussed in \cite{prosab}, and derived, in a proper way, 
in \cite{babush}. After that, the same equation has been obtained in
\cite{hammond,singh,lamme}, and the derivation of higher order 
corrections through the Foldy-Wouthuysen transformation with torsion 
has been done in \cite{rysh}. In \cite{lamme} the Foldy-Wouthuysen 
transformation has been applied to derive Pauli equation. 

The starting point is the 
action (\ref{diraconly}) of Dirac fermion in external 
electromagnetic and torsion fields.
$$
S = \int d^4 x\,\{\,i\bar{\psi}\gamma^\mu(\partial_\mu
+i\eta\gamma_5\,S_\mu+ie \,A_\mu) \psi+m\bar{\psi}\psi\,\}\,.
$$
The equation of motion for the fermion $\psi$ can be 
rewritten using the standard representation of the Dirac matrices
(see, for example, \cite{beresteckii})
$$
\beta = \gamma^0 = \left(\matrix{1 &0\cr 0 &-1\cr} \right)
\,,\,\,\,\,\,\,\,\,\,\,\,\,\,\,\,\,\,\,\,\,\,\,
\vec{\alpha} = \gamma^0 \vec{\gamma} =
\left(\matrix{0 &\vec{\sigma}\cr \vec{\sigma} & 0\cr} \right)\,.
$$
Here, as before, $\gamma_5 = -i\gamma^0 \gamma^1 \gamma^2 \gamma^3$.
The equation has the form
\beq
i \hbar\frac{\partial \psi}{\partial t} = \Big[ c\vec{\alpha}\vec{p}-
e\vec{\alpha}\vec{A} - \eta \vec{\alpha}\vec{S}\gamma_5
+ e\Phi + \eta \gamma_5 S_0 + m c^2 \beta \Big]\,\psi\,.
\label{eqDirac}
\eeq
Here, the dimensional constants $\hbar$ and $c$ were 
taken into account, and we denoted
$$
A_\mu = (\Phi, \vec{A})\; ,\;\;\; \; \; \; \; \; \; \; \; \; 
S_\mu = (S_0, \vec{S})\,.
$$

Following the simplest procedure of deriving the non-relativistic
approximation  \cite{beresteckii} we write
\beq
\psi\, = \,\left(\matrix{\varphi\cr\chi\cr}\right)\,
e^{ \frac{imc^2t}{\hbar}}\,.
\label{divide}
\eeq
Within the non-relativistic approximation
$\chi \ll \varphi$. From the equation (\ref{eqDirac}) follows:
\beq
(i\hbar\frac{\partial}{\partial t}-
\eta_1 \vec{\sigma}\cdot \vec{S} - e\Phi )\varphi
= (c\vec{\sigma}\cdot\vec{p} - e\vec{\sigma}\cdot \vec{A} 
- \eta S_0)\chi
\label{non1}
\eeq
and
\beq
(i\hbar\frac{\partial}{\partial t}
- \eta \vec{\sigma}\cdot\vec{S} - e\Phi + 2mc^2 ) \chi
= (c\vec{\sigma}\cdot\vec{p} - e\vec{\sigma}\cdot\vec{A} 
- \eta S_0)\ph =0\,.
\label{non2}
\eeq
At low energies, the term $\,2mc^2\chi\,$ in the {\it l.h.s.} of
(\ref{non2}) is dominating. Thus, one can disregard other terms
and express $\chi$ from (\ref{non2}). Then, in the leading order in
$\frac{1}{c}$ we meet the equation:
\beq
i \hbar\frac{\partial \varphi}{\partial t} =
\Big[\,\eta \vec{\sigma} \vec{S} + e\Phi
+ \frac{1}{2mc^2\chi} (c\vec{\sigma}\cdot\vec{p}
- e\vec{\sigma}\cdot\vec{A} - \eta S_0)^2 \,\Big]\, \varphi\,.
\label{non3}
\eeq
The last equation can be easily written in the Schr\"{o}edinger form
\beq
i\hbar\frac{\partial \varphi}{\partial t} =
\hat{H} \varphi\,,
\label{non4}
\eeq
with the Hamiltonian
\beq
\hat{H} = \frac{1}{2m} \vec{\pi}^2 + B_0 + \vec{\sigma}\cdot\vec{Q}\,,
\label{Hamiltonian}
\eeq
where
$$
\vec{\pi} = \vec{p} - \frac{e}{c}\vec{A}
-\frac{\eta_1}{c}\vec{\sigma}S_0\,,
$$
$$
B_0 = e \Phi  - \frac{1}{mc^2}\eta^2 S_0^2\,,
$$
\beq
\vec{Q} = \eta\vec{S} + \frac{\hbar\,e}{2mc}\,\vec{H}\,.
\label{non5}
\eeq
Here, $\vec{H} = rot\vec{A}$ is the magnetic field strength.
This equation is the analog of the Pauli equation for the
general case of external torsion and electromagnetic fields.

The above expression for the Hamiltonian can be
compared to the standard one, which contains only  the 
electromagnetic terms. Some torsion-dependent terms resemble the 
ones with the magnetic field. At the same time, the term
$\,\,- (\eta_1 S_0\,/\,{mc})\,\vec{p}\cdot\vec{\sigma}\;$ does 
not have the analogies in quantum electrodynamics.

%%%%%%%%%%%%%%%%%%%%%%%%%%%%%%%%%%%%%%%%%%%%%%%%%%%%%%%%%%%%%
\section{Foldy-Wouthuysen transformation with torsion}
%%%%%%%%%%%%%%%%%%%%%%%%%%%%%%%%%%%%%%%%%%%%%%%%%%%%%%%%%%%%%

One can derive the next to the
leading order corrections to the non-relativistic approximation 
(\ref{non4}) in the framework of the Foldy-Wouthuysen 
transformation with torsion \cite{rysh}.
The initial Hamiltonian of the Dirac spinor in 
external electromagnetic
and torsion fields can be presented in the form:
\beq
H = \be m + {\cal E} + {\cal G}\,,
\label{inham}
\eeq
where
\beq
 {\cal E} = e\,\Phi + \eta\,\ga_5\,{\vec \al}\,{\vec S}
\,\,\,\,\,\,\,\,\,\,\,\,\,\,\,\,\, {\rm and}
\,\,\,\,\,\,\,\,\,\,\,\,\,\,\,\,\,
{\cal G} = {\vec \al}
\, \left({\vec p} - e {\vec A} \right) - \eta\,\ga_5\,S_0
\label{evenodd}
\eeq
are even and odd parts of the expression. From this instant,
if this is not indicated explicitly, we shall 
use the conventional units $c={\hbar} = 1$.

Our purpose is to find a unitary transformation which separates ``small''
and ``large'' components of the Dirac spinor. In other words, we need to
find a Hamiltonian which is block-diagonal in the new representation.
We use a conventional prescription (see, for example, \cite{BD}):
\beq
H' = e^{i{\cal S}}
\,\left( H - i \,\partial_t\right)\,e^{-i{\cal S}}\,,
\label{gener}
\eeq
where ${\cal S}$ has to be chosen in an appropriate way. We shall try
to find ${\cal S}$ and $H'$ in a form of the weak-relativistic expansion,
and thus start by taking ${\cal S}$ to be of order $1/m$ (with
${\hbar} = c = 1$). Then, to the usual accuracy, we arrive at the
standard result
$$
H' = H + i \,\left[{\cal S},H \right] -
\frac12\,\left[{\cal S}, \left[{\cal S},H \right] \right] -
\frac{i}{6}\,\left[{\cal S}, \left[{\cal S}, \left[{\cal S},H
\right]  \right] \right] +
\frac{1}{24}\,\left[{\cal S}, \left[{\cal S}, \left[{\cal S},
\left[{\cal S},H \right] \right]  \right] \right] -
$$
\beq
- {\dot {\cal S}}
- \frac{i}{2}\,\left[{\cal S}, {\dot {\cal S}} \right] +
\frac{1}{6}\,\left[{\cal S}, \left[{\cal S}, {\dot {\cal S}} \right]\right]
+ ...
\label{expan}
\eeq
One can easily see that ${\cal E}$ and ${\cal G}$ given above
(anti)commute with $\be$ in a usual way
\beq
{\cal E}\,\be = \be\, {\cal E}
,\,\,\,\,\,\,\,\,\,\,\,\,\,\,\,\,\,\,\,
{\cal G} \,\be = - \be\, {\cal G}
\label{commute}
\eeq
and therefore one can safely use the standard prescription for
the lowest-order approximation:
\beq
{\cal S} = - \frac{i}{2m}\,\be \, {\cal G}\,.
\label{anzats}
\eeq
This gives
\beq
H' = \be m + {\cal E}' + {\cal G}'\,,
\label{1ham}
\eeq
where ${\cal G}'$ is of order $1/m$. Now one has to perform
second Foldy-Wouthuysen transformation with
${\cal S}' = - \frac{i}{2m}\,\be {\cal G}'$. This leads to the
\beq
H'' = \be m + {\cal E}' + {\cal G}''
\label{2ham}
\eeq
with ${\cal G}'' \approx 1/m^2$; and then a third
Foldy-Wouthuysen transformation with
${\cal S}'' = - \frac{i}{2m}\,\be {\cal G}''$ removes odd operators
in the given order of the non-relativistic
expansion, so that we finally obtain the usual result
\beq
H''' = \be\,\left( m + \frac{1}{2m}\,{\cal G}^2 -
\frac{1}{8m^3}\,{\cal G}^4 \right) + {\cal E}
- \frac{1}{8m^2}\,\left[{\cal G}, \left(
\left[{\cal G}, {\cal E}\right] + i\,{\dot {\cal G}} \right)\right]\,.
\label{usual}
\eeq
Substituting our ${\cal E}$ and ${\cal G}$ from (\ref{evenodd}),
after some algebra we arrive at the final form of the Hamiltonian
$$
H''' = \be \left[ m + \frac{1}{2m}
\left( {\vec p}
- e {\vec A} - \eta S_0 {\vec \si} \right)^2 
- \frac{1}{8m^3}\,{\vec p}^{\,4} \right] + e\Phi
- \eta \left( {\vec \si}\cdot{\vec S} \right)
- \frac{e}{2m} {\vec \si} \cdot {\vec H} -
$$
\vskip 0.5mm
$$
- \frac{e}{8m^2}\,\left[ div {\vec E} +
i{\vec \si}\cdot {rot}{\vec E}
+ 2 {\vec \si}\cdot\,[{\vec E} \times {\vec p}] \right]
+ \frac{\eta}{8m^2}\,\Big\{
\, {\vec \si}\cdot\nabla {\dot S}_0 - \Big[p_i\,,\,\Big[p^i\,,\,
({\vec \si}\cdot{\vec S})\Big]_{+}\,\Big]_{+} +
%\right.
$$
\vskip 0.5mm
\beq
%\left.
+ 2\, rot{\vec S}\cdot {\vec p} 
- 2i \left({\vec \si}\cdot\nabla\right)
({\vec S}\cdot  {\vec p})
- 2i\,({\na}{\vec S})\,\left({\vec \si}\cdot{\vec p}
\right) \Big\}\,,
\label{separado}
\eeq
\vskip 0.5mm
\noindent
where we have used standard notation for the anticommutator
$\,\left[A,B \right]_+ = AB + BA$. As usual, $\vec{E}$ denotes 
the strength of the external electric field
$\,{\vec E} = -\frac{1}{c}\frac{\pa {\vec A}}{\pa t}
- {\rm grad}\,\Phi.$
One can proceed in the same way and get separated Hamiltonian
with any given accuracy in $\,1/m$.

The first five terms of (\ref{separado})
reproduce the Pauli-like equation with torsion (\ref{Hamiltonian}).
Other terms are the next-to-the-leading order
weak-relativistic corrections and torsion-dependent corrections 
to the Pauli-like equation (\ref{non4}).
In those terms we follow the system of approximation which is
standard for the electromagnetic case \cite{BD}; that is we keep 
the terms linear in interactions.
One can notice that for the case of the constant torsion and
electromagnetic fields one can achieve the exact
Foldy-Wouthuysen transformation \cite{Nikitin}. We do not reproduce
this result here, because torsion (if it exists) is definitely weak
and the the leading order approximation \cite{rysh} is certainly 
the most important one.

Further simplifications of (\ref{separado}) are possible if we are
interested in constant torsion. This version of torsion can be
some kind of relic cosmological field or it can be generated by the
vacuum quantum effects. In this case we have to keep only
the constant components of the pseudovector $S_\mu$. Then the effects
of torsion will be:
i) a small correction to the potential energy of the spinor field,
which sometimes looks just like a correction to the mass, and
ii) the appearance of a new gauge-invariant spin-momentum 
interaction term in the Hamiltonian.

%%%%%%%%%%%%%%%%%%%%%%%%%%%%%%%%%%%%%%%%%%%%%%%%%%%%%%%%%%%%%%%%%%%%
\section{Non-relativistic particle in the external torsion field}
%%%%%%%%%%%%%%%%%%%%%%%%%%%%%%%%%%%%%%%%%%%%%%%%%%%%%%%%%%%%%%%%%%%%

In this section we start from the 
simple derivation of the action for the non-relativistic particle.
This action will be used later on to test the more general 
relativistic expression.

If we consider (\ref{Hamiltonian}) as the Hamiltonian operator 
of some quantum particle, then the corresponding classical 
energy has the form
\beq
H = \frac{1}{2m} \vec{\pi}^2 + B_0 + \vec{\sigma}\cdot\vec{Q} \,,
\label{energ}
\eeq
where $\vec{\pi}, B_0, \vec{Q}$ are defined by (16) and
$\vec{\pi} = m \vec{v}$. Here $\vec{v} = \dot{\vec{x}}$
is the velocity of the particle. The expression for the canonically 
conjugated momenta $\vec{p}$ follows from  (\ref{energ}).
\beq
\vec{p} 
= m\vec{v}+\frac{e}{c}\,\vec{A}+\frac{\eta}{c}\,\vec{\sigma}S_0\,.
\label{moment}
\eeq
One can consider the components of the vector $\vec{\sigma}$ as
internal degrees of freedom, corresponding to spin.

Let us perform the canonical quantization of the theory. For this, we
introduce the operators of coordinate $\hat{x}_i$, momenta $\hat{p}_i$ and
spin $\hat{\sigma}_i$ and demand that they satisfy the equal - time 
commutation relations of the following form:
\beq
\left[\hat{x}_i, \hat{p}_j\right] 
= i\hbar \;\delta_{ij}\;,\;\;\;\;\;\;\;\;\;\;
\left[\hat{x}_i, \hat{\sigma}_j \right] =
\left[\hat{p}_i, \hat{\sigma}_j\right] = 0\;, \;\;\;\;\;\;\;\;\;
\left[\hat{\sigma}_i,\hat{\sigma}_j \right] =
2i\,\varepsilon_{ijk}\; \hat{\sigma}_k\;.
\label{kom}
\eeq
The Hamiltonian operator $\hat{H}$ which corresponds to the energy 
(\ref{energ}) can be easily constructed in terms of the operators 
$\hat{x}_i, \hat{p}_i, \hat{\sigma}_i$. From it we may write the 
equations of motion
$$
i\hbar \frac{d\hat{x}_i}{dt} = \left[\hat{x}_i, \hat{H} \right],
$$
$$
i\hbar \frac{d\hat{p}_i}{dt} = \left[\hat{p}_i, \hat{H} \right],
$$
\beq
i\hbar \frac{d\hat{\sigma}_i}{dt} = \left[\hat{\sigma}_i, 
\hat{H} \right]\,.
\label{eq-ca}
\eeq
After the computation of the commutators in (\ref{eq-ca}) we arrive at 
the explicit form of the operator equations of motion. Now, we can omit
all terms which vanish when $\hbar \rightarrow \; 0$. Thus we obtain
the non-relativistic, quasi-classical equations of motion for the 
spinning particle in
the external torsion and electromagnetic fields. Note that the operator
ordering problem is irrelevant because of the 
$\hbar \rightarrow \; 0$ limit. The straightforward calculations
lead to the equations \cite{babush}:
\beq
\frac{d\vec{x}}{dt} = \frac{1}{m} \left( \vec{p} - \frac{e}{c}\vec{A}
- \frac{\eta}{c}\,\vec{\sigma}\,S_0 \right) = \vec{v},
\label{e1}
\eeq
\vskip 0.5mm
$$
\frac{d\vec{v}}{dt} = e\vec{E} + \frac{e}{c}\left[ \vec{v}\times\vec{H} \right]
- \eta\nabla\left(\vec{\sigma}\cdot \vec{S} \right) +
$$
\beq
+ \frac{\eta}{c} \Big[\, \left(\vec{v}\cdot\sigma\right) \nabla S_0
- \left(\vec{v}\cdot \nabla S_0 \right)\vec{\sigma}
- \frac{d S_0}{d t}\,\vec{\sigma} \,\Big]  
+ \frac{\eta^2}{mc^2}\; \nabla (S_0^2)
+ \frac{\eta}{c}\,S_0\,\Big[ \vec{\si}\,\times\, \vec{R}\,\Big]\,,
\label{e2}
\eeq
and
\beq
\frac{d\vec{\sigma}}{dt} = \left[\vec{R}\times\vec{\sigma}\right]
\,,\,\,\,\,\,\,\,\,\,\,\,\,\,\,\,\,{\rm where}
\,\,\,\,\,\,\,\,\,\,\,\,\,\,\,\,\,
\vec{R} = \frac{2\eta}{\hbar}\left[ \vec{S}
- \frac{1}{c}\vec{v}S_0 \right]
+ \frac{e}{mc}\vec{H}\,.
\label{e3}
\eeq
Equations (\ref{e1}) - (\ref{e3}) contain the torsion - dependent 
terms which are similar to the magnetic terms, and also some terms 
which have a qualitatively new form. 

%%%%%%%%%%%%%%%%%%%%%%%%%%%%%%%%%%%%%%%%%%%%%%%%%%%%%%%%%%%%%
\section{Path-integral approach for the relativistic particle
   with torsion}
%%%%%%%%%%%%%%%%%%%%%%%%%%%%%%%%%%%%%%%%%%%%%%%%%%%%%%%%%%%%%

In this section, we are going to construct a path integral
representation for a propagator of a massive spinning particle
in external electromagnetic and torsion fields. 
The consistent method of constructing the path integral
representation has been developed by Berezin and Marinov
\cite{BerMa75}. Various aspects of the
Berezin-Marinov approach were investigated in the consequent
publications (see, for example, \cite{Casal76}). In this
section we shall follow Ref. \cite{gegi} where the path 
integral representation has been generalized for the background
with torsion. It was demonstrated in \cite{FraGi91,Gitma97},
that a special kind of path integral representations for
propagators of  relativistic particles  allow one
to derive gauge invariant pseudoclassical actions
for the corresponding particles. 
Let us remark that in \cite{FraSh92} some path integral 
representation for massive spinning particle in the presence
of the torsion was derived using the perturbative approach to
path integrals.

First, we consider the path integral representation of the 
 scalar field  propagator in external torsion $S_\mu$ and
electromagnetic $A_\mu$ fields. As we already know, the scalar
field interacts with torsion non-minimally, and this interaction is
necessary for the renormalizability of scalar coupled to the 
fermions. Therefore,  the Klein-Gordon equation in
external electromagnetic and torsion fields has the form
\beq
\left[\,{\hat{\cal P}}^2 + m^2 +\xi\,S^2 \right]\,\ph(x) = 0\,,
\label{scal}
\eeq
where ${\cal P}_\mu = i\pa_\mu - e\,A_\mu,\,\,S^2=S_\mu S^\mu$
and $\,\xi\,$ is an arbitrary non-minimal parameter.
This is the
very same parameter which was called $\xi_4$ in Chapter 2.
Since in this Chapter there are no other $\xi$, it is reasonable
to omit index $4$, exactly as we omitted the index in $\eta_1$.

Our consideration is very similar to the one presented in
\cite{Gitma97} for the torsionless case.
The propagator obeys the equation
\beq
\left[\,{\hat{\cal P}}^2 + m^2 +\xi\,S^2\right]\,D^c(x,y)
= -\de(x,y)\,.
\label{scal-pr}
\eeq
The Schwinger representation for the propagator is
$$ %%%% \beq
D^c(x,y)\, = \,<x| {\hat D}^c |y>\,.
$$ %%% \label{Sch-sc}
%%%%%% \eeq
Here $| x \rangle$ are eigenvectors for some Hermitian
operators of coordinates $X^\mu$ and the corresponding 
canonically conjugated momenta  operators are $P_\mu$. 
Then, the following relations hold:
\begin{eqnarray}
&&X^\mu|x\rangle = x^\mu |x\rangle\,, \,\,\,\,\,\,\,\,
    \langle x | y \rangle = \delta^4(x-y)\,,
                \,\,\,\,\,\,\,\,
\int|x\rangle\langle x|dx = I\,,   \nonumber \\
&& \left[P_\mu,X^\nu \right]_- = - i
\delta_\mu^\nu
\,, \,\,\,\,\,\,\,\,
P_\mu|p\rangle = p_\mu |p\rangle\,,
\,\,\,\,\,\,\,\, \langle p | p' \rangle =
\delta^4(p-p')\,,\nonumber \\
&&\int|p\rangle\langle p|dp = I\,,
\,\,\,\,\,\,\,\, \langle x |P_\mu| y \rangle = -
i\partial_\mu\delta^4(x-y) \,,\,\,\,\,\,\,\,\,
\langle x | p \rangle = \frac{1}{(2\pi)^2}e^{ipx}\,,
\nonumber \\ &&
\left[\Pi_\mu,\Pi_\nu \right]_{-} = - ieF_{\mu\nu}(X)
\,, \,\,\,\,\,\,\,\,
\Pi_\mu = - P_\mu - eA_\mu(X)\,,\,\,\,\,\,\,
F_{\mu\nu}(X)= \pa_\mu A_\nu-\pa_\nu A_\mu   \; .
\label{algebra}
\end{eqnarray}
We can write (\ref{scal-pr}) in an operator way
$$
{\hat F}{\hat D}^c={\hat 1}\,,\,\,\,\,\,\,\,\,\,\,\,\,\,\,\,
{\rm where}\,\,\,\,\,\,\,\,\,\,\,\,\,\,
{\hat F}=m^2 +\xi\,S^2 -{\Pi}^2\,,
$$
and use the Schwinger proper-time representation:
\beq
{\hat D}^c = {\hat F}^{-1}
= i\int_0^{\infty}\,e^{-i\la({\hat F}-i\ep)}\,\la\,,
\label{Sch-inv}
\eeq
where $\ep\to 0$ at the end of calculations. For the massive theory
$\ep$-term can be included into the mass, and that is why we do not
write it in what follows. Indeed, for the massless case $\ep$ is
important for it stabilizes the theory in the IR domain. In principle,
for the constant torsion and $\xi S^2 > 0$, torsion term can
stabilize the proper time integral even in the massless case. It
proves useful to denote ${\hat {\cal H}}={\hat F}\cdot\la$, and rewrite
the previous expression for the Green function (\ref{Sch-inv})
$$
{\hat D}^c = {\hat D}^c(x_{out},x_{in}) = i\,\int_0^\infty\,\,
<x_{out}|e^{-i{\hat {\cal H}} (\la)}|x_{in}>\,d\la=
$$
\beq
=i\,\lim_{N\to\infty}\int_0^\infty d\la_0\,
\int_{-\infty}^{+\infty} dx_1\,...\,dx_N\, d\la_1\,...\,d\la_N\,
\prod_{k=1}^{N}<x_k|e^{-i{\hat {\cal H}} (\la_k)/N}|x_{k-1}>\,
\de(\la_k - \la_{k-1})\,,
\label{green}
\eeq
where $x_0=x_{in},\,\,x_N=x_{out}$. For large enough $N$ one can
approximate 
\beq
<x_k|e^{-i{\hat {\cal H}} (\la_k)/N}|x_{k-1}> \approx
<x_k|1 - \frac{i}{N}\,{\hat {\cal H}} (\la_k)|x_{k-1}>\,.
\label{matr}
\eeq
Introducing the Weyl symbol $\,{\cal H} (\la,x,p)\,$
of the operator $\,{\hat {\cal H}} (\la)$ as \cite{Berez80}
$$
{\cal H} (\la,x,p) = \la\,(m^2  + \xi\,S^2
-{\cal P}^2)\,,\,\,\,\,\,\,\,\,\,\,\,
{\cal P}_\mu = -p_\mu -eA_\mu\,,
$$
we can express each of (\ref{matr}) in terms of the Weyl symbols
in the middle points ${\bar x}_k = (x_k+x_{k-1})/2$:
\beq
<x_k|e^{-i{\hat {\cal H}} (\la_k)/N}|x_{k-1}> \, \approx \,
\int \frac{dp_k}{(2\pi)^4}\,\exp\;
\Big\{\,\Big[\,p_k (x_k-x_{k-1})
- \frac{ {\cal H}(\la_k,{\bar x}_k,p_k)}{N}\,\Big]\,\Big\}\,.
\label{matr1}
\eeq
Then the expression for the propagator becomes
$$
{\hat D}^c = i\,\lim_{N\to\infty}\int_0^\infty d\la_0\,
\int_{-\infty}^{+\infty}\,\prod_{k=1}^N\,
dx_k\, d\la_k\,\frac{dp_k}{(2\pi)^4}\,\frac{d\pi_k}{2\pi}\,
\times
$$
\beq
\times\,\exp\;
\Big\{\,\Big[\,p_k (x_k-x_{k-1})
- \frac{ {\cal H}(\la_k,{\bar x}_k,p_k)}{N} + \pi_k(\la_k-\la_{k-1})
\,\Big]\,\Big\}\,,
\label{greenfun}
\eeq
where we have also used the Fourier representation for the
delta-functions $\,\,\de(\la_k - \la_{k-1})\,\,$ of (\ref{green}).
The Eq. (\ref{greenfun}) is the definition of the Hamiltonian
path integral for the propagator of scalar particle
\beq
{\hat D}^c = i\,\int_0^\infty d\la_0\,
\int_{x_{in}}^{x_{out}} Dx \int D\la \int Dp D\pi
\,\exp
\Big\{i\int_0^1 d\tau
[\la({\cal P}^2 - m^2 - \xi S^2) + p{\dot x} + \pi{\dot \la}]\Big\}.
\label{pathpart}
\eeq
The integral is taken over the path
$x^\mu(\tau),\,p_\mu(\tau),\,\la(\tau),\,\pi(\tau)$ with fixed
ends $x(0)=x_{in}, x(1)=x_{out},\,\la(0)=\la_0$. Integrating over
the momenta $p_\mu$, one arrives at the Lagrangian form of the
path integral representation (where we substituted $\la = \th/2$
in order to achieve the conventional form)
\beq
{\hat D}^c = \frac{i}{2}\,\int_0^\infty d\th_0\,
\int_{x_{in}}^{x_{out}} Dx \,\int_{\th_0} D\th M(\th) \int D\pi
e^{i\int_0^1 L d\tau}\,,
\label{pathLag}
\eeq
where the Lagrangian of the scalar particle has the form
\beq
L = -\frac{{\dot x}^2}{2\th}
- \frac{\th}{2}\,(m^2 + \xi S^2) - ex_\mu A^\mu
+ \pi{\dot \th}\,.
\label{scalar-L}
\eeq
It is easy to see that the presence of torsion does not make any
essential changes in the derivation of the particle action. The
result may be obtained by simple replacement $\,m^2 \to m^2+\xi\,S^2$.
\vskip 3mm

Let us now consider the path integral representation for the 
propagator of spinning particle. We shall follow the original paper 
\cite{gegi}, where the technique of Refs. \cite{FraGi91} was 
applied to the case of the spinning particle on the torsion and 
electromagnetic background. One can also consult
\cite{gegi} for the further list of references on the subject.

Consider the causal Green function $\Delta^c(x,y)$ of the equation
of motion corresponding to the action (\ref{diraconly}).
$\Delta^c(x,y)$ is the propagator of the spinor particle.
\begin{equation}
\label{b1}
 \left[\,\ga^\mu\,\left(\hat{{\cal P}}_\mu
+ \eta\ga^5 S_\mu \right) - m\,\right]\Delta^c(x,y)= -
\delta^4(x-y)\;.
\end{equation}
It proves useful to introduce, along with
$\ga^0,\,\ga^1,\,\ga^2,\,\ga^3,\,\ga^5$,
another set of the Dirac matrices
$\Ga^0,\,\Ga^1,\,\Ga^2,\,\Ga^3,\,\Ga^4$. These matrices are
defined through the relations
\beq
\Ga^4 = i\ga^5\;\,\,\,\,\, \;\; \Ga^\mu =\Ga^4 \ga^\mu\,.
\label{b3}
\eeq
It is easy to check that the matrices $\Ga^n,\,\,\;n = 0,1,..,4$,
form a representation of the Clifford algebra in 5-dimensions:
\beq
\left[ \Ga^n, \Ga^m \right]_+ =
2\,\eta^{nm}\;,\;\;\;\;\;\;\; \eta_{nm} = {\rm
diag}(1,-1,-1,-1,-1)\,.
\label{bbb3}
\eeq
In order to proceed, we
need a homogeneous form of the operator $(\Delta^c)^{-1}$.
Therefore, we perform the $\Ga^4$-transformation for $\Delta^c(x,y)$.
$$
\tilde{\Delta}^c(x,y) = \Delta^c(x,y)\Gamma^4\,.
$$
The new propagator $\tilde{\Delta}^c(x,y)$ obeys the equation
\begin{equation}
\label{b2}
 \left[\,\Gamma^\mu\,\left(\hat{{\cal P}}_\mu
- i\eta \,\Gamma^4 S_\mu \right) - m\Gamma^4\right]
\tilde{\Delta}^c(x,y)= \delta^4(x-y),
\end{equation}
Similar to the scalar case, we present $\,\tilde{\Delta}^c(x,y)\,$ 
as a matrix element of an operator $\tilde{\Delta}^c$.
%%% \begin{equation}
$$ %%% \label{bb4}
\tilde{\Delta}^c(x,y)
= \langle x | \tilde{\Delta}^c | y
\rangle\;.
$$ %%% \end{equation}
All further notations are those of (\ref{algebra}).
The formal solution for the operator $ \tilde{\Delta}^c$ is
$$
\tilde{\Delta}^c = {\wh {\cal F}}^{-1}\;,\;\;\;\;\;\;\;\;
{\wh {\cal F}}=\Pi_\mu\Gamma^\mu -
m \Gamma^4 - i\eta\Gamma^\mu\Gamma^4 S_{\mu}\;.
$$
The operator ${\wh {\cal F}}$ may be written in an
equivalent form, \begin{equation}\label{b5}
{\wh {\cal F}}=\Pi_\mu\Gamma^\mu -
m \Gamma^4-
\frac{i}{6}\,\eta\epsilon_{\mu\nu\alpha\beta}S^{\mu}\Gamma^{\nu}
\Gamma^\alpha\Gamma^\beta \;,
\end{equation}
using the following formula
$$
\Ga_\mu \Ga^4 = \frac16\,\ep_{\mu\nu\al\be}\,
\Ga^\nu\Ga^\al\Ga^\be\; ,\;\;\;\;\;\;\;  \ep_{0123}=1\;.
$$
The last relation is important, for it replaces the product of
an even number of $\Ga$'s for the product of an odd number of $\Ga$'s.
Together with the $\Ga^4$-transformation of the propagator,
this helps us to provide the homogeneity of the equation,
so that ${\wh {\cal F}}$ becomes purely fermionic operator.
Now, ${\wh {\cal F}}^{-1}$ can be presented by means of an
integral \cite{FraGi91}:
\beq
{\wh {\cal F}}^{-1} = \int_0^\infty \,
d\lambda \int  e^{i[\lambda({\wh {\cal F}}^2
+i\epsilon) +\chi {\wh {\cal F}}]} d\,\chi\,.
\label{superproper}
\eeq
Here $\,\lambda,\,\chi\,$ are the parameters of even and odd
Grassmann parity. Taken together, they can be considered as
a super-proper time \cite{FraGi91}. Indeed, $\lambda$ commutes and
$\chi$ anticommutes with ${\wh {\cal F}}$:
$$
\left[\la\,,\,{\wh {\cal F}}\right]=0\,,
\,\,\,\,\,\,\,\,\,\,\,\,\,\,\,\,\,\,\,\,\,\,\,\,\,\,\,\,\,\,\,\,\,
\left[\chi\,,\,{\wh {\cal F}}\right]_{+}=0\,.
$$
Calculating ${\wh {\cal F}}^2$ we find
\beq\label{b6}
{\wh {\cal F}}^2 = \Pi^2 - m^2 - \eta^2 S^2
 - \frac{ie}{2}\,F_{\mu\nu}\,\Ga^\mu\Ga^\nu
 + \hat{K}_{\mu\nu}\,\Ga^\mu\Ga^\nu
+ \eta\,\pa_\mu S^\mu \,\Ga^0\Ga^1\Ga^2\Ga^3    \,,
\eeq
where
\beq\label{b7}
\hat{K}_{\mu\nu} = \frac{i\eta}{2}\,
\left[\Pi^\al\,,\,S^\be\right]_+\ep_{\al\be\mu\nu}
\,,\;\;\;\;\;\;\;\;\;\;
\Pi^2=P^2+e^2A^2+e\left[P_{\mu},A^{\mu}\right]_+ \,\;.
\eeq
Thus we get the integral representation for the propagator:
$$
\tilde{\Delta}^c =\int_0^\infty \, d\lambda
\int d\chi\, {\rm exp}\,\left[\,-i\hat{\cal H}(\lambda,\chi)\,\right] 
\,\,,
$$
where
\begin{eqnarray*}
&&\hat{{\cal H}}(\lambda,\chi) = \lambda \left(
m^2+\eta^2 S^2 - \Pi^2 +\frac{ie}{2}\,F_{\mu\nu}\Gamma^\mu\Gamma^\nu -
\hat{K}_{\mu\nu}\Gamma^\mu\Gamma^\nu
\right.   \\
&&\left.-\eta\,\pa_\mu S^\mu\,\Ga^0\Ga^1\Ga^2\Ga^3
\right) - \chi\left(
\Pi_\mu\Gamma^\mu -
\frac{im\eta}{6}\,\epsilon_{\kappa\mu\nu\alpha}S^{\kappa}
\,\Ga^4\Gamma^{\mu}\Gamma^\nu\Gamma^\alpha\right)\,.
\end{eqnarray*}
The Green function $\tilde{\Delta}^c(x_{\rm out},x_{\rm in})$   
has the form: \begin{equation}\label{b8}
\tilde{\Delta}^c(x_{\rm out},x_{\rm in}) =\int_0^\infty \,
d\lambda \int \langle x_{\rm out} | e^{-i\hat{\cal
H}(\lambda,\chi)}|x_{\rm in} \rangle d\chi\,\,.
\end{equation}

Now we are going to represent the matrix element entering 
in the expression (\ref{b8}) by means of a path integral
\cite{FraGi91,gegi}. The calculation goes very
similar to the scalar case. We  write, as
usual, $\,e^{-i\hat{{\cal H}}} = \left(e^{-i\hat{{\cal H}}/N}
\right)^N\,$, then insert $(N-1)$
identities $\int|x\rangle\langle x|dx = I$ and
introduce $N$ additional integrations over $\lambda$ and $\chi$
\begin{eqnarray}
\label{b9}
&& \tilde{\Delta}^c(x_{\rm out},x_{\rm in}) =  \lim_{N\rightarrow
\infty} \int_0^\infty d\, \lambda_0
\int \prod_{k=1}^{N}\langle x_k\,|
e^{-\frac{i}{N}\,\hat{{\cal H}}(\lambda_k,\chi_k)\Delta}|\,x_{k-1}
\rangle \\
&&
\times\,\delta(\lambda_k-\lambda_{k-1})\,\delta(\chi_k-\chi_{k-1})
\,\,d\chi_0\, dx_1\, ...\, d x_{N-1}
\,\,d\lambda_1 ... d\, \lambda_N d\,\,\chi_1 ... d\, \chi_N\,,
\nonumber 
\end{eqnarray}
where $x_0=x_{\rm in}$, $x_N=x_{\rm out}$. Using the approximation
(\ref{matr}), and introducing the Weyl symbol
 ${\cal H}(\lambda,\chi,x,p)$ of the symmetric operator $\hat{{\cal H}}$
\begin{eqnarray}
\label{b11}
&&{\cal H}(\lambda,\chi,x,p) =  \lambda \left(m^2+ \eta^2 S^2
- {\cal P}^2 + \frac{ie}{2}F_{\mu\nu}\Gamma^\mu\Gamma^\nu
-K_{\mu\nu}\Gamma^\mu\Gamma^\nu
\right.  \nonumber \\
&&\left.
-\eta\pa_\mu S^\mu\,\Ga^0\Ga^1\Ga^2\Ga^3
\right) - \chi\left(
{\cal P}_\mu\Gamma^\mu -
\frac{im\eta}{6}\epsilon_{\kappa\mu\nu\alpha}S^{\kappa}
\,\Ga^4\Gamma^{\mu}\Gamma^\nu\Gamma^\alpha
\right)\;,
\nonumber
\end{eqnarray}
with
$\,K_{\mu\nu} = - \eta{\cal P}^\al S^\be\ep_{\al\be\mu\nu}$,
we can express the matrix elements (\ref{matr}) in terms of
the Weyl symbols at the middle point ${\bar x}_k$. Then
(\ref{matr}) can be replaced by the expressions
\begin{equation}
\label{b13}
\int \frac{d\,p_k}{(2\pi)^4}
\exp i \left[ p_k\,\frac{x_k-x_{k-1}}{\De \tau}
- {\cal H}(\lambda_k,\chi_k, \overline{x}_k,p_k)
\right]\Delta\tau\,,
\end{equation}
where $\Delta\tau = 1/N$. Such expressions with different values 
of $k$ do not commute  due to the $\Gamma$-matrix structure and, 
therefore, have to be replaced into (\ref{b9}) in such a way that 
the numbers $k$ increase from the right to the left. For the two 
$\delta$-functions, accompanying each matrix element in the 
expression (\ref{b9}), we use the integral representations
$$
\delta(\lambda_k-\lambda_{k-1})
\,\delta(\chi_k-\chi_{k-1}) \,=\, \frac{i}{2\pi}
\int \,d\, \pi_k d\, \nu_k\, {\rm exp}\,\Big\{\,
i\left[ \pi_k\left(\lambda_k-\lambda_{k-
1}\right)+ \nu_k\left(\chi_k   -\chi_{k-1}
\right)\right]\,\Big\}
\;,
$$
where $\nu_k$ are odd variables. Then we attribute
to the $\Gamma$-matrices in (\ref{b13}) an index $k$. 
At the same time we attribute to all quantities 
the ``time'' $\tau_k$ according to the index $k$ they have, 
$\tau_k=k\Delta\tau $. Then, $\tau \in [0,1]$. Introducing 
the T-product, which acts on $\Gamma$-matrices, it is possible 
to gather all the expressions, entering in (\ref{b9}), in one 
exponent and postulate that at equal times the $\Gamma$-matrices 
anticommute. Finally, we arrive at the 
propagator
\begin{eqnarray}
\label{b14}
&&\tilde{\Delta}^{c}(x_{\rm
out},x_{\rm in}) =  {\rm T}\int_0^\infty \, d\lambda_0
\int
d\chi_{0}\int_{x_{in}}^{x_{out}}Dx \int Dp
\int_{\lambda_0}D\lambda \int_{\chi_0}D\chi\int D\pi \int
D\nu \nonumber \\
&&\times\exp \left\{i\int_0^1 \left[ \lambda \left({\cal P}^2 -
m^2 - \eta^2S^2 -\frac{ie}{2}F_{\mu\nu}  \Gamma^\mu\Gamma^\nu
+K_{\mu\nu}\Gamma^\mu\Gamma^\nu + \eta\pa_\mu
S^\mu\,\Ga^0\Ga^1\Ga^2\Ga^3 \right)
\right.\right.  \nonumber \\
&&\left.\left. + \chi\left(
{\cal P}_\mu\Gamma^\mu -
\frac{im\eta}{6}\,\epsilon_{\kappa\mu\nu\alpha}S^{\kappa}
\,\Ga^4\Gamma^{\mu}\Gamma^\nu\Gamma^\alpha\right)
+ p\dot{x} + \pi\dot{\lambda} + \nu\dot{\chi}\right]d\tau \right
\} \,, \end{eqnarray}
where   $x(\tau)$,  $p(\tau)$,  $\lambda(\tau)$,  $\pi(\tau)$,
are even and $\chi(\tau), \;\nu(\tau)$ are odd functions.
The boundary conditions are $\,\,\,x(0)=x_{\rm in}$,
$\,\,\,x(1)=x_{\rm out}$, $\,\,\,\lambda (0) = \lambda_0$,
$\,\,\,\chi(0) = \chi_0$. The operation of T-ordering
acts on the $\Gamma$-matrices which formally 
depend on the time $\tau$. The expression (\ref{b14}) can be
transformed as follows:
\begin{eqnarray*}
&& \tilde{\Delta}^{c}(x_{\rm out},x_{\rm in}) =
\int_0^\infty\,d\lambda_0 \int d\chi_{0}\int_{\lambda_0}D\lambda
\int_{\chi_0}D\chi  \int_{x_{in}}^{x_{out}}Dx
\int Dp     \int D\pi \int D\nu \times \\
&& \exp \left\{i\int_0^1 \left[
\lambda\left( {\cal P}^2 - m^2 - \eta^2 S^2 
 - \frac{ie}{2}F_{\mu\nu}
\frac{\delta_l}{\delta \rho_\mu}\frac{\delta_l}{\delta
\rho_\nu} +K_{\mu\nu}\frac{\delta_l}{\delta \rho_\mu}
\frac{\delta_l}{\delta \rho_\nu}
+\eta\pa_\mu S^\mu\, \frac{\delta_l}{\delta \rho_0}
\frac{\delta_l}{\delta \rho_1}
  \frac{\delta_l}{\delta \rho_2}
\frac{\delta_l}{\delta \rho_3}  \right)\right.\right. \\
&&+\left.\left. \chi\left(
{\cal P}_\mu \frac{\delta_l}{\delta \rho_\mu}
- m \frac{\delta_l}{\delta \rho_4}
-\frac{i\eta}{6}\epsilon_{\kappa\mu\nu\alpha}S^{\kappa}
\frac{\delta_l}{\delta \rho_\mu}
\frac{\delta_l}{\delta \rho_\nu}
 \frac{\delta_l}{\delta \rho_\alpha}     \right)
p\dot{x} + \pi\dot{\lambda} +
\nu\dot{\chi}\right]d\tau\right\} \\
&&\times{\rm T}\left.\exp
\int_0^1\rho_n(\tau)\Gamma^n d\tau \right|_{\rho=0},
\end{eqnarray*}
where five  odd sources $\rho_n(\tau)$ are introduced. They
anticommute with the $\Gamma$-matrices by definition. One can
represent the quantity ${\rm T}\exp \int_0^1 \rho_n(\tau)\Gamma^n
d\tau$  via a  path integral over odd trajectories
\cite{FraGi91},
\begin{eqnarray}
\label{bb15}
&&{\rm T}\exp \int_0^1\rho_n(\tau)\Gamma^n d\tau  =
\exp\left(i\Gamma^n \frac{\partial_l}{\partial\Theta^n}   \right)
\int_{\psi(0)+\psi(1)=\Theta}\exp \left[ \int_0^1 \left(
\psi_n\dot{\psi}^n - 2i\rho_n\psi^n\right) d\tau \right.\nonumber
\\
&&
+ \left.\left.\psi_n(1)\psi^n(0)\right]{\cal D}
\psi\right|_{\Theta=0}\;,
\end{eqnarray}
with the modified integration measure
$$
{\cal D}\psi=D\psi\left[\int_{\psi
(0)+\psi (1)=0}
D\psi \exp\left\{\int^{1}
_{0}\psi_{n}\dot{\psi}^{n}d\tau\right\}\right]^{-1}\,.
$$
Here $\Theta^n$ are
odd variables, anticommuting with the $\Gamma$-matrices, and
$\psi^{n}(\tau)$ are odd trajectories of integration. These trajectories
satisfy the boundary conditions indicated below the signs of
integration. Using (\ref{bb15}) we get the Hamiltonian path
integral representation for the propagator:
\begin{eqnarray}\label{b16}
\tilde{\Delta}^{c}(x_{\rm out},x_{\rm in}) &=&\exp\left(i\Gamma^n
\frac{\partial_l}{\partial\Theta^n} \right)\int_0^\infty \, d\lambda_0
\int d\chi_{0}\int_{\lambda_{0}}D\lambda
\int_{\chi_{0}}D\chi \int_{x_{in}}^{x_{out}}Dx \int Dp \int D\pi \int
D\nu \nonumber \\
&\times&\int_{\psi(0)+\psi(1)=\Theta} {\cal D}\psi \exp \left\{i\int_0^1
\left[ \lambda\left({\cal P}_\mu+\frac{i}{\lambda}\psi_\mu\chi
+d_\mu\right)^2
-\lambda \left(m^2 + \eta^2 S^2\right)\right.\right. \nonumber \\
&+&\left.\left.
2i\lambda eF_{\mu\nu}\psi^\mu\psi^\nu
+ 16\lambda\eta\,\pa_\mu S^\mu\, {\psi^0} {\psi^1}
 {\psi^2} {\psi^3}+2i\chi\left(m\psi^4
+ \frac{2}{3}\psi^\mu d_\mu\right)\right.\right. \nonumber \\
&-&\left.\left. i\psi_n\dot{\psi}^n + p\dot{x}
+ \pi \dot{\lambda} +\nu \dot{\chi}\right] d\tau
+ \left.\psi_n(1)\psi^n(0) \right\}\right|_{\Theta=0} \;,
\end{eqnarray}
where
\beq
d_\mu=-2i\eta\,\epsilon_{\mu\nu\alpha\beta}S^\nu\psi^\al\psi^\be\;.
\label{dmu}
\eeq

Integrating over the momenta, we get the Lagrangian path integral
representation for the propagator,
\begin{eqnarray}
\label{b17}
&&\tilde{\Delta}^{c}(x_{\rm out},x_{\rm in})=\exp\left(i\Gamma^{n}
\frac{\partial_{\ell}}{\partial \Theta^{n}}\right)
\int_{0}^{\infty}d\th_{0}
\int d\chi_{0}\int_{\th_{0}} {\cal M}(\th)D\th
\int_{\chi_{0}}D\chi
\int_{x_{in}}^{x_{out}}Dx
\int D\pi \int D\nu
\nonumber \\
&&\times \int_{\psi(0)+\psi(1)=\Theta} {\cal D}\psi
\, \exp\left\{i\int_{0}^{1}\left[-\frac{z^{2}}{2\th}
-\frac{\th}{2}M^{2} -\dot{x}_\mu\left(eA^\mu-d^\mu\right)
+i\th e\,F_{\mu \nu}\psi^{\mu}\psi^{\nu} \right.\right.
\nonumber \\
&&\left.\left. +i\chi\left(m\psi^4+\frac{2}{3}\psi^{\mu}d_\mu \right)
-i\psi_{n}\dot{\psi}^{n}+\pi \dot{\th}+\nu \dot{\chi}\right]d\tau
+ \left.\psi_{n}(1)\psi^{n}(0)\right\}\right|_{\Theta=0}\;,
\end{eqnarray}
where $\th=2\la$ and the measure ${\cal M}(\th)$ has the form:
\begin{equation}\label{b18}
{\cal M}(\th)=\int {\cal D} p \exp \left[
\frac{i}{2}\int_0^1 \th p^2 d\,\tau \right]\;,
\end{equation}
and
\beq
M^2 = m^2+\eta^2S^2 - 16\eta\,\pa_\mu S^\mu\,{\psi^0} {\psi^1}
 {\psi^2} {\psi^3}\,,\;\;\,\,\,\,\;\,\,\,\,
z^\mu=\dot{x}^\mu+i\chi\psi^\mu\;.
\label{Mz}
\eeq
The discussion of the role of the measure (\ref{b18}) can be found in
\cite{FraGi91}.

The exponential in the integrand (\ref{b17}) can be considered as an
effective non-degenerate Lagrangian action of a spinning particle
in electromagnetic and torsion fields. It consists of two principal
parts. The term
$$
S_{\rm GF} = \int_0^1\left(\pi \dot{\th} + \nu \dot{\chi}
\right)d\,\tau,
$$
can be treated as a gauge fixing term corresponding to the gauge 
conditions $\,\dot{\th} = \dot{\chi} = 0$.
The other terms can be treated as a gauge 
invariant action of a spinning particle. It has the form
\begin{eqnarray}
\label{c1}
S &=& \int_0^1 \left[-\frac{z^{2}}
{2\th}-\frac{\th}{2}M^{2} -\dot{x}_\mu\left(eA^\mu-d^\mu\right)
+i\th e F_{\mu \nu}\psi^{\mu}\psi^{\nu} \right. \nonumber \\
&&\left.+i\chi\left(m\psi^4+\frac{2}{3}\psi^{\mu}d_\mu\right)
-i\psi_{n}\dot{\psi}^{n}\right]d\tau\;.
\end{eqnarray}
where
$\,z^\mu,\,\, M^2,\,$ and $\, d_\mu$ have been defined 
in (\ref{Mz}) and (\ref{dmu}).
The action (\ref{c1}) is a generalization of the
Berezin-Marinov action \cite{BerMa75,Casal76} to the background
with torsion. One can easily verify that (\ref{c1}) is invariant
under reparametrizations:
$$ %%% \beq
\de x = {\dot x}\xi\,,\,\,\,\,\,\,\,
\de \th = \frac{d(\th\xi)}{dt}\,,\,\,\,\,\,\,\,
\de \psi^n = {\dot \psi}\,^n\xi \,\,\,\,\,\,
(n=0,1,2,3,4)\,,\,\,\,\,\,\,\,\,\,\,\,
\de\chi = \frac{d(\chi\xi)}{dt}\,.
$$ %%% \label{repara}
%%%%%% \eeq
I could establish the explicit form of the local supersymmetry
transformations, which generalize the ones for the
Berezin-Marinov action, only in the linear in torsion approximation.
These transformations have exactly the same form as in the case
without torsion (see, for example, \cite{GitSa93}):
$$
\de x^\al = i\psi^\al\ep \,,\,\,\,\,\,\,\,\,\,\,\,
\de \th = i\chi\ep \,,\,\,\,\,\,\,\,\,\,\,\,
\de \psi^\al = \frac{1}{2\th}\,\Big({\dot x}^\al + i\chi\psi^\al\Big)\ep
\,\,(\al=0,1,2,3)\,,
$$
$$ %%%% \beq
\de\chi = {\dot \ep}\,,\,\,\,\,\,\,\,\,\,\,\,\,\,\,
\de \psi^5 = \Big[\frac{m}{2} - \frac{i}{m\th}\,\psi^5\,
(\,{\dot \psi}^5 -\frac{m}{2}\,\chi\,)\,\Big]\ep\,,
$$ %%%% \label{susy}
%%%%%%% \eeq
with $\ep=\ep(\tau)$.
The demonstration of supersymmetry is technically nontrivial,
and in particular one needs the identity:
$$
\psi^\al\psi^\be\psi^\mu\psi^\nu\,\vp_{\al\be\mu\la}\,\pa_\nu S^\la=
\frac14\,
\psi^\al\psi^\be\psi^\mu\psi^\nu\,\vp_{\al\be\mu\nu}\,\pa_\la S^\la\,.
$$
In the general case one can establish the supersymmetry of the action
through the structure of the Hamiltonian constraints in the
course of quantization \cite{gegi}.
\vskip 3mm

Let us analyze the equations of motion for the 
theory with the action (\ref{c1}).
These equations contain some unphysical variables, that are
related to the reparametrization and supersymmetry invariance.
One can choose  the gauge conditions $\chi=0$ and $\th=1/m$ to 
simplify the analysis. Then we need only two equations
\begin{eqnarray}
\label{c3}
\frac{\delta_r S}{\delta\psi^\alpha} &=& 2i\dot{\psi}_\alpha
- 2i\th e\,F_{\al\be}\psi^\beta - \frac{i}{e}\,\dot{x}_\al\chi
+ \frac{2i}{3}\,\chi d_\al
+ 4i\eta\,\vp_{mu\nu\al\be}\,\dot{x}^\mu S^\nu\psi^\be -
\nonumber \\
&& - \frac{8\eta}{3}\,\chi\vp_{mu\nu\al\be}\,\psi^\mu S^\nu\psi^\be
- \frac{4\eta \th}{3}\,\pa_\la S^\la
\vp_{\mu\nu\al\be}\,\psi^\mu \psi^\nu\psi^\be\,,    \\
%%%%%%%%%%%%%%%%%%%%%%%%%%%%%%%%%%%%%%%%%%%%%%
\label{c5}
\frac{\delta S}{\delta x^\alpha} &=&
\frac{d}{d\tau}\left(\frac{\dot{x}_\alpha}{\th}\right)
+ e\dot{x}^\beta F_{\beta\alpha}
+ i\th e F_{\mu\nu,\alpha}\psi^\mu\psi^\nu
+ e\dot{x}_\mu\pa_\al A^\mu
+ \frac{d}{d\tau}\,\left(
\frac{i}{\th}\,\psi_\al \chi - eA_\al + d_\al \right) \nonumber \\
&&+ \eta \th S^\mu\pa_\al S_\mu
- 8\eta \th\, (\pa_\al\pa_\mu S^\mu)\psi^0 \psi^1 \psi^2 \psi^3
- \dot{x}_\mu  (\pa_\al d^\mu)
- \frac{2i}{3}\,\chi\psi_\mu  (\pa_\al d^\mu) = 0 \label{c6}\,.
\end{eqnarray}

Now, in order to perform the nonrelativistic limit we define the 
three dimensional spin vector $\,\vec{\sigma}\,$ as \cite{BerMa75}:
\beq
\si_k = 2i\,\epsilon_{kjl}\psi^l\psi^j \,,\,\,\,\,\,\,\,
\psi^j\psi^l = \frac{i}{4}\,\epsilon^{kjl}\si_k
 \,,\,\,\,\,\,\,\,
\dot{\psi^j}\psi^l = \frac{i}{4}\,\epsilon^{kjl}\dot{\si}_k \,,
\label{c7}
\eeq
and consider
\begin{equation}
\psi^0 \approx 0 \,,\,\,\,\,\,\,
\dot{x}^0 \approx 1 \,,\,\,\,\,\,\,
\dot{x}^i \approx v^i = \frac{dx^i}{dx^0}\,,
\label{c8}
\end{equation}
as a part of the nonrelativistic approximation. Furthermore, we use
standard relations for the components of the stress tensor:
$$
F_{0i} = - E_i = \pa_0 A_i - \pa_i A_0 \,\,\,\,\,\,\,\,\,\,\,
{\rm and} \,\,\,\,\,\,\,\,\,\,\,
F_{ij} = \epsilon_{ijk}\, H^k\,.
$$
Substituting these formulas into (\ref{c6}) and (\ref{c3}),
and disregarding the terms of higher orders in the external 
fields, we arrive at the equations:
\begin{eqnarray}\label{c9}
&&m \,\dot{\vec {v}} = e\vec{E} + \frac{e}{c}\left[
\vec{v}\times\vec{H} \right]
- \eta\nabla\left(\vec{\sigma}\cdot \vec{S} \right)
- \frac{\eta}{c}\,\frac{d S_0}{dt}\,\vec{\sigma}
+ \frac{\eta}{c}\, \left(\vec{v}\cdot\sigma\right) \na S_0 +
... \nonumber \\
&& {\dot {\vec{\sigma}}} = \left[\,\Big(\frac{e}{mc}\,\vec{H}
+ \frac{2\eta}{\hbar}\,\vec{S} - \frac{2\eta S_0}{c\hbar}\,\vec{v}
\Big)\,\times\,\vec{\sigma}\,\right] \,.
\end{eqnarray}
They coincide perfectly  with the classical equations of motion
(\ref{e2}), (\ref{e3}) obtained from the generalized Pauli equation.
This correspondence 
confirms our interpretation of the action (\ref{c1}). Additional
arguments in favor of this interpretation were obtained in
the framework of canonical quantization \cite{gegi}.
The quantization leads us back to the Dirac equation on torsion and
electromagnetic background. Therefore, we have all reasons to 
consider (\ref{c1}) as the correct expression for the action of 
spin-$1/2$ particle in the external background of $A_\mu$ and $S_\mu$
fields.

%%%%%%%%%%%%%%%%%%%%%%%%%%%%%%%%%%%%%%%%%%%%%%%%%%%%%%%%%%%%%%%%%
\section{Space-time trajectories for the spinning and spinless
  particles in an external torsion field}
%%%%%%%%%%%%%%%%%%%%%%%%%%%%%%%%%%%%%%%%%%%%%%%%%%%%%%%%%%%%%%%%%

In this section, we shall consider several particular examples of
motion of spinless and spinning particles in an external torsion field.
Let us start from the scalar particle with the action (\ref{scalar-L}).
For the sake of simplicity we do not consider electromagnetic field.
Using the gauge condition $\pi=0$, and replacing the solution for the
auxiliary field $\,\th\,$ back into the action, we cast it in the form
\beq
S = - \int_0^1\,
d \tau\,\sqrt{(m^2+\xi\,S_\mu S^\mu) \,\,{\dot x}^2}\,.
\label{scalar-L1}
\eeq
It is obvious, already from the action (\ref{scalar-L}),
that the role of the constant torsion axial vector $S_\mu=const$
is just to change
the value of the mass of the scalar particle $m^2\to m^2+\xi S^2$.
Indeed, this is true only until the metric is flat. As far as we
take a curved metric, the square $S_\mu S^\mu$ depends on it, even
for a constant torsion. Let us suppose that $\,S_\mu S^\mu\,$
is coordinate dependent. 
Performing the variation over
the coordinate $x^\al$, after some algebra we arrive at the equation
of motion
$$ %%%  \beq
{\ddot x}_\al \,=\, \frac{{\dot x}^2}{2}\,\,\cdot\,\,
\frac{\xi\, \pa_\al (S^\mu S_\mu)}{m^2+\xi S^2}\,.
$$ %%% \label{scaleq1}
%%%%%% \eeq
Taking into account that the square of the $4$-velocity is
constant, ${\dot x}^2=1$, we obtain
\beq
{\ddot x}_\al \,=\, \frac12\,\pa_\al\,{\rm ln}\,\left|\,
1\,+\,\frac{\xi S^2}{m^2}\,\right|\,.
\label{scale2}
\eeq
Thus, in case of the non-constant $\,S_\mu S^\mu\,$ the motion 
of particle corresponds to the additional four-acceleration 
(\ref{scale2}), and the torsion leads to the potential of the 
form 
$$
V_{S}\, =\, - \, \ln \sqrt{|m^2+\xi S^2|/m^2}\,.
$$ 
Indeed, this potential
is constant for the scalar minimally coupled to torsion $\xi=0$.
However, as we have seen in the previous Chapter, for any scalar
which interacts with spinors, the non-minimal interaction $\xi\neq 0$
is nothing but a consistency condition. So, if torsion would
really exists, and if its square is not constant, the force producing
(\ref{scale2}) should exist too. At the same time, 
the only scalar which is supposed to couple to the fermions
through the Yukawa  interaction, is the Higgs. Since
the Higgs mass is supposed to be, at least, of the order of $100\,GeV$,
all experimental manifestations of the Higgs particle are possible
only at the high energy domain. Since torsion (if exists) is a very weak
field, there are extremely small chances to observe torsion through the 
acceleration (\ref{scale2}).
\vskip 3mm

Let us now discuss the spin $1/2$ case. Here we follow, mainly,
Ref. \cite{rysh}.
The physical degrees of freedom of the particle are its coordinate
${\vec {x}}$ and its spin ${\vec {\si}}$. For the sake of simplicity
we shall concentrate on the non-relativistic case and
consider the motion of a spinning particle in a space with
 constant axial torsion
$S_\mu = (S_0, {\vec {S}})$, but without electromagnetic field.
In this case the equations of motion (\ref{e2}), (\ref{e3}) have the
form:
$$ %%% \beq
\frac{d{\vec {v}}}{dt}
= - \eta \,{\vec {S}}\,({\vec {v}}\cdot {\vec {\si}})
- \frac{\eta S_0}{c}\,\frac{d{\vec {\si}}}{dt}\,,
$$ %%% \label{eq1}
%%%%%% \eeq
\beq
\frac{d{\vec {\si}}}{dt} =
+ \frac{2\eta}{\hbar}\,\left[{\vec {S}}\times {\vec {\si}}\right]
- \frac{2\eta S_0}{\hbar c}\,\left[{\vec {v}}\times {\vec {\si}}\right]\,.
\label{eq2}
\eeq
Consider first the case when $\,S_0 = 0\,$ so that only
${\vec {S}}$ is present. Since $\,{\vec S} = const$,
we can safely put $\,S_{1,2}=0$. The solution for the spin
can be easily found to be
\beq
{\si}_3 = {\si}_{30} = const,\,\,\,\,\,\,\,\,\,\,\,\,
\si_1 = \rho\,\cos \left( \frac{2\eta S_3t}{\hbar}\right)
,\,\,\,\,\,\,\,\,\,\,\,\,
\si_2 = \rho\,\sin \left( \frac{2\eta S_3t}{\hbar}\right)\,,
\label{sol-spin}
\eeq
where $\rho = \sqrt{\si_{10}^2 + \si_{20}^2}$.
For the first two components of the velocity we have
$\,\,v_1=v_{10}=const,\,\,\,v_2=v_{20}=const,\,\,$
 but the solution for $v_3$ turns out to be complicated. For  
$\si_3 \neq 0$ the solution is
$$
v_3(t) = \left[\, v_{30}\, + \,
\frac{\rho\hbar\,\left(\si_3 v_{10}\hbar -
2mv_{20}\right)}{4m^2 + \hbar^2\si_3^2}\,\right]
\;e^{-\frac{\eta S_3 \si_3}{m}\,t}\, -
$$
\beq
-\,\frac{\rho\hbar}{4m^2 + \hbar^2\si_3^2}\;\left[
\, \left(\si_3 v_{10}\hbar - 2mv_{20}\right)
\,\cos \left( \frac{2\eta S_3t}{\hbar}\right)
\,+\, \left(\si_3 v_{20}\hbar + 2mv_{10}\right)
\,\sin \left( \frac{2\eta S_3t}{\hbar}\right) \,\right]\,.
\label{sol-skor2}
\eeq
Physically, such a solution means i) precession of the spin around the
direction of ${\vec S}$ and ii) oscillation of the particle velocity
in this same direction accompanied (for $\si_3\neq 0$ )
by the exponential damping of the initial velocity in
this direction. We remark that the value of the
relic torsion field should be very weak so that very precise
experiments will be necessary to measure these (probably extremely
slow) precession, oscillation and damping.

Consider another special case $\,{\vec {S}} = 0\,$
and $\,S_0\neq 0\,$, which is the form of the torsion field 
motivated by the isotropic cosmological model 
(\ref{nashi_starye_dobrye_dostijenia}). The equations of motion 
have a form \footnote{The analysis of the equations
(\ref{equa3}) in our paper \cite{rysh} was wrong.
I am very grateful to Luiz Garcia de Andrade who found
this mistake and noticed me about it.}:
$$
\frac{d{\vec {v}}}{dt} = \frac{2\eta^2 S_0^2}{c \hbar}\,
\left[{\vec {v}}\times {\vec {\si}}\right]\,,
$$
\beq
\,\,\,\,\,\,\,\,\,
\frac{d{\vec {\si}}}{dt} = - \frac{2\eta S_0}{\hbar}\,
\left[{\vec {v}}\times {\vec {\si}}\right]\,.
\label{equa3}
\eeq
In order to analyze these equations we notice, that the
squares $\,\,\vec {v}^{\,2}\,,\,\,\,\,{\vec {\si}}^{\,2}\,\,$ and 
the product $(\vec {v}\cdot  {\vec {\si}})$ are integrals of motion. 
Consequently, the magnitudes of the two vectors, and the angle 
between them do not change during the motion. Another obvious 
integral of motion is the linear combination
\beq
\vec {w} = \frac{2\eta S_0}{\hbar}\,\Big(\,{\vec {v}}
+  \frac{\eta S_0}{c}\,{\vec {\si}}\,\Big)\,.
\label{freq}
\eeq
Therefore, the evolution of ${\vec {v}}$ and ${\vec {\si}}$ 
performs such that the plane of two vectors is rotating around the
constant vector ${\vec {w}}$. By elementary means one can check
that the frequency of this rotation is exactly $w=|{\vec {w}}|$,
so that the period is $\,T=2\pi/w$.
In this case, the magnitude and direction of the precession of spin and
acceleration of the particle depends on the magnitude and
mutual orientation of its spin and velocity. It is interesting that
in the case $[\vec {\si}\,,\,\vec {v}]=0$ both spin and velocity
are constants, but as far as $\vec {\si}\,$ and $\,\vec {v}$
are not exactly parallel, the period of precession depends only on
the magnitude of the vector $\vec {w}$. In other words, both vectors
may be infinitesimally non-parallel, and the frequency of the
precession will not be infinitesimal (but the amplitude of the 
precession will). The last observation is that, for the gas of 
particles with random orientation of velocities, their precession
in the $S_0$ field would be also random. We note that the 
possibility of the accelerating motion of the spinning particles 
in an external torsion field has been discussed also in 
\cite{stoeger,singh}.

The torsion field is supposed to act on the spin of particles
but not on their angular momentum \cite{hehl, stoeger}. Therefore a
motion like the one described above will occur for individual electrons
or other particles with spin as well as for macroscopic bodies with
fixed spin orientation but it does not occur for
the (charged or neutral) bodies with random orientations of spins.

%%%%%%%%%%%%%%%%%%%%%%%%%%%%%%%%%%%%%%%%%%%%%%%%%%%%%%%%%%%%%%%%%%%%%
\section{Experimental constraints for the constant background
torsion}
%%%%%%%%%%%%%%%%%%%%%%%%%%%%%%%%%%%%%%%%%%%%%%%%%%%%%%%%%%%%%%%%%%%%%

In the previous and present Chapters we are considering the approach 
in which torsion is purely background field. Thus, we shall describe
only those possible effects which do not need propagating 
torsion. A brief discussion of the possible experimental 
manifestations of the propagating torsion will be given in 
the next Chapter. In principle, the background torsion may 
produce two kinds of effects: the change of the trajectory 
for the particles with (or even without - for the case of 
Higgs particle) spin, or the change of the spectrum due 
to the torsion-dependent terms in the Dirac equation. 
The possible experiments with the motion of particles
are quite obvious. Let the electromagnetic field to be absent.
Then, according to the  results of the previous section,
the interaction with torsion twists the particle trajectory. 
Then, any charged particles may be the source of 
electromagnetic radiation. The structure of the radiation 
provides the opportunity to look for torsion effects.
However, for a very feeble torsion, the electromagnetic
radiation (as a second order effect) will be very weak
and this way of detecting torsion is not really promising.

Let us now comment on the spectroscopy 
part, using the non-relativistic approximation. We shall 
follow the consideration of \cite{babush}.  
Consider the Schroedinger equation (\ref{non4}) with the 
Hamiltonian operator (\ref{Hamiltonian}) without 
electromagnetic field. 
It is evident that the effect of a torsion field can modify
the particles spectrum. Some modifications are similar 
to the ones which arise in the electromagnetic field.
At the same time, another modifications
are possible due to the qualitatively new term
$\;\,\frac{\eta}{mc}\,S_0\, \vec{p}\cdot\vec{\sigma}\;\,$ in 
(\ref{Hamiltonian}).

It is natural to suppose that torsion is feeble enough and 
therefore one can consider it as a perturbation. This 
perturbation
might lead to the splitting of the known spectral lines and 
hence one can, in principle, derive an upper bound for 
the background torsion using the spectral analysis experiments
\footnote{Indeed, torsion effects will compete with the 
relativistic corrections and with the fine structure 
effects coming from QED. Therefore, this our consideration
has mainly pedagogical purpose. At the end of the section 
we shall briefly present the modern limits on torsion 
coming from the complete studies.}.

One can expect the splitting of the spectral lines for 
the hydrogen atom (such a splitting has been also
discussed in \cite{gaha} for torsion coupled to massive
electrodynamics). Consider the constant torsion 
$\;S_\mu = const\;$ and estimate possible modifications
of the spectrum. In this particular case the 
Hamiltonian operator is
$$ %%% \beq
\hat{H} = \frac{1}{2m}\hat{\pi}^2 
+ \eta\, \hat{\vec{S}}\cdot \hat{\vec{\sigma}}  
- \frac{\eta}{2mc}\,(\hat{S}_0\,\hat{\vec{p}} 
+ \hat{\vec{p}}\,\hat{S}_0 )\cdot \hat{\vec{\sigma}}\,,
$$ %%% \label{hammi}
%%%%%% \eeq
where
$$
\vec{\pi} = \vec{p} - \frac{e}{c}\vec{A} \,.   
$$
In the framework of the non-relativistic approximation 
$|\vec{p}| \ll
mc$ and hence the second $S_0$ dependent term in the brackets 
can be omitted. The remaining term 
$\,\,\eta\vec{S}\cdot\vec{\sigma}\,\,$ admits the standard
interpretation and gives the contribution $\pm \eta S_3$ into the
spectrum. Thus, if the $S_3$ component of the torsion tensor is not
equal to zero, the energy level is splitted into two sub-levels with
the difference $2 \eta S_3$. If now, the week transversal magnetic
field is switched on then the cross between the new levels will arise
and the energy absorption takes place at the magnetic field frequency 
$w = \frac{2\eta}{S_3}$. Note that the situation is typical for the
magnetic resonance experiments, however in the present case the effect
arises due to the torsion, but not to the magnetic field effects.
\vskip 3mm

%%%%%%%%%%%%%%%%%%%%%%%%%%%%%%%%%%%%%%%%%%%%%%%%%%%%%%%%%%%%%%%
There were several attempts to draw numerical bounds on the
background torsion using known experiments and the
torsion corrections to the Schr\"{o}edinger or Dirac equation.
One can distinguish
two approaches. One of them is more traditional, it does not
really distinguish between purely background and propagating
torsion. It is sufficient to suppose that the torsion mass is
dominating over the possible kinetic terms. In this case the
effect of torsion is to provide the contact spin-spin interactions.
As an examples of works done in this direction one can mention
\cite{hehl} (one can find there more references) and \cite{hammond},
where the Pauli-like equation (\ref{non4}), (\ref{non5})
has been applied together with the Einstein-Cartan action for 
torsion. The most recent and complete
upper bound for torsion parameters from the contact interactions
have been obtained in \cite{betor}. We shall present the
corresponding results in the next Chapter after discussing the
problem of torsion mass and propagating torsion.

An alternative way is to suppose the existence, in our part of the
Universe, of some constant torsion axial vector $S_\mu$, and to
look for its possible manifestations. The most
recent publication with the analysis of this possibility and the
derivation of the corresponding upper bound for torsion is
\cite{lamme}, where the constraints on the space-time torsion
were obtained from the data on the Hughes-Drever experiments
on the basis of the Pauli-like equation. 
The limit on the magnitude of the space component
of the antisymmetric contorsion has been obtained
in \cite{lamme} using the experimental data 
concerning violation of Lorentz invariance \cite{chupp}.

Besides these papers devoted to the search of the torsion effects,
there were, in the last decades, numerous publications on the same
subject, but without explicit mentioning the word ``torsion''. 
These works were devoted to the search for the Lorentz
and CPT violations coming from various odd terms in the Dirac
equation. The most popular form (there are others) of such an
insertion is the $b_\mu$ axial vector. The modified form of the
Dirac equation is
\beq
\Big(\,i\,\ga^\mu\,D_\mu-\ga_5\,\ga^\mu\,b_\mu-m\,\Big)\,\psi=0\,,
\label{b-field}
\eeq
where $D_\mu =\pa_\mu - i\,A_\mu$. It is easy to see that the
$b_\mu$ is nothing but the normalized axial vector of torsion
$\,b_\mu=\eta\,S_\mu$ and, fortunately, we have the possibility
to use the corresponding data on the limits coming from the
CPT and Lorentz anomalies. There is no reason to repeat the
details of the existing numerous reviews on the subject (see, 
for example, \cite{bluhm,kosto} and references therein),
so we shall just present the main results.
The violation of the Lorentz and CPT symmetries occurs because
$b_\mu$ is a constant vector with the fixed space component.
Consequently, any Lorentz boost breaks the form of the Dirac
equation. The limits on the magnitude of the $b_\mu$ fields
come from the studies of
neutral-meson oscillations in the kaon system,
experimental test with leptons and barions using Penning traps,
comparative spectroscopy of the hydrogen and
anti-hydrogen atoms, measurements of muon properties,
clock-comparison Hughes-Drever type experiments, 
observation of the anomaly in the behaviour of the spin-polarized
torsion pendulum and tests with the spin-polarized solids
\cite{blukos}. The overall limits on $\,\left|b\right|\,$
differ and depend on the type of experiment. In particular, these
limits are different for different fermions. 
These limits are typically from $\,10^{-25}\,GeV$
to $\,10^{-30}\,GeV$, so that the universal phenomenological
bound, valid for all fermion species, is between $\,10^{-27}\,GeV\,$
and $\,10^{-30}\,GeV$. If we really associate $\,b_\mu$ vector 
with torsion, and remember the renormalization-group
based arguments (see section 3.4) about the universality 
of the fermion-torsion interaction, the estimates for different
fermions can be put together and we arrive at the total universal
limit $\,\left|b\right| < 10^{-30}\,GeV$.
Thus, the limits derived from numerous laboratory experiments, are 
very small. They leave no real chance to use torsion for the 
explanation of physical phenomena \cite{doma} like the anomaly in 
the polarization of light coming from distant galaxies 
\cite{nodrol}. The same concerns the creation of particles by 
external torsion field \cite{rump} and the helicity flip for the 
solar neutrino which could be, in principle, induced by torsion
\cite{hammond2}.
%%%%%%%%%%%%%%%%%%%%%%%%%%%%%%%%%%%%%%%%%%%%%%%%%%%%%%%%%%%%%%%%%%%%%%%%
%%%  \newpage
\vskip 10mm

%%%%%%%%%%%%%%%%%%%%%%%%%%%%%%%%%%%%%%%%%%%%%%%%%%%%%%%%%%%%%%%%%%%%%%%%
%%%%%%%%%%%%%%%%%%%%%%%%%%%%%   CHAPTER 5  %%%%%%%%%%%%%%%%%%%%%%%%%%%%%
%%%%%%%%%%%%%%%%%%%%%%%%%%%%%%%%%%%%%%%%%%%%%%%%%%%%%%%%%%%%%%%%%%%%%%%%
\chapter{The effective quantum field theory approach for the 
dynamical torsion}
%%%%%%%%%%%%%%%%%%%%%%%%%%%%%%%%%%%%%%%%%%%%%%%%%%%%%%%%%%%%%%%%%%%%%%%%
%%%%%%%%%%%%%%%%%%%%%%%%%%%%%%%%%%%%%%%%%%%%%%%%%%%%%%%%%%%%%%%%%%%%%%%%

The theoretical description of any new field, including torsion, must 
have two important elements: the interaction of this field with the well 
established matter fields and the proper dynamics of the new field. 
From Quantum Field Theory point of view, any classical description may
be considered as an approximation to some complete theory including quantum 
effects. Following this line, one has to construct the theory of torsion 
in such a way that it would pass the necessary test of consistency as a 
quantum theory. As we have seen in the previous two Chapters, the 
interaction of torsion with matter does not lead to any difficulty, 
until we consider torsion as a purely background field. However, this 
semi-classical theory is definitely incomplete if we do not attempt to 
formulate torsion dynamics. The simplest approach is just to postulate the 
Einstein-Cartan theory (\ref{EC}) as a torsion action. As we have already 
learned in Chapter 2, in this case torsion does not propagate and leads 
only to the contact spin-spin interactions. Furthermore, since the torsion 
mass is of the Planck order of magnitude, such a contact interaction 
is suppressed, at low energies, by the Planck mass. As a result there 
are very small chances to observe torsion at low energies. The main 
purpose of the present Chapter is to follow the recent papers 
\cite{betor,guhesh} where we have discussed an alternative possibility 
for a smaller torsion mass. We start the Chapter by making a short 
account of the previous works on the dynamical torsion, and 
then proceed by applying the ideas of effective quantum field theory 
to the formulation of torsion dynamics. We shall mainly 
concentrate on the theoretical aspects, and provide only a short 
review of the phenomenological bounds on the torsion mass and 
couplings. The interested reader is referred to the second paper
in Ref. \cite{betor} for further phenomenological details. 

%%%%%%%%%%%%%%%%%%%%%%%%%%%%%%%%%%%%%%%%%%%%%%%%%%%%%%%%%%%%%%%%%%%%%%%%
\section{Early works on the quantum gravity with torsion}
%%%%%%%%%%%%%%%%%%%%%%%%%%%%%%%%%%%%%%%%%%%%%%%%%%%%%%%%%%%%%%%%%%%%%%%%

Since the early days (see \cite{hehl} for a review) torsion
has been considered as an object related to quantum theory. Thus, it is
natural to discuss propagating torsion in terms of Feynman diagrams, 
Green functions and $S$-matrix instead of using the dynamical equations.

As any other propagating field, torsion must satisfy the condition of 
unitarity. So, it is natural that the attempts to construct the theory
of the propagating torsion started from the study of the constraints 
imposed by the unitarity \cite{nev1,nev2,seznie,sez,nev3}. 
The initial motivation of  \cite{nev1,nev2,seznie,sez} was to construct 
the theory of quantum gravity which would be both unitary and 
renormalizable. So, let us briefly describe the general situation in 
Quantum Gravity (see \cite{susy-enc} for more extensive review). 
It is well known that the program of quantizing General Relativity
met serious difficulty, because this quantum theory is non-renormalizable 
by power counting. If taking only the superficial  logarithmic 
divergences of the diagrams into account, the
dimension (number of derivatives of the metric) of the $n$-loop
counterterms is $\,d = 2+2n$. Then, 
with every new order of the loop expansion the dimension of the
counterterms grows up and hence one needs an infinite number 
of the renormalization conditions to extract a single prediction of 
the theory in the high energy region. The non-renormalizability of 
quantum General Relativity becomes apparent already at the one-loop
level for the case of gravity coupled to matter \cite{hove,dene}
and at two-loop level for the pure gravity \cite{2loop}. The
situation in supergravity (see, e.g. \cite{super})
is better in the sense that the on-shell divergences do not 
show up at second ($N=1$ case) or even higher (perhaps seventh 
for $N=8$ supergravity) loops. However, the supersymmetry does not
solve the principal problem, and all known versions of  
supergravity generalizations of General Relativity are expected to
be non-renormalizable. 

At the same time, it is fairly simple to construct renormalizable 
theory of the gravitational field by adding the fourth-derivative 
terms \cite{stelle}
\beq
S_{HD}\, = \,\int d^4x\sqrt{-g}\,
\left(\,\alpha R_{\mu\nu\rho\sigma}R^{\mu\nu\rho\sigma}
+ \beta R_{\mu\nu}R_{\mu\nu} + \gamma R^2 +\de\,{\Box}\,R
\,\right)
\label{highQG}
\eeq 
into the classical action. It is better to write the above 
expression in another basis, so that the algebraic properties 
of the terms become more explicit. Let us use (\ref{basis}), 
so that the eq. (\ref{highQG}) can be written as 
\beq
S_{HD}\, = \,\int d^4x\sqrt{-g}\,
\left(\,a_1\, C^2 + a_2\, E + a_3\, R^2 + a_4\,{\Box}\,R
\,\right)\,
\label{highQGmod}
\eeq 
It is well known (see, e.g. \cite{stelle,book}), that
the contributions of the terms of eq. (\ref{highQGmod})
to the propagator of the gravitational perturbations are very
different. This propagator can be divided into irreducible 
parts through introducing the projectors to the spin-2, spin-1
and spin-0 states. It turns out that the spin-1 
states can be completely removed by the gauge fixing
(of course, the Faddeev-Popov ghosts must be taken into account). 
Furthermore, the $C^2$ term contributes to the gauge fixing 
independent spin-2 part of the propagator, while the $R^2$-term
contributes only to the spin-0 part and does not contribute to 
the spin-2 part. The term $\int E$ does 
not contribute to the propagator at all, even if we change
the dimension of the space-time from $\,4\,$ to $\,n$. 
In $\,n=4\,$ this term is topological, so it is supposed not 
to affect the vertices either (algebraically, the situation is not 
so simple \cite{kimber}, but there are no indications of the 
non-trivial quantum effect of this term). 
At the same time, for $n\neq 4$ all three terms contribute 
to all vertices:
to the interactions of the metric components of all spins. 

Now, suppose we take a theory with the action 
\beq
S_{t}\, = \,\int d^4x\sqrt{-g}\,
\left[\,-\frac{1}{\ka^2}\,R 
+ a_1\, C^2 + a_2\, E + a_3\, R^2 \,\right]\,.
\label{highQGtot}
\eeq 
with $a_1\neq 0$ and $a_3\neq 0$. In this case the propagators
of all components, including ghosts \cite{book}, behave like 
$k^{-4}$ in the high energy domain. Similarly, there are vertices 
proportional to the fourth, second and zero power of the momenta.
In this situation, the power counting shows that the superficial 
degree of divergence of the IPI
diagrams does not depend on the  loop order, and the maximal 
possible dimension of the logarithmic counterterms is $4$. 
Taking the locality and general covariance of the counterterms 
\cite{stelle,votyu84} into account, we see that they have the 
same form as the classical action  
(\ref{highQGtot}). Thus, the theory is renormalizable. 

Still, it is not perfect. The spin-2 sector of the propagator 
has the form
\beq
G^{(2)}(k) \sim \frac{1}{k^2(k^2 + m^2_2)}
= \frac{1}{m_2^2}\,\left[\frac{1}{k^2} 
- \frac{1}{k^2 + m^2_2} \right]\,,
\label{prop2}
\eeq
where $m_2 \sim 1/\ka^2$ is the mass of the non-physical ghost. 
Another massive pole exists in the spin-$0$ sector, but its
position depends on the gauge fixing while the spin-2 part 
(\ref{prop2}) is gauge independent. Excluding the non-physical 
particles from the physical spectrum, one breaks unitarity 
\cite{stelle}. The situation described above is quite general, 
because further modifications of the action do not change the 
situation. For instance, introducing more higher derivative 
terms into the action can not provide unitarity \cite{highderi}.
It is quite obvious that the situation can not be improved 
by changing the dynamical variables, because this operation 
can not change the position of the massive gauge independent 
pole of the propagator. 

Some observation is in order. If we put, in the action (\ref{highQGtot}),
the coefficient $a_1=0$, the theory will not be renormalizable 
(despite it still possesses higher derivatives). The point is that
such a theory contains high derivative vertices of the interaction 
of spin-2 states with themselves and with other states, but the
spin-2 part of the propagator behaves like $G^{(2)}(k) \sim k^{-2}$. 
It is easy to see that the power counting in this theory would 
be even worst than in quantum General Relativity. Thus, the 
renormalizability of quantum gravity theory with high derivatives
crucially depends on the $k^{-4}$ UV behaviour of the spin-2
propagator. In turn, since the spin-2 sector always contains 
a $k^{-2}$ part due to the Einstein term, the (\ref{prop2})
structure of the propagator looks as an unavoidable consequence 
of the renormalizability. Of course, these considerations do 
not have absolute sense, one can try to look for an action which 
could solve the problem of quantum gravity. 

The original idea of \cite{nev1} was to use the first order formalism
and to establish the combination 
of the $R^2_{\,..\,}$-terms which could preserve the renormalizability 
and, simultaneously, provide the absence of the massive unphysical
pole. As it was already explained in section 2.5, the first order 
formalism treats the affine connection as an object independent 
on the metric. Anyhow, the connection can always be  
presented as a sum of the Christoffel symbol and some additional 
tensor which includes torsion and non-metricity. Furthermore,
if one is interested in the propagator of perturbations on 
the flat background, it is possible to classify all the fields by 
the representations of the Lorentz group. Indeed, the non-metricity 
tensor and $q^\al_{\,\cdot\,\be\ga}$-component of torsion both
contain spin-2 states. At least one of these 
states should be massless in order to provide the correct Newtonian 
limit. Then, if there are other, massive spin-2 states, the 
general structure of the complete spin-2 propagator should be 
equivalent to (\ref{prop2}). Of course, the word ``equivalent''
in the last statement can not be understood as an identity.  
For instance, in the first order formalism 
the propagator may be free of high derivatives at all. The 
equivalence signifies that the {\it r.h.s.} of (\ref{prop2})
will be restored, in the linearized theory, after elimination
of an extra (one can call them auxiliary) fields. 

The first attempt to find the renormalizable and ghost-free
theory of Ref. \cite{nev1} did not include torsion, which was  
implemented later in Refs. \cite{nev2,seznie,sez,nev3}. 
In all these papers, the gravitational field has been 
parametrized by the vierbein $e_\mu^a$ and spinor connection
$w^\mu_{ab}$, but as we have discussed in Chapter 2, 
these variables are equivalent to another ones -- 
($g_{\mu\nu},\,T^\al_{\,\cdot\,\be\ga}$): if one does not 
introduce a non-metricity. 
In principle, one could include, as we mentioned above, all 
curvature and torsion depending terms (there are 168 of them
\cite{chris}) plus all possible terms depending on the 
non-metricity. The technical part of the original papers 
\cite{nev1,nev2,seznie,sez} is quite cumbersome and we 
will not reproduce it here. The final result has been achieved 
for the theories with torsion but without non-metricity. The 
number of ghost-free $R^2_{\,..\,}$-type actions has been 
formulated, but all of them 
are non-renormalizable. Obviously, these actions resemble 
(\ref{highQGtot}) with $a_1=0$, for they possess high 
derivatives in the vertices but not in the propagators. 
If one performs the loop calculation in such theory, 
the ghost-free structure of the propagator will be immediately 
destroyed. Therefore, at the quantum level the unitarity of 
the classical $S$-matrix can be spoiled, if the theory is 
not renormalizable. The conclusion is that, for the consistent
quantum theory, one needs both unitarity and renormalizability.
  
In the next sections we will not focus on the problem of 
quantum gravity. Instead, we shall concentrate on the possibility
to have a theory of the propagating torsion, which should 
be consistent at the quantum level. The application of the ideas of the 
effective field theories enables one to weaken the requirement  
of renormalizability (see, e.g. \cite{weinberg,dogoho}), 
but even in this case
we shall meet serious problems and limitations in constructing
the theory of the propagating torsion. As to the formulation 
of consistent quantum gravity, the string theory is, nowadays,
the only one visible candidate. 

%%%%%%%%%%%%%%%%%%%%%%%%%%%%%%%%%%%%%%%%%%%%%%%%%%%%%%%%%%%%%%%%%%%%%%%%
\section{General note about the effective approach to torsion}
%%%%%%%%%%%%%%%%%%%%%%%%%%%%%%%%%%%%%%%%%%%%%%%%%%%%%%%%%%%%%%%%%%%%%%%%

In order to construct the action of a propagating torsion, 
we have to apply two requirements: unitarity and renormalizability.
The problem of renormalizability is casted in another form if we
consider it in the content of effective field theory 
\cite{weinberg,dogoho}.
In the framework of this approach one has to start with the action
which includes all possible terms satisfying the symmetries of the
theory. Usually, such an action contains higher derivatives at least in
a vertices. However, as far as one is interested in the lower
energy effects, those high derivative vertices are suppressed by the
great massive parameter which should be introduced for this purpose.
Then, those vertices and their renormalization are not visible and
effectively at low energies one meets renormalizable and unitary theory.
The gauge invariance of all the divergences is guaranteed by the
corresponding theorems \cite{volatyu} and thus this scheme may be
applied to the gauge theories including even gravity \cite{effgrav,don}.
Within this approach, it is important that the lower-derivative
counterterms have the same form as the terms included into
the action. This condition, together with the symmetries and the
requirement of unitarity, may help to construct the effective field
theories for the new interactions such as torsion.

If one starts to formulate the dynamical theory for torsion in this
framework, the sequence of steps is quite definite. First, one has
to establish the field content of the dynamical torsion theory and
the form of the interactions between torsion and other fields. 
Then, it is necessary to take into account the symmetries 
and formulate the action in such a way that the resulting theory
is unitary and renormalizable as an effective field theory.
Indeed, there is no guarantee that all these requirements are consistent
with each other, but the inconsistency might indicate that some
symmetries are lost or that the theory with the given
particle content is impossible. In the next sections we consider 
how this scheme works for torsion, and then compare the situation 
with that of effective low-energy quantum gravity 
(see, e.g. \cite{don}).

Among the torsion components
(\ref{irr}) $S_\mu,\,T_\mu,\,q^\al_{\,\cdot\,\be\ga}$, only 
$S_\mu$ is really important, because only this 
ingredient of torsion couples to spinors in a minimal way. 
Therefore, in what follows we shall restrict the consideration 
to the axial vector $S_\mu$ which parameterizes the
completely antisymmetric torsion.

%%%%%%%%%%%%%%%%%%%%%%%%%%%%%%%%%%%%%%%%%%%%%%%%%%%%%%%%%%%%%%%%%%%%%%%%
\section{Torsion-fermion interaction again:
Softly broken symmetry associated with torsion
and the unique possibility for the low-energy torsion action}
%%%%%%%%%%%%%%%%%%%%%%%%%%%%%%%%%%%%%%%%%%%%%%%%%%%%%%%%%%%%%%%%%%%%%%%%

In this section, we  consider the torsion-spinor system 
without scalar fields. Thus we start 
from the action (\ref{diraconly}) of the Dirac spinor nonminimally 
coupled to the vector and torsion fields
$$
S_{1/2}= i\,\int d^4x \,\, {\bar \psi}\, \left[
\,\ga^\al \,\left( {\pa}_\al + ieA_\al 
+ i\,\eta\,\ga_5\,S_\al\,\right) - im \,\right]\,\psi\,.
$$
First one has to establish its symmetries. At this stage we 
consider the vector field $A_\mu$ as an abelian one but later 
we will focus on the vector fields of the SM which are nonabelian. 
The symmetries include the usual gauge transformation (\ref{trans1}), 
and also softly broken symmetry (\ref{trans}):
$$
 \psi' = \psi\,e^{\ga_5\be(x)}
,\,\,\,\,\,\,\,\,\,\,\,\,\,\,
{\bar {\psi}}' = {\bar {\psi}}\,e^{\ga_5\be(x)}
,\,\,\,\,\,\,\,\,\,\,\,\,\,\,
S_\mu ' = S_\mu - {\eta}^{-1}\, \pa_\mu\be(x)\,.
$$
The massive term is not invariant under the last transformation.

The symmetries of the theory have serious impact on the renormalization 
structure. In particular, since the symmetry under (\ref{trans}) is softly
broken, it does not forbid massive counterterms in the torsion sector and 
hence  $S_\mu$ has to be a massive field.
Below we consider the torsion mass as a free parameter which should be 
defined on a theoretical basis, and maybe also subject of 
experimental restrictions.

As far as torsion is taken as a dynamical field,
one has to incorporate it into the SM along with other vector 
fields. Let us discuss the form of the torsion action in the framework 
of the effective approach -- that is focusing on the low-energy 
effects. The higher derivative terms may be 
included into the action, but they are not seen at low energies.
Thus, we restrict the torsion action by the lower-derivative terms
and arrive at the expression:
\beq
S_{tor} = \int d^4 x\,\left\{\, 
-a\,S_{\mu\nu}S^{\mu\nu} + b\,(\pa_\mu S^\mu)^2
+ M_{ts}^2\,S_\mu S^\mu\,\right\}\,,
\label{geral1}
\eeq
where $\,S_{\mu\nu} = \pa_\mu S_\nu - \pa_\mu S_\nu\,$ and $a,b$
are some positive parameters. The action (\ref{geral1}) contains both
transverse vector mode and the longitudinal vector 
mode where the last one is equivalent to the scalar \cite{carroll}.
In particular, in the $a=0$ case only the scalar mode,
and for $b=0$ only the vector mode propagates.
It is well known \cite{vector} (see also \cite{carroll} for the
discussion of the theory (\ref{geral1})) that in the unitary
theory of the vector field the longitudinal and transverse
modes can not propagate simultaneously
\footnote{One can easily check this without even looking into 
the textbooks: the kinetic terms for the transverse vector and 
for the ``longitudinal'' scalar, coming from the 
$\,\,a\,(\pa_\mu S^\mu)^2$-term, have opposite signs, while 
both fields share massive term. As a result one of these fields 
must have, depending on the signs of $a,b,M_{ts}^2$, either 
negative mass or negative kinetic energy. Hence, the theory with 
both components always includes either ghost or tachyon.}, and 
therefore one has to choose  either $a$ or $b$ to be zero.

In fact the only correct choice is $b=0$. 
In order to see this one has to reveal
that the symmetry (\ref{trans}), which is spoiled by the massive terms
only, is preserved in the renormalization of the dimensionless coupling
constants of the theory (at least at the one-loop level). 
In other words, the divergences and
corresponding local counterterms, which produce the dimensionless
renormalization constants, do not depend on the dimensional parameters
such as the masses of the fields. This structure of renormalization 
resembles the one in the Yang-Mills theories with spontaneous
symmetry breaking. As we shall see later, 
even the $b=0$ choice is not free of problems, but at least 
they are not related to the leading one-loop effects, as in the
opposite $a=0$ choice. Thus, the only one possible 
torsion action is given by Eq. (\ref{geral}) with $b=0$.
In order to illustrate this, we remind that the divergences
coming from fermion loop are given by the expression (\ref{contra}),
which is in a perfect agreement with the transformation 
(\ref{trans}). Namely, the one-loop divergences contain $S_{\mu\nu}^2$ 
and the massive term while the $\,\left(\pa_\nu S^\nu\right)^2$ 
term is absent.

It is well-known that the fermion loop gives rise, in the theory
(\ref{diraconly}), to axial anomaly. But, as we have 
discussed in section 3.8, the problem 
of anomaly does not spoil our attempts to implement torsion 
into the fermion sector of the SM. 
And so, the only possible form of the torsion action which 
can be coupled to the spinor field (\ref{diraconly}) is
\beq
S_{tor} = \int d^4 x\,\left\{\, -\frac14\,S_{\mu\nu}S^{\mu\nu}
+ M_{ts}^2\, S_\mu S^\mu\,\right\}\,.
\label{action}
\eeq
In the last expression we put the conventional coefficient $\,-1/4$
in front of the kinetic term. With respect to the renormalization
this means that we (in a direct analogy with QED) can remove the 
kinetic counterterm by the renormalization of the field $S_\mu$ 
and then renormalize the parameter $\eta$ in the action 
(\ref{diraconly}) such that the combination $\eta S_\mu$ is the 
same for the bare   and renormalized quantities. Instead one can 
include $1/\eta^2$ into the kinetic term of (\ref{action}), that 
should lead to the direct renormalization of
this parameter while the interaction of torsion with spinor has
minimal form (\ref{dirac2}) and $S_\mu$ is not renormalized.
Therefore, in the case of a propagating torsion the difference
between minimal and nonminimal types of interactions is only a
question of notations on both classical and quantum levels.

%%%%%%%%%%%%%%%%%%%%%%%%%%%%%%%%%%%%%%%%%%%%%%%%%%%%%%%%%%%%%%%%%%%%%%%%
\section{Brief review of the possible torsion effects in 
high-energy physics}
%%%%%%%%%%%%%%%%%%%%%%%%%%%%%%%%%%%%%%%%%%%%%%%%%%%%%%%%%%%%%%%%%%%%%%%%

As we have already seen, spinor-torsion interactions enter the Standard 
Model as interactions of fermions with a new axial vector field 
$S_{\mu}$. Such an interaction is characterized 
by the new dimensionless parameter  -- coupling constant $\eta$. 
Furthermore, the mass of the torsion field $M_{ts}$ is unknown, 
and its value is of crucial importance for the possible experimental
manifestations of the propagating torsion and finally for the 
existence of  torsion at all (see the discussion in the last 
sections of this Chapter and in the Chapter 6). In the present section 
we consider $\,\eta\,$ and $\,M_{ts}\,$ as arbitrary parameters and 
review the limits on their values from the known experiments 
\cite{betor}. Later on we shall see that the consistency of the 
fermion-torsion system can be achieved for the heavy torsion 
only, such that $M_{ts} >> M_{fermion}$. However, we shall follow 
\cite{betor} where one can find the discussion of the ``light'' 
torsion with the mass of the order of 1 GeV. 

The strategy of \cite{betor} was to use known experiments 
directed to the search of the new interactions. One can regard torsion 
as one of those interactions and obtain the limits for the torsion 
parameters from the data which already fit with the phenomenology. 

Torsion, being a pseudo-vector particle interacting with fermions, 
might change different physical observables. For instance, 
this specific type of
interaction might lead to the forward-backward asymmetry. The last has 
been precisely measured at the LEP $e^+e^-$ collider, so the upper 
bounds for torsion parameters may be set from those measurements.  
One can consider two different cases: 
i) torsion is much heavier than other particles of the SM and 
$\,\,$ 
ii) torsion has a mass comparable to that of other particles. 
In the last case one meets a propagating particle which must be 
treated 
on an equal footing with other constituents of the SM.  Contrary to 
that, the very heavy torsion leads to the effective contact 
four-fermion interactions. Let us briefly review all mentioned 
possibilities.
\vskip 3mm

i) {\large\it Forward-backward asymmetry.}
Any parity violating interactions eventually give rise to the space
asymmetry and could be revealed in the  
forward-backward asymmetry of the particle scattering. 
Axial-vector type interactions of torsion with matter fields is
this case of interactions. But the  source 
of asymmetry also exists in the SM electroweak interactions  
because of the presence of the $\gamma_\mu\gamma_5$ structure in
the  interactions of $Z$- and $W$-bosons with  fermions.
The interactions between $Z$-boson and fermions can be written in
general form as:
\beq
L_{Zff}= -\frac{g}{2\, cos\theta_W}
\sum_i \,{\bar {\psi}}_i\,\gamma^\mu(g^i_V-g^i_A\gamma^5) \psi\, 
Z_\mu\,,
\label{zff}
\eeq
where, $\theta_W$ is Weinberg angle, $g=e/sin\theta_W$
 ($e$ - positron charge); 
and the vector and axial couplings are:
\begin{eqnarray}
g_V^i&\equiv& t_{3L}(i)- 2 q_i\, sin^2\theta_W,\\
g_A^i&\equiv& t_{3L}(i).
\label{ga}
\end{eqnarray}
Here $t_{3L}$ is the weak isospin of the fermion
and $i$ has the values $+1/2$ for $u_i$ and $\nu_i$ while it is 
$-1/2$ for $d_i$ and $e_i$. Here
$i=1,2,3\,$ is the index of the fermion generation and
$q_i$ is the charge of the $\psi_i$ in  units of charge of positron.

The forward-backward asymmetry for $e^+e^- \rightarrow l^+l^-$ 
is defined as 
\begin{equation}
A_{FB}\equiv \frac{\sigma_F-\sigma_B}{\sigma_F+\sigma_B},
\label{asym}
\end{equation}
where $\sigma_F(\sigma_B)$ is the cross section for $l^-$
to travel forward(backward) with 
respect to electron direction. Such an asymmetries are measured
 at LEP \cite{ewdata}.

In the SM, from asymmetries one derives the ratio $g_V/g_A$ of 
vector and axial-vector couplings, but the presence of torsion could 
change the result. In fact, the measured electroweak parameters 
are in a good agreement with the
theoretical predictions and hence one can establish the limits on the
torsion parameters based on the experimental errors. The contribution 
from torsion exchange diagrams has been calculated 
in \cite{betor}. From those calculations one can establish the limits 
on $\eta$ and $M_{ts}$ taking into account the mentioned error of 
the experimental measurements. Deviations of the asymmetry from SM
predictions would be an indication of the presence of the additional 
torsion-like type axial-vector  interactions. The analysis of
\cite{betor} shows that the electron $A^e_{FB}$ asymmetry is the best 
observable among others asymmetries to look for torsion.
The details of the corresponding analysis can be easily found in 
\cite{betor} and we will not reproduce them here. 
\vskip 3mm

ii) {\large\it Contact interactions}.
Since the torsion mass comparable to the mass of the fermions
leads to problems (which will be explained in the next sections
of this Chapter), it is especially important for us to consider 
the case of heavy torsion. Since
the massive term dominates over the covariant kinetic
part of the action, the last can be disregarded. Then the total 
action leads to the algebraic equation of motion for $\,S_\mu$. 
The solution of this equation can be substituted back into
the action and thus produce the contact four-fermion
interaction term
\beq
{\cal L}_{int} = - \frac{\eta^2}{M_{ts}^2}\,
({\bar \psi}\ga_5\ga^\mu\psi)\,({\bar \psi}\ga_5\ga_\mu\psi)
\label{contact}
\eeq

As one can see the only quantity which appears in this 
approach is the ratio ${M_{ts}}/{\eta}\,$ and therefore for the 
very heavy torsion field the phenomenological consequences 
depend only on this single parameter. As it was mentioned above, the
axial-vector type interactions would give rise to the forward-backward
asymmetry which have been precisely measured in the
$e^+e^- \rightarrow l^+l^- (q\bar{q})$  scattering
(here $l=(\tau,\mu,e)$ stands for the leptons and $q$ for quarks)
at LEP collider with the center-mass energy near the Z-pole.
Due to the resonance production of Z-bosons the statistics is good
(several million events) and it allowed to measure electroweak (EW)
parameters with high precision. There are several experiments
from which the constraints on the contact four-fermion interactions
come \cite{betor}:
\vskip 1mm

\noindent
1)Experiments on polarized electron-nucleus scattering:
\noindent
SLAC e-D scattering experiment \cite{slac}, 
Mainz e-Be scattering
experiment \cite{mainz} and bates e-C scattering experiment 
\cite{bates};
\vskip 1mm

\noindent
2)Atomic physics parity violations measures \cite{apv}
electron-quark coupling that are different from those tested at
high energy experiment provides alternative constraints on new physics
(see also section 4.6).

\vskip 1mm
\noindent
3) $e^+e^-$ experiments  - SLD, LEP1, LEP1.5 and LEP2 (see for example
\cite{lep,opal,l3,aleph,lang});
\vskip 1mm
\noindent
4)Neutrino-Nucleon DIS experiments -- CCFR collaboration obtained 
a model independent constraint on the
effective $\nu\nu q q$ coupling \cite{ccfr}.
\vskip 1mm

Consider the limits on the contact interactions induced by the
torsion. The contact four-fermion interaction may be described by the
Lagrangian~\cite{eich} of the most general form:
\begin{equation}
L_{\psi'\psi'\psi\psi}=g^2\sum_{i,j=L,R}\sum_{q=u,d}\frac{\epsilon_{ij}}
{(\Lambda_{ij}^{\epsilon})^2}\,
(\bar \psi'_i\gamma_\mu \psi'_i)\,(\bar \psi_j\gamma^\mu \psi_j)
\label{cont}
\end{equation}
Subscripts $i,j$ refer to different fermion helicities:
$\, \psi^{(')}_i= \psi^{(')}_{R,L}= (1\pm\gamma_5)/2\cdot \psi^{(')}\,$;
where $ \psi^{(')}$ could be quark or lepton; $\,\,\Lambda_{ij}\,$
represents the mass scale of the exchanged new particle; coupling 
strength
is fixed by the relation: $\,g^2/4\pi=1$,
the sign factor  $\,\epsilon_{ij}=\pm 1\,$
allows for either constructive or destructive interference with
the SM $\gamma$ and $Z$-boson exchange amplitudes.
The formula (\ref{cont}) can be successfully used for the study of the
torsion-induced contact interactions because it includes an axial-axial
current interaction (\ref{contact}) as a particular case.

Recently, the global study of the
electron-electron-quark-quark($eeqq$) interaction sector of the
SM have been done using data from all mentioned 
experiments \cite{global}. For the axial-axial
$\,eeqq\,$ interactions~(\ref{cont}) takes the form (we put $g^2=4\pi$):
\beq
L_{eeqq}=-\frac{4\pi}{{(\Lambda_{AA}^{\epsilon})^2}}
(\bar e\gamma_\mu\gamma_5 e)(\bar q\gamma^\mu \gamma_5 q)
\label{contaa}
\eeq
The limit for the contact axial-axial $eeqq$ interactions comes
from the global analysis of Ref.~\cite{global}:
\beq
\frac{4\pi}{\Lambda_{AA}^2} < 0.36\mbox{ TeV} ^{-2}
\label{glob}
\eeq
Comparing the parameters of the effective contact four-fermion
interactions of general form
(\ref{contaa}) and contact four fermion interactions induced by
torsion (\ref{contact}) we arrive at the following relations:
\beq
\frac{\eta^2}{M_{ts}^2}=\frac{4\pi}{{\Lambda_{AA}}^2}
\label{rel}
\eeq
From (\ref{glob}) and (\ref{rel}) one gets the following limit
on torsion parameters:
\beq
\frac{\eta}{M_{ts}}<0.6\mbox{ TeV}^{-1}\; \Rightarrow \;
M_{ts}>1.7\mbox{ TeV }\cdot\eta\,.
\label{globfin}
\eeq
The last relation puts rigid phenomenological constraints 
on the torsion parameters $\eta,\,M_{ts}$, and one can also 
take into account that the modern scattering-based analysis 
can not be relevant for the masses beyond the $3\,TeV$. 
It is easy to see that the bound (\ref{globfin}) is incompatible 
with the one established in \cite{hammond}, because in this
paper the torsion mass has been taken to be of the Planck order 
of magnitude. 

An additional restriction 
can be obtained from the analysis of the TEVATRON data 
but since they concern mainly the light torsion, we will not 
give the details here. In \cite{betor}, one can find the 
total limits on torsion from an extensive variety of experiments
(see also \cite{extra-dim} for the consequent analysis
concerning the torsion coming from small extra dimensions).

In conclusion, we learned that torsion, if exists, may produce 
some visible effects, while the phenomenological analysis put 
some limits on the torsion parameters $\eta,\,M_{ts}$. Of course, 
these limits will improve if the experimental data and(or) the 
theoretical
derivation of observables in the SM become more precise. In case
one detects some violation of the phenomenological bounds, 
one can suppose that this is a manifestation of some new physics.
This new physics can be torsion or something else (GUT, 
supersymmetry, higher dimensions and so on). In each case one has
to provide the consistent quantum theory for the corresponding 
phenomena. Therefore, the relevance of phenomenological 
considerations always depends on the formal field-theoretical 
investigation. As we shall see in the next sections, in the case 
of torsion the demands of the theory are more restrictive than 
the phenomenological limits. 

%%%%%%%%%%%%%%%%%%%%%%%%%%%%%%%%%%%%%%%%%%%%%%%%%%%%%%%%%%%%%%%%%%%%%%%%%
\section{First test of consistency: loops in the fermion-scalar 
systems break unitarity}
%%%%%%%%%%%%%%%%%%%%%%%%%%%%%%%%%%%%%%%%%%%%%%%%%%%%%%%%%%%%%%%%%%%%%%%%

To this point, we have considered only interaction between torsion
and spinors. Now, in order to implement torsion into the SM, one
has to include scalar and Yukawa interactions. 
When introducing scalar field we shall follow the
same line as in the previous section and try first to construct the
renormalizable theory. Hence, the first thing to do is to analyze
the structure of the possible divergences. The divergent diagrams
in the theory with a dynamical torsion include, in particular, all
such diagrams with external lines of torsion and internal lines
of other fields. Those grafs are indeed the same one meets in 
quantum field theory on a classical torsion background. Therefore,
one has to include into the action all terms which were necessary
for the renormalizability when torsion was a purely external field.
All such terms are already known from our investigation of quantum
field theory on an external torsion background. Besides the
nonminimal interaction with spinors, one has to introduce the 
nonminimal interaction $\,\,\ph^2 S^2\,\,$ between scalar field
and torsion as in (\ref{scalar1-nm}) 
and also the terms which played the
role of the action of vacuum (see, e.g., (\ref{scalardivs})
or (\ref{1d})) in the form
\beq
S_{tor} = \int d^4x\,\left\{\, - \frac14\,S_{\mu\nu}S^{\mu\nu}
+ M_{ts}^2\, S_\mu S^\mu 
- \frac{1}{24}\,\ze \left(S_\mu S^\mu\right)^2
\,\right\} + {\rm surface\, terms}\,.
\label{act}
\eeq
Here $\,\ze\,$ is some new parameter, and the coefficient
$\,\,{1}/{24}\,\,$ stands for the sake of convenience only. The 
necessity of the $\left(S_\mu S^\mu\right)^2$
term in the classical action follows from the fact that such a
term emerges from the scalar loop with a divergent coefficient. 
Then, if not included into the classical action, it will
appear with 
infinite coefficient as a quantum correction. On the other hand, 
by introducing this term 
into the classical action we gain the possibility to remove 
the corresponding divergence renormalizing the coupling $\ze$. 

So, if one implements torsion into the complete SM including the 
scalar field, the total action includes the following new terms: 
torsion action  (\ref{act}) with the self-interacting term, and 
nonminimal interactions between torsion and spinors (\ref{diraconly}) 
and scalars (\ref{scalar1-nm}). However, at the quantum level, such 
a theory suffers from a serious difficulty.

The root of the problem is that the Yukawa interaction term
$\,h\ph{\bar {\psi}}\psi$ is not invariant under the transformation
(\ref{trans}). Unlike the spinor mass, the Yukawa constant $h$ is
massless, and this noninvariance may affect the
renormalization in the massless sector of the theory. In particular,
the noninvariance of the Yukawa interaction causes the necessity
of the nonminimal scalar-torsion interaction in 
(\ref{scalar1-nm}) which,
in turn, requires an introduction of the self-interaction term in
(\ref{act}). Those terms do not pose any problem at the one-loop
level, but already at the second loop one meets two dangerous diagrams
presented at Fig. 3.

%%% \vskip 1mm
%%%%%%%%%%%%%%%%%%%%%%%%%%%%%%%
\noindent
\begin{picture}(120,120)(0,0)
\Line(0,23)(25,23)
\Line(0,27)(25,27)
\Line(0,23)(0,27)
\BCirc(50,25){27}
\BCirc(50,25){23}
\Line(75,23)(100,23) 
\Line(75,27)(100,27)
\Line(100,23)(100,27) 
\Line(25,23)(75,23) 
\Line(25,27)(75,27) 
\Vertex(75,25){3} \Vertex(25,25){3} 
\end{picture}
$\,\,\,\,\,\,\,\,\,\,$ $\,\,\,\,\,\,\,\,\,\,$
%%%%%%%%%%%%%%%%%%%%%%%%%%%%%%%
\begin{picture}(120,120)(0,0)
\Line(0,23)(25,23)
\Line(0,27)(25,27)
\Line(0,23)(0,27)
\BCirc(50,25){25}
\Line(75,23)(100,23) 
\Line(75,27)(100,27)
\Line(100,23)(100,27) 
\Line(25,23)(75,23) 
\Line(25,27)(75,27) 
\Vertex(75,25){3} \Vertex(25,25){3} 
\end{picture}  
\vskip 3mm
\noindent
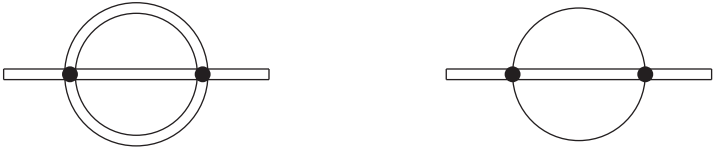
{\small \it Figure 3.  Two-loop diagrams corresponding to 
torsion self-interaction and torsion-scalar interaction.}
\vskip 3mm
%%%%%%%%%%%%%%%%%%%%%%%%%%%%%%%%%%%%%%%%%%%%%%%%%%%%%%%%%%

These diagrams are divergent and they can lead
to the appearance of the $\,\left(\pa_\mu S^\mu\right)^2\,$-type
counterterm. No any symmetry is seen which forbids these divergences.
Let us consider, following \cite{betor}, the diagrams presented 
at Fig. 3 in more details. Using the
actions of the scalar field coupled to torsion (\ref{scalar1-nm}) 
and
torsion self-interaction (\ref{act}), we arrive at the following
Feynman rules:
\vskip 4mm

\noindent
i) Scalar propagator: $\,\,\,\,\,\,\,\,\,\,\,\,
\,\,\,\,\,\,\,\,\,\,\,\,\,\,\,\,\,\,\,\,\,\,\,\,\,\,\,\,\,\,\,\,
\,\,\,G(k) = \frac{i}{k^2+M^2}\,\,\,\,\,\,
{\rm where}\,\,\,\,\,\,M^2=2M_{ts}^2$,
\vskip 2mm

\noindent
ii) Torsion propagator:  $\,\,\,\,\,\,\,\,\,\,\,\,\,\,\,\,\,\,\,\,\,\,
\,\,\,\,\,\,\,\,\,\,\,\,\,\,\,\,\,\,\,\,\,\,
\,\,\,\,\,\,D_\mu^{\,\nu}(k) = \frac{i}{k^2+M^2}
\,\left(\de_\mu^{\,\nu} + \frac{k_\mu k^{\nu}}{M^2} \,\right)\,\,,$
\vskip 2mm

\noindent
iii) The $\ph^2 S^2$ - vertex: $\,\,\,\,\,\,\,\,\,\,\,\,
\,\,\,\,\,\,\,\,\,\,\,\,\,\,\,\,\,\,\,\,\,\,
V^{\mu\nu}(k,p,q,r)\, = \,- \,2 i\xi\,\eta^{\mu\nu}\,\,,$
\vskip 2mm

\noindent
iv) Vertex of torsion self-interaction: 
$\,\,\,\,\,\,\,\,\,V^{\mu\nu\al\be}(k,p,q,r)\, =
 \,\frac{i\ze}{3}\,g_{(4)}^{\mu\nu\al\be}$
\vskip 4mm

\noindent
where $g_{(4)}^{\mu\nu\al\be} = g^{\mu\nu}g^{\al\be} +
g^{\mu\be}g^{\al\nu} + g^{\mu\al}g^{\nu\be}$ and $k,p,q,r$ denote the
outgoing momenta.

The only one thing that we would like to check is the violation of 
the transversality in the kinetic 2-loop counterterms.
We shall present the calculation in some details because it is
quite instructive.
To analyze the loop integrals we have used dimensional
regularization and in particular the formulas from 
\cite{leibr,hela}.
It turns out that it is sufficient to trace the
$\frac{1}{\vp^2}$-pole, because even this leading divergence
requires the longitudinal counterterm.
The contribution to the mass-operator of torsion from the second
diagram from Fig. 3 is given by the following integral
\beq
\Pi^{(2)}_{\al\be}(q)\,=\, -\,2\, \xi^2\,
\int \frac{d^{n} k}{(2\pi)^4} \frac{d^{n} p}{(2\pi)^4}\,
\frac{  \,\eta_{\al\be}\, + \,M^{-2}\, (k-q)_\al\,(k-q)_\be\,}{(p^2+
M^2)[(k-q)^2+M^2][(p+k)^2 + M^2]}
\label{int1}
\eeq
First, one has to notice that (as in any local quantum field theory) 
the counterterms needed to subtract 
the divergences of the above integrals are local expressions, 
hence the divergent part of the above integral is finite polynomial 
in the external momenta $q^\mu$. Therefore, in order to extract 
these divergences one can expand the factor in the
integrand into the power series in $q^\mu$:
\beq
\frac{1}{(k-q)^2+M^2} =
\frac{1}{k^2+M^2}\,\left[1 + \frac{-2k\cdot q+q^2}{k^2+M^2}\right]^{-1}
=\frac{1}{k^2+M^2} \,\sum_{l=1}^{\infty}(-1)^l\,
\left(\frac{-2k\cdot q+q^2}{k^2+M^2}\right)^l
\label{expand}
\eeq
and substitute this expansion into (\ref{int1}). It is easy to see that
the divergences hold in this expansion till the order $\,l=8$. On the
other hand, each order brings some powers of $\,q^\mu$. 
The divergences of the above integral may be canceled only by adding 
the counterterms which include high 
derivatives~\footnote{This is a consequence of the
 power-counting non-renormalizability
of the theory with massive vector fields.}.
To achieve the renormalizability one has
 to include these high derivative terms into the action (\ref{act}). 
However, since we are aiming to construct the effective (low-energy)
field theory of torsion, the effects of the higher derivative terms are
not seen and their renormalization is not interesting for us. All we 
need are the second derivative counterterms. Hence, for our purposes
the expansion (\ref{expand}) can be cut at $l=2$ rather that at $l=8$
and moreover only $O(q^2)$ terms should be kept. Thus, one arrives 
at the known integral \cite{hela} 
$$
\Pi^{(2)}_{\al\be}(q) = -6\xi^2\,q^2\,\eta_{\al\be}\,
\int \frac{d^{n} k}{(2\pi)^4}
\frac{d^n p}{(2\pi)^4}\,\frac{1}{p^2+M^2}
\,\frac{1}{(k^2+M^2)^2}\,\frac{1}{(p+k)^2+M^2} + \,...\,=
$$
\beq
= -\frac{12\,\xi^2}{(4\pi)^4\,(n-4)^2}\,q^2\,\eta_{\al\be} 
\,+\,({\rm lower}\,\,{\rm poles})\, + \, ({\rm higher} \,\, {\rm
derivative} \,\, {\rm terms}).
\eeq
Another integral looks a bit more complicated, but its derivation 
can be done in a similar way.
The contribution to the mass-operator of torsion from the first
diagram from Fig. 3 is given by the integral
$$
\Pi^{(1)}_{\al\la}(q)=- \frac{\ze^2}{108}\,
g^{(2)\al\rho\si\be}\,g^{(2)\la\tau\mu\nu}\,
\int \frac{d^n k}{(2\pi)^4}\,\int \frac{d^n p}{(2\pi)^4}\,
\frac{1}{k^2+M^2}
\left(\eta_{\tau\be} + \frac{k_\tau k_\be}{M^2}\right)\times
$$
\beq
\times
\left(\eta_{\rho\mu} +
\frac{(p-q)_\rho (p-q)_\mu}{M^2}\right)\,\frac{1}{p^2+M^2}
\,\left( \eta_{\si\nu} +
\frac{(p+k)_\si (p+k)_\nu }{M^2}\right)\,\frac{1}{(k+q)^2+M^2 }
\label{int3}
\eeq
Now, we perform the same expansion (\ref{expand}) and, disregarding
lower poles, finite contributions and higher derivative 
divergences arrive at the following leading divergence
\beq
\Pi^{(1)}_{\al\la}(q)=
- \frac{\ze^2}{(4\pi)^4\,(n-4)^2}\,q^2\,\eta_{\al\la} +\, ...\,.
\label{int4}
\eeq

%%%%%%%%%%%%%%%%%%%%%%%%%%%%%%%%%%%%%%%%%%%%%%%%%%%%%%%%%%%%%%%%%%%%%
Thus we see that both diagrams from Fig. 3 really give 
rise to the longitudinal kinetic counterterm and no any simple
cancellation of these divergences is seen.

In order to 
understand the situation better let us compare it with the one that
takes place for the usual abelian gauge transformation
(\ref{trans1}). In this case, the symmetry is not violated by the
Yukawa coupling, and (in the abelian case) the $\,A^2\ph^2\,$
counterterm is impossible because it violates gauge invariance.
The same concerns also the self-interacting $A^4$ counterterm.
The gauge invariance of the theory on quantum level is controlled
by the Ward identities.
In principle, the noncovariant counterterms can show up, but they
can be always removed, even in the non-abelian case,
by the renormalization of the gauge parameter
and in some special class of (background) gauges they are 
completely forbidden. 
Generally, the renormalization can be always performed in a
covariant way \cite{volatyu}.

In the case of the transformation (\ref{trans})
if the Yukawa coupling is inserted there are no reasonable gauge
identities at all. Therefore, in the theory of torsion field coupled
to the SM with scalar field there is a conflict between
renormalizability and unitarity. The action of the renormalizable
theory has to include the $\,\left(\pa_\mu S^\mu\right)^2\,$ term,
but this term leads to the massive ghost.
This conflict between unitarity and renormalizability reminds 
another
one which is well known -- the problem of massive unphysical ghosts
in the high derivative gravity \cite{stelle}. The difference is that 
in our case, unlike higher derivative gravity,
the problem appears due to the non invariance with respect
to the transformation (\ref{trans}). We shall proceed with the
discussion of this analogy in section 5.8.

%%%%%%%%%%%%%%%%%%%%%%%%%%%%%%%%%%%%%%%%%%%%%%%%%%%%%%%%%%%%%%%%%%%%%%%%
\section{Second test: problems with the quantized
fermion-torsion systems}
%%%%%%%%%%%%%%%%%%%%%%%%%%%%%%%%%%%%%%%%%%%%%%%%%%%%%%%%%%%%%%%%%%%%%%%%

The problem of consistency of the fermion-torsion system has been 
studied in \cite{guhesh}. Since the result of this study had crucial 
importance for the theoretical possibility of torsion, we shall 
present many details of the investigation of \cite{guhesh} here. 

In order to understand the source of the problems, let us first 
write the fermion action with torsion using the Boulware-like
parametrization. For pedagogical reasons, we first consider the usual
vector case, that is, repeat the transformation of \cite{bou}.
The action of  the massive 
vector field $V_\mu$, in original variables, has the form:
\beq
S_{m-vec} = \int d^4x\,\left\{\,
- \frac14\,V_{\mu\nu}V^{\mu\nu}
+ \frac12\,M^2\,V_\mu V^\mu +
i \,{\bar \psi}\, [\,\ga^\al \,( {\pa}_\al
- i\,g\,V_\al) - im \,]\,\psi\,\right\}\,,
\label{mvec}
\eeq
where $V_{\mu\nu}=\pa_\mu V_\nu - \pa_\nu V_\mu$
and, after the change of the field variables \cite{bou}:
\beq
\psi = \exp \left\{ \frac{ig}{M}\cdot\ph \right\}\cdot\chi
\,,\,\,\,\,\,\,\,\,\,\,\,\,\,\,\,\,
{\bar {\psi}} = {\bar {\chi}}\cdot
\exp \left\{ -\frac{ig}{M}\cdot\ph \right\}
\,,\,\,\,\,\,\,\,\,\,\,\,\,\,\,\,\,
V_\mu = V_{\mu}^{\bot} - \frac{1}{M}\,\pa_\mu\ph\,,
\label{bu1}
\eeq
the new scalar, $\ph$, is completely factored out:
\beq
S_{m-vec} = \int d^4x\,\left\{\,
-\frac14\,\left(V^{\bot}_{\mu\nu}\right)^2
+ \frac12\,M^2\,V^{\bot}_\mu V^{\bot\mu}
+ \frac12\,\pa^\mu\ph\pa_\mu\ph
+ i\,{\bar \chi}\,[\ga^\al \,( {\pa}_\al
+ i\,g\,V^{\bot}_\al) ]\,\chi \,\right\}.
\label{masvec}
\eeq  
Let us now consider the fermion-torsion system given by the action
\footnote{For the sake of simplicity we consider a single spinor only.
Since torsion is an abelian massive vector, the results can not depend 
on the gauge group.}
\beq
S_{tor-fer} = \int d^4x\,\left\{\,
- \frac14\,S_{\mu\nu}S^{\mu\nu}
+ \frac12\,M^2\,S_\mu S^\mu
+ i\,{\bar \psi}\, [\,\ga^\al \,( {\pa}_\al
+ i\,\eta\,\ga_5\,S_\al) - im \,]\,\psi\,\right\}\,.
\label{geral}
\eeq
The change of variables, similar to the
one in (\ref{bu1}), has the form:
\beq
\psi = \exp\left\{\frac{i\eta}{M}\,\gamma^5\,\ph\right\}\,\chi
\,,\,\,\,\,\,\,\,\,\,\,\,\,\,\,\,\,
{\bar {\psi}} =
{\bar {\chi}}\, \exp\left\{ \frac{i\eta}{M}\,\gamma^5\,\ph\right\}
\,,\,\,\,\,\,\,\,\,\,\,\,\,\,\,\,\,
S_\mu = S_{\mu}^{\bot} - \frac{1}{M}\,\pa_\mu\ph\, ,
\label{bu}
\eeq
where $\,S_{\mu}^{\bot}\,$ and $\,S^{\|}_\mu = \pa_\mu\ph\,$ are the
transverse and longitudinal parts of the axial vector respectively, the
latter being equivalent to the pseudoscalar $\ph$. One has to notice that,
contrary to (\ref{bu1}), but in full accordance with (\ref{trans}),
the signs of both the exponents in (\ref{bu}) are the same.
In terms of the new variables, the action becomes
$$
S_{tor-fer} = \int d^4x\,
\left\{\,
-\frac14\,S^{\bot}_{\mu\nu}S^{{\bot}\mu\nu}
+ \frac12\,M^2\,S^{\bot}_\mu S^{\bot\mu} +
\right.
$$
\beq
\left.
+ i\,{\bar \chi}\, [\,\ga^\al \,( {\pa}_\al
+ i\,\eta\,\ga_5\,S^{\bot}_\al\,)
- im\cdot e^{ \frac{2i\eta}{M}\,\gamma^5\,\ph}]\,\chi
+ \frac12\,\pa^\mu\ph\pa_\mu\ph \,\right\}\,,
\label{heral}
\eeq
where $\,S^{\bot}_{\mu\nu} =
\pa_\mu S^{\bot}_\nu - \pa_\mu S^{\bot}_\nu = S_{\mu\nu}\,$.
Contrary to the vector case (\ref{bu1}), for the torsion axial 
vector (\ref{heral}) the scalar mode does not decouple.
Moreover, the interaction has the following unpleasant features:
\vskip 1mm

\noindent
i) it is Yukawa-type, resembling the problems with the ordinary scalar.

\noindent
ii) besides, the interaction is exponential, which makes almost certain 
that the model is not power-counting renormalizable.
\vskip 1mm

On quantum level, the symmetry (\ref{trans}) manifests itself through 
the Ward identities for the Green functions. The analysis of the 
Ward identities arising in the fermion-torsion system confirms
that the new scalar mode does not decouple and that one can not 
control the $\ph$-dependence on quantum level \cite{guhesh}. 
However, at first sight, there is a hope that the above properties
would not be fatal for the theory. With respect to the point (i), 
one can guess that the only result of the non-factorization, which
can be dangerous for the consistency of the effective quantum theory,
would be the propagation of the longitudinal mode of the torsion, and this
does not directly follow from the non-factorization of the scalar
degree of freedom in the classical action. On the other hand,
(ii) indicates the non-renormalizability, which might mean just the
appearance of higher-derivative divergences. But, this does not 
matter within the effective approach. Thus, a more detailed analysis
is necessary. In particular, the one-loop and especially two-loop 
calculations in the theory (\ref{geral}) may be especially helpful.
%%%%%%%%%%%%%%%%%%%%%%%%%%%%%%%%%%%%%%%%%%%%%%%%%%%%%%%%%%%%%%%%%%%%%%
\vskip 3mm

The {\it one-loop calculation} in the theory (\ref{geral})
can be performed using the generalized Schwinger-DeWitt technique
developed by Barvinsky and Vilkovisky \cite{bavi}, 
but the application of this
technique here is highly non-trivial. The problem has been solved in
\cite{guhesh}. First of all, we notice that the derivation of 
divergences in the purely torsion sector is not necessary, since
the result is (\ref{contra}), that is the same as for the spinor 
loop on torsion background. Of course, one has to disregard all
curvature-dependent terms in (\ref{contra}), since now we work on 
the flat metric background.

%%%%%%%%%%%%%%%%%%%%%%%%%%%%%%%%%%%%%%%%%%%%%%%%%%%%%%%%%%%%%%%%%%

Now, we are in a position to start the complete calculation of 
divergences. The use of the background field method supposes the
shift of the field variables into a background and a
quantum part. However, in the case of the (axial)vector-fermion
system, the simple shift of the fields leads to an enormous volume of
calculations, even for a massive vector. Such a calculation becomes
extremely difficult for the axial massive vector (\ref{geral}).
That is why, in \cite{guhesh}, we have invented a simple trick 
combining the background field method with the Boulware transformation
(\ref{bu}) for the quantum fields. 

According to the method of \cite{guhesh}, one has to  
divide the fields into background $(S_\mu,\psi,{\bar \psi})$
and quantum $(t^{\bot}_\mu,\ph,\chi,{\bar \chi})$ parts, according to 
$$
\psi \to \psi ^{'}=e^{i\frac{\eta}{M}\ga _5\ph}\cdot(\psi +\chi)\,,
$$
$$
\bar{\psi} \to \bar{\psi}^{'}=
(\bar{\psi}+\bar{\chi})\cdot e^{i\frac{\eta}{M}\ga _5\ph}\,,
$$
$$
S_{\mu} \to S_{\mu}^{'}
=S_{\mu}+t^{\perp}_{\mu}-\frac{1}{M}\partial _{\mu}\ph\,.
$$
The one-loop effective action depends on the quadratic (in quantum
fields) part of the total action:
$$
S^{(2)}=\frac{1}{2}\int d^4x \,\left\{ t^{\perp}_{\mu}
(\,\Box + M^2\, )t^{\perp\mu} + \ph \,( - \Box )\,\ph
+ t^{\perp}_{\mu}\,(\, -2\eta\bar{\psi}\ga ^{\mu}\ga _5\, )\,\chi
+ \ph \,( - \frac{4m\eta ^2}{M^2}\,\bar{\psi}\psi )\,\ph +
\right.
$$
\beq
\left.
+ \bar{\chi}(\, -2\eta\ga ^{\nu}\ga^5\psi\, )\,t^{\perp}_{\nu}
+ \bar{\chi}\,(\frac{4im\eta}{M}\,\ga^5\psi )\,\ph
+\ph\, (\frac{4i\eta m}{M}\,\bar{\psi}\ga^5 )\,\chi +
\bar{\chi}\,(2i\ga ^{\mu}D_{\mu}+2m\,)\chi \right\} \,.
\label{bili}
\eeq
Making the usual change of the fermionic variables
$\,\chi = -\frac{i}{2}(\ga ^{\mu}D_{\mu}+im)\ta\,$,
and substituting $\,\ph \to i\ph$, we arrive at the following
expression:
$$
S^{(2)} \,=\,\frac{1}{2}\,\int d^4x \,
\left(\matrix{t^{\perp}_{\mu} & \ph & {\bar \chi} \cr}\right)
\cdot \,{\hat {\bf H}}\,\cdot
\left(\matrix{t^{\perp}_{\nu} \cr \ph \cr \tau \cr}\right)\,,
$$
where the Hermitian bilinear form $\,{\hat {\bf H}}\,$
has the form
\beq
{\hat {\bf H}} = \left( \begin{array}{ccc}
\th^{\mu\nu}(\Box + M^2) & 0 & \th ^{\mu}\mbox{}_{\be}
(L^{\be\al}\partial _{\al} + M^{\be}) \\
0 & \Box + N & A^{\al}\partial _{\al} + B \\
P_{\be}\th ^{\be\nu} & Q & \hat{1}\Box + R^{\la}\partial _{\la}+ \Pi
\end{array}\right)\,,
\label{2h}
\eeq
$\th^{\mu}\mbox{}_{\nu}=\de^{\mu}\mbox{}_{\nu}-\pa^{\mu}
\frac{1}{\Box}\pa_{\nu}$ being
the projector on the transverse vector states.
The elements of the matrix operator (\ref{2h}) are defined
according to (\ref{bili}). 
$$
R^\mu =  2\eta\si ^{\mu\nu}\ga _5 S_{\nu}
\,,\,\,\,\,\,\,\,\,\,\,\,\,\,\,
\Pi = i\eta\ga _5(\partial _{\mu}S^{\mu})
+\frac{i}{2}\eta\ga ^{\mu}\ga ^{\nu}\ga _5 S_{\mu\nu}
+\eta^2 S_{\mu}S^{\mu} + m^2\,.
$$
$$
L^{\al\be}=-i\eta \bar{\psi}\ga _5\ga ^{\al}\ga ^{\be}
\,,\;\;\;\;\;\;\;\;\;\;\;\;\;\;
M^{\be} = \eta ^2\bar{\psi}\ga ^{\be}\ga ^{\al}S_{\al}+
\eta m\bar{\psi}\ga _5\ga ^{\be}\,,
$$
$$
A^{\al} = 2i\eta\frac{m}{M}\bar{\psi}\ga _5\ga ^{\al}
\,,\;\;\;\;\;\;\;\;\;\;\;\;\;\;
B = 2\eta ^2\frac{m}{M}\bar{\psi}\ga ^{\be}S_{\be}-
2\eta \frac{m^2}{M}\bar{\psi}\ga _5\,,
$$
\beq
N = 4\eta ^2 \frac{m}{M^2}\bar{\psi}\psi
\,,\;\;\;\;\;\;\;
P^{\be} = -2\eta\ga ^{\be}\ga _5\psi
\,,\;\;\;\;\;\;\;
Q = -4\eta\frac{m}{M}\ga _5 \psi\,.
\label{blocks}
\eeq

The operator ${\hat {\bf H}}$ given above might look like the
minimal second order operator ($\,\Box + 2h^\la\na_\la + \Pi$); 
but, in fact, it is not minimal because of the projectors
$\th^{\mu\nu}$ in the axial vector- axial vector
$\,t^{\perp}_{\mu}$ -- $t^{\perp}_{\nu}\,$ sector.
That is why one cannot directly
apply the standard Schwinger-DeWitt expansion to derive the
divergent contributions to the one-loop effective action,
and some more sophisticated method is needed.
Such a method, which can be called the generalized 
Schwinger-DeWitt technique in the transverse space,  
has been developed in \cite{guhesh}, where the reader 
can find the complete details concerning the one-loop 
calculations in both theories (\ref{masvec}) and 
(\ref{geral}). Let us, for the sake of brevity,
present only the final result for the one-loop divergences
in the theory (\ref{geral}):
$$
\Ga^{(1)}_{div} \,= \,- \,\frac{\mu^{n-4}}{\varepsilon}\,
\int d^n x \,\left\{ \,-\frac{2\eta^2}{3}\,S_{\mu\nu}S^{\mu\nu}
+ 8m^2\eta^2\,S^\mu S_\mu - 2m^4 + \frac{3}{2}\,M^4 +
\right.
$$
\beq
\left.
+ \Big( \,\frac{8\eta^2\,m^3}{M^2} 
- 6\,\eta^2m\,\Big) \bar{\psi}\psi 
+8\,\frac{\eta^4\,m^2}{M^4}\,(\bar{\psi}\psi)^2 +
4i\,\frac{\eta^2\,m^2}{M^2}\,\bar{\psi}\ga^\mu D^{*}_\mu \psi 
\,\right\}\,.
\label{result}
\eeq

It is interesting to notice that the above expression
(\ref{result}) is not
gauge invariant. One can trace the calculations back in order to 
see that the non-invariant
terms come as a contribution of the scalar $\ph$
(see, e.g., eq.(\ref{heral})). Thus, the non-invariant divergences
emerge because there are variables, in which  
the violation of the symmetry (\ref{trans}) is not soft. 

Consider the one-loop renormalization and the corresponding
renormalization group. 
The expression (\ref{result}) shows that the theory (\ref{geral})
is not renormalizable, but the new type of divergences are 
suppressed if we suppose that the torsion has very big mass
$m \ll M_{ts}$ (remember our notation $\,\,M^2_{ts}=2M^2$). 
Let us, for a while, take this relation as a 
working hypothesis. Then, the relations between bare
and renormalized fields and the coupling $\eta$ follow from
(\ref{result}):
$$
S_\mu^{(0)} = \mu^{\frac{n-4}{2}}\,S_\mu\,
\left(\, 1 + \frac{1}{\ep}\cdot\frac{8\eta^2}{3}\,\right)
\,,\,\,\,\,\,\,\,\,\,\,\,\,\, \psi^{(0)} =
\mu^{\frac{n-4}{2}}\,\psi\, \left(\, 1 +
\frac{1}{\ep}\cdot\frac{2\eta^2m^2}{M^2}\,\right)\,,
$$
\beq
\eta^{(0)} = \mu^{\frac{4-n}{2}}\, \left( \,\eta -
\frac{1}{\ep}\cdot \frac{8\eta^3}{3}\cdot \left[1+
\frac{m^2}{M^2}\right]\,\right)\,. \label{renor}
\eeq
Similar
relations for the parameter $\,{\tilde \la} =
\frac{M^4}{m^2}\,\la\,$ of the $\,\la (\bar \psi\,\psi)^2\,$-
interaction, have the form: \beq {\tilde \la}^{(0)} =
\mu^{4-n}\,\left[ {\tilde \la} + \frac{16\eta^4}{\ep} -
\frac{8{\tilde \la} \eta^2 m^2}{M^2\,\ep}\right]\,.
\label{reno}
\eeq
These relations lead to a renormalization group equation
for $\eta$, which contains a new term proportional to $(m / M)^2$:
\beq (4\pi)^2\,\frac{d\eta^2}{dt} = \frac{8}{3}\,[1+
\frac{m^2}{M^2}]\,\eta^4\,,\,\,\,\,\,\,\,\, \,\,\,\,\,\, \eta(0) =
\eta_0\,. \label{rg} \eeq Indeed, for the case $m \ll M$ and in
the low-energy region, this equation reduces to the one presented
in \cite{betor} (that is identical to the similar equation of
QED). In any other case, the theory of torsion coupled to the
massive spinors is inconsistent, and equation (\ref{rg}) is 
meaningless.

One can also write down the renormalization group equation for the
parameter $\,{\tilde \la}\,$ defined above. Using (\ref{reno}), we
arrive at the following equation:
\beq
(4\pi)^2\,\frac{d {\tilde \la} }{dt} = 16\,\eta^4\,.
\label{psi4}
\eeq
This equation confirms the
lack of a too fast running for this parameter. Indeed, all this
renormalization group consideration has meaning only under the
assumption that $\,\,m \ll M$. In order to provide the one-loop 
renormalizability, it is necessary to add the 
$({\bar \psi}\,\psi)^2$-term to the classical action. However, as 
we have learned, in Chapter 3, on the example of the Nambu-Jona-Lasinio 
model, such a term does not 
affect the one-loop renormalization of other 
sectors of the action. However, it becomes extremely important 
for the two-loop contributions. 
%%%%%%%%%%%%%%%%%%%%%%%%%%%%%%%%%%%%%%%%%%%%%%%%%%%%%%%%%%%%%%%%%%%%%%%%
\vskip 3mm

The {\it two-loop calculation} is the crucial test for 
the consistency of the fermion-torsion effective theory. 
The point is that there are no one-loop diagrams which 
include any non-symmetric vertices and can contribute to the 
dangerous longitudinal term in the torsion action. However, there 
are 2-loop diagrams contributing to the
propagator of the axial vector, $\,S_\mu$. The question  
is whether there are longitudinal divergences 
$(\pa_\mu S^\mu)^2$-type at the two-loop level. Then, 
it is reasonable to start
from the diagrams which can exhibit $\,1/\ep^2$-divergences.

\noindent
%%%%%%%%%%%%%%%%%%%%%%%%%%%%%%%
\begin{picture}(160,160)(0,0)
\Line(0,23)(25,23)
\Line(0,27)(25,27)
\Line(0,23)(0,27)
\DashCArc(50,25)(25,0,360){1} %%% {1} ???
\DashCArc(100,25)(25,0,360){2} %%% {2} ???
\Line(125,23)(150,23)
\Line(125,27)(150,27)
\Line(150,23)(150,27)
\Vertex(25,25){3} 
\Vertex(75,25){2} 
\Vertex(125,25){3}
\end{picture}
$\,\,\,\,\,\,\,\,\,\,\,\,\,\,\,\,$
%%%%%%%%%%%%%%%%%%%%%%%%%%%%%%%
\noindent
\begin{picture}(120,120)(0,0)
\Line(0,23)(25,23)
\Line(0,27)(25,27)
\Line(0,23)(0,27)
\DashCArc(50,25)(25,0,360){1} %%% {1} ???
\DashCArc(50,75)(25,0,360){2} %%% {2} ???
\Line(75,23)(100,23)
\Line(75,27)(100,27)
\Line(100,23)(100,27)
\Vertex(25,25){3} 
\Vertex(75,25){3} 
\Vertex(50,50){2}
\end{picture}
\\
%%%%%%%%%%%%%%%%%%%%%%%%%%%%%%%  2 line
\noindent
\begin{picture}(120,120)(0,0)
\Line(0,23)(25,23)
\Line(0,27)(25,27)
\Line(0,23)(0,27)
\DashCArc(50,25)(25,0,360){1} %%% {1} ???
\Line(75,23)(100,23)
\Line(75,27)(100,27)
\Line(100,23)(100,27)
%%%%%%%
\Line(48,0)(48,50)
\Line(52,0)(52,50)
\Vertex(25,25){3} 
\Vertex(75,25){3} 
\Vertex(50,50){3}
\Vertex(50,0){3}
\end{picture}
$\,\,\,\,\,\,\,\,\,\,\,\,\,\,\,\,$
%%%%%%%%%%%%%%%%%%%%%%%%%%%%%%%
\noindent
\begin{picture}(120,120)(0,0)
\Line(0,23)(25,23)
\Line(0,27)(25,27)
\Line(0,23)(0,27)
\DashCArc(50,25)(25,0,360){1} %%% {1} ???
\Line(75,23)(100,23)
\Line(75,27)(100,27)
\Line(100,23)(100,27)
\DashLine(25,25)(75,25){4}
%%   CArc(50,25)(25,0,180){1} %%% {2} ???
%%%  CArc(50,25)(12,0,180){1} %%% {2} ???
\Vertex(25,25){3} 
\Vertex(75,25){3} 
%%%%%%%
\Line(36,44)(64,44)
\Line(31,40)(69,40)
\Vertex(33,42){3} 
\Vertex(67,42){3} 
\end{picture}
\\
\\
\noindent
{\small\it Figure 4. 
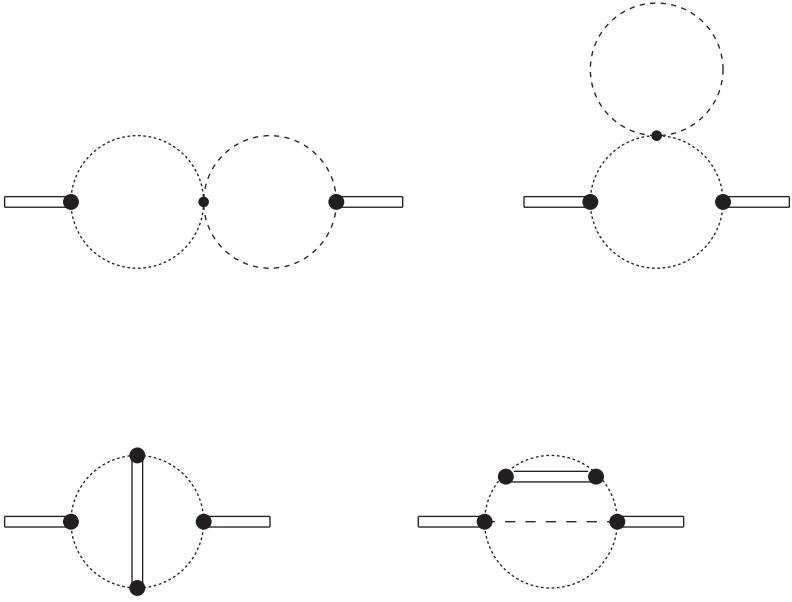
Two-loop ``dangerous'' diagrams, which can contribute to the
$(\partial_\mu S^\mu)^2$-counterterm. The first two give 
non-zero contributions and the second two can not cancel them.}
\vskip 5mm
\vskip 1mm

The leading $\,1/\ep^2$-two-loop divergences of the
mass operator for the axial vector $\,S_\mu\,$ come from two
distinct types of diagrams: the ones with the
$\,({\bar \psi}\psi)^2\,$-vertex
and the ones without this vertex. As we shall see,
the most dangerous diagrams are those with 4-fermion interaction.
As we have seen above, this kind of interaction
is a remarkable feature of the axial vector
theory, which is absent in a massive vector theory.
Now, we shall calculate divergent $\,1/\ep ^2\,$-contributions 
from two diagrams with the $\,({\bar \psi}\psi)^2\,$-vertex, 
using the expansion (\ref{expand});  in the
Appendix B of \cite{guhesh} this calculation has been checked 
using Feynman parameters.

Consider the first diagram of Figure 4.
It can be expressed, after making some commutations
of the $\ga$-matrices, as
\beq
\Pi^1_{\mu\nu}  =
- \la\eta ^2 \; tr\;\left\{ I_{\nu} \cdot I_{\mu}
\right\},
\label{produ1}
\eeq
where $\,\la\sim \frac{m^2}{M^4}\,$ is the coupling of the
four-fermion vertex,
the trace is taken over the Dirac spinor space and
\beq
I_{\nu}(p) = \int\frac{d^np}{(2\pi)^n}
\frac{p\sla -m}{p^2-m^2}\,\, \ga _{\nu}\,\,\ga _5\,\,
\frac{p\sla -q\sla -m}{(p-q)^2-m^2}\, .
\label{I}
\eeq
Following \cite{betor}, we can perform the expansion
\beq
\frac{1}{(p-q)^2-m^2}
=\frac{1}{p^2-m^2}\,\sum_{l=0}^{\infty}\,(-1)^l\,
\left( \frac{-2p\, .\, q+q^2}{p^2-m^2}\right)^l 
\label{expans}\,.
\eeq
Now, as in the previous section, one can omit the powers 
of $q$ higher than $2$. Besides, 
when performing the integrations, we trace just the divergent
parts, thus arriving (using the integrals from \cite{hela})
at the expressions:
$$ %%%%%   \beq
I_{\nu}=\frac{i}{\ep}\,\,
\left\{ -\frac{1}{6}q^2\ga _{\nu}-2m^2\ga _{\nu}-
\frac{1}{6}\ga _{\al}\ga _{\nu}\ga _{\be}q^{\al}q^{\be}
+ mq_{\nu}\right\}
+\,\, ... ,
$$ %%%%%  \label{resultI}
%%%%%%%%  \eeq
where the dots stand for the finite and higher-derivative divergent
terms. Substituting this into (\ref{produ1}), we obtain the
leading divergences of the diagram:
\beq
\Pi^{1,div}_{\mu\nu}= -\frac{\la\eta^2}{\ep ^2} \,\,
\left\{+16m^4\,\eta _{\mu\nu}+\frac{28}{3}\,m^2\,q_{\mu}q_{\nu}
-\frac{16}{3}\,m^2\,q^2\eta _{\mu\nu}\right\}
+\,\, ...
\label{resultPi}
\eeq
This result shows that the construction of the first diagram
contains an $\,1/\ep^2\,$-longitudinal counterterm.

\vskip 2mm
%%%%%%%%%%%%%%%%%%%%%%%%%%%%%%%%%%%%%%%%%%%%%%%%

The contribution of the second two-loop diagram of Fig. 4 
to the polarization operator,
$\,\Pi^2_{\mu\nu}\,$, is written, after certain transformations,
in the following way:
\beq
\Pi^2_{\mu\nu}= - \la\eta^2 \;\,tr
\,\left\{ I_{\nu\mu}\cdot J  \right\}\,,
\label{produ2}
\eeq
where
\beq
I_{\nu\mu}=\int\frac{d^np}{(2\pi)^n}\,
\frac{p\sla -m}{p^2-m^2}\,\, \ga _{\nu}\,\,
\frac{p\sla -q\sla +m}{(p-q)^2-m^2}\,\, \ga _{\mu}\,\,
\frac{p\sla -m}{p^2-m^2}
\eeq
and
\beq
J=\int\frac{d^np}{(2\pi)^n}\,\frac{k\sla -m}{k^2-m^2}\,.
\eeq
It proves useful to introduce the following definitions:
$$
A_{\al\nu\be\mu\rho}=
\ga _{\al}\ga _{\nu}\ga _{\be}\ga _{\mu}\ga _{\rho}\,,
$$$$
B_{\al\nu\mu\be}=
- q^{\rho}\,\ga_\al\ga_\nu\ga_\rho\ga_\mu\ga_\be
+ m\,(\ga_\al\ga_\nu\ga_\mu\ga_\be
-\ga_\nu\ga_\al\ga_\mu\ga_\be - \ga_\al\ga_\nu\ga_\be\ga_\mu)
$$$$
C_{\al\nu\mu} = m^2\,(\ga_\nu\ga_\al\ga_\mu
- \ga_\nu\ga_\mu\ga_\al - \ga_\nu\ga_\al\ga_\mu) +
m q^\be\,(\ga_\nu\ga_\be\ga_\mu\ga_\al
+ \ga_\al\ga_\nu\ga_\be\ga_\mu)\,.
$$
$$
D_{\nu\mu}=-m^2q^{\be}\,\ga_\nu\ga_\be\ga_\mu
+m^3\,\ga_\nu\ga_\mu\,.
$$
Then, the first integral can be written as
\beq
I_{\nu\mu}=\int\frac{d^np}{(2\pi)^n}\,\,
\frac{A_{\al\nu\be\mu\rho}\cdot p^{\al}p^{\be}p^{\rho}
+B_{\al\nu\mu\be}\cdot p^{\al}p^{\be}
+C_{\al\nu\mu}p^{\al}+D_{\nu\mu}}
{(p^2-m^2)^2\left(\, (p-q)^2-m^2\,\right)}.
\label{I2}
\eeq
Using the expansion (\ref{expans}), and disregarding higher
powers of $q$, as well as odd powers of $p$ in the numerator
of the resulting integral, one obtains, after using the
integrals of \cite{hela}:
$$ %%%% \beq
I_{\nu\mu}=\frac{i}{\ep}\; \left\{\;
\frac{1}{4}B^{\al}\mbox{}_{\nu\mu\al}+\frac{1}{12}
(A^{\al}\mbox{}_{\nu\al\mu\rho}+A^{\al}\mbox{}_{\nu\rho\mu\al}
+A_{\rho\nu\al\mu}\mbox{}^{\al})q^{\rho}\;\right\}+\; ...
$$ %%%%  \eeq
which gives, after some algebra,
\beq
I_{\nu\mu} = \frac{i}{\ep}\; \left\{ \;
m\,\ga_\mu\ga_\nu + 3m \eta_{\mu\nu}
-\frac{2}{3}\ga_{\rho}q^{\rho}\eta_{\mu\nu}
+ \frac{1}{3}\ga _{\mu}q_{\nu}+
\frac{1}{3}\ga _{\nu}q_{\mu} \right\} +\; ...
\eeq
The divergent contribution to $J$ is
$$ %%% \beq
J=-\frac{i}{\ep}\; m^3 +\; ...
$$ %%% \eeq
Now, the calculation of (\ref{produ2}) is straightforward:
\beq
\Pi_{\mu\nu}= \frac{\la\eta^2}{\ep^2}\;
\,8\,m^4\,\eta_{\mu\nu}\,\, + \,\,\; ...
\label{result2}
\eeq
As we see, this diagram does not contribute to the kinetic
counterterm
(with accuracy of the higher-derivative terms), and hence the
cancellation of the contributions to the longitudinal
counterterm coming from $\,\Pi^1_{\mu\nu}\,$ do not take place.

One has to notice that other two-loop
diagrams do not include the 
$\,({\bar \psi}\psi)^2\,$-vertex. Thus, even if those
diagrams contribute to the longitudinal counterterm,
the cancellation with $\,\Pi^1_{\mu\nu}\,$ should require
some special fine-tuning between $\,\la\,$ and $\,\eta\,$.
In fact, one can prove, without explicit calculation, that
the remaining two-loop diagrams of Fig. 4 
do not contribute to the longitudinal $\,1/\ep^2\,$-pole.
In order to see this, let us notice that the leading
(in our case  $\,1/\ep^2\,$) divergence may be obtained
by consequent substitution of the contributions from
the subdiagrams by their local divergent components.
Since the local counterterms produced by the
subdiagrams of the last two graphs of Fig. 4
are minus the one-loop expression
(\ref{result}), the corresponding divergent vertices
are $\,1/\ep\,$ factor classical vertices. Hence, in the
leading $\,1/\ep^2\,$-divergences of the last two 
diagrams of Fig. 4, one meets again the same expressions as
in (\ref{result}). The result of our consideration is, therefore,
the non-cancellation of the $\,1/\ep^2\,$-longitudinal
divergence (\ref{resultPi}). This means that the theory
(\ref{geral}), without additional restrictions
on the torsion mass, like $\,m \ll M\,$, is inconsistent at the
quantum level.

%%%%%%%%%%%%%%%%%%%%%%%%%%%%%%%%%%%%%%%%%%%%%%%%%%%%%%%%%%%%%%%%%%%%%%
\section{Interpretation of the results: do we have a chance to meet
propagating torsion?}
%%%%%%%%%%%%%%%%%%%%%%%%%%%%%%%%%%%%%%%%%%%%%%%%%%%%%%%%%%%%%%%%%%%%%%

Obviously, we have found a very pessimistic answer about the 
possibility of a
propagating torsion. Torsion is not only helpless in constructing 
the renormalizable quantum gravity, as people thought 
\cite{nev1,seznie}, but it can not be renormalizable by itself. 
Moreover, even if we give up the requirement of power counting
renormalizability and turn to the effective approach, the theory 
remains inconsistent. In this situation one can try the 
following: i) Invent, if possible, some restriction on the 
parameters of the theory, for which the problems disappear.
ii) Analyze the approach from the very beginning in order to 
look for the possible holes in the analysis. 
Let us start with the first option, and postpone the second one
for the next Chapter.

The problems with the non-renormalizability and with the
violation of unitarity become weakened if one input
severe restrictions on the torsion mass, which has to be much 
greater than the mass of the heaviest fermion (say, t-quark, 
with a mass about 175 $GeV$), and continue to use an effective 
quantum field theory approach, investigating 
the low-energy observables only. This approach implies the
existence of a fundamental theory which is valid at higher
energies. 

Hence, in order to have a propagating torsion,
one has to satisfy a double inequality:
\beq
m_{fermion} \ll M_{torsion} \ll M_{fundamental}\,.
\label{double}
\eeq
Usually, the fundamental scale is associated with the Planck 
mass, $M_{P} \approx 10^{19}\,GeV$. Indeed, the identification 
of the torsion mass with the Planck scale means, for 
$M_{torsion}\approx M_P$, that at the
low (much less then $M_P$) energies, when the kinetic and other 
terms are negligible, we come back to the Einstein-Cartan theory
(\ref{EC}). As we have readily noticed, in this theory torsion 
is not propagating, but it can mediate contact interactions. 
These interactions are too weak for the high energy experiments
described in section 5.4, but maybe, in future, they can be 
detected in very precise experiments like the ones in 
the atomic systems.

It is important to remark that, in principle, 
the effective quantum field theory approach may be used only 
at the energies
essentially smaller than the typical mass scale of the
fundamental theory. If the mass of torsion is comparable
to the fundamental scale $M_P$, all the torsion degrees of 
freedom should be described directly in the framework of the 
fundamental theory. For instance, in the low-energy effective 
actions of the available versions of string theory torsion 
enters with the mass $\,\,1/\al^{\prime}$, which is conventionally 
taken to be of the Planck order. But, other degrees of freedom
associated to torsion also have a mass of the same order.
Thus, it is unclear why one can take only this ``static''
mode and neglect infinite amount of others with the same huge 
mass. Later on, in Chapter 6, we shall discuss torsion coming 
from the string theory, and see that the standard approach 
to string forbids propagating torsion at all. Then, the 
Einstein-Cartan theory (\ref{EC}) becomes some kind of 
universal torsion theory for the low-energy domain. 

On the other hand, the relation (\ref{double}) still leaves a
huge gap in the energy spectrum, which is not completely
covered by the present theoretical consideration. In other 
words, (\ref{double}) might be inconsistent with the string 
theory, but it does not contradict the effective approach.  
Of course, this gap cannot be closed by any experiment, because
the mass of torsion is too big. Even the restrictions coming
from the contact experiments \cite{betor} achieve only the
region $\,M < 3 \,TeV\,$. And that is not really enough to 
satisfy (\ref{double}) for all the fermions
of the Standard Model. It is clear that the existence of
fermions with masses many orders of magnitude 
larger than $m_t$ (like the
ones which are expected in many GUT's or SSM) can close the gap
on the particle spectrum and ``forbid'' propagating torsion.

%%%%%%%%%%%%%%%%%%%%%%%%%%%%%%%%%%%%%%%%%%%%%%%%%%%%%%%%%%%%%%%%%%%%%
\section{What is the difference with metric?}
%%%%%%%%%%%%%%%%%%%%%%%%%%%%%%%%%%%%%%%%%%%%%%%%%%%%%%%%%%%%%%%%%%%%%

The situation with torsion is similar to the one with quantum 
gravity. In both cases, there is a conflict between renormalizability 
and unitarity. In the case of quantum gravity, there are
models which are unitary (General Relativity and its supersymmetric 
generalizations) but non-renormalizable, and other models, with 
higher derivatives, which are renormalizable but not unitary.  

For the case of torsion, there are models which are unitary 
at the tree level
(take, for instance, all versions described in \cite{sez}, or 
simply our action (\ref{action})), but non-renormalizable. 
At the same time, it is not difficult to formulate 
renormalizable theory of torsion. For that one has to take,
say, the general metric-torsion action of \cite{chris}, which 
will provide renormalizable metric-torsion theory. The same 
action of \cite{chris}, but with the flat metric 
$g_{\mu\nu}=\eta_{\mu\nu}$ will give a renormalizable theory of 
the torsion alone. However, such a theory will not be unitary, 
because both transverse and longitudinal components of the axial 
vector $S_\mu$ will propagate. 

In some sense, this analogy is natural, because both metric
and torsion are geometric characteristics of the space-time
manifold rather than usual fields. Therefore, one of the
options is to give up the quantization of these two
fields and consider them only as a classical background.
This option cures all the problems at the same time, and 
leaves one a great freedom in choosing the model for metric,
torsion and other gravity components: there are no constraints
imposed by quantum theory anymore. Indeed, there are very 
small number of the shortcomings in this point of view 
\cite{susy-enc}. The most important of them is the 
quantum-mechanical inconsistency of the systems composed 
by quantum and classical constituents (see, for example, 
\cite{wald}). 

If one does not accept this option, it is only possible to
consider both metric and torsion as effective low-energy
interactions resulting from a more fundamental theory like
string. There is, however, a great difference between metric 
and torsion. The effective field theory permits the long-distance
propagation of the metric waves, because metric 
has massless degrees of freedom. Indeed, there may be other 
degrees of freedom, with the mass of the Planck order, which 
are non visible at low energies. But, the massless degrees 
of freedom ``work'' at very long distances, and provide 
the long-range gravitational force. Then, the study of an 
effective quantum field theory for the metric
does not meet major difficulties \cite{don,weinberg}. 
In case of torsion, the
massless degrees of freedom are forbidden, because the symmetry 
(\ref{trans}) is violated by the spinor mass. 

It might happen, that some new symmetries will be discovered, which 
make the consistent quantum theory of the propagating torsion 
possible. However, in the framework of the well-established 
results, the remaining possibilities are that torsion does not
exist as an independent field, or it has a huge mass, or it is 
a purely classical field which should not be quantized. Indeed,
in the first two cases there are no chances of detecting torsion
experimentally. 

\vskip 10mm
%%% \newpage
%%%%%%%%%%%%%%%%%%%%%%%%%%%%%%%%%%%%%%%%%%%%%%%%%%%%%%%%%%%%%%%%%
%%%%%%%%%%%%%%%%%%%%%%%%%%%%%%%%%%%%%%%%%%%%%%%%%%%%%%%%%%%%%%%%%%
\chapter{Alternative approaches: induced torsion}
%%%%%%%%%%%%%%%%%%%%%%%%%%%%%%%%%%%%%%%%%%%%%%%%%%%%%%%%%%%%%%%%%%

$\,\,\,\,\,\,$
As we have seen in the previous Chapter, the formulation of the
theory of propagating torsion meets serious difficulties. 
In this situation it is natural to remind that some analog of 
the space-time torsion is generated in string theory, and then 
try to check whether this is consistent with the situation 
in the effective field theory. 
On the other hand, we know that usual metric gravity can be 
induced not only in string theory, but also through the quantum 
effects of the matter fields (see, for example, \cite{adler}
for a review and further references). This 
short Chapter is devoted to a brief description of these two 
approaches: torsion induced in string theory and torsion induced 
by the quantum effects of matter fields. 

%%%%%%%%%%%%%%%%%%%%%%%%%%%%%%%%%%%%%%%%%%%%%%%%%%%%%%%%%%%%%%%%%%
\section{Is that torsion induced in string theory?}
%%%%%%%%%%%%%%%%%%%%%%%%%%%%%%%%%%%%%%%%%%%%%%%%%%%%%%%%%%%%%%%%%%

The covariant (super)string action has a geometric interpretation 
as a two-dimensional nonlinear sigma-model. Then, the completeness 
of the spectrum (it can be seen as the requirement of the 
correspondence between string and the sigma-model) requires an 
additional Wess-Zumino-Witten term
to be included. The action of a bosonic string has the form
$$
S_{str} = \int {d^2}z \sqrt{|h|} \Big\{ {1\over {2 \alpha'}}
\,h^{ab} g_{\mu\nu}(X)\,
\partial_a X^\mu\, \partial_b X^\nu + {1\over {\alpha'}}\,
{\varepsilon^{ab}\over {\sqrt {|h|}}}\,\,b_{\mu\nu}(X)\,
\partial_a X^\mu\, \partial_b X^\nu +
$$
\beq
+ \,\Phi(X)\,\,^{(2)}R + T(X)\,\Big\} \,. 
\label{string}
\eeq
Here $\mu,\nu = 1,2,...,D\,;\,\,\,\,\,\,\, {\rm and}\,\,\,\,\,\,\,     
a,b = 1,2\,.\,\,\,\,\,\,\, g_{\mu\nu}(X)$ is the background
metric, $b_{\mu\nu}(X)$ is the antisymmetric tensor background 
field, $\Phi(X)$ is the
dilaton field, $T(X)$ is the tachyon background field, which we
do not consider in what follows\footnote{Tachyon does not pose
a problem for the superstring.}. $^{(2)}R$ is the 
two-dimensional curvature, $X^\mu=X^\mu(z)$ are string coordinates.
The parameter $\,\alpha^\prime\,$ may be considered as 
the parameter of the
string loop expansion. Since the dimension of $\,\alpha^\prime\,$
is the inverse of the mass, usually $\,\alpha^\prime\,$ is 
associated with the inverse square of the Planck mass $\,1/M^2_P$.
Indeed, this choice is strongly motivated by the common belief 
that the (super)string theory is the theory which should unify 
all the interactions including gravitational, in one 
quantum theory. 

The $X^\mu$ are the dynamical fields 
defined on the world sheet. At the same time they are 
coordinates of the $D$-dimensional space with the metric 
$g_{\mu\nu}$. Initially, the geometry of this $D$-dimensional 
space is not known, it is generated by the quantum effects 
of the two-dimensional theory. In the sigma-model approach
(see, for example, \cite{hall,tseytlin} for the general 
review and \cite{ketov} for the technical introduction 
and complete list of results of the higher-loop calculations
of the string effective action) the effective $D$-dimensional 
action of the 
background fields $\,g_{\mu\nu}(X), \,b_{\mu\nu}(X),\,\Phi(X)\,$ 
appears as a result of the imposition of the Weyl invariance
principle at each order of the perturbative 
expansion of the $2$-dimensional 
effective action. For example, the dilaton term in (\ref{string})
is not Weyl invariant, but it has an extra factor of
$\,\alpha^\prime\,$. Therefore, after integration over the
fields $X^\mu$ we find that this classical term contributes to 
the trace of the Energy-Momentum Tensor and that this 
contribution is of the first order in $\al^\prime$. Other similar 
contribution comes from the one-loop effects, including the 
renormalization of the composite operators in the $<T_\mu^\mu>$
expression \cite{frts85,CFMP,ketov}. 
Requesting the cancellation of two contributions we get 
a set of the one-loop effective equations: conditions on the
background fields $\,g_{\mu\nu}(X), \,b_{\mu\nu}(X),\,\Phi(X)\,$. 
The corresponding action is interpreted as a low-energy effective 
action of string.

%%%%%%%%%%%%%%%%%%%%%%%%%%%%%%%%%%
The one-loop effective equations
do not directly depend on the antisymmetric field $\,b_{\mu\nu}(X)$,
but rather on the stress tensor 
\beq
H_{\mu\nu\la}=\pa_{[\mu}\,b_{\nu\la]}\,.
\label{stress}
\eeq
The covariant calculation 
includes the expansion in Riemann normal coordinates   
\cite{hon,osborn,ketov}), and the coefficients of the 
expansion for $b_{\mu\nu}$ depend only on $H_{\mu\nu\la}$
and its derivatives, but not directly on the $\,b_{\mu\nu}\,$
components. 
After all, the geometry of the $D$-dimensional space is
characterized by three kinds of fields: metric $g_{\mu\nu}$,
dilaton $\,\Phi\,$ and completely antisymmetric 
field $\,H_{\mu\nu\la}$, which satisfies the constraint 
$\,\vp^{\mu\nu\al\be}\,\pa_\al\,H_{\nu\al\be}=0\,\,$ coming 
from (\ref{stress}). In principle, the field $H_{\mu\nu\la}$ 
can be interpreted as a space-time torsion. 
In order to understand better the correspondence with our 
previous treatment of torsion, let us parametrize the 
$H_{\mu\nu\la}$ field by the axial vector 
$S^\mu = \vp^{\mu\nu\al\be}\,H_{\nu\al\be}$. Then the 
constraint $\,\vp^{\mu\nu\al\be}\,\pa_\al\,H_{\nu\al\be}\,$ 
gives $\,\pa_\al S^\al=0$, and we arrive at the conclusion 
that the string-induced torsion has only transverse 
axial vector component -- exactly the same answer which is 
dictated by the effective 
field theory approach (see the beginning of section 5.3). 

In fact, on the way to this interpretation 
$\,H_{\mu\nu\la}\sim T_{\mu\nu\la}\,$
one has to check only one thing: how 
this field interacts to the fermions. Indeed, we are
interested in the answer after the
compactification into the $4$-dimensional space-time.
However, compactification can not change the general form 
of the interaction. If we suppose that the fermions interact 
to the 
$H_{\mu\nu\la}$ field through the axial current, then 
$H_{\mu\nu\la}$ can be identified as the space-time torsion 
$T_{\mu\nu\la}$. One can notice that this is the form of 
the interaction which looks quite natural from the point of 
view of dimension and covariance. 
This is not the unique possible choice, indeed. If we 
consider the tensor $b_{\mu\nu}$ as an independent field
(see, e.g. \cite{soeg}), the 
situation becomes quite different. In four dimensions, 
the antisymmetric field is dual to the axion - specific 
kind of scalar field \cite{ogpolu} (see also \cite{book}). 
In a conventional supergravity theory, which might be 
taken as the low-energy limit of the superstring, the axion 
field is present and 
the formulation of its quantum theory does not pose any 
problem (in a sharp contrast to torsion!). 
Of course, the interaction of axion with fermions
is quite different from the one of (\ref{dirac2-nm}), so 
in this case the antisymmetric field can not be associated 
to the space-time torsion. 

On the other hand, the 
renormalizable and local supergravity theory is not the
unique possible choice. There are some arguments \cite{coll+kost}
that the non-local effects can play a very important role
in the string effective actions, and one of the 
popular phenomenological models for these effects are
related to the $b_\mu$ field, 
which we already mentioned by the end of section 4.6.
The $b_\mu$ field is a close analog of torsion, and therefore
the available predictions of the string theory are not 
completely definite. 
 
%%%%%%%%%%%%%%%%%%%%%%%%%%
Let us now discuss the form of the string-induced action for 
torsion. At the one-loop level, this is some kind of the 
metric-dilaton-torsion action considered in section 2.4.
As an example of the string-induced action we 
reproduce the one for the bosonic string
\beq
S_{eff}\,\, \sim\,\,  \int {d^D}x \sqrt {|g|} \,e^{-2\Phi}
\,\left[\, - R - 2\,{\cal D}^2\Phi 
+\frac13\,H_{\la\al\be}\,H^{\la\al\be}\,\right]\,,
\label{string-eff}
\eeq
where one can put $\Phi=const$ for the analysis of the 
torsion dynamics.
Since the torsion square enters the expression 
in the linear combination with the Ricci scalar,
the torsion mass equals to the square of the Planck 
mass. This is, again, in a perfect agreement with 
our results about the effective approach to the
propagating torsion. The expressions 
similar to (\ref{string-eff}) show up for all known 
versions of string theory. For superstring the one-loop
result is exactly the same \cite{zach}. For the heterotic
string the torsion term is the same as in (\ref{string-eff}), 
the difference concerns only the dilaton. No version of string 
is known, which would produce the zero torsion mass in the 
low-energy effective action. Perhaps, this is more than a 
simple coincidence, for the massless torsion should lead 
to  problems in the effective framework.

The propagation of torsion can be caused by the higher 
loop string corrections to the low-energy effective action.
It was notices by Zweibach in 1985, that the definition of the 
higher-order corrections includes an ambiguity related to the 
choice of the background fields \cite{zwei,dere,tse,den4}. 
In particular, one can perform such a reparametrization, 
that the propagators of all three 
fields: metric, torsion and dilaton do not depend on 
higher derivatives. For the theory with torsion the corresponding
analysis has been done in Ref. \cite{den4}.
The general form of the transformation is
\beq
g_{\mu\nu} \to g_{\mu\nu} + \de g_{\mu\nu}\,,\,\,\,\,\,\,\,\,\,\,\,\,\,\,
b_{\mu\nu} \to b_{\mu\nu} + \de b_{\mu\nu}\,,\,\,\,\,\,\,\,\,\,\,\,\,\,\,
\Phi \to \Phi + \de \Phi\,,
\label{repa}
\eeq
where
$$
\de g_{\mu\nu} = x_1R_{\mu\nu}+x_2Rg_{\mu\nu}+x_3H^2_{\mu\nu}
+ x_4\na_\mu\na_\nu \Phi +x_5\na_\mu\Phi\na_\nu \Phi+
$$
$$
\,\,\,\,\,\,\,\,\,\,\,\,\,\,
\,\,\,\,\,\,\,\,\,\,\,\,\,\,\,\,\,\,\,\,\,\,\,\,\,\,\,\,
+ x_6H^2g_{\mu\nu}+x_7g_{\mu\nu}\na^2\Phi+x_8(\na\Phi)^2+\,...\,,
$$
\vskip 2mm
$$
\de b_{\mu\nu} = x_9\na_\la H^\la_{\,\,\cdot\,\mu\nu}
+ x_{10} H^\la_{\,\,\cdot\,\mu\nu}\na_\la\Phi 
\,\,\,\,\, + \,\,\,\,\,\,\,\,\,\,...\,,
$$
\beq
\de \Phi = x_{11}R+x_{12}H^2+x_{13}\na^2\Phi+x_{14}(\na \Phi)^2
\,\,\,\,\, + \,\,\,\,\,\,\,\,\,\,...\,.
\label{repa in details}
\eeq
In these transformations, the coefficients 
$x_{1}, \,x_2,\, ...\,x_{14},\,...\,$ are chosen in such a 
way that the high derivative terms contributing to the 
propagators disappear. The transformation can be continued
(in \cite{den4} the proof is presented until the third 
order in $\alpha^\prime$, but the statement is likely to hold at 
any order, see for example, \cite{highderi}). 

The motivation to make a transformation (\ref{repa}) is to 
construct the low-energy theory of gravity free of the 
high derivative ghosts. At the same time, one can make several 
simple observations:
\vskip 2mm

i) At higher orders the transformation is not uniquely 
defined, at least if we request just a ghost-free
propagator. For instance, let us take a 
$R_{\mu\nu}R^\mu_{\al}R^{\al\nu}$-term. This term can be 
easily removed by some transformation similar to (\ref{repa}),
despite it is innocent -- it does not contribute to the 
propagator of the metric perturbations. Usually all such 
terms are removed in order to simplify the practical 
calculations, but the validity of this operation is 
not obvious.
\vskip 2mm

ii) If we are not going to quantize the effective theory, the
necessity of the whole (\ref{repa}) is not clear. It is 
well known, that in many cases the physically important 
classical solutions are  due to higher derivatives 
(for example, inflation may be achieved in this way 
\cite{star,marot}).
\vskip 2mm

iii) If we are going to quantize the effective theory (see 
corresponding examples in \cite{don} for metric and 
\cite{betor,guhesh} for torsion), one has to repeat the 
reparametrization (\ref{repa}) after calculating any loop 
in the effective quantum field theory. This operation 
becomes a necessary component of the whole effective
approach, and it substitutes the standard consideration 
of Ref. \cite{effgrav,don}. Of course, since the 
effective approach cares about lower derivatives
only, the physical consequences of two definitions must
be the same.  
\vskip 2mm

Indeed, the transformation (\ref{repa}) kills the 
torsion kinetic term, so that the torsion action consists 
of the mass and interaction terms. 
In the second loop, for the bosonic and heterotic strings, 
the interaction terms of the $H^4$ and $RH^2$-types emerge
(see, for example, \cite{ketov}). These terms can 
not be removed by (\ref{repa}). For the superstring, the 
2 and 3-loop contributions cancel, and at the 4-loop 
level the torsion terms can be ``hidden'' inside the 
$R_{....}^4$-type terms. It is unlikely that this 
can be done at higher loops but, since we are interested 
in the torsion dynamics, it has no real importance. 
The conclusion is that string 
torsion does not propagate, and that in this respect 
the effective low-energy action of string agrees with
the results of the effective quantum field theory. 

%%%%%%%%%%%%%%%%%%%%%%%%%%%%%%%%%%%%%%%%%%%%%%%%%%%%%%%%%%%%%%%%%%
\section{Gravity with torsion induced by quantum effects 
of matter}
%%%%%%%%%%%%%%%%%%%%%%%%%%%%%%%%%%%%%%%%%%%%%%%%%%%%%%%%%%%%%%%%%%

Despite induced torsion looks as an interesting possibility, 
it seems there are not many publications on this issue (except
\cite{buodsh,anhesh} devoted to the anomaly-induced action
with torsion). One can notice, however, that
the action of gravity with torsion may be induced in the very same 
manner as the action for gravity without torsion. Let us repeat
the consideration typical for the standard approach (see, for 
example, \cite{adler}), but with torsion. 

One has to suppose that there are gravity fields: metric and 
torsion, which couple to quantized matter fields. 
For instance, it can be the non-minimal interaction described
in section 2.3. The non-minimal interaction is vitally important, 
for otherwise the theory is non-renormalizable. The geometry of 
the space-time is described by some vacuum action, depending 
on the metric and torsion, but this action is, initially, not 
defined completely. The exact sense of the last statement will 
be explained in what follows. Now, the action of gravity with 
torsion $S_{grav}\left[g_{\mu\nu},T^\al_{\,\cdot\,\be\ga}\right]$ 
is a result of the quantum effects of matter
on an arbitrary curved background. This means the following 
representation for this action (compare to (\ref{Z}))
\beq
e^{iS_{grav}\left[g_{\mu\nu},\,T^\al_{\,\cdot\,\be\ga}\right]}
= \int {\cal D}\Phi\,
e^{iS_{matter}
\left[\Phi,\,g_{\mu\nu},\,T^\al_{\,\cdot\,\be\ga}\right]}\,.
\label{indu1}
\eeq

Here, $\Phi$ means the whole set of non-gravitational fields: 
quarks, gluons, leptons, scalar and vector 
bosons and their GUT analogs, gauge ghosts etc. Consequently, 
$S_{matter} = S_{matter}
\left[\Phi,g_{\mu\nu},T^\al_{\,\cdot\,\be\ga}\right]$
is the action of the fields $\Phi$. 
After these non-gravitational fields are integrated out, the 
remaining action can be interpreted as an action of gravity 
and the solutions of the dynamical equations following from
this action will define the space-time geometry. Then, the 
geometry depends on the quantum effects of matter, including 
spontaneous symmetry breaking which occurs at different scales,
phase transitions, loop corrections, non-perturbative effects
and scale dependence governed by the renormalization group. 

An important observation is in order. As we have already 
indicated in Chapter 3, the action must also 
include a vacuum part. For the massive theory one
has to introduce not only the $R^2, R\cdot T^2, R\na T,
T^2\na T$ and $T^4$-terms, but also $R$ and 
$T^2_{\,..}$-type terms, similar to the
ones in the Einstein-Cartan action. Then, the lower derivative
action of gravity can not be completely induced. The induced 
part will always sum up with the initial vacuum part which 
is subject of an independent renormalization.   
Of course, for the massless theory, one can choose the
$R^2_{\,..}$-type vacuum action, and then the low-energy
term will be completely induced through the dimensional 
transmutation mechanism (see section 3.5).

Let us now comment on the induced part of the $S_{grav}$. 
One can introduce
various reasonable approximations and evaluate the action
(\ref{indu1}) in the corresponding framework. One can list 
the following approaches in deriving the one-loop 
effective vacuum action: 

 {\it i}) The anomaly-induced action (see \cite{anhesh} and section 
3.6 of this report 
for the best available result in gravity with torsion). 
In some known cases this looks as the best approximation
\cite{star,buodsh,anju,balbi}. Indeed, in all 
cases, except \cite{buodsh,anhesh}, the theory without 
torsion has been considered. Usually, the anomaly-induced action 
has been treated as a quantum correction to the Einstein action. 
The advantage of the anomaly-induced action is that it includes 
the non-local pieces. This is especially important for the 
cosmological applications (see, e.g. the solution 
(\ref{nashi_starye_dobrye_dostijenia}) as an example of 
such application).

 {\it ii}) The induced action of gravity with torsion which 
emerges as a result of the phase transition induced by torsion 
or curvature (see section 2.5 for the case with torsion). 
We remark that the derivation of the effective action can be 
continued beyond the effective potential, so that the 
next terms in the derivative expansion could be taken into
account. For gravity without torsion this has been 
done in \cite{wolf} (see also \cite{book}), and it is 
technically possible to realize similar calculus for the 
theory with torsion. Then, after the phase transition, 
in the critical point one meets not only the non-minimal 
version (\ref{indu}) of the Einstein-Cartan theory, but also 
the next order corrections, including high derivative 
terms, terms of higher order in torsion and curvature,
the ones described in \cite{chris}, etc.  
The common property of all these terms is locality. 
The expansion parameter will be the inverse square 
of the Planck mass, therefore
these terms will be negligible at low energies. 

 {\it iii}) Alternatively one can simply take the contributions
of the free fields and take into account the higher orders of 
the Schwinger-DeWitt expansion (\ref{a2..}). Starting from 
$\,{\rm tr}\,{\hat a}_3(x,x)$, all these terms will be
finite and they are indeed contributions to $S_{grav}$.
In general, these terms are not very much different from
the ones described in {\it ii}). There is a possibility 
to sum the Schwinger-DeWitt expansion, but this has been  
achieved \cite{avram} only for the especially simple backgrounds 
without torsion. 

{\it iv}) The non-local terms can be taken into account  
using the generalized 
Schwinger-DeWitt technique. Such a calculations have 
been performed by Vilkovisky et al (see \cite{vilk} for 
the review and further references). This method seems more 
appropriate for the low-energy region, for the non-localities
are not directly related to the high-energy behaviour of the 
massive fields (as it is in the {\it i}) case). The calculations
for torsion gravity has not been performed yet. It may 
happen, that the non-local effects are relevant for torsion
(as they are, perhaps, for gravity), and then the effective 
field theory approach does not give full information. In the
content of string theory similar possibility led to the 
consideration of the (already mentioned) $b_\mu$-field, 
which strongly resembles torsion. 

{\it v)} Another way of deriving the non-local piece is to 
take the renormalization-group improved action \cite{maroto-thes}.
The corresponding corrections are obtained through the 
replacement of $\,1/\vp$, in the expression for vacuum 
divergences, by the $\,\,{\rm ln} ({\Box}/\mu^2)$. 
Indeed, these terms can be relevant only in the high 
energy region.

All mentioned possibilities concern the definition of the
torsion action, but one can make a stronger question about 
inducing the torsion itself. In principle, the axial current 
and the interactions similar to the torsion-fermion one,
can be induced in a different ways. One can mention, in this
respect, the old works on the anomalous magnetic field
\cite{Nikitin}, recent works on the anomalous effects of 
the spinning fluid \cite{volovik} and the torsion-like 
contact interactions coming from extra dimensions 
\cite{extra-dim}. The phenomenological bounds on the 
contact interaction derived in the last work are similar 
to the ones of \cite{betor}. 

\vskip 10mm
%%%  \newpage
%%%%%%%%%%%%%%%%%%%%%%%%%%%%%%%%%%%%%%%%%%%%%%%%%%%%%%%%%%%%%%%%%%
%%%%%%%%%%%%%%%%%%%%%%%%%%%%%%%%%%%%%%%%%%%%%%%%%%%%%%%%%%%%%%%%%%
\chapter{Conclusions} 
%%%%%%%%%%%%%%%%%%%%%%%%%%%%%%%%%%%%%%%%%%%%%%%%%%%%%%%%%%%%%%%%%%

We have considered various aspects of the space-time torsion. 
There is no definite indications, 
from experiments, whether torsion exists or not, but it is
remarkable that purely theoretical studies can put
severe limits of this hypothetical field which would be an 
important geometrical characteristic of the space-time.   
The main point is that the consistency of the propagating torsion 
is extremely restricted. Indeed, there 
is no any problem in writing the classical action for torsion,
and this can be done in many ways. Also, the quantum
field theory on classical torsion background can be successfully 
formulated, and we presented many results in this area of research. 
However, without the dynamical theory for the torsion itself the 
description of this phenomena is incomplete, and one can only 
draw phenomenological upper bound for the background torsion
from known experiments.
 
The serious problems show up when one demands the consistency of 
the propagating torsion theory at the quantum level
\cite{betor,guhesh}. Then, we have 
found that there is no any theory of torsion which could be 
simultaneously unitary and renormalizable. Moreover, there is 
no theory which can be consistent even as 
an effective theory, when we give up the requirement of the 
power counting renormalizability. The only one possibility is to 
suppose that torsion has a huge mass -- much greater than the
mass of the heaviest fermion. Then, if we assume that torsion
couples to {\large\it all} fermions, its mass has to greatly
exceed the $TeV$ level. Hence, torsion can not propagate
long distances and can only produce contact spin-spin 
interactions. The necessity of a huge torsion mass can 
explain the weakness of torsion and difficulties in its 
observation. Up to our knowledge, this is the first example
of rigid restrictions on the geometry of the space-time,
derived from the quantum field theory principles.

In this review, we avoided to discuss some, technically obvious, 
possibilities -- like a spontaneous symmetry breaking 
which would provide, 
simultaneously, the mass to all the fermions and to the torsion. 
The reason is that such an approach would require torsion
to be treated as a matter field, and in particular to be 
related to some internal symmetry group -- like $SU(2)$, for
example. Besides possible problems with anomalies, this 
would mean that we do not consider, anymore, torsion as part 
of gravity, but instead take it as a matter field. And that 
certainly goes beyond the scope of the present review, 
devoted to the field-theoretical investigation of the 
space-time torsion. 

Thus, the only solution is to take a string-induced or 
matter-induced torsion. As we have seen in the previous 
Chapter, both approaches do not give any definite 
information. First of all, the torsion induced in string 
theory has a mass of the Planck order. Thus, it can not
be really treated as an independent field, but rather as
a string degree of freedom, integrated into the complete
string spectrum. Furthermore, if we take supergravity as
a string effective theory, there is no torsion, but 
rather there is an axion - scalar field with quite
different interaction to fermions. Equally, the investigation 
of the matter-induced torsion offers some possibilities 
concerning torsion action, but in all available cases 
the torsion mass is of the Planck order.  
Anyhow, the possibility of induced torsion remains, and 
it is still interesting to look for more precise 
bounds on a background torsion and to develop the 
formal aspects of quantum field theory on curved 
background with torsion.
 
%%%%%%%%%%%%%%%%%%%%%%%%%%%%%%%%%%%%%%%%%%%%%%%%%%%%%%%%%%
%% \newpage
\vskip 6mm\vskip 6mm\vskip 6mm

%%%%%%%%%%%%%%%%%%%%%%%%%%%%%%%%%%%%%%%%%%%%%%%%%%%%%%%%%
%%%%%%%%%%%%%%%%%%%%%%%%%%%%%%%%%%%%%%%%%%%%%%%%%%%%%%%%%
\noindent
{\bf Acknowledgments.}
\vskip 3mm

First of all, I would like to express my great thanks to all 
colleagues with whom I collaborated in the study of torsion
and especially to I.L. Buchbinder, A.S. Belyaev, D.M. Gitman, 
J.A. Helayel-Neto, L.H. Ryder and G. de Berredo Peixoto. 
I am very grateful to  M. Asorey, T. Kinoshita, 
I.B. Khriplovich, I.V. Tyutin, L. Garcia de Andrade and 
S.V. Ketov for discussions.
I wish to thank J.A. Helayel-Neto and G. de Oliveira Neto 
for critical reading of the manuscript. 
Also I am indebted to all those
participants of my seminars about torsion, who asked 
questions.

I am grateful to the CNPq (Brazil) for permanent support
of my work,
and to the RFFI (Russia) for the support of the group 
of theoretical physics at Tomsk Pedagogical University
through the project 99-02-16617.

%%%%%%%%%%%%%%%%%%%%%%%%%%%%%%%%%%%%%%%%%%%%%%%%%%%%%%%%%
\vskip 6mm
%%%%%%%%%%%%%%%%%%%%%%%%%%%%%%%%%%%%%%%%%%%%%%%%%%%%%%%%%
%%%%%%%%%%%%%%%%%%%%%%%%%%%%%%%%%%%%%%%%%%%%%%%%%%%%%%%%%
\newpage

%%%%%%%%%%%%%%%%%%%%%%%%%%%%%%%%%%%%%%%%%%%%%%%%%%%%%%%%%%

%%%%%%%%%%%%%%%%%%%%%%%%%%%%%%%%%%%%%%%%%%%%%%%%%%%%%%%%%%%%%%

\end{document}